\pdfoutput=1
\documentclass[12pt, twoside, a4paper, openany]{book}

\usepackage{longtable}
\usepackage[T1]{fontenc}
\usepackage{titlesec, blindtext, color}
\usepackage{multicol}
\definecolor{gray75}{gray}{0.75}
\newcommand{\hsp}{\hspace{20pt}}
\titleformat{\chapter}[hang]{\Huge\bfseries}{\thechapter\hsp\textcolor{gray75}{|}\hsp}{0pt}{\Huge\bfseries}
\usepackage{a4wide}
\usepackage{graphicx}
\usepackage{natbib}
\usepackage{caption}
\usepackage{pdfpages}
\usepackage{fancyhdr}
\usepackage{lastpage}
\usepackage{pdfpages}
\usepackage{url}
\usepackage{float}
\usepackage{setspace}
\usepackage{enumerate}
\usepackage{epstopdf}
\usepackage{afterpage}
\usepackage{minitoc}
\usepackage{fancyhdr}
\usepackage{afterpage}
\usepackage{lscape}
\usepackage{amsmath}
\usepackage{amsfonts}
\usepackage{txfonts}
\usepackage{wasysym}
\usepackage[font={small}]{caption}
\usepackage{hyperref}

\usepackage[top=3.5cm, bottom=3.5cm, outer=2.5cm, inner=3.0cm, headsep=14pt]{geometry}

\fancypagestyle{plain}
\usepackage{indentfirst}

\newcommand{\kms}{\mrm{\, km \, s}^{-1}}
\newcommand{\ms}{\mrm{\, m \, s}^{-1}}
\newcommand{\cms}{\mrm{\, cm \, s}^{-1}}
\def \gtsima{$\; \buildrel > \over \sim \;$}
\def \ltsima{$\; \buildrel < \over \sim \;$}
\def \gtrsim{\lower.5ex\hbox{\gtsima}}
\def \lesssim{\lower.5ex\hbox{\ltsima}}

\def\ms{\,m\,s$^{-1}$}         
\def\cms{\hbox{\,cm\,s$^{-1}$}}       
\def\kms{\hbox{\,km\,s$^{-1}$}}       
\def\vsini{\hbox{$v$\,sin\,$i_\star$}}      
\newcommand{\Msun}{$M_{\odot}$}             
\def\Rsun{\hbox{$R_{\odot}$}}

\def \e{\times 10^}
\newcommand{\vrad}{$v_{\rm rad}$} 
\newcommand{\teff}{$T_{\rm eff}$}
\newcommand{\logg}{$\log g$}
\newcommand{\vmic}{$v_{\rm micro}$}
\newcommand{\vmac}{$v_{\rm macro}$}
\newcommand{\feh}{[Fe/H]}
\newcommand{\MJ}{{\it M}$_{\rm J}$}
\newcommand{\ME}{{\it M}$_\oplus$}
\newcommand{\RE}{{\it R}$_\oplus$}
\newcommand{\RJ}{{\it R}$_{\rm J}$}

\newcommand{\sn}{SNR}
\newcommand{\ff}{\textit{FF'}}
\newcommand{\logrhk}{$\log R'_{\rm{HK}}$}

\newcommand{\corot}{CoRoT}
\newcommand{\pastis}{\texttt{PASTIS}}
\newcommand{\vwa}{\texttt{VWA}}
\newcommand{\sme}{\texttt{SME}}
\newcommand{\ares}{\texttt{ARES}}
\newcommand{\tmcalc}{\texttt{TMCalc}}
\newcommand{\macula}{\texttt{macula}}
\newcommand{\mmercury}{\texttt{mercury}}
\newcommand{\ksint}{\texttt{KSint}}
\newcommand{\soap}{\texttt{SOAP}}
\newcommand{\ebop}{\texttt{EBOP}}
\newcommand{\amax}{$\alpha_{\mathrm{max}}$}
\newcommand{\tmax}{$t_{\mathrm{max}}$}
\newcommand{\tlife}{$t_{\mathrm{life}}$}
\newcommand{\ingress}{$\mathcal{I}$}
\newcommand{\egress}{$\mathcal{E}$}

\def\apjl{ApJ}%
\def\apjs{ApJS}%
\def\ao{Appl.~Opt.}%
\def\aap{A\&A}%
\begin{document}

\thispagestyle{empty}
\begin{center}
\begin{figure}[!ht]
\centering
\includegraphics[scale=0.043]{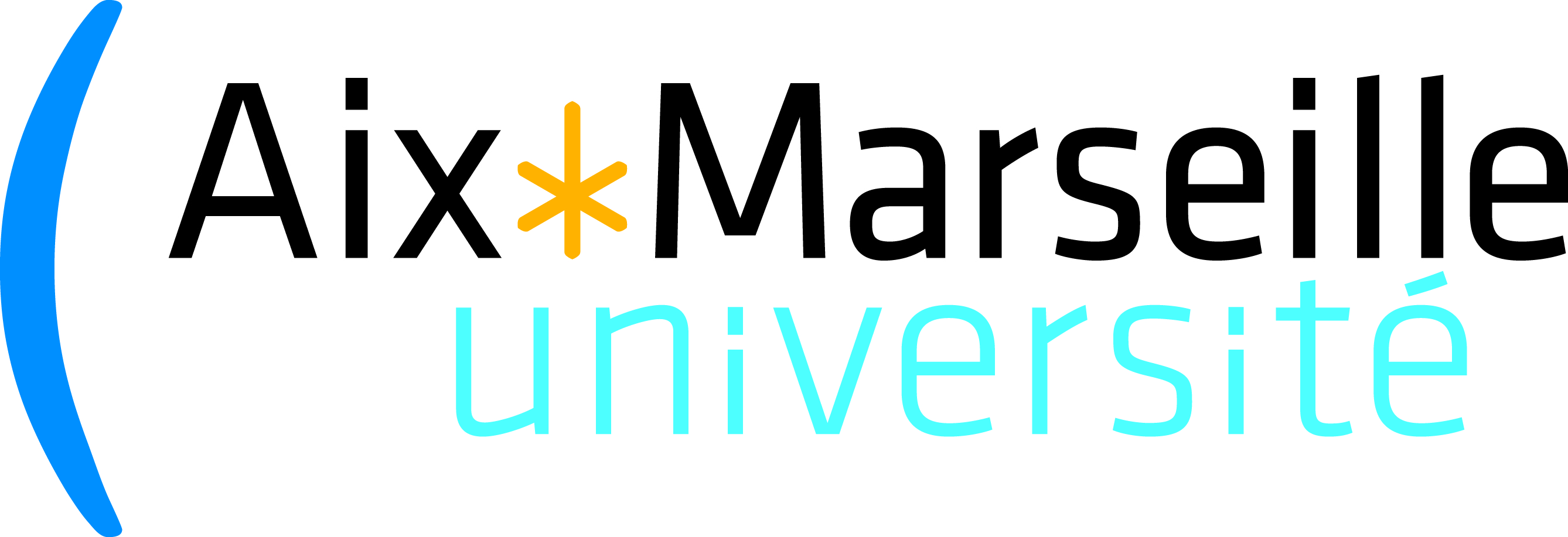}
\hspace{4cm}
\includegraphics[scale=0.5]{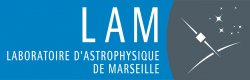}
\end{figure}
\vspace{0.5cm}
\large{Aix-Marseille Universit\'e\\
ECOLE DOCTORALE 352 - Physique et Sciences de la Mati\`ere\\}
\large{Specialit\'e : Astrophysique et Cosmologie\\}
\vspace{1 cm}
\Large{Th\`ese de Doctorat\\
\large{Present\'ee par Giovanni BRUNO\\}
\textbf{\LARGE {Characterization of transiting exoplanets:\\ 
analyzing the impact of the host star on the planet parameters}}\\
\vspace{0.5 cm}
\large{Th\`ese dirig\'ee par Magali DELEUIL\\
Soutenue publiquement le 21 Octobre 2015 \`a Marseille}
}
\end{center}
\large{\textbf{Jury:}\\

\begin{tabular}{ll}
\textit{Rapporteurs :} & \\
Antonino F. LANZA & INAF - Osservatorio Astrofisico di Catania\\
Don POLLACCO & University of Warwick\\
\textit{Examinateurs :}  & \\
Pascal BORD\'E &  Laboratoire d'Astrophysique de Bordeaux \\
Magali DELEUIL & Laboratoire d'Astrophysique de Marseille\\
Malcolm FRIDLUND & Sterrewacht Leiden\\
Guillaume H\'EBRARD & Institut d'Astrophysique de Paris /\\
 & Observatoire de Haute-Provence\\
Alessandro SOZZETTI & INAF - Osservatorio Astrofisico di Torino
\end{tabular}}
\vspace{1 cm}
\begin{center}
\normalsize{Laboratoire d'Astrophysique de Marseille\\
P\^ole de l'\'Etoile Site de Ch\^ateau-Gombert\\
38, rue Fr\'ed\'eric Joliot-Curie\\
13388 Marseille cedex 13\\
FRANCE}
\end{center}
\normalsize
\clearpage{\pagestyle{empty}\cleardoublepage}

\pagenumbering{roman}
\fancyhead[LO]{\itshape Contents}
\fancyhead[RE]{\itshape Contents}
\dominitoc
\tableofcontents
\clearpage
\fancyhead[LO]{\itshape List of figures}
\fancyhead[RE]{\itshape List of figures}
\listoffigures
\clearpage
\fancyhead[LO]{\itshape List of tables}
\fancyhead[RE]{\itshape List of tables}
\listoftables

\pagenumbering{arabic}
\fancyhead[LE,RO]{\thepage}
\fancyhead[LO]{\itshape\rightmark}
\fancyhead[RE]{\itshape\leftmark}
\chapter{Introduction}

\section{The importance of measuring precise planet parameters}
\begin{figure}[htb]
\centering
\includegraphics[scale = 0.37]{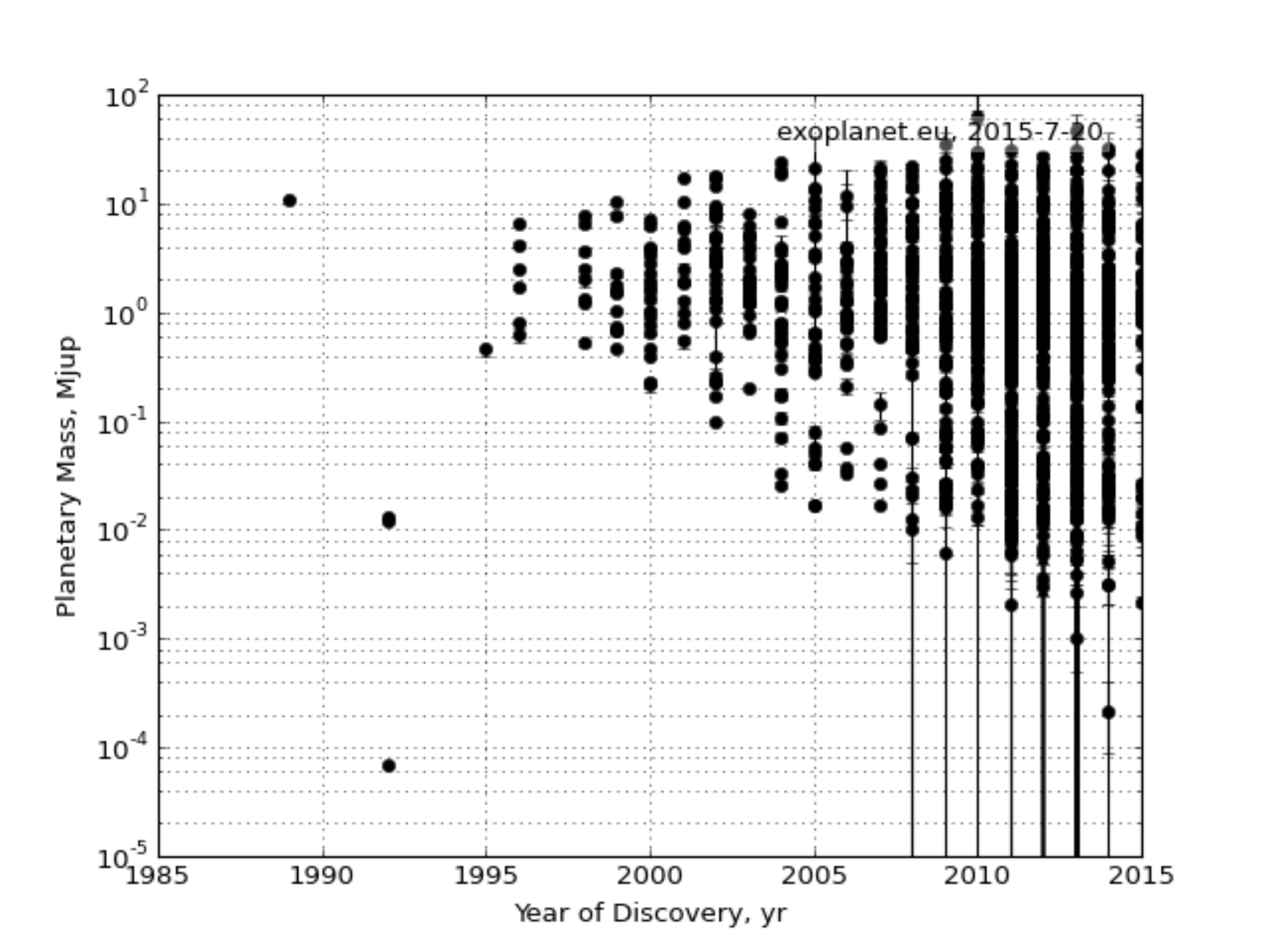}
\includegraphics[scale = 0.37]{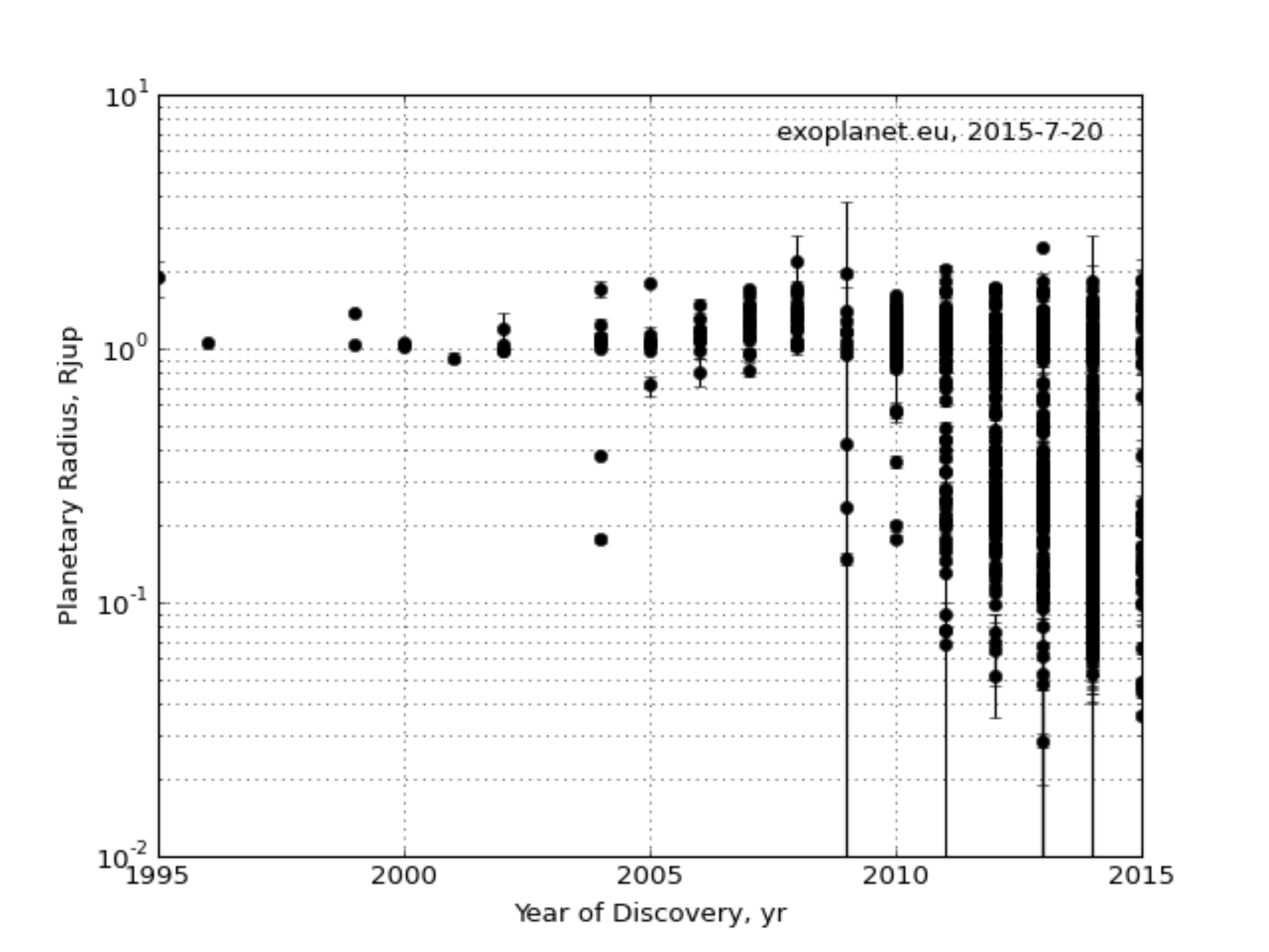}
\caption[Masses and radii of the planets discovered by year]{Masses (left) and radii (right) of the planets discovered by year. From \texttt{http://exoplanet.eu/}.}
\label{yearR}
\end{figure}

Since the first observation of a multi-planetary system around a pulsar \citep{wolszczan1991} and of a Jupiter-mass planet around a solar-type star \citep{mayor1995}, the exoplanet field has advanced with an ever increasing pace. At the time of writing, 1942 validated planets are archived in \texttt{exoplanet.eu} \citep{schneider2011}. Figure \ref{yearR} represents the masses and radii that have become accessible up to this day, giving an idea of the refinement in detection techniques and methodologies that have been achieved in the last twenty years. 

\begin{figure}[htb]
\centering
\includegraphics[scale = 1.0]{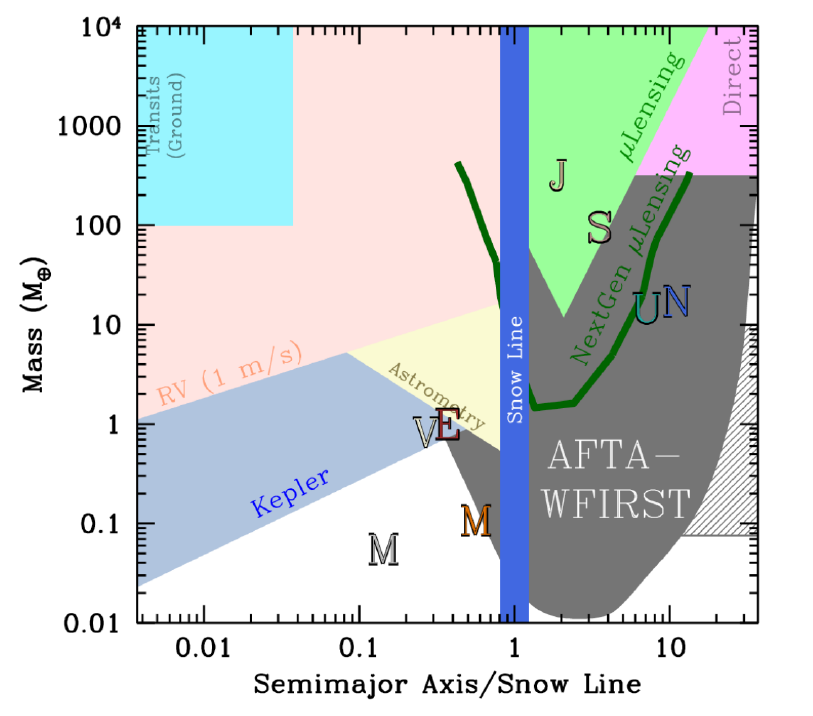}
\caption[Mass-separation diagram with the sensitivity regions of RVs, transit method, microlensing, direct imaging, astrometry, and timing]{Mass-separation diagram with the sensitivity regions of RVs, transit method, microlensing, direct imaging, astrometry, and timing. The separation is expressed in units of the snow line, calculated as $2.7 \, M_\star/M_\odot$. The letters indicate the locations of the Solar System planets. The AFTA/WFIRST mission, covering a large part of the long-period planet region of the plot, is still in its development phase. Adapted from \cite{wright2013}.}
\label{wright}
\end{figure}

To test theories of planetary system formation and evolution, a statistical picture of the planet properties needs to be drawn. Statistically significant inferences can only be derived for large samples. Hence, it is crucial to detect as many planets as possible, and to look for planets that are as diverse as possible. Assessing properties of planets of varying nature requires the precise measure of the planetary and orbital parameters, as well as of the stellar parameters.\\
No single detection technique, be it radial velocities (RVs), transits, microlensing, direct imaging, astrometry, or timing, is able to provide the full set of planetary parameters. Despite their complementarity, different techniques are not generally applied to the same sample of stars, so that a large number of planets remain with only a part of their main parameters measured. Moreover, different techniques have a different range of sensitivity to the parameters. Figure \ref{wright} is one way of representing this: it shows a mass-orbital separation diagram, with the sensitivity domain of each technique highlighted. It can be seen that some regions of the plot are still out of reach of the techniques available today: the region of planets with mass $\lesssim 1 M_\oplus$ with more than a few days of orbital period, and a large part of the region of long-period planets. This latter region, as shown in the plot, will be covered by the AFTA/WFIRST mission \citep{spergel2013}, which is still in its development phase.

\begin{figure}[!tbt]
\centering
\includegraphics[scale = 0.58]{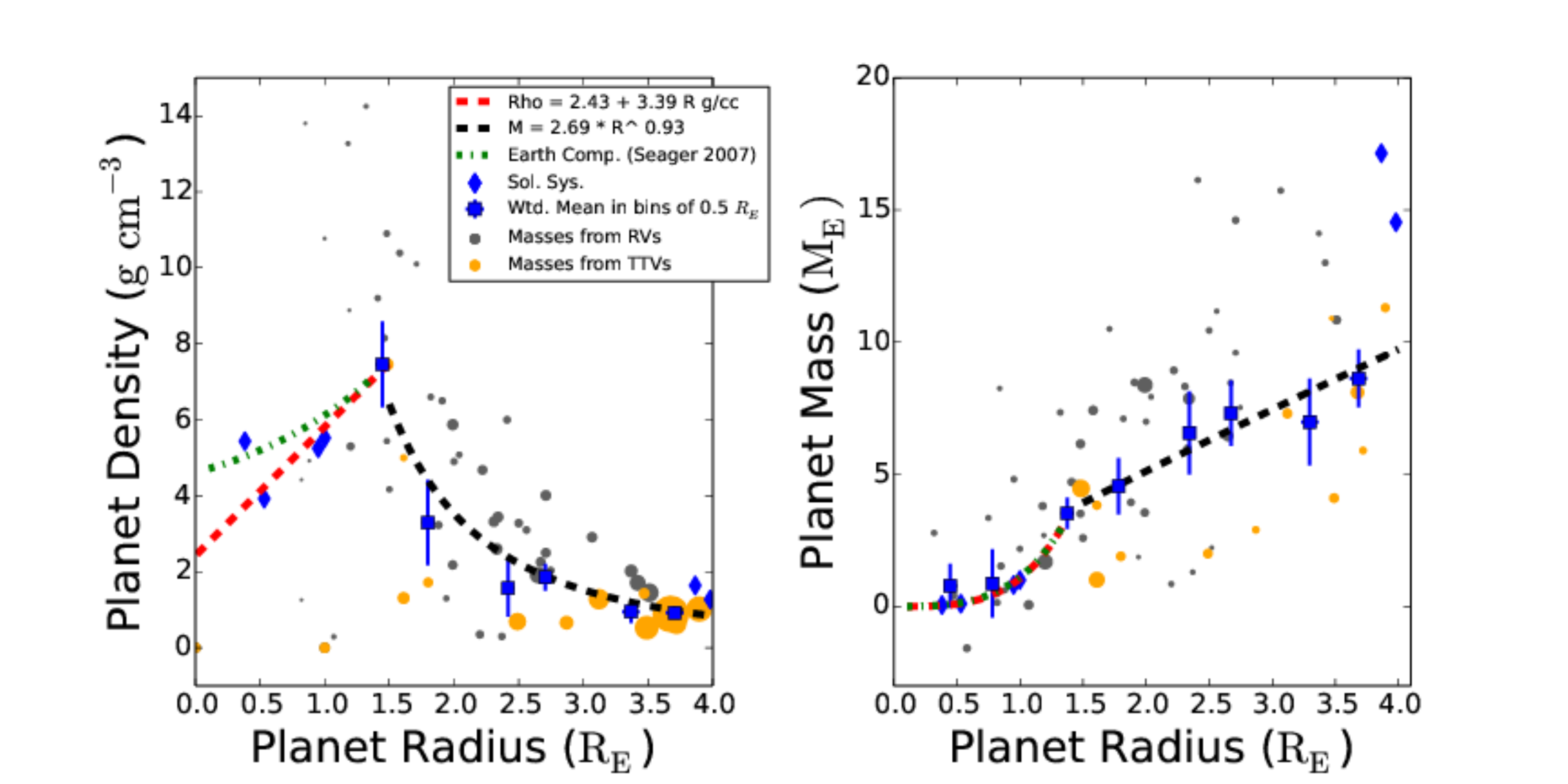}
\includegraphics[scale = 0.9]{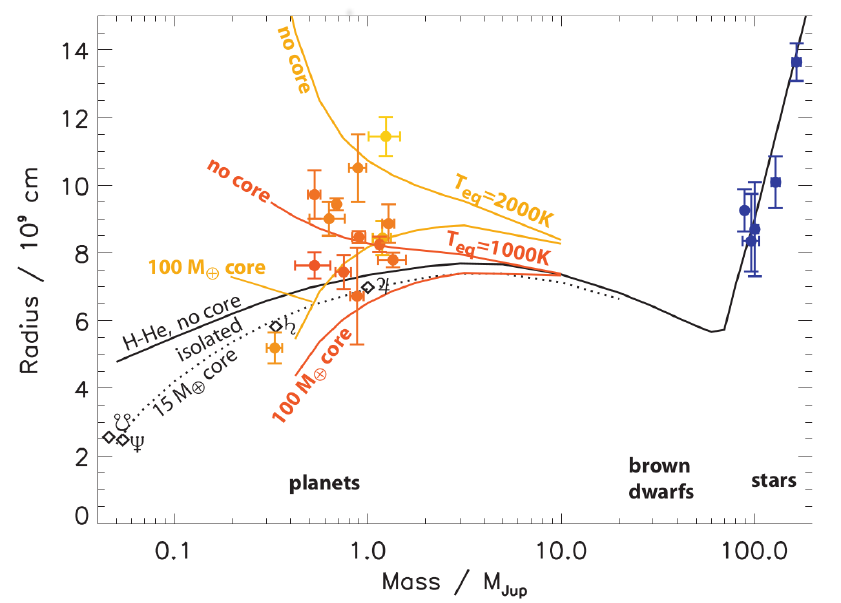}
\caption[Curves of the mass-radius relationships for different compositions]{$Left:$ curves of the mass-radius relationship for 8 Gyr-old low-mass planets, for different compositions. Some planets are indicated. In particular, those of the Solar System are labeled with their initial letter (V, E, U and N, corresponding to Venus, Earth, Neptune and Uranus, respectively). Lower-case letters indicate the six planets detected in the Kepler-11 system. The color code indicates the measured planetary temperatures. From \cite{lissauer2011b}. $Right:$ the same relationship for giant planets. The different curves converge as mass increases, passing from the planet, across the brown dwarf, to the stellar regime. From \cite{fressin2007}.}
\label{mr_comp}
\end{figure}

\begin{figure}[!htb]
\centering
\includegraphics[scale = 1.0]{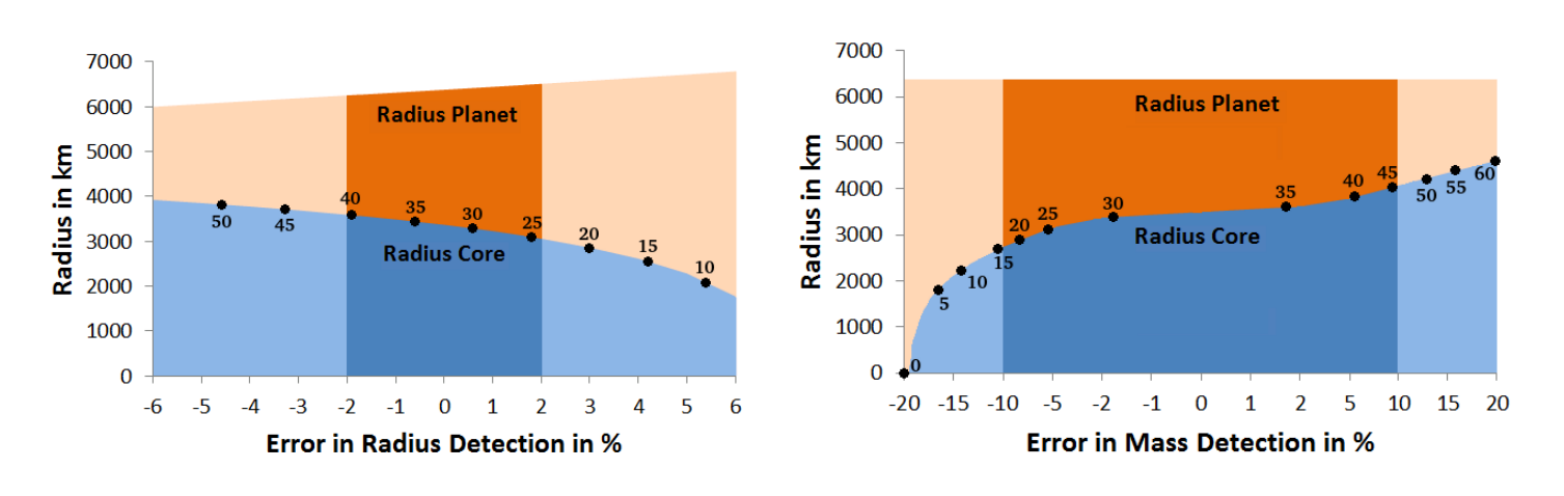}
\caption[Radius of a small-size planet and its core depending on the uncertainty in the  planet radius and mass]{$Left$: Radius of a small-size planet and its core depending on the uncertainty in the  planet radius, assuming a fully differentiated iron core and silicate mantle. $Right$: the same for the uncertainty on the planet mass. The calculations are based on the mass-radius relationship of \cite{wagner2011}. The width of the $x$-axes corresponds to present-day observational limits; the darker regions indicate the expected uncertainties for PLATO 2.0 ($2\%$ and $10\%$ in planet radius and mass, respectively). From \cite{rauer2014}.}
\label{coreunc}
\end{figure}

The work of this PhD dissertation is focused on problematics related to the RV and transit methods. These techniques yielded the largest number of discoveries. Both of them are more sensitive to close-in planets. For RVs, this is because one complete orbit must be covered to robustly measure the orbital period; for transits, because short periods mean that many transits can be observed, and therefore their signal-to-noise ratio (\sn) increased.\\ 
At the time of writing, 608 planets have been detected by RVs. This technique allows for the measurement of the $m_p \sin i$ of a planet, i.e. its mass times the sine of the orbital inclination, provided that the mass of the star is known. Indeed, the RV semi-amplitude $K$ is related to the system parameters through
\begin{equation}
K = \frac{8.95 \cms}{\sqrt{1-e^2}}\frac{m_p \sin i}{M_\oplus} \left( \frac{M_\star + m_p}{M_\odot} 
\right)^{-2/3} \left(\frac{P}{\mathrm{yr}} \right)^{-1/3},
\label{semiamp}
\end{equation}
where $e$ is the orbital eccentricity, $P$ the planet orbital period, and $M_\star$ the host star mass. This equation shows that RVs are more sensitive to massive, short period planets around low-mass stars.\\
The transit technique quickly gained the primacy in the number of discovered planets, after the launch of the Kepler space telescope \citep{borucki2008}. It is important to note that space-borne transit surveys were pioneered by the CoRoT mission \citep{baglin2006}, thus opening a new era with respect to previous ground-based transit surveys. The transit technique counts 1214 confirmed planets at the time of writing. Thanks to this method,  the size of a planet $r_p$ relatively to the one of its transited star $R_\star$ is measured, as
\begin{equation}
\frac{\Delta F}{F} = 8.41 \cdot 10^{-5} \left( \frac {r_p}{R_\oplus} \right)^2 
\left( \frac{R_\star}{R_\odot}\right)^{-2}
\label{transit}
\end{equation} 
\citep{mandel2002}, where $\Delta F$ is the drop in the observed stellar flux due to the transit, and $F$ the flux received from the non-transited star. This equation shows that the transit method is more sensitive to large planets, or to planets around small stars. Also, the transit method allows for the measurement of the planetary impact parameter $b$ as
\begin{equation} 
b \equiv \frac{a}{R_\star}\cos i  = \left\{ \frac{ (1 - \sqrt{\Delta F})^2 
 - [\sin^2 (t_{F} \pi/P)/\sin^2 (t_{T} \pi / P)] (1 + \sqrt{\Delta F})^2}
{1 - [ \sin^2(t_{F} \pi /P) /\sin^2(t_{T} \pi/P)]}
\right\}^{1/2}
\label{impact}
\end{equation}
\citep{seager2003}, where $a$ is the orbital semi-major axis, $t_F$ the transit duration completely inside ingress and egress, and $t_T$ the total transit duration. In this way, the modulation by $\sin i$ of the planet mass, unavoidable in RVs, is removed. Because of this, and because RVs is one of the methods to validate the planetary nature of transiting candidates\footnote{Both RVs and transits are affected by cases of false positives. Grazing transits in a system of two main-sequence stars, their dilution by a third star (either bound to the system or in the foreground of the target), and transits of a giant by a main-sequence star can all mimic a planetary transit in front of a main-sequence star \citep[e.g.][]{brown2003,cameron2012}. Stellar activity, on its side, can create cases of false positives for both RVs and transits \citep[e.g.][]{queloz2001,barros2013}.}, RVs and transits are often coupled.\\
As explained, when RVs and transit method are combined, the mass and radius of an exoplanet can be univocally determined. With these two parameters, the planet bulk density can be determined, and the modeling of its internal structure is possible \citep[e.g.][]{guillot2014}. As equation \ref{semiamp} and \ref{transit} show, however, a planet mass and radius are only known as a function of the same stellar parameters. Stellar masses and radii are usually determined by measuring the stellar atmospheric parameters by spectroscopy, and by combining them to stellar evolutionary models. This method implies a wide variety of problems. The precision and the accuracy on the stellar parameters are crucial, especially for small size planets.\\
Figure \ref{mr_comp} shows the mass-radius relationship for low-mass and for giant planets. Different curves correspond to different chemical compositions: the error bars for a given planet's parameters are crucial to determine on which curve this planet lies. Indeed, a precision of 1\% is required on the planet radius for the modeling of super-Earths \citep{wagner2011}. Instead, typical uncertainties on the stellar mass and radius derived from spectroscopy are of $5\%$ \citep{wright2011}. Moreover, systematic errors for stellar models can reach up to $10\%$ \citep{boyajian2012}. For Earth-like planets, typical current uncertainties for radius and mass are $\sim6\%$ and $\sim20\%$, respectively \citep{rauer2014}. This leads to uncertainties of 30 to 50\% in mean density. Figure \ref{coreunc} shows the internal structure of an Earth-like planet with a fully differentiated iron core and silicate mantle (after \citealp{noack2014}). The uncertainty on the radius of the planet core varies as a function of the uncertainty on the planet radius and mass. For an Earth-like planet one can see that, with present observational limits, it is difficult to determine the size of the planet core \citep{rauer2014}.

The importance of spectral analysis goes beyond measuring the host star mass and radius. For large enough \sn, the chemical composition of the star can be determined. This allows for the study of correlations between the stellar and the planetary properties. Using spectroscopy, a ``planet-metallicity correlation'' was observed, that is the fact that metal-rich stars with $0.7 < M_\star/M_\odot < 1.2$ are more likely to host giant planets within 5 AU \citep{santos2004,fischer2005}. It was also observed that sub-Neptune planets (with $r_p < 4 \, R_\oplus$) do not follow the same pattern \citep{buchhave2012}.\\
Additionally, spectroscopy is used to study star-planet interactions. Thanks to spectral analysis, it was found that the chromospheric activity index of the stars \logrhk\ \citep{noyes1984} is correlated with both the temperature inversions in the atmospheres of transiting Hot Jupiters \citep{knutson2010}, and with the Hot Jupiters surface gravity \citep{hartman2010}. This contradicts current models of planetary evolution, and will be discussed in more detail in this work.\\
Furthermore, spectroscopy is employed for stellar age determination. Indeed, the atmospheric parameters allow the determination of the age of a star through the modeling of the star in the Hertzsprung-Russel diagram. Other age indicators are also derived from spectroscopy. An example is the level of lithium depletion. This is a marker of the age of solar-type stars \citep{soderblom1993}, and challenges current stellar models. Thanks to spectroscopy, controversial indications have been found that star-planet interactions enhance stellar Li depletion \citep{baumann2010,figueira2014li}. Finally, spectroscopy yields the measurement of the rotational velocity of a star. This parameter decreases with the age and the magnetic activity level of the star \citep{skumanich1972}, because of angular momentum loss due to the stellar wind \citep{weber1967}. Stellar color index and stellar rotation velocity can be combined in an empirical law to estimate stellar ages, called  gyrochronology \citep[e.g.][]{barnes2007}. The study of stellar rotation of transited stars questioned the validity of gyrochronology \citep[e.g.][]{brown2011,angus2015,maxted2015} and fostered the development of models of tidal spin-up of transited host stars by Hot Jupiters \citep[e.g.][and references therein]{goodman2009,lanza2010tides,ferrazmelo2015}.\\

The precision of transits and RVs has remarkably improved in the last few years. The Doppler precision improved from about 10 \ms\ in 1995 to 3 \ms\ in 1998, and to about 1 \ms\ in 2005, when HARPS was commissioned \citep{mayor2003}. This allowed for the detection of $\sim 1 M_\oplus$ close-in planets \citep{dumusque2012}. To detect Earth analogs at a $\sim 1$ AU orbital separation, however, requires a precision of about 10 \cms, as equation \ref{semiamp} shows. To get there, technological improvements will not be sufficient. The magnetic activity of the star, manifesting as starspots, p-mode variations, and variable granulation, is superposed on the planet signal. \cite{meunier2013} estimated the contribution of plages in the RV signal received from the Sun. They found that the peak-to-peak RV amplitude they produce is about 8 \ms. Unless stellar noise is modeled or disentangled, an Earth analog around a moderately active star such as the Sun is beyond our reach. However, we do not need to look for Earth analogs to be limited by stellar activity. \corot-7 A, hosting the first observed transiting super-Earth \citep{leger2009}, shows RV variations due to activity which are 3 to 12 times larger than those due to its planets \corot-7 b and c \citep{queloz2009}.\\
Similarly, stellar activity limits transit surveys. Kepler has achieved the best precision for transit surveys so far: a median photometric precision of 29 ppm with 6.5 hour cadence on $V =12$ stars \citep{fischer2014}. By comparison, the best precision in ground-based transit surveys is of 0.47 mmag at 80-second cadence, obtained on a $V = 12.7$ star \citep{johnson2009}. The photometric precision of Kepler allowed for the detection of sub-Mercury sized planets \citep{barclay2013} in light curves not affected by stellar activity. When a planet orbits a strongly active star, the precision on the planet parameters is severely reduced. Classic examples can be found among the planets discovered by \corot. We can refer again to \corot-7 A, whose stellar density was found to be much lower when derived from the fit of the transits than from spectroscopy. \corot-2 is a Hot Jupiter whose measured radius changes by more than 3\%, according to the way stellar activity in the light curve is dealt with \citep{alonso2008,czesla2009,gillon2010,guillot2011}. Hence, stellar activity affects the measure of the whole range of planet sizes, becoming more severe as observations try to reach the realm of Earth-sized planets and exomoons \citep{kipping2014m}.\\
For the characterization of very low-mass objects, photometry can be exploited thanks to dynamical modeling. Systems with at least two planets are likely to present variations of the orbital periods due to the dynamical interactions of the planets \citep{miraldaescude2002,holman2005,agol2005}. The fit of these transit timing variations (TTVs) can take advantage of the precise measure of the transit times offered by high-precision transit surveys. However, this technique is affected by the distortions of the transits due to starspots. Moreover, starspots can produce apparent TTVs \citep[e.g.][]{alonso2009,barros2013}, making this method inapplicable.

\section{Plan of this work}

The precise measure of the stellar and planetary parameters has been the main drive of this PhD. The project was oriented in two main directions. The first one was spectral analysis, to which chapter \ref{chapspectro} is dedicated. Within the PASI team\footnote{The exoplanet team at the Astrophysics Laboratory of Marseille.}, I took part in campaigns of characterization of Kepler giant transiting candidates, or validated giant planets. A regular ground-based follow-up of selected Kepler giant candidates is carried out by PASI team members and collaborators mainly with the SOPHIE spectrograph at the Haute Provence Observatory \citep[OHP,][]{perruchot2008}. The follow-up allows for the study of the Kepler giant planet population, as well as the study of the false positive probability of Kepler candidates \citep{bouchy2011,santerne2012,bouchy2013}. I took in charge the spectral analysis of the stars of nine systems, resulting in the characterization of their companions. Most of the stars belong to the main sequence, and have spectral type from F to K. Examples of stars which have just entered the giant phase were found. Also, our follow-up program led to the discovery of two brown dwarfs.\\
As part of my spectral analysis studies, I was involved in two other projects: one related to the physics of low-mass stars, and an other to star-planet interactions. For the first, I took in charge the spectral analysis of twenty-one CoRoT and Kepler stars, which are the primary members of binary systems whose secondary objects are low-mass stars. By measuring the mass and the radius of the low-mass secondary objects, I attempted to help in constraining the mass-radius relationship of long-period low-mass stars, which is still poorly estimated by observations. For the second, I measured the \logrhk\ of thirty-one weakly and non-active stars observed with SOPHIE, belonging to the CoRoT sample, and to the SuperWASP \citep{colliercameron2007} and HATNet \citep{bakos2007} ground-based transit surveys. This was aimed at extending the sample of the work of \cite{hartman2010}, who found a correlation between stellar activity and the surface gravity of Hot Jupiters on a sample of thirty-nine stars.\\
For most stars, I made use of the SOPHIE spectra which result from the RV observations. I performed the analysis starting from the data processed by the SOPHIE pipeline \citep{bouchy2009}. One of the main difficulties in spectral analysis, disregarding the peculiarities of each star, is the data reduction phase. This phase involves the correction for instrumental effects and precedes any proper spectral normalization. Without a robust data reduction and spectral normalization, the derived stellar parameters can be affected by systematic errors. This issue is more serious for low-\sn\ spectra. With the acquired expertise, I studied issues related to the SOPHIE spectra in the low-\sn\ regime. I determined the \sn\ regime in which spectra can be safely used to measure the stellar parameters, and confirmed that the co-addition of single exposures fixes the problems of low-\sn\ spectra.\\

The participation to the campaign of planet characterization led me to take in charge the analysis of a two-planet Kepler system, to which chapter \ref{chapttvs} is dedicated. When they are in mean-motion resonance (MMR), planets in multiple systems show important TTVs. As mentioned before, these can be exploited to measure, or refine, the mass of the planets. The planets of the studied system, Kepler-117, are out of resonance. In spite of this, they present significant TTVs \citep{steffen2010,mazeh2013}. An important part of this PhD was dedicated to the study of this system and to the development of a technique to improve the precision on all the system parameters by taking advantage of TTV dynamical fitting.\\ 
  
The development of techniques to correct the data for the stellar noise is critical for the exploitation of \corot\ and Kepler data, as well as in the perspective of the future transit surveys CHEOPS \citep{broeg2013} and PLATO 2.0 \citep{rauer2014}. Moreover, as discussed previously, such techniques are also needed for RV observations. The second main direction of this PhD, therefore, was the development of a method to model and fit the signal of stellar activity in transits and RVs. This is the topic of chapter \ref{chapactivity}. To that purpose, I studied numerical and analytic methods to model starspots. I chose and studied in detail three of them: two for photometry, and one for RVs. Such methods were applied to fit and correct for the activity component in different data sets, by exploiting the information content of the entire light curve. I explored the advantages they offer and their limitations. I used the test case of \corot-2, to apply and benchmark a method to model starspots and transits at the same time. Thanks to spot modeling, I have been able to constrain the transit and spot parameters in a way that would not have been possible with a standard data reduction.

\clearpage
\chapter{Spectral analyses}\label{chapspectro}
\minitoc

\bigskip
Spectral analysis is one of the main techniques used to characterize exoplanet host stars. My participation in programs of exoplanet characterization consisted mainly in the spectral analysis of their host stars. A part of this PhD was also dedicated to the study of the performance of the SOPHIE spectrograph at low signal-to-noise ratio. I contributed also to two ongoing studies, related to the physics of low-mass stars and to star-planet interactions.

\section{What is the importance of stellar parameters?}\label{why}
\begin{figure}[!htb]
\centering
\includegraphics[width = 0.6 \textwidth]{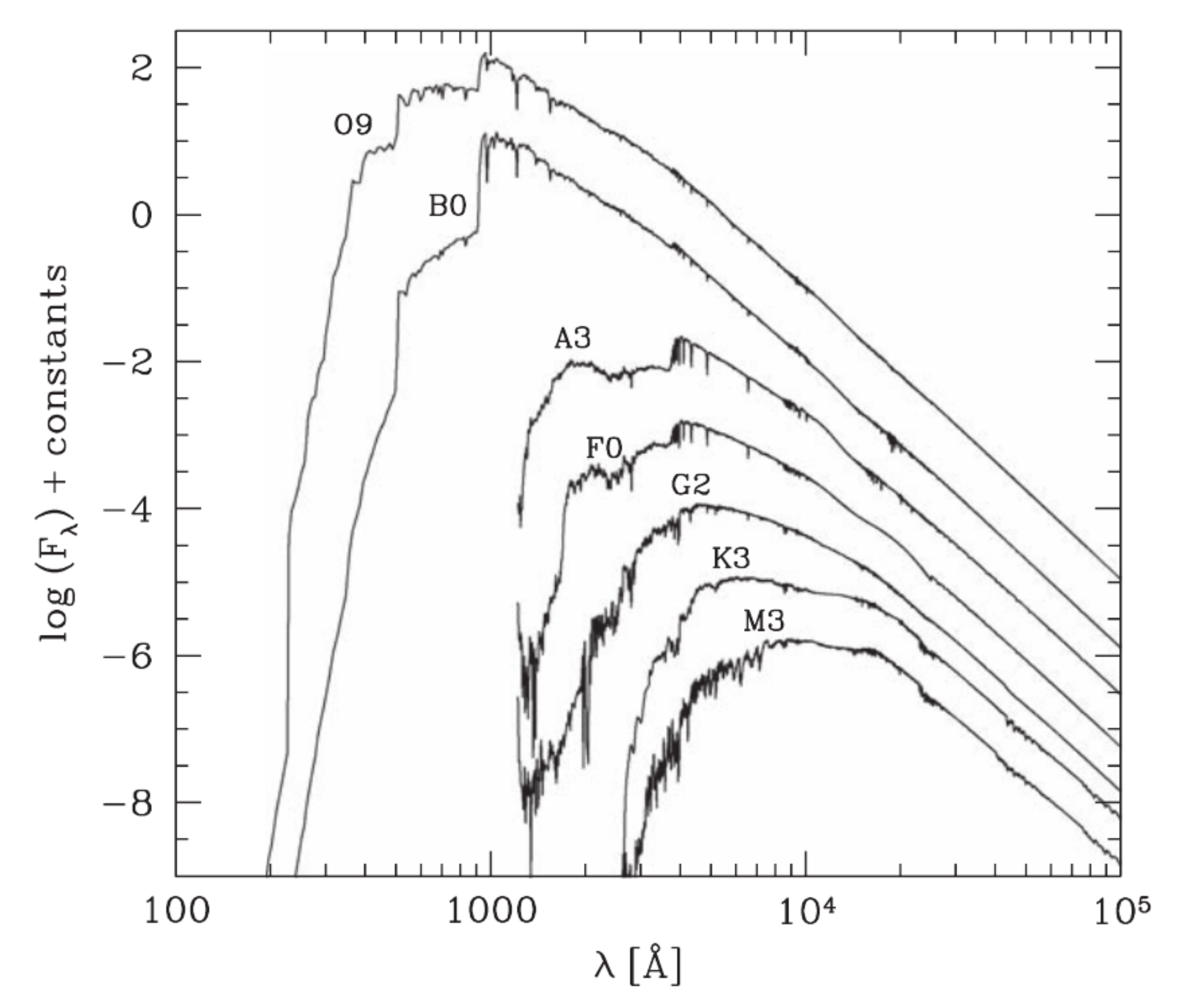}
\caption[SED for varying spectral types]{SED for varying spectral types. The parameters $f_\lambda$ is the flux per angstrom, and a constant is arbitrarily added to each spectrum for clarity. From \cite{mo2010}.}
\label{seds}
\end{figure}

The mass and radius of a planet are fundamental parameters which allow for the modeling of its internal structure \citep[and references therein]{guillot2014}. Their measurement relies on the knowledge we have of the host star's equivalent parameters. Indeed, the transit and radial velocities methods only provide the relationship between the planet and the host star in terms of their respective mass and radius. Characterizing an exoplanet requires, therefore, to determine the mass and the radius of its host star (or stars) as accurately as possible.\\
The mass ($M_\star$), the radius ($R_\star$), and the age of a star are called its fundamental parameters. In particular, the radius varies with the wavelength and the evolutionary stage of the star, and is defined as the depth of formation of the continuum, which in the visible is approximately constant for all stars \citep{gray2005}.\\
For the determination of the main parameters of a star, there are both direct and indirect methods. Direct methods require particularly favorable conditions, such  as the star being particularly close to us or being in a binary stellar system. These conditions are fulfilled by stars with calibrated flux, measured angular diameters, or eclipsing binaries \citep[][and references therein]{smalley2005}. This is, however, very rarely the case, and is indeed the exception in exoplanet science. For most stars, these measurements are not available, and indirect methods need to be used.\\ 
Indirect methods rely on the determination of the atmospheric parameters, which are the effective temperature (\teff), surface gravity (\logg), and metallicity, and on the deduction of the mass and radius in a model-dependent way.  With indirect methods, the stellar metallicity must be taken into account, as it has an important impact on the other parameters. The iron-to-hydrogen abundance ratio, \feh, is used as a proxy of the overall metallicity of the star, because of the large number of available spectral lines of iron. However, it may not represent it correctly, as other elements like C, N, and O are more abundant. These elements are seldom employed, because their abundance is more difficult to measure \citep{mucciarelli2013}\footnote{In the following, the symbol \feh\ will be used for both the iron abundance and the overall stellar metallicity, notwithstanding the difference between the two.}.\\
Indirect methods can be either photometric or spectroscopic. Photometric methods rely on calibrations of magnitude (color) differences, on the modeling of the infrared flux of a star \citep{blackwell1977,blackwell1980}, or on the fit of model atmosphere fluxes to the Spectral Energy Distribution (SED) observations. In figure \ref{seds}, SEDs for various spectral types are plotted. An observed SED depends on \teff, \logg\, \feh, stellar radius and interstellar extinction. 

Whenever a good quality spectrum is available, spectroscopic methods are to be preferred, as they are less affected by systematic errors and uncertainties. Moreover, spectroscopy allows for the measurement of the radial velocity of the star, the elemental abundances for the stellar photosphere, the projected equatorial rotational velocity (\vsini), and information about the magnetic activity of a star. Its range of application, therefore, goes beyond the determination of the atmospheric parameters for the purpose of exoplanet characterization.

\begin{figure}[!htb]
\centering
\includegraphics[width = 0.5 \textwidth]{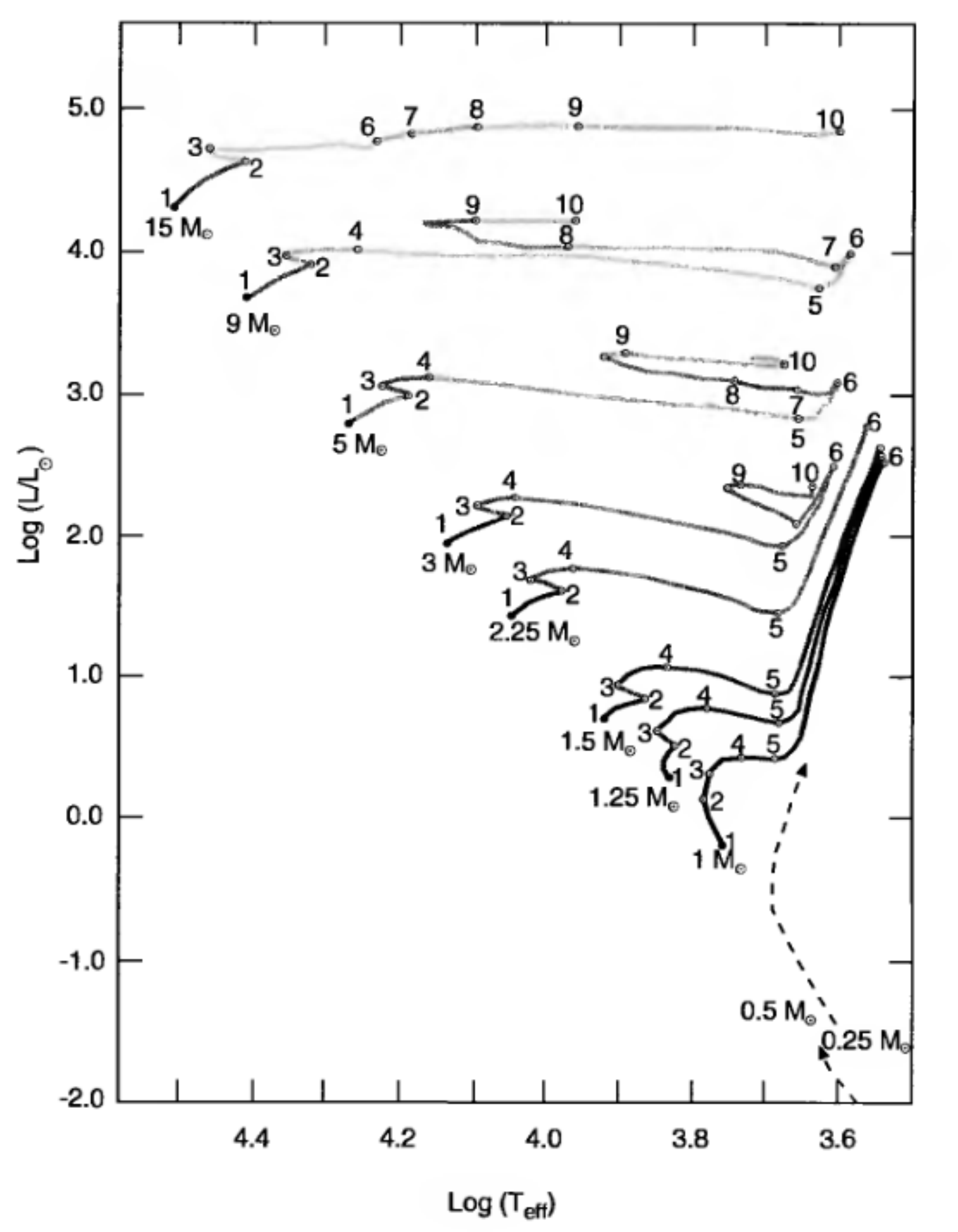}
\caption[Evolutionary paths in the H-R diagram]{Evolutionary paths in the H-R diagram for stars of different masses (as marked) up to the stage of He burning in a shell. The different phases in a star evolution are indicated by the numbers: 1-2, main sequence; 2-3, overall contraction; 3-5, H burning in thick shell; 5-6, shell narrowing; 6-7, red giant branch; 7-10, core He burning; 8-9, envelope contraction. From \cite{prialnik2009}.}
\label{hrdiagram}
\end{figure}

\begin{figure}[htb]
\centering
\includegraphics[width = 0.5 \textwidth]{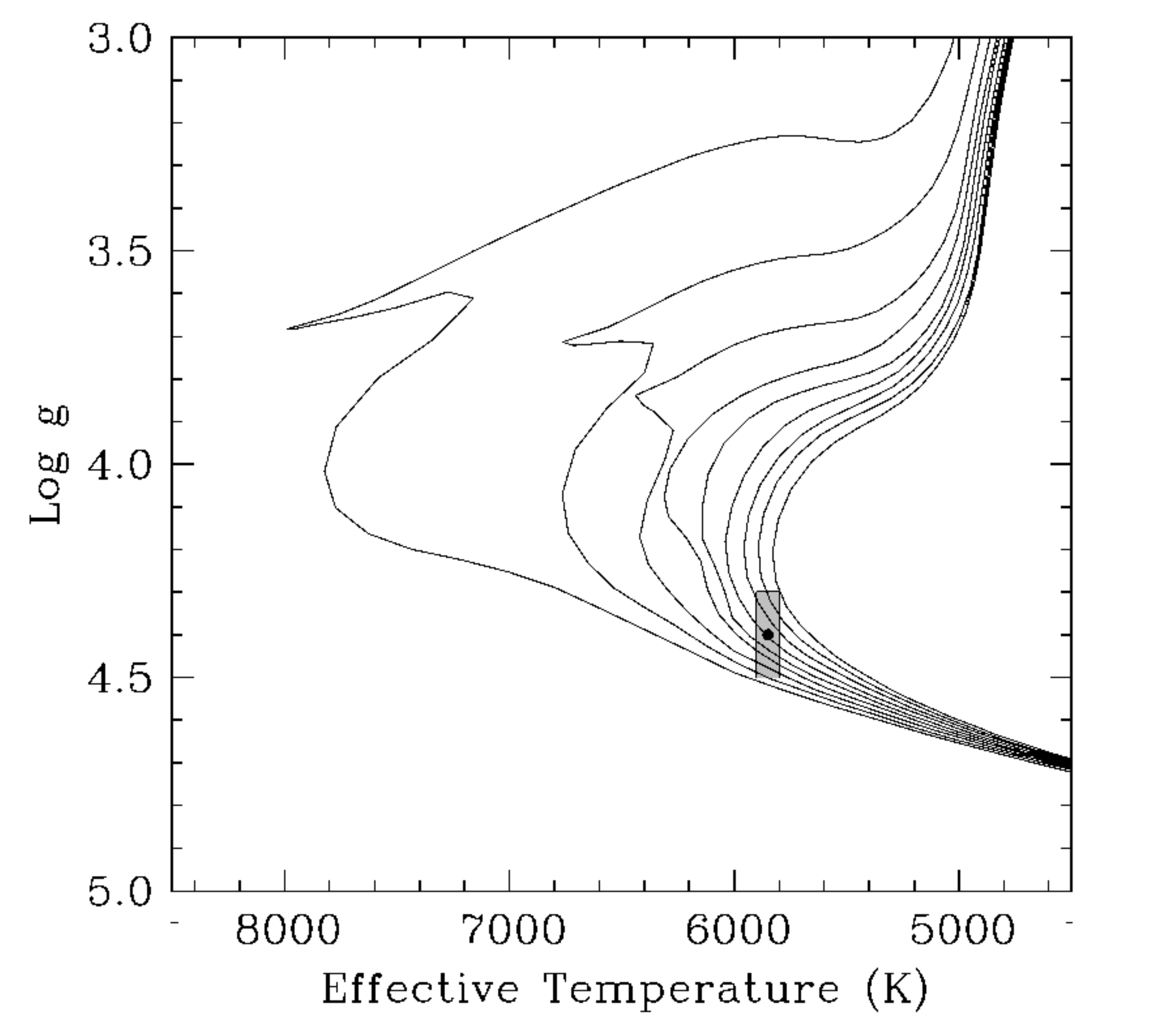}
\includegraphics[width = 0.485 \textwidth]{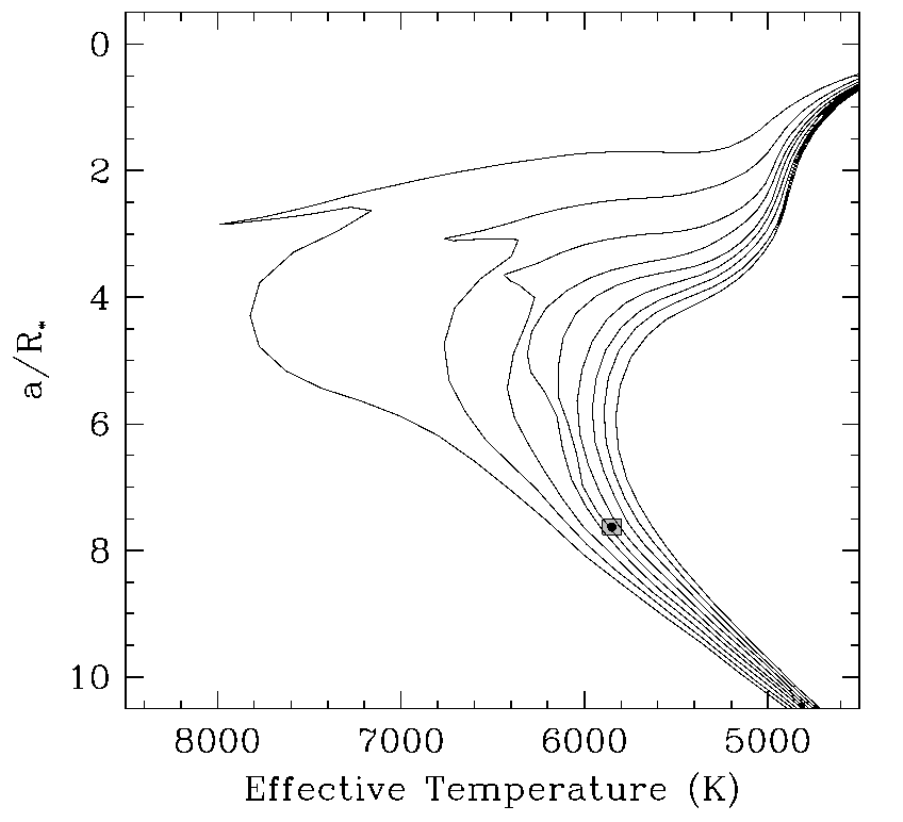}
\caption[Model isochrones]{Model isochrones from the Yonsei-Yale series by \cite{yi2001} and \cite{demarque2004}, for ages from 1 to 9 Gyr (left to right), for a given \teff\ and \feh. On the $x$-axis the \teff, on the $y$-axis the spectroscopically determined \logg\ (left) and the $a/R_\star$ derived from the transit fit (right). The observational constraints are five times tighter in the second case. From \cite{sozzetti2007}.}
\label{isochrones}
\end{figure}

When the atmospheric parameters are derived, the star is placed in a Hertzsprung-Russel (H-R) diagram and its position is compared to a grid of stellar evolutionary tracks. The tracks take as input the stellar mass and the metallicity of a star, and interpolate the evolution of luminosity, \teff, \logg, stellar radius, mass loss and stellar density at successive ages from stellar evolution models. Figure \ref{hrdiagram} shows a set of evolutionary paths on the H-R diagram for stars of different masses; the different phases in a star's life are highlighted. Given the measured atmospheric parameters of a star, its mass, radius, and age are therefore obtained by maximizing the likelihood function  $\mathrm{e}^{-\chi^2/2}$, where 
\begin{equation}
\chi^2 = \left[ \left( \frac{\Delta \mathrm{[Fe/H]}}{\sigma(\mathrm{[Fe/H]})} \right)^2 + \left( \frac{\Delta \mathrm{T}_\mathrm{eff}}{\sigma(\mathrm{T}_\mathrm{eff})} \right) ^2 + \left( \frac{\Delta (M_\star^{1/3}/R_\star)}{\sigma(M_\star^{1/3}/R_\star)} \right)^2 \right].
\end{equation}
In this equation, $\Delta$ indicates the difference between the measured and the modeled values, and $\sigma$ their uncertainty.\\
Different parameter spaces can be used for the stellar modeling in the H-R diagram. Indeed, for the special case of transiting planets, other sources of information than the spectra are usually exploited. This is due to the difficulty of determining \logg\ on low signal-to-noise (\sn) spectra, and to the degeneracy between the atmospheric parameters, in particular between \logg\ and \feh. The semi-major axis-to-stellar radius ratio, $a/R_\star$, can be derived with high precision from the photometric transits, as
\begin{equation}
\frac{a}{R_\star} = \left\{ \frac{(1 + \sqrt{\Delta F})^2 - b^2[1 - \sin^2(t_T\pi/P)]}{\sin^2(t_T\pi/P)} \right\}^{1/2}
\end{equation}
\citep{seager2003}, where the same notation as in equation \ref{impact} is used. This quantity can be converted into the stellar density $\rho_\star$, which is another way of expressing \logg\ \citep{sozzetti2007}. This is done by rearranging Kepler's third law:
\begin{equation}
\frac{M_\star}{R_\star^3} = \frac{4 \pi^2}{GP^2}\left( \frac{a}{R_\star} \right)^3 - \frac{m_p}{R_\star^3},
\end{equation}
where $M_\star$ is the mass of the star, $m_\mathrm{p}$ the mass of the planet, $G$ Newton's constant, and $P$ the orbital period (the quantities are expressed in cgs units).\\
In figure \ref{isochrones}, the variation of \logg\ (left panel) and $a/R_\star$ (right panel) is plotted versus \teff, for different stellar ages and a limited range of \teff.

\section{Methods}\label{specmeth}

In spectroscopy, there are two major strategies to measure the stellar atmospheric parameters \teff, \logg, and \feh: the detailed inspection of individual spectral lines, and a fit of the observed spectrum with a grid of models \citep{smalley2005,gray2005}.\\
In the actual fit of a spectrum, other phenomena broaden the line profile: the projected equatorial rotational velocity (\vsini), and the convective velocity fields described by the macroturbulence (\vmac) and microturbulence (\vmic) parameters. It can be observed that most of the stellar evolution models do not take into account the stellar rotation and its impact on the stellar parameters along its evolution. However, as most of the exoplanet host stars are slow rotators (that is, their \vsini\ is lower than 10 \kms), this detail can be neglected.\\
In what follows, I describe the methods that have been used for the spectral analysis used in this work, including their advantages and disadvantages.

\subsection{Equivalent width analysis}\label{ewa}
\begin{figure}[htb]
\centering
\includegraphics[width = 0.4 \textwidth]{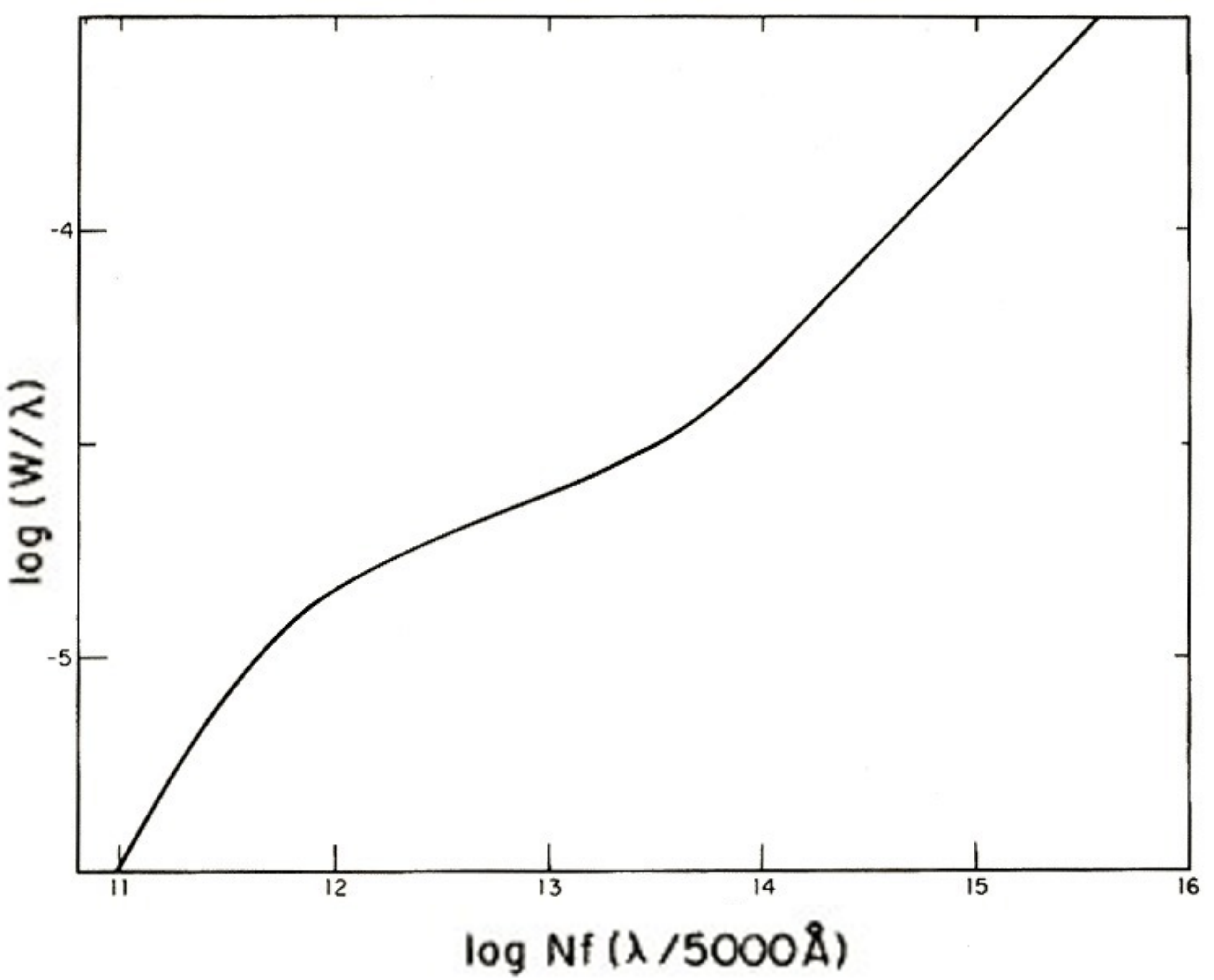}
\caption[A generic curve of growth]{A generic curve of growth (EW vs. abundance). From \cite{aller1971}.}
\label{cog}
\end{figure}

One of the main methods to derive stellar parameters is the diagnostic of metal lines. Three parameters are usually inspected for each line: equivalent width (EW), excitation potential and element abundance. The EW is fitted on the observed spectrum, and the excitation potential is known from atomic physics. The abundance is deduced through a ``curve of growth'' that describes the increase of the line strength with element abundance. A generic curve of growth is represented in figure \ref{cog}. Its relationship to the line parameters is expressed as
\begin{equation}
\log \left( \frac{w}{\lambda} \right) = \log C + \log A + \log g_n f \lambda - \theta_\mathrm{ex} \chi -\log \kappa_\nu,
\label{wlam}
\end{equation}
where $\lambda$ is the wavelength, $C$ is a term related to the ionization state of the element and to the temperature of the source, $A$ is the abundance of the considered element with respect to hydrogen, $g_n$ the statistical weight for the transition, $f$ the oscillator strength, $\theta_\mathrm{ex}= 5040/T$ ($T$ being the temperature of the source), $\chi$ the excitation potential for the transition, and $\kappa_\nu$ the continuum absorption coefficient. Finally, $w$ indicates the integral over the line profile $w$. This latter is defined by 
\begin{equation}
w = \frac{\mathcal{F}_c - \mathcal{F}_\nu}{\mathcal{F}_c},
\end{equation}
where $\mathcal{F}_c$ is the flux in the continuum, and $\mathcal{F}_\nu$ the flux across the spectral line.

Three parts can be recognized in a curve of growth: the first one, characterized by $\mathrm{EW} \propto A$, corresponds to weaker lines and is more suitable for EW analysis. The second and the third parts are characterized by $\mathrm{EW} \propto \sqrt{\ln A}$ and by (approximately) $\mathrm{EW} \propto \sqrt{A}$, respectively.\\ 
As shown by equation \ref{wlam}, different atmospheric parameters produce different curves of growth. By searching for an agreement between the elemental abundances of a given element in different stages of ionization, for the different excitation potentials of the lines, and for their EW, a consistent measurement of the atmospheric parameters is obtained. The use of different ionized species is based on the ionization-equilibrium method, requiring that different ionization states of the same species have the same abundance (Saha equation). The minimization of the abundance-excitation potential correlation corresponds to the requirement of excitation equilibrium (Boltzmann equation). The minimization of the abundance-EW correlation requires the adjustment of \vmic.\\
This methodology assumes a condition of local thermodynamical equilibrium (LTE). Non-LTE corrections may apply for possible departures from this condition, for example for metal-poor, low-gravity, or hot (\teff$\gtrsim 6300$ K) stars.\\

\begin{figure}[!bth]
\centering
\includegraphics[width = 1.0 \textwidth]{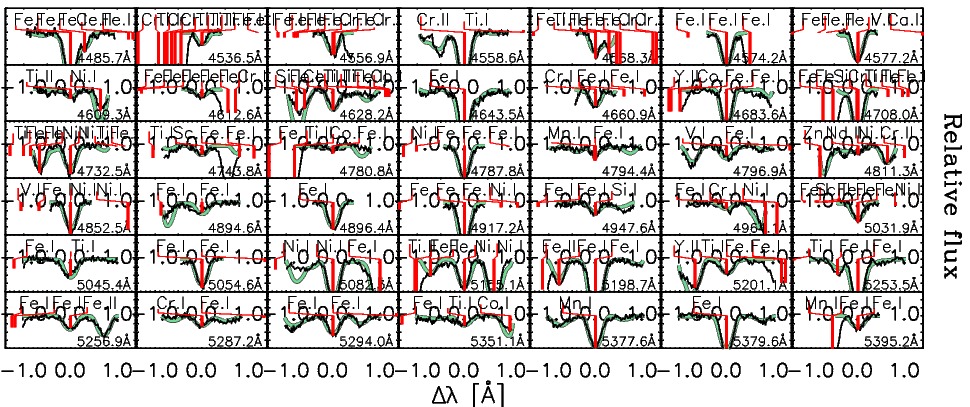}
\caption[A sample of the lines proposed by \vwa]{A sample of the lines proposed by \vwa\ on the basis of the rejecting criteria imposed by the user. The observed (black) and the calculated model (green) are shown. For each line, the wavelength and the atomic species are indicated.}
\label{vwaexam}
\end{figure}

The Versatile Wavelength Analysis package, or \vwa\ \citep[][and references therein]{bruntt2012}, intensively used in this work, adopts this approach. \vwa\ works with a normalized spectrum. To begin with, a set of some hundreds of non-blended spectral lines needs to be selected. With \vwa, the spectral lines included in the VALD database \citep{piskunov1995vald,kupka1999vald} can be selected. Selecting hundreds of spectral lines is possible only in the ideal case, and one often needs to accept a compromise due to the number, width, and depth of the spectral lines (changing with the spectral type of the star) and quality of the spectrum. As discussed above, these spectral lines should be weak metallic lines, for which the EW depends linearly on the abundance. Then, using a first guess on \teff, \logg, \feh, \vmic, \vmac, and \vsini, synthetic line profiles are computed and fitted to the observed lines, deriving the EW of the observed ones. The observed and the synthetic spectral lines are visually inspected, and those which are too blended or badly fitted (often because of a wrong $\log g_n$ and $f$) can be rejected. The visual inspection is performed through plots like the one of figure \ref{vwaexam}. 

After this, a series of models is iteratively computed by changing the values of \teff, \logg, and \vmic. For a given element, the parameters are adjusted until the correlation between the derived abundances of two differently ionized species and the excitation potentials of the lines are minimized, together with the correlation between the abundances of two different species and the lines' equivalent widths.

\begin{figure}[p]
\centering
\includegraphics[width = 0.8 \textwidth]{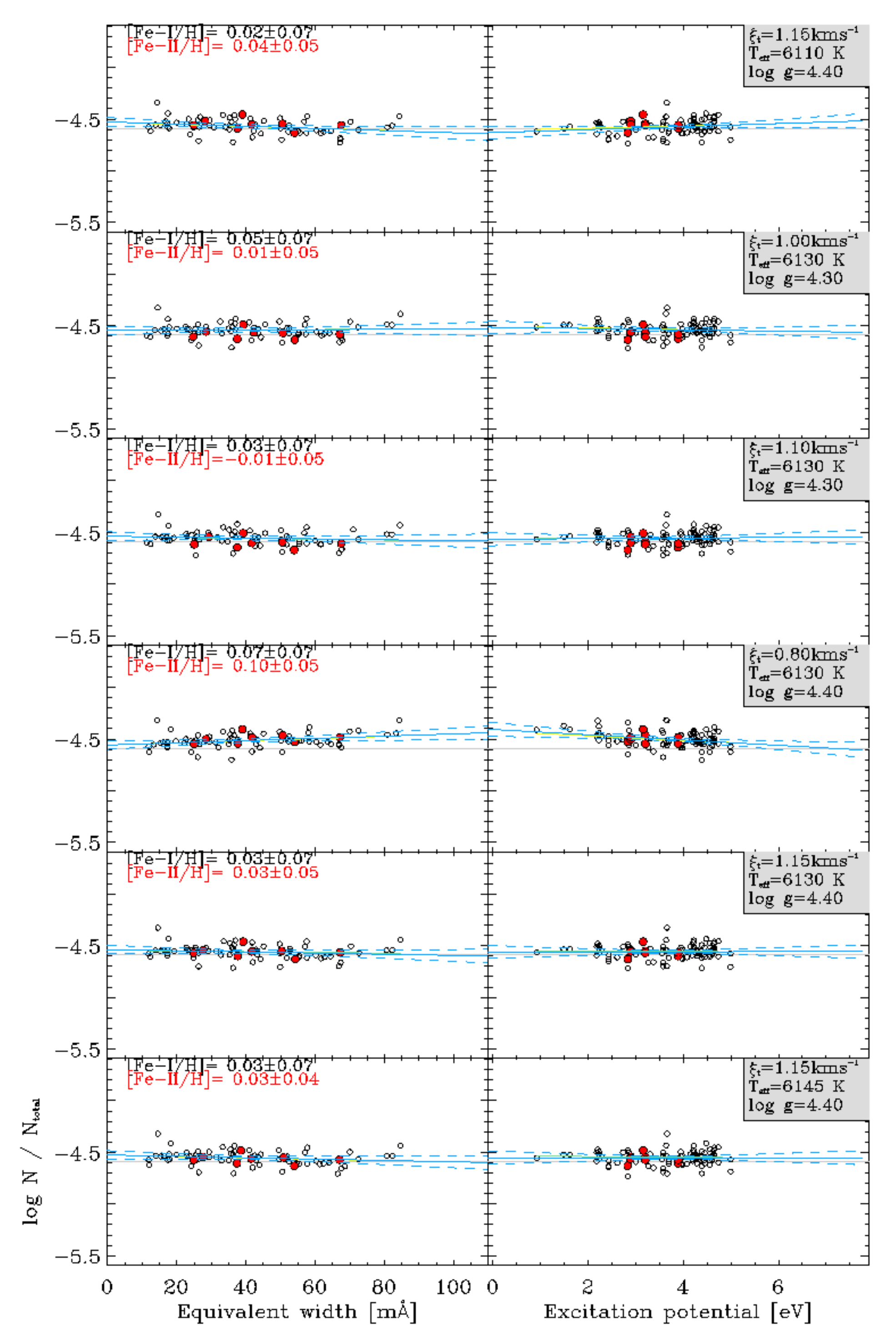}
\caption[Correlation minimization in \vwa]{Abundances of FeI and FeII (open and red circles, respectively) versus equivalent width and excitation potential, of a spectrum analyzed with \vwa. Each plot represents the results obtained with different choices of \teff, \logg, and \vmic. Also indicated is the solar Fe abundances (thin horizontal line)
and a linear fit with 95\% confidence limits indicated by the solid and dashed lines.}
\label{vwares}
\end{figure}

\begin{figure}[!htb]
\centering
\includegraphics[width = 0.8 \textwidth]{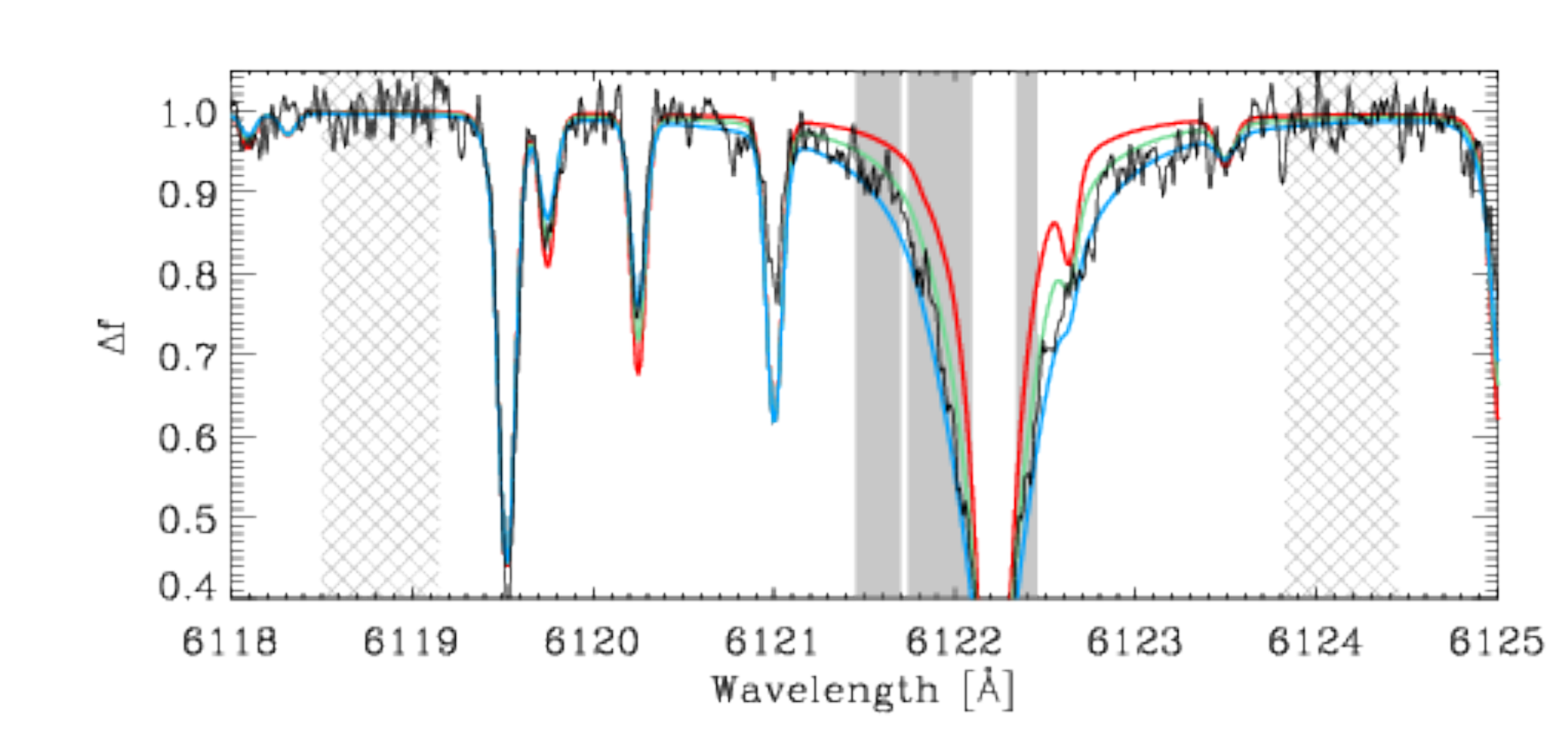}
\caption[The pressure-sensitive CaI line at 6122 \AA]{The pressure-sensitive CaI line at 6122 \AA, from a generic spectrum. Three synthetic spectra, varying only in \logg, are fitted to the line profile. A larger \logg\ means a wider line profile. The grey regions indicate the wavelength intervals used for the fit.}
\label{Ca6122}
\end{figure}

In most cases, only iron lines are used for these minimizations: the calculation of the abundance of the other metallic elements is robust only for very high quality spectra. Moreover, iron lines are often more abundant than lines for other elements. Figure \ref{vwares} shows the correlations that are inspected, and the differences caused by changing the values of the parameters separately. From the plots, it can also be also that the parameters are degenerate, as different combinations produce similar correlations.\\
After \teff, \logg, \feh, and \vmic\ are derived, the measure of \logg\ is checked by inspecting pressure-sensitive lines such as CaI at 6122 and 6162 \AA, the MgIb triplet, and NaID lines. Figure \ref{Ca6122} shows the fits of one of these lines. Finally, the couple \vsini - \vmac\ is measured by fitting a rotational profile to a set of metallic lines. In case of important discrepancies with the initially guessed \vsini, the analysis is repeated, beginning with the selection of the lines.\\
The domain of validity of \vwa\ is mostly for stars with \vsini $\lesssim 15 \kms$. For larger rotational velocities, the rotational component becomes dominant in the broadening of the spectral lines, inducing a strong blending between the lines; therefore, the derived equivalent widths and abundances are not reliable anymore. Moreover, at low \sn, the number of isolated and non-blended spectral lines becomes insufficient.\\ 

Another tool dedicated to EW analysis is the combination of the codes \ares\ and \tmcalc\  \citep{sousa2007,sousa2012}. \ares\ is a code for the determination of the EW of spectral lines. With \tmcalc, the results from \ares\ are used to calculate \teff\ and \feh, from a calibration for F, G, and K type stars. In particular, \teff\ is calibrated on a set of published line depth ratios for several spectral lines of different elements \citep{sousa2010}. Thanks to this approach, any abundance dependence is avoided \citep{gray2005}. For \feh, the calibration is based on the same set of lines, over which the polynomial equation 
\begin{equation}
\mathrm{EW} = c_0 + c_1 \mathrm{[Fe/H]} + c_2 T_{\mathrm{eff}} + c_3 \mathrm{[Fe/H]}^2 + c_4 T_{\mathrm{eff}}^2 + c_5 T_{\mathrm{eff}}\mathrm{[Fe/H]}
\end{equation}
is solved for \feh. Both the \teff\ and the \feh\ calibrations have been obtained from a sample of 451 stars, observed with HARPS, with a resolution $R \sim 110000$ and \sn\ between $\sim 70$ and $\sim 2000$, 90\% of them having \sn$\, > 200$. For a simulated solar spectrum with \sn $\, < 25$, \teff\ deviations from the correct parameters are reported in \cite{sousa2012}.\\
\ares\ offers the advantage of performing a local normalization of the continuum flux around each spectral line. For this, it requires various parameters, each of them needing to be adjusted by the user: 1) the width of the spectral window around a line inside which the continuum is fitted; 2) the coefficients of a polynomial function, called $rejt$, which fits the local continuum; 3) the width (in pixels) of the spectral window around each spectral line over which a smoothing function is applied before the fit of the EW; 4) the minimum separation between the lines whose EW is fitted (used to avoid the selection of blended lines); 5) the minimum EW for a line to be reported in the results that \tmcalc\ will use (to limit the use of spectral lines that are too noisy). The user is supposed to visually inspect the quality of the fit and to tune the parameters of the code until a correct normalization is performed.\\
\tmcalc\ has the advantage of being a very fast computation tool, which allows for the systematic and automatic analysis of a large quantity of spectra. However, it is limited by the fact that ti is calibrated only for F, G, and K slowly-rotating stars. 

\subsection{Fit of synthetic spectra}\label{synthsp}
\begin{figure}[htb]
\centering
\includegraphics[width = 0.6 \textwidth]{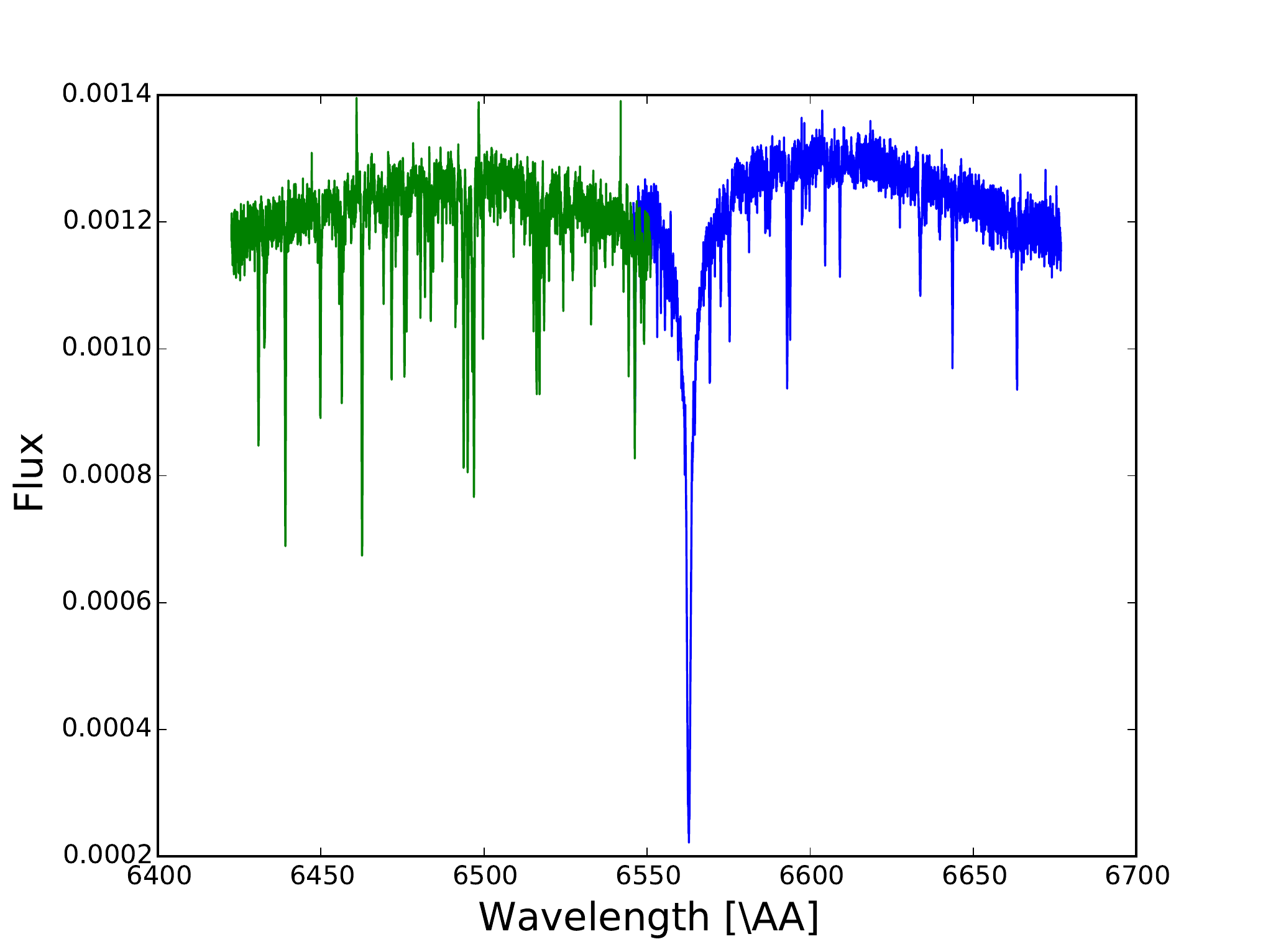}
\caption[The H$_\alpha$ line of a HARPS spectrum, previous to normalization]{The H$_\alpha$ line of a HARPS spectrum, previous to the normalization. The colors indicate the two consecutive echelle orders in which the line lies. The wings of H$_\alpha$ extend over tens of angstroms, creating a problem in the normalization of the spectrum.}
\label{halpha}
\end{figure}

In this approach, the observed spectrum is compared to a grid of synthetic spectra, computed as a function of the atmospheric parameters and of the lines atomic properties.\\
The package Spectra Made Easy (\sme) \citep[][and references therein]{valenti2005}, used in this work, adopts this approach. Provided a normalized spectrum, \sme\ calculates the spectral model that best fits the observed spectrum, for spectral windows chosen by the user. The code makes use of the VALD database, as does \vwa. For the fit, a non-linear least-squares algorithm is used. \sme\ lets the user choose which parameters to fit or to fix, or to solve for all the atmospheric parameters. This allows one, for example, to individually fit the profiles of Balmer lines, solving for \teff, and then, using the \teff\ estimate, to fit \logg\ on pressure-sensitive lines such as CaI at 6122 and 6162 \AA, the MgIb triplet, and the NaID lines. After several fits with varying starting values, the standard deviation of the results gives the internal uncertainties on the derived parameters (Fridlund, private communication). Fixing some parameters, moreover, is helpful when the spectrum has a low \sn. Large \vsini\ can also be measured in this way.\\
The advantages of this method come at the expenses of a high sensitivity to the normalization of the spectrum. This is particularly important for Balmer lines. The large wings of these lines, that extend for tens of angstroms, make the normalization difficult and tricky. This is particularly true for echelle spectrographs, like SOPHIE and HARPS, as the Balmer lines can fall over two adjacent orders. Figure \ref{halpha} shows a typical H$_\alpha$ line of a main sequence star, before the normalization.

\subsection{Data reduction}\label{spectrumtreat}
The spectra I analyzed were obtained with the SOPHIE spectrograph at the \textit{Observatoire de Haute Provence} (OHP), HARPS at ESO \citep{mayor2003}, HARPS-N at TNG \citep{cosentino2012}, and ESPaDOns at CFHT \citep{donati2003}. Table \ref{tabinstr} presents the main characteristics of these instruments.

\begin{table}[!ht]
\centering
\begin{tabular}{lccc}
 \hline
Instrument & Wavelength range [\AA] & Resolution ($\lambda/\Delta \lambda$) \\
\hline
SOPHIE/OHP & 3972-6943 & 75000/40000 (HR/HE mode)$^{(a)}$ \\
HARPS/ESO & 3780-6910 & 115000/80000 (HAM/EGGS mode)$^{(b)}$  \\
HARPS-N/TNG & As HARPS& As HARPS \\
ESPaDOns/CFHT & 3700/10500 & 81000/68000 (three observing modes)\\
\hline
\end{tabular}
\caption[Main characteristics of the spectrographs used to obtain the spectra analyzed in this work.]{Main characteristics of the spectrographs used to obtain the spectra analyzed in this work. All the instruments are echelle spectrographs. $^{(a)}$: High Resolution/High Efficiency mode. $^{(b)}$: High Accuracy Mode/High Efficiency mode (probably dubbed EGGS to remain in the food analogy).}
\label{tabinstr}
\end{table}

\begin{figure}[htb]
\centering
\includegraphics[width = 1.0 \textwidth]{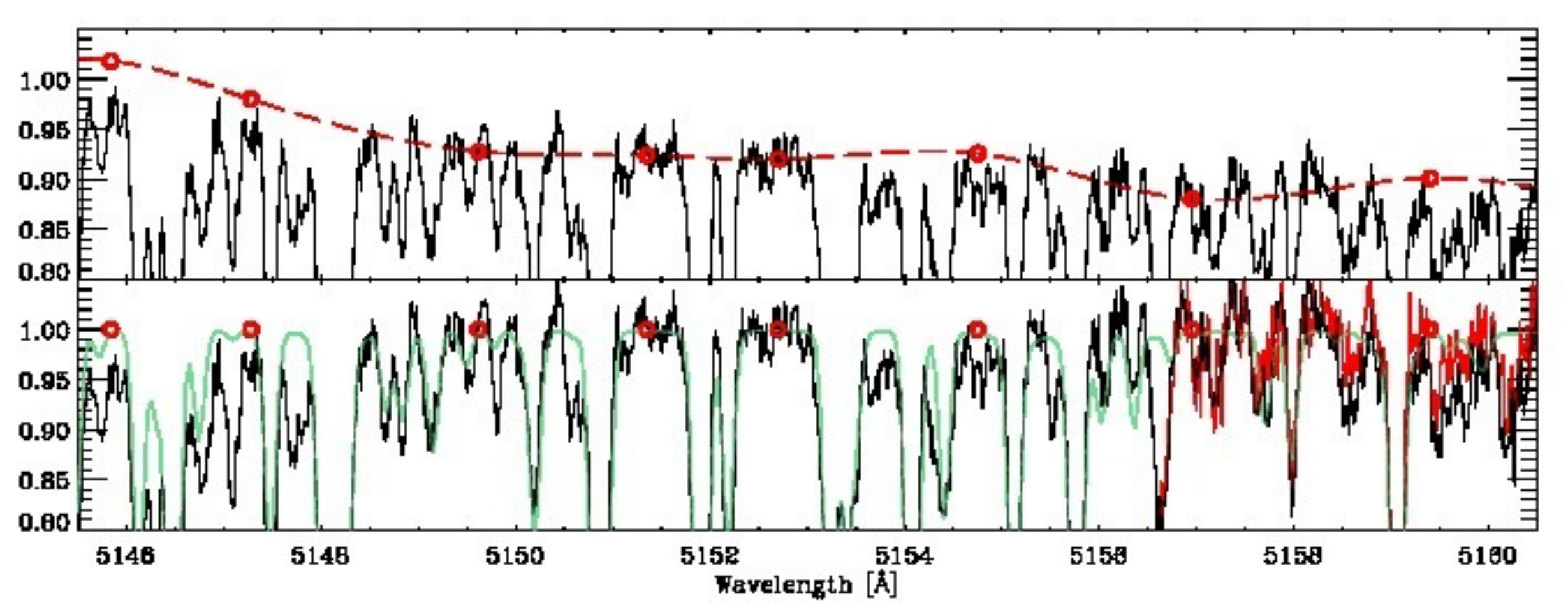}
\caption[Spectrum normalization in \vwa]{A section of a generic spectrum during the normalization with \texttt{rainbow}. In the upper panel the observed spectrum, with the selected continuum points marked with red circles: the red dotted lines is their spline fit. In the lower panel, the normalized spectrum with the template coloured in green. On the right, the adjacent echelle order is shown in red. On the upper left the fit can be improved: the sensitivity to the choice of the points can be seen. A new choice is then preferable.}
\label{rainbow}
\end{figure}

The spectra were reduced following the same methodology, independently to the instrument they were obtained with. The reduction started with the spectra provided by the instrument pipeline. If enough exposures were available, I made a first selection, by rejecting the spectra with \sn$\lesssim 10$ and those affected by the moonlight. The flux coming from the fiber exposed to the sky was subtracted from the flux of the fiber pointed towards the target. The spectra were then set to the rest frame, and corrected for the blaze function and the cosmic rays. Each  order was uniformly rebinned and automatically roughly normalized. Then, all the spectra were co-added order-by-order.\\
Subsequently, I performed a manual normalization with \texttt{rainbow}, a tool included for this purpose in \vwa. This requires the selection of points on the observed spectrum to fit with a spline function, which is used to divide the observed flux. The typical interface of \texttt{rainbow} is shown in figure \ref{rainbow}. Finally, all the orders were concatenated in a single spectrum ready to be analyzed.

\subsection{Limitations}
Spectral analyses lead to a number of uncertainties, which are briefly described below.
\begin{enumerate}[-]
\item The measure of the atmospheric parameters depends on the atomic line properties. One can choose to use laboratory values for all lines, accepting error bars of 10-20\% \citep{sousa2012}, or perform differential analysis based on a well-known star, such as the Sun \citep[e.g.][]{bruntt2004}, which has the disadvantage of limiting the analysis to stars similar to the calibrator. Corrections on the oscillator strengths of the lines can be applied, as well \citep[see, e.g.,][]{bruntt201023}. In general, erroneous values of the atomic parameters of a given line will result in wrong fits, regardless of the values of the atmospheric parameters used for the fit. Usually, lines with wrong atomic parameters are identified by visual inspection (allowed for by the codes used in this work) and are excluded from the line list used for the analysis. Hence, a human intervention is needed for the critical selection of the spectral lines to be used, or the spectral windows to be fitted. As a consequence, the analyses are very time-consuming.
\item The easiest element to study in spectral analysis is iron, because it shows the largest number of lines in a spectrum. Ideally, other metallic elements can be included in the line list to make the analysis more robust, but this is possible only for very high \sn\ spectra. For a spectrum with an \sn\ of $\sim 100$, a line list can include some hundreds of FeI lines and around 15 FeII lines or more. However, blends among lines cause the exclusion of a large number of lines, reducing the number of available FeII lines. If an EW analysis is performed, this will affect the reliability of the measure of all the atmospheric parameters. For this reason, it is important to include in the line list some pressure-sensitive lines such as Ca at 6122 and 6162 \AA, the NaID line, and the MgIb triplet. In this way, the atmospheric parameters derived with the EW method can additionally be checked by the synthetic spectra fit.
\item All the methods presented here require a normalized spectrum: no automatic procedure is completely satisfactory, and a human intervention is necessary. Another uncertainty related to data reduction, which is usually not included in the total error budget, is therefore introduced. This is particularly important for the fit of \teff\ though Balmer lines (section \ref{synthsp}), as the slope of the wings of these lines is highly affected. In the analyses I carried out by fitting synthetic spectra to the observed ones, I could obtain differences of $\sim 100$ or 200 K, by changing the normalization on H$_\alpha$.\\
The issue of normalization affects the determination of all the atmospheric parameters, especially if the EW analysis method is used. It is difficult to quantify to what extent the uncertainty on normalization affects the uncertainty on the atmospheric parameters. To be conservative, in the analyses carried out for this work, I took into account the internal uncertainties of the methods estimated in previous studies. For \vwa, these were derived by \cite{bruntt201023}: a systematic offset of $-40 \pm 20$ K in \teff\ and systematic errors of 50 K and 0.05 dex in \teff\ and \logg, respectively. \ares\ + \tmcalc\ provide the uncertainties based on the dispersion of the lines' ratios used for the analysis. When I used these codes, I referred to these uncertainties. For \sme, I relied on the internal uncertainties estimated by \cite{valenti2005}: 44 K for \teff\ and 0.06 dex for \logg. All the errors were added quadratically to the uncertainties obtained from the analysis of the spectra.
\item Both the EW analysis and the synthetic spectra fits are affected by the correlations between the atmospheric parameters. In particular, [Fe/H] affects the \teff\ and \logg\ determination. \cite{torres2012} found that the correlations between \teff, \feh, and \logg\ derived by the curve of growth method are much weaker than those derived by spectral fitting. However, when the \sn\ of a spectrum decreases, one may need to fix some parameters, and therefore choose the fit of synthetic spectra.
\item The \logg\ is a particularly difficult parameter to precisely measure, and it can lead to errors of the order of 20\% for stellar mass and up to 100\% on the radius \citep{torres2012}. To avoid this, it is advisable to fix the \logg\ to the most accurate value that can be derived from light curve modeling, thanks to the measure of the stellar density. For the case of WASP-13, \cite{chew2013} showed that, if the pressure-sensitive lines NaID and the MgIb triplet are included in the analysis,  then correct $M_\star$ and $R_\star$ are derived even without constraining \logg. However, as this is only an individual case, it is always good practice to constrain \logg\ thanks to $\rho_\star$ to avoid systematic errors in the determination of $M_\star$ and $R_\star$. This was done for all the stars analyzed during this work, as they were all part of photometric surveys.
\end{enumerate}

\section{Kepler planets}\label{kepstars}
After the release of the first Kepler candidates, a collaboration that mostly involves PASI\footnote{The exoplanet team at the Astrophysics Laboratory of Marseille.} team members initiated a ground based program for the follow-up of Kepler giant candidates \citep{bouchy2011,santerne2012}. The purpose was to establish the planetary nature of the transiting candidates, to determine the false positive rate for the giant planets, and to characterize the mass and radius of the validated planets.\\
The candidates are selected from the most up-to-date list of the Kepler Objects of Interest (KOI), reported in \cite{mullally2015}. The selection criteria for the candidates, which are still currently followed-up with SOPHIE, are: 1) the magnitude of the transited object K$_p \leq 14.7$, which corresponds to the limit of sensitivity of SOPHIE; 2) a transit depth between 0.4\% and 3\%, compatible with the signal expected from a giant planet; 3) period less than 400 days, so that at least three transits will have been detected during Kepler prime mission. Regular observations are carried out with SOPHIE, sometimes in combination with the other spectrographs mentioned in section \ref{spectrumtreat}.\\

\begin{figure}[!htb]
\centering
\includegraphics[width = 0.6 \textwidth]{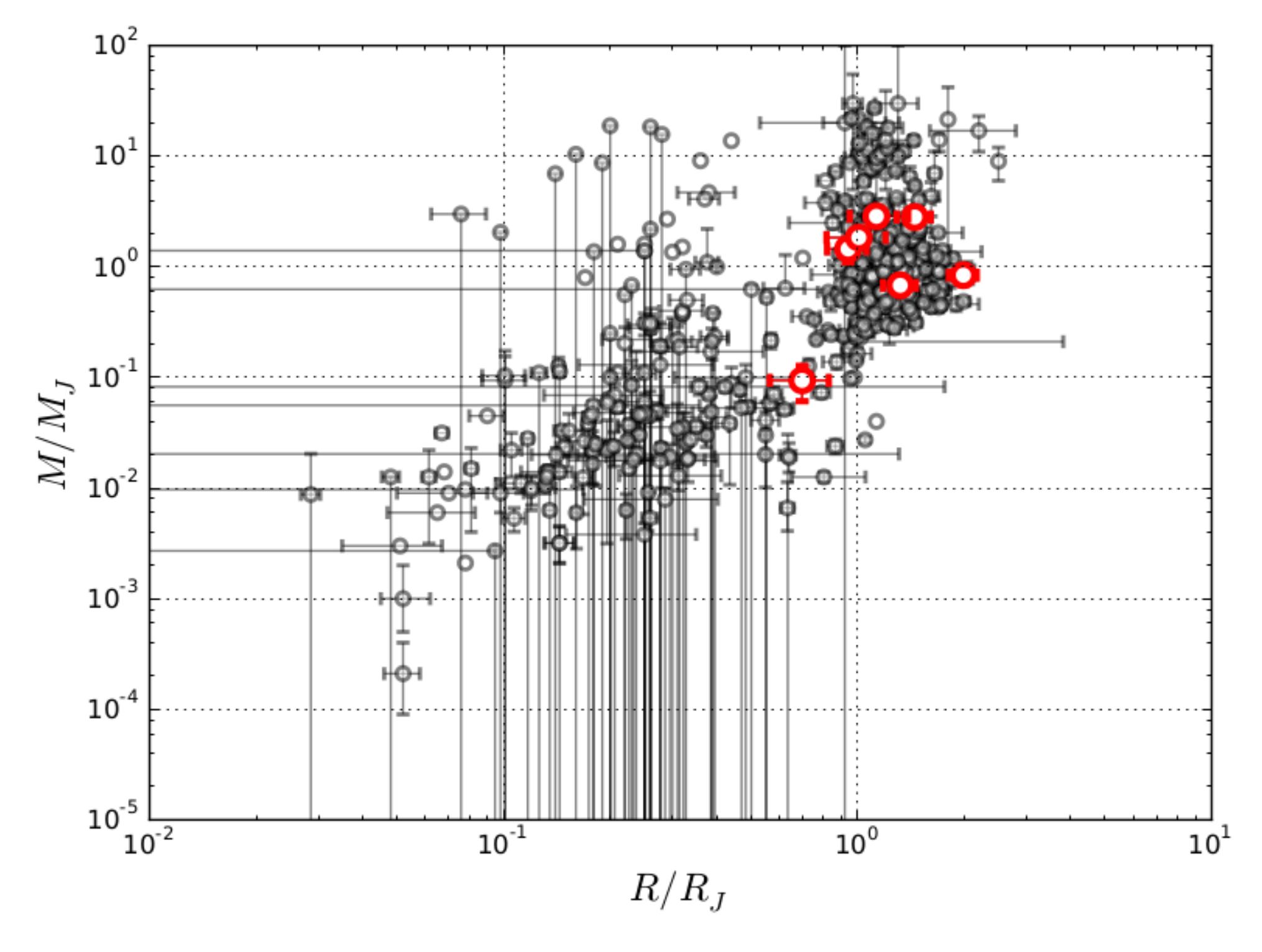}
\includegraphics[width = 0.6 \textwidth]{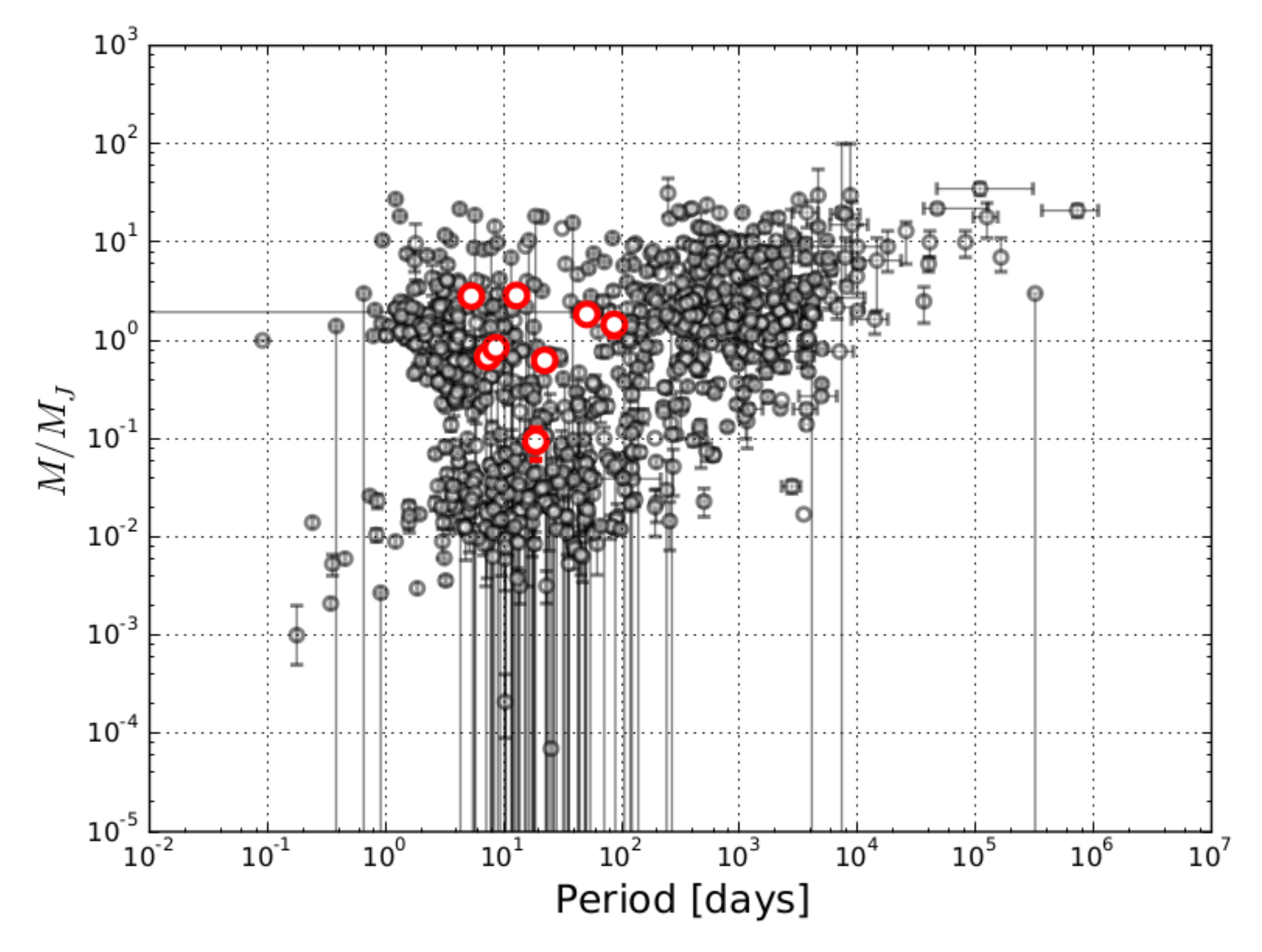}
\includegraphics[width = 0.6 \textwidth]{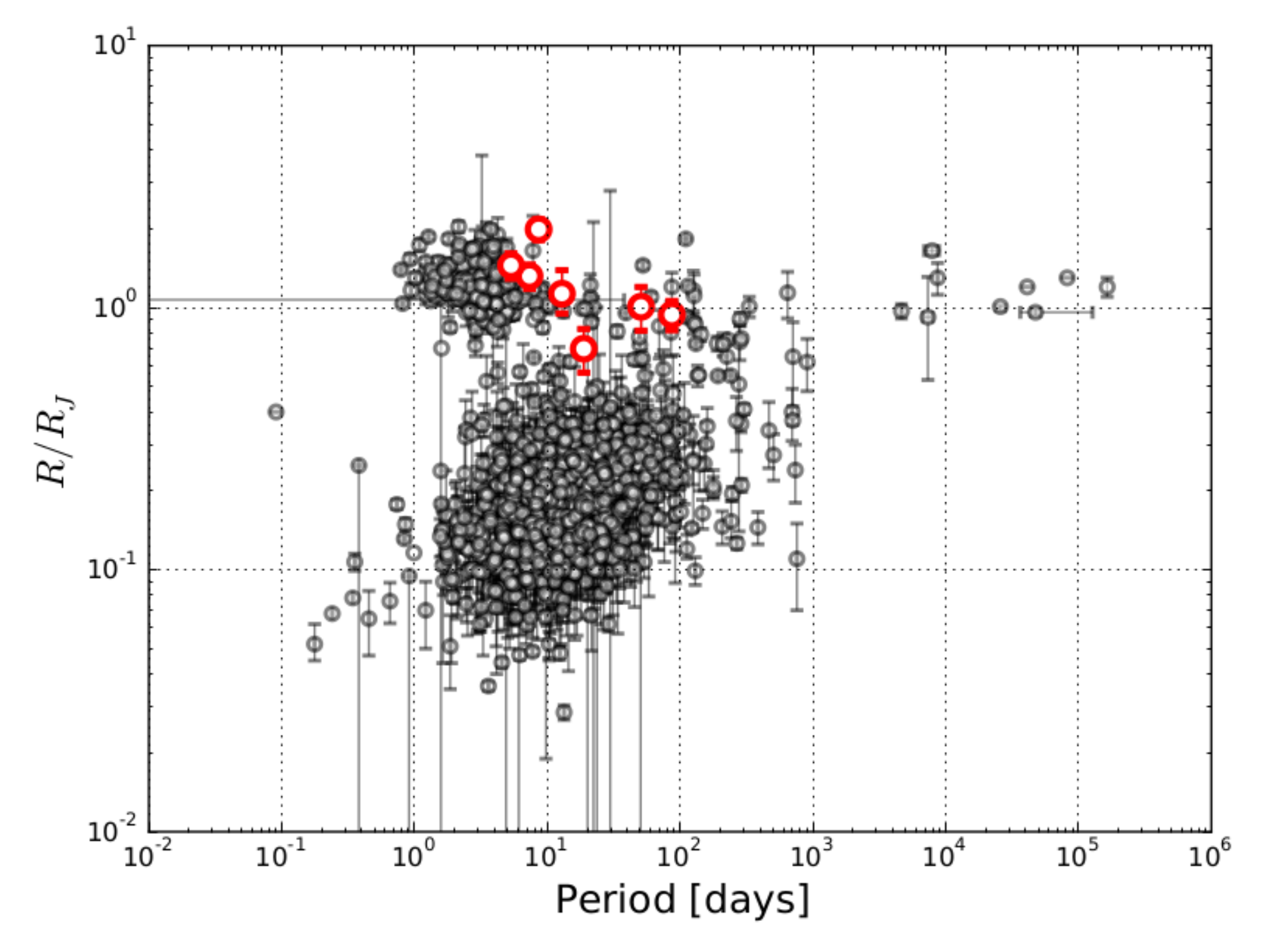}
\caption[Mass-radius, mass-period, and radius-period diagrams for the exoplanets known at the time of writing]{Mass-radius (top), mass-period (center), and radius-period (bottom) diagrams for the exoplanets known at the time of writing. The Kepler planets whose stars have been analyzed in this PhD are indicated in red. The two discovered brown dwarfs (KOI-205 b and KOI-415) are not represented.}
\label{plotkepmr}
\end{figure}

During this PhD, I analyzed the spectra of nine Kepler planet host stars. The analyses were performed with \vwa\ and \sme, which I used to derive the atmospheric parameters of the stars. Then, the derivation of the stellar masses and radii was made by constraining the \logg\ thanks to the $\rho_\star$ deduced from the light curve.\\
These analyses resulted in the characterization of nine planets and two brown dwarfs. The list of the published objects, with their main parameters, is given in tables \ref{keptab} and \ref{keptab2}. The main properties of the systems are outlined below.\\
Figure \ref{plotkepmr} shows these planets in mass-radius, mass-period, and radius-period diagrams.\\
The results of these works, and the complete relative system studies, are presented in a series of papers in \aap\ with title \textit{SOPHIE velocimetry of Kepler transit candidates}. As an example, two of these publications are reported at the end of this chapter. 

\begin{landscape}
\begin{table}[!htb]
\centering
\scalebox{0.8}{\begin{tabular}{lcccccll}
 \hline
Name & \sn\ at 550 nm & \teff\ [K] & \logg\ [dex] & \feh\ [dex] & \vsini\ [\kms] & Planets & Ref.\\
\hline
KOI-205 & 90 & $5210 \pm 70$ & $4.65 \pm 0.07$ & $0.27 \pm 0.14$ & $2 \pm 1$ & KOI-205 b & \cite{diaz2013}\\
Kepler-74 (KOI-200) & 75 & $6050\pm110$ & $4.2 \pm 0.1^{(a)}$& $0.34 \pm 0.14$ & $5 \pm 1$ & Kepler-74 b & \cite{hebrard2013}\\
KOI-415 & 120 & $5810 \pm 80 $ &$4.5 \pm 0.2$ & $-0.24 \pm 0.11$ & $1 \pm 1$ & KOI-415 b & \cite{moutou2013}\\
Kepler-88 (KOI-142) & 180 &$5460 \pm 70 $ & $4.6 \pm 0.2$ & $0.25 \pm 0.09$ & $2 \pm 1$ &  Kepler-88 c & \cite{barros2014}\\
KOI-1257 & 270 & $5540 \pm 90$ & $4.30 \pm 0.15$ & $0.26 \pm 0.10$ & $4 \pm 2$ & KOI-1257 b & \cite{santerne2014}\\
Kepler-117 (KOI-209) & 130 & $6260 \pm 80$ & $4.40 \pm 0.11$ & $0.25 \pm 0.09$ & $2 \pm 1$ & Kepler-117 b, c & \cite{bruno2015}\\
Kepler-433 (KOI-206) & 90 & $6340 \pm 140$ & $4.0 \pm 0.3$ & $0.06 \pm 0.19 $ & $11 \pm 1$ & Kepler-433 b & \cite{almenara2015}\\
Kepler-434 (KOI-614) & 50 & $5970 \pm 100 $ & $4.22 \pm 0.10$ & $0.35 \pm 0.15$ & $3 \pm 1$ & Kepler-434 b & \cite{almenara2015}\\
Kepler-435 (KOI-680) & 250 & $6090 \pm 110$ & $3.5 \pm 0.1$ & $-0.17 \pm 0.10$ & $6 \pm 1$ & Kepler-435 b & \cite{almenara2015}\\
\hline
\end{tabular}}
\caption[The stars analyzed for the Kepler follow-up program and the corresponding characterized planets]{The stars analyzed for the Kepler follow-up program and the corresponding characterized planets. $^{(a)}$: Derived from the stellar density fitted on the transits.}
\label{keptab}
\end{table}

\begin{table}[!htb]
\centering
\scalebox{0.8}{\begin{tabular}{lccc}
 \hline
Name & $R/R_J$ & $M/M_J$ & $P$ [days]\\
\hline
KOI-205 b & $0.807\pm0.022$ & $39.9 \pm 1.0$ & $11.7201248 \pm 2.1\cdot 10^{-6}$ \\
Kepler-74 b &$1.32 \pm  0.14$ & $0.68\pm  0.09$ & $7.340718 \pm  1\cdot 10^{-6}$ \\
KOI-415 b & $0.79^{+0.12}_{-0.07}$ & $62.14 \pm2.69$ & $166.78805 \pm 0.00022$ \\
Kepler-88 c$^{(a)}$ & - & $0.76^{+0.32}_{-0.16}$ & $22.10\pm0.25$\\
KOI-1257 A b&$ 0.94\pm0.12$ & $1.45\pm0.35$&  $86.647661\pm   3.4\cdot 10^{-5}$\\
Kepler-117 b&$ 0.697 \pm 0.132$ & $0.094\pm 0.033$ & $18.795952 \pm 2.7\cdot 10^{-5}$ \\
Kepler-117 c&$ 1.008 \pm 0.191$ & $1.84 \pm0.18$ & $50.790412\pm 3.9\cdot 10^{-5} $ \\
Kepler-433 b   & $1.45 \pm 0.16$ &$  2.82\pm0.52$ & $5.33408384 \pm 1.1\cdot 10^{-6}$\\
Kepler-434 b  & $  1.13^{+0.26}_{-0.18}$ & $2.86 \pm0.35$ &$12.8747099\pm  5\cdot 10^{-6}$ \\
Kepler-435 b   &$   1.99 \pm 0.18$ &$0.84 \pm 0.15$ &$8.6001536 \pm 1.8\cdot 10^{-6}$\\ 
\hline
\end{tabular}}
\caption[Main parameters for the planets of table \ref{keptab}]{Main parameters for the planets of table \ref{keptab}. $^{(a)}$: Only characterized by RVs, therefore without a measure for the radius. The mass indicates the minimum planet mass constrained by RVs.}
\label{keptab2}
\end{table}
\end{landscape}

\clearpage

\paragraph{Planets}
\begin{itemize}
\item \textbf{Kepler-74.} Orbiting this star, Kepler-74 b is a Jupiter-like planet on an eccentric orbit, lying at the frontiers between regimes where tides can explain circularization and tidal effects are negligible. It is therefore interesting for tidal evolution scenarios. Kepler-74 b, together with Kepler-75 b, was one of the first planets detected and characterized through the synergy between SOPHIE and HARPS-N/TNG.
\item \textbf{Kepler-88.} This star hosts the planet Kepler-88b, also known as the ``king of transit variations'', because of the large amplitude of the transit timing variations (TTVs) it presents ($\simeq 12$ h). TTVs and transit duration variations of $\simeq 5$ min, in phase with TTVs, led to the prediction of a non-transiting companion, Kepler-88 c, close to the 2:1 resonance with planet b \citep{nesvorny2013}. In \cite{barros2014}, for the first time, a radial velocity (RV) confirmation of a non-transiting planet detected by TTVs was presented. 
\item \textbf{KOI-1257.} This star is a member of a binary system and hosts the warm giant planet KOI-1257 b (equilibrium temperature $511 \pm 50$ K). The Kepler transit light curve, the SOPHIE RVs, the line bisector, the full-width half maximum (FWHM) variations, and the SED were consistently fitted with a Bayesian method. Notwithstanding the need of future observations to confirm the result, the properties of the binary system were constrained, and the star orbited by the planet determined.
\item \textbf{Kepler-117.} This system is made up of an F8-type main sequence star, hosting two planets presenting TTVs. Chapter \ref{chapttvs} is entirely dedicated to this system, which required specific developments.
\item \textbf{Kepler-433.} The planet KOI-206 b is a hot Jupiter, with a Jupiter-like density. Its radius ($1.13^{+0.26}_{-0.18}$ \RJ) is particularly large and challenging for planetary evolution models. If the transit depth had been overestimated because of starspots, an unlikely big ($> 59^\circ$) polar spot with a typical contrast of 0.67 would have been needed. Complementary information about the activity of the star, e.g. from the CaII H \& K lines, could not be obtained because of the low \sn\ of the SOPHIE spectra in the spectral domain of these lines.
\item \textbf{Kepler-434.} The high-density giant planet KOI-614 b is one of the few giant planets with a period longer than ten days. It has an equilibrium temperature of $1000 \pm 45$ K and therefore is at the boundary between ``hot'' and ``warm'' Jupiters. Moreover, its distance from the host star makes it an important candidate for the test of migration theories. Migration and circularization would require a heavy companion, but the RV residuals of our fit excluded the presence of a companion of more than $2.0$ \MJ\ on orbital periods less than 200 days, at the $3 \sigma$ level. However, conclusions about more distant companions could not be made with the available data.
\item \textbf{Kepler-435.} This star was the largest transited star known at the time of writing the paper, and its hosted planet Kepler-435 b was one of the largest and least dense observed at that time. Standard planetary evolution models were not able to reproduce its characteristics, but an improvement in the equation of state is expected to successfully model its inflated radius \citep{militzer2013}.\\
The host star is in a phase of rapid stellar evolution, and is forecasted to engulf the planet within approximately the next 260 Myr. 
\end{itemize}

\paragraph{Brown dwarfs}
\begin{itemize}
\item \textbf{KOI-205}. This K0 main sequence star is the first star of this type known to host a brown dwarf (KOI-205 b). Moreover, this brown dwarf was the smallest known at the time of discovery, and its proximity to the host star ($P < 15$ d) is challenging to explain on the basis of the current scenarios of formation and migration of massive objects around late-type stars.
\item \textbf{KOI-415.} This evolved solar-type star hosts a long-period  ($\simeq 167$ d) brown dwarf in an eccentric orbit. Its measured properties are compatible with a 10 Gyr, low-metallicity and non-irradiated object.
\end{itemize}

\section{Low-mass binaries}\label{lowm}

\begin{figure}[!htb]
\centering
\includegraphics[width = 0.6 \textwidth]{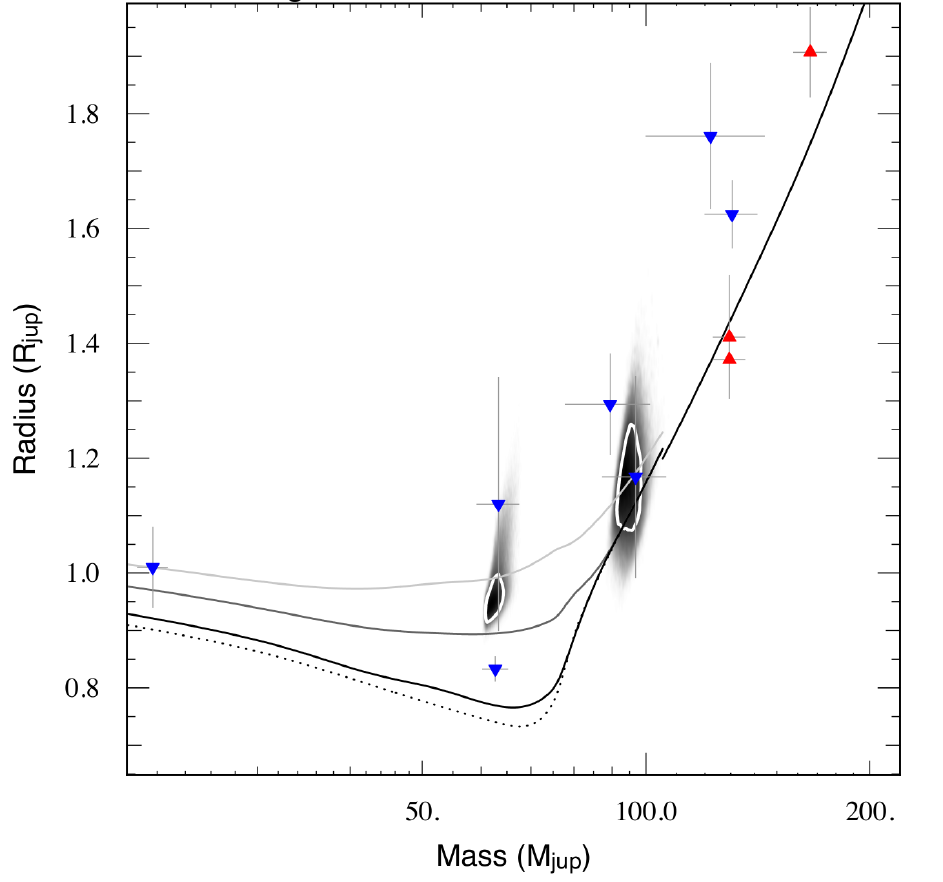}
\caption[The mass-radius theoretical relationship for M dwarfs]{The mass-radius theoretical relationship (black line) for M dwarfs. The lines represent the isochrones of \cite{baraffe1998,baraffe2003} for different ages: from bottom to top, 10 Gyr, 5 Gyr, 1 Gyr, 0.5 Gyr. Blue and red points and shaded regions show some known systems. Courtesy of G. Montagnier.}
\label{mrlow}
\end{figure}

While the empirical knowledge of the mass-luminosity relation for M dwarfs has improved in the last years \citep{delfosse2000,segransan2000,xia2008}, the mass-radius relation for low-mass stars remains a controversial topic. Figure \ref{mrlow} shows the mass-radius relationship from the isochrones of \cite{baraffe1998,baraffe2003}. 

As discussed in section \ref{why}, the model-independent determination of stellar masses and radii is rarely possible. Double-line eclipsing binaries allow one to overcome this limitation \citep{torres2010,kraus2011}. By photometry and RVs, the masses and radii of these objects are known at a precision better than 1\% in the most favourable cases. However, most of these binaries orbit with periods less than three days. This is due to selection effects: binaries with small separations have longer eclipses, which can be better sampled. By tidal synchronization, the bodies composing these systems are expected to rotate fast, increasing their magnetic activity, and enlarging their radii \citep{mullan2001}. Indeed, the components of these binary systems have larger radii by up to 10\% than what stellar evolution models predict. To assess whether this is a peculiar characteristic of these systems or a shortcoming of the theory, longer-period systems are needed to better calibrate the models.\\
Long period systems or isolated M stars can be observed with long-based interferometry, in order to constrain the stellar radii by a few percent \citep{demory2009}. Unfortunately, this technique is suitable only for the closest, brightest stars. The eclipsing binaries detected for transiting planet searches, instead, offer plenty of M dwarfs orbiting F and G dwarfs as a by-product of the observations.\\
A follow-up program of CoRoT and Kepler eclipsing binaries was therefore carried out with SOPHIE between 2012 and 2013. The aim was to determine the fundamental parameters of the secondary object (the M star) by characterizing the primary (an F or a G star). The primaries were selected between the stars that had already been observed with SOPHIE in the planet candidates follow-up program, to be able to constrain the mass of the secondary. All the selected objects show an RV variation compatible with a low-mass star ($70 < m_{\rm{companion}}/M_{\mathrm{J}} < 300$ ), a V magnitude $m_V < 15$, and a non-grazing transit. Their list is presented in table \ref{lowmtab}.\\
I analyzed the spectra of these stars with \sme, because of their low \sn. In the cases where the quality of the spectrum was too low to obtain a reliable measurement of the atmospheric parameters, the SED was used. For this latter, we referred to the web archives of APASS\footnote{\texttt{http://www.aavso.org/apass}.}, 2MASS \citep{sktutskie2006,cutri2003}, and WISE \citep{wright2010}. Moreover, some of the stars are fast rotators (\vsini\ $> 20$ \kms), therefore their parameters are very problematic to derive.\\
This work is still in progress, and the full list of masses and radii for the low-mass objects is not yet available. 

\begin{table}[htb]
\centering
\begin{tabular}{lccccccc}
 \hline
Name & \sn & \teff\ [K] & \logg\ [dex] & \feh\ [dex] & \vsini\ [\kms] \\
\hline
IRa01\_E2\_0203 & 175 & $4960 \pm 70$ & $2.90 \pm 0.07$ & $-0.28 \pm 0.08$ & $ 3 \pm 1$ \\
SRa02\_E2\_1065 & 44 & $6560 \pm 180$& $4.3\pm0.3$ & $-0.27 \pm 0.19$	& $6\pm2$ \\		
LRa02\_E2\_0122 & 140 & $6240\pm100$	&$3.70 \pm 0.21$	& $-0.21 \pm 0.12$	& $7\pm1$ \\
SRa04\_E2\_0335 & 176 & $6490 \pm100$ & $4.3 \pm 0.1$ & $0.21 \pm 0.07 $	& $\sim 65$ \\
SRa02\_E2\_0486 & 45 & $6030 \pm 100$ & $4.2 \pm 0.25$	& $-0.24 \pm 0.1$	& $6 \pm 1$ \\
LRa03\_E2\_0269 & 111 & $6000 \pm 100$ & $4.0 \pm 0.2$	& $-0.12 \pm 0.1 $ & $10 \pm 1$ \\
LRa01\_E2\_2249 & 39 & $5400 \pm 100 $& $4.0 \pm 0.12$ & $> 0$ & - \\
IRa01\_E2\_2430 & 69&	$\sim 5000$ &-	& - & $\sim 70$\\
SRc02\_E2\_3977 & 45 &	$6190 \pm 100$ &$\sim1.9$	& -  & $13 \pm 1$\\
SRa02\_E2\_0893 & 72	&$\sim6500$ &$\sim2.5$& -	& $\sim50$\\
SRa02\_E2\_0749 & 39	&$6100\sim70$&	2.6 (Ca6122)& - & $\sim27$ \\
LRc02\_E2\_2154 & 35 & $5900 \pm 30$& -	& $0.4 \pm 0.1$ & -\\
LRc03\_E2\_1126 & 85&	$5450 \pm300$&- &$-0.41\pm0.38$ & $22\pm1$\\
LRc02\_E1\_0981 &70&	$5500 \sim140$&	-	&$-0.25\pm 0.29$ & $\sim17$ \\
IRa01\_E1\_1158 & 63&	$\sim 7200 $& - &	$< 0$ & $\sim20$ \\
LRc02\_E2\_1207 & 58&	$6610$&	$4.2\pm 0.2$	& $-0.69\pm0.38$	 & $5\pm1$\\
LRc07\_E2\_0108 & 100	&$\sim 6500$ & $>4$& -& $\sim 70$ \\
LRc09\_E2\_0548 & 129	&$5470\pm150$&	$3.4\pm 0.3$ & $0.21 \pm 0.1$ & $20\pm3$ \\
KOI-552 & - &5990&	-&	$-0.48\pm0.15$&$18\pm1$ \\
KOI-554 & 66 &	$6110\pm130$&$4.3 \pm 0.11$	&$-0.01\pm0.13$&	$23\pm2$\\
KOI-1230 & 175 & $4980\pm70$& $2.90 \pm0.07$&$-0.18\pm0.10$&$3\pm1$\\
\\
\hline
\end{tabular}
\caption[Targets of the search for low-mass stars among the CoRoT and Kepler candidates and derived parameters]{Targets of the search for low-mass stars among the CoRoT and Kepler candidates and derived parameters. Empty cells indicate that a spectroscopic derivation was not robust, so that an SED needs to be used. Such fits will be performed together with the fit of the stellar mass and radius, not complete yet, and is therefore not indicated.}
\label{lowmtab}
\end{table}

\section{Activity index of SOPHIE stars}\label{logrhk}

\subsection{Context}\label{contrhk}
\begin{figure}[htb]
\centering
\includegraphics[width = 0.6 \textwidth]{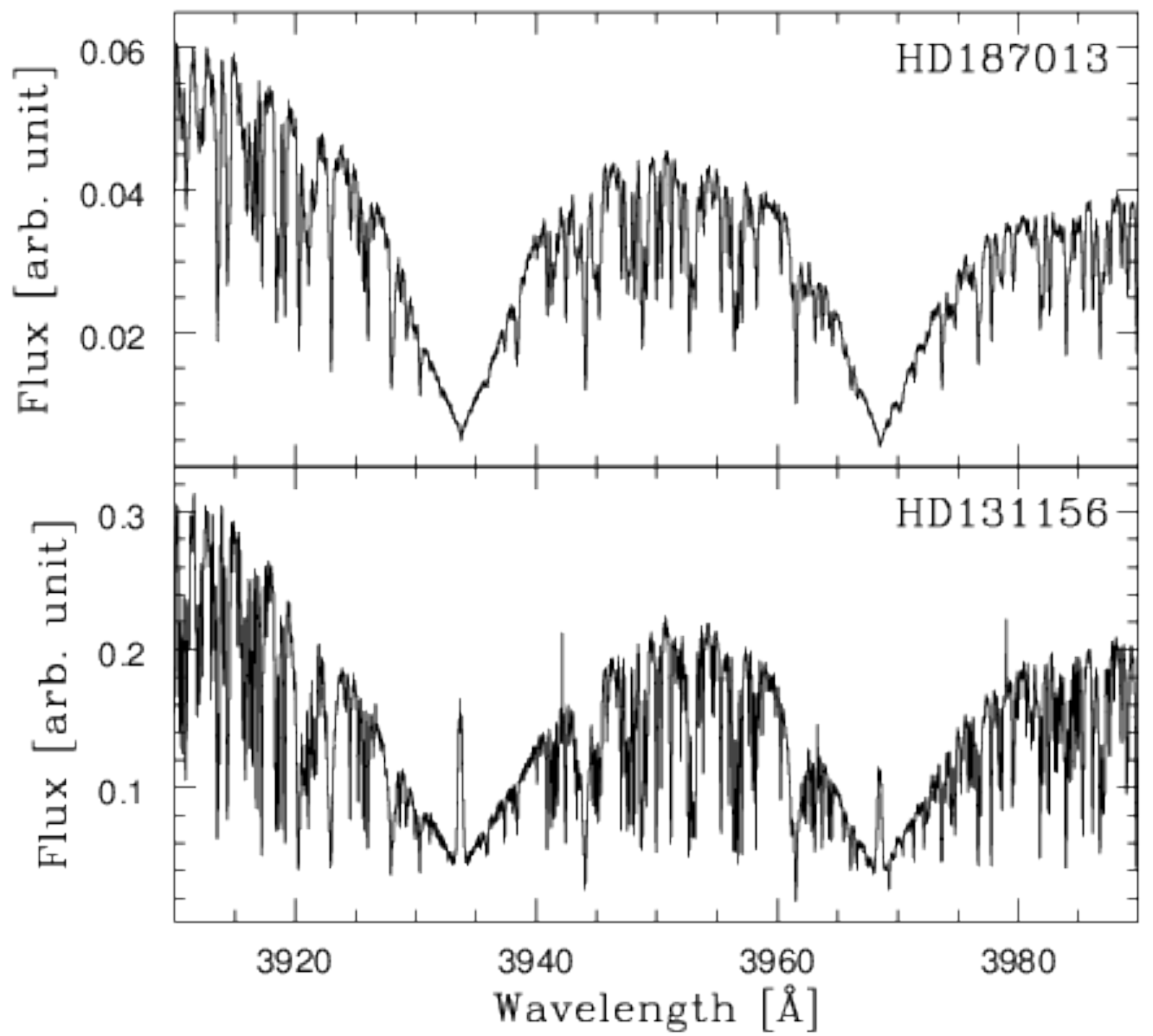}
\caption[The spectral region around the CaII  H and K lines for the SOPHIE spectrum]{The spectral region around the CaII  H and K lines for the SOPHIE spectrum of a non-active (top) and an active (bottom) star. From \cite{boisse2010}.}
\label{calcium}
\end{figure}

The Calcium II H and K lines, at 3933.68 and 3968.49 \AA, are among the deepest and broadest absorption lines in the spectrum of Sun-like stars. They were first observed in the solar spectrum by Fraunhofer in 1814. They are resonant lines whose cores are formed in the chromosphere, so that they can be employed as probes of the chromospheric structure and properties \citep{linsky1970}. In particular, they are are widely used indicators of magnetic activity, of convective envelopes and of the presence of spots and plages on the stellar surface. Figure \ref{calcium} shows these lines for a non-active and an active star.

\begin{figure}[htb]
\centering
\includegraphics[width = 0.6 \textwidth]{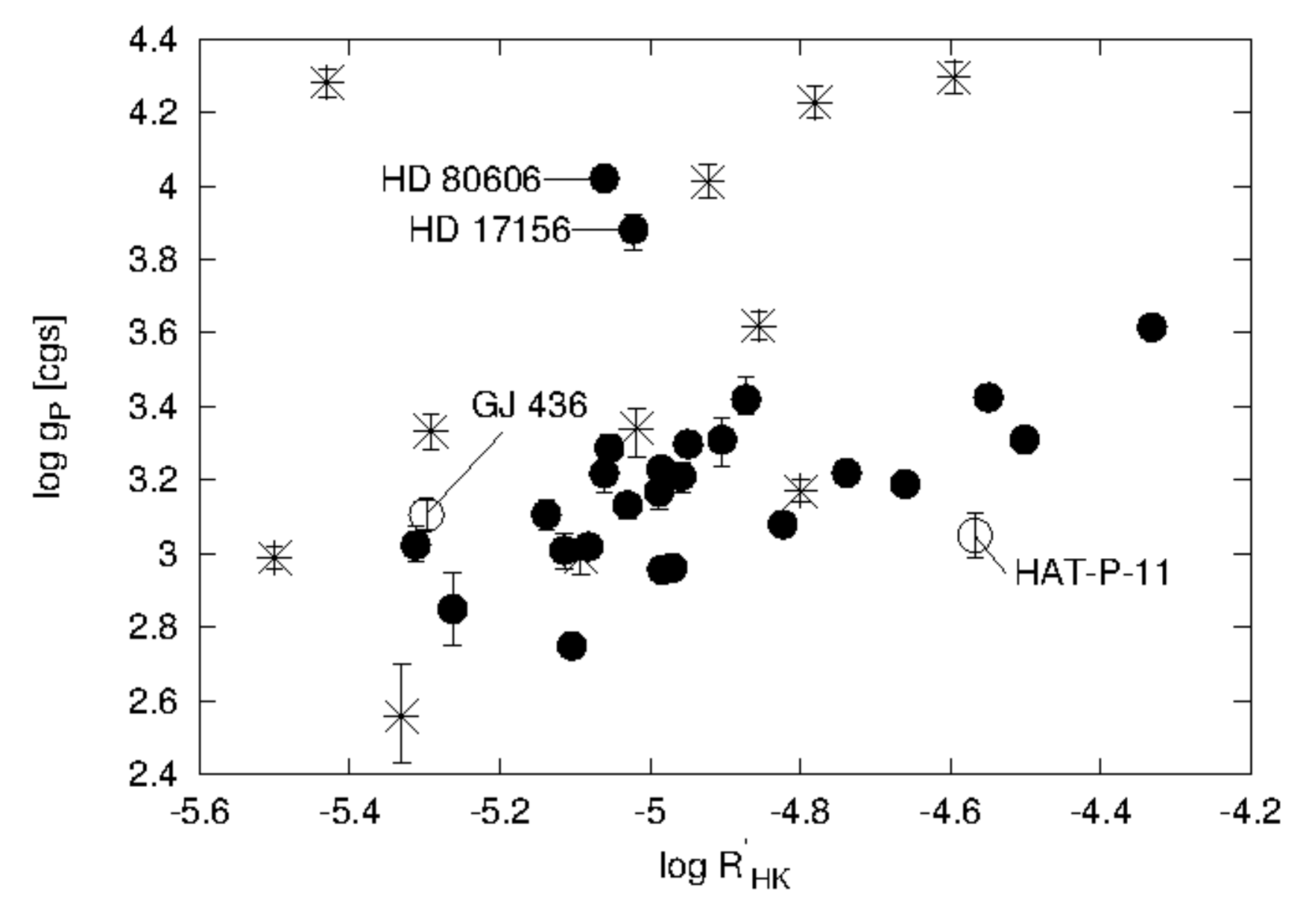}
\caption[The correlation \logrhk - \logg$_{\rm{p}}$ from \cite{hartman2010}]{The correlation \logrhk - \logg$_{\rm{p}}$ from \cite{hartman2010}. Filled circles show planets with mass $> 0.1$ \MJ\ orbiting host stars with 4200 K $<$ \teff\ $<$ 6200 K, crosses show planets with mass $>$ 0.1 \MJ\ orbiting host stars with \teff\ $<$ 4200 K or \teff\ $>$ 6200 K (out of the calibration range of \logrhk\ used by the authors), and open circles show two hot Neptunes with mass $< 0.1$ \MJ\ , which might be expected to have different atmospheric properties than more massive planets. The labeled planets have semi-major axes $>$ 0.1 AU, and their properties are less likely to be influenced by the stellar flux.}
\label{hartman}
\end{figure}

Chromospheric activity is generated by the stellar magnetic dynamo, whose intensity scales with the rotational velocity of the star \citep{kraft1967,noyes1984,montesinos2001}. Both chromospheric activity and rotation rate, moreover, have been observed to decay with age \citep{wilson1963,skumanich1972,soderblom1983,soderblom1991,mamajek2008}, as an effect of a mass loss in a magnetized wind \citep{schatzman1962,weber1967,mestel1968}.\\
The CaII H and K lines present emission cores in active stars, indicating that the source function (that is, the emission-to-absorption coefficient ratio) in the chromosphere is larger than in the photosphere. \cite{wilson1968} defined an index, called $S$-value, as the ratio between the emitted flux in the center of the lines and the continuum flux. Historically, this has been converted to the Mount Wilson system \citep{vaughan1978,vaughan1980,duncan1991}. From the $S$-value, a chromospheric activity index is defined, and called \logrhk\ (see section \ref{methlogr} for the definition).\\
\cite{knutson2010} presented evidence of a correlation between the presence of temperature inversions in the atmosphere of transiting hot Jupiters and the chromospheric activity level of their host stars. \cite{hartman2010} found also a positive correlation between the surface gravity of the transiting hot Jupiters, \logg$_{\rm{p}}$, and the activity level of their host stars, with a 99.5\% confidence. This correlation is shown in figure \ref{hartman}. 

If one assumes that the activity of a star decreases with age, then this correlation means that planets expand with age, instead of contracting \citep[e.g.][]{fortney2007}. This contradicts the models planetary structure evolution. After having discarded a selection effect as a cause of the correlation, these authors propose different explanations: 1) the stellar insolation on the planetary radius could be larger than what is assumed, and thus counterbalance the gravitational contraction of the planet; 2) mass loss from the hydrogen atmosphere, for planets irradiated by strong UV fluxes \citep[e.g.][]{lecavelier2010}; 3) tidal spin-up of the convection zones of stars or magnetic star-planet interactions \citep{shkolnik2009}, implying that the \logrhk\ does not necessarily decrease with the age of a star. Recently, \cite{lanza2014} proposed another possible interpretation based on the evaporation of plasma from the close-in planets. The debate on the most likely scenario is still ongoing.\\
 \cite{figueira2014} extended the previous sample of stars by a factor three, finding significant correlations, but with lower Spearman's rank correlation coefficients than in \cite{hartman2010}. We added 31 transiting stars observed with SOPHIE to the sample of \cite{figueira2014}, to increase the number of stars from which the \logrhk-\logg$_{\rm{p}}$ correlation can be studied. The sample includes both active and non-active stars.

\subsection{Method}\label{methlogr}
I used a script written by I. Boisse, which calculates the \logrhk\ index for SOPHIE spectra. The definition of \cite{noyes1984} was used:
\begin{equation}
R^{'}_{\rm{HK}} = 1.340 \cdot 10^{-4} C_{cf}S,
\end{equation}
where 
\begin{equation}
\log C_{cf} (B-V) = 1.13 (B-V)^3 - 3.91 (B-V)^2 + 2.84 (B-V) -0.47,
\end{equation}
and the $B-V$ color index refers to the $UBV$ magnitude system \citep{johnson1953,nicolet1978}.
This relation is available only for $0.43 \leq B-V \leq 1.2$ and is based on the $S$-value of Mount Wilson, which was calibrated for SOPHIE by \cite{boisse2010}. The $S$ value is calculated as
\begin{equation}
S = \frac{H + K}{B + V},
\end{equation}
where $H$ and $K$ are the flux values measured in two triangular 1.09 \AA-FWHM windows centered on the Ca lines, and $B$ and $V$ represent the value of the continuum on the sides of the lines, with the flux measured in 20 \AA-wide windows respectively centered on 3900 and 4000 \AA.\\
The determinant variable is the $B-V$. For homogeneity, I computed this value from stellar models, depending on the \teff, \logg, and \feh\ of the stars. I referred to the mean values reported in literature for these parameters. The relative uncertainties, as well as the internal uncertainties of the models, were not taken into account.\\
In a second step, I quantified the impact of the reddening $E(B-V)$ on the derived \logrhk. I added the reddening values reported in the EXODAT archive\footnote{\texttt{http://cesam.oamp.fr/exodat/}.} \citep{meunier2007, meunier2009, deleuil2009} to the theoretical $B-V$ for the CoRoT stars, while I referred to the values reported in the SIMBAD astronomical database\footnote{\texttt{http://simbad.u-strasbg.fr/simbad/}.} for the others.

\subsection{Analyzed spectra}
I examined the CoRoT, the WASP \citep{colliercameron2007}, and the HAT \citep{bakos2007} objects observed for the follow-up with SOPHIE. Several exposures were available for each target, each with a different \sn. I considered only the stars observed at least three times. The spectra with \sn\ $< 3$ in the echelle order containing the Ca lines were rejected. All the spectra were obtained in the HE mode of SOPHIE. The list of targets is reported in table \ref{tabrhk}.

\begin{table}[!htbp]
\centering
\caption[The stars whose \logrhk\ has been measured]{The stars whose \logrhk\ has been measured. From left to right: name of the star, \sn\ of the co-added spectrum in the inspected echelle order, theoretical $B-V$ from the stellar models, corresponding \logrhk, $(B-V)_0 + E(B-V)$ (for CoRoT stars) or observed $B-V$ (for HAT-WASP stars), corresponding \logrhk, and reference paper for the stellar parameters. The cases in which non-reliable reddened values were obtained are indicated as ``-'' (section \ref{resrhk}).}
\scalebox{0.95}{\begin{tabular}{lcccccl}
 \hline
Star &   \sn  &  $(B-V)_0$  &  \logrhk & $B-V$ & \logrhk$_{\rm{, red}}$ & Ref.\\
\hline		       
CoRoT-2	   &	18.37  &0.80  &$-4.32  \pm 0.07 $&  1.05 & $-4.55 \pm 0.14$ & \cite{alonso2008} \\
CoRoT-3	   &	20.62  &0.47  &$-4.44  \pm 0.11 $&  0.97 & $-4.80 \pm 0.16$ & \cite{deleuil2008} \\
CoRoT-4	   &	23.43  &0.64  &$-4.46  \pm 0.16 $&  0.74 & $-4.53 \pm 0.15$ & \cite{aigrain2008} \\
CoRoT-5	   &	17.69  &0.61  &$-4.45  \pm 0.26 $&  0.66 & $-4.48 \pm 0.25$ &  \cite{rauer2009} \\
CoRoT-9	   &	14.39  &0.80  &$-4.44  \pm 0.24 $&  0.90 & $-4.55 \pm 0.22$ & \cite{deeg2010} \\
CoRoT-11   &	12.78  &0.54  &$-4.39  \pm 0.08 $&  1.14 & $-5.01 \pm 0.05$ & \cite{gandolfi2010} \\
CoRoT-18   &	14.69  & 0.85 &$ -4.32 \pm 0.11 $&  1.20 & $-4.73 \pm 0.12$ & \cite{hebrard2011} \\
CoRoT-19   &	22.71  &0.66  &$-4.34  \pm 0.14 $&  1.16 & $-4.93 \pm 0.14$ & \cite{guenther2012} \\
CoRoT-20$^{(a)}$   &	15.05  &0.72  &$-4.46  \pm 0.28 $&  - & -&  \cite{deleuil2012} \\	
HAT-P-42   &	26.92  &0.79  &$-4.80  \pm 0.14 $&  0.67 & $-4.72 \pm 0.15$ & \cite{boisse2013} \\
HAT-P-43   &	19.31  &0.82  &$-4.51  \pm 0.15 $&  0.76 & $-4.52 \pm 0.11$ & \cite{boisse2013} \\	
WASP-1	   &	26.73  &0.62  &$-4.77  \pm 0.17 $&  - &  -  & \cite{simpson2011}\\
WASP-3	   &	44.53  &0.55  &$-4.85  \pm 0.06 $&  - & - & \cite{pollacco2008} \\
WASP-10	   &	26.27  &1.15  &$-4.62  \pm 0.08 $&  - & - &  \cite{christian2009}  \\	
WASP-11	   &	15.25  &1.09  &$-4.89  \pm 0.06 $&  1.00 & $-4.76 \pm 0.06$ & \cite{west2009}  \\
WASP-13	   &	46.52  &0.73  &$-5.15  \pm 0.08 $&  0.89 & $-5.13 \pm 0.06$ & \cite{skillen2009}\\
WASP-14	   &	114.44 &0.52  &$-5.05  \pm 0.13 $&  0.46 & $-5.05 \pm 0.12$ & \cite{joshi2009} \\
WASP-21	   &	62.84  &0.66  &$-4.91  \pm 0.11 $&  0.54 & $-4.84 \pm 0.11$ & \cite{bouchy2010}\\
WASP-37    &	24.98  &0.67  &$-4.70  \pm 0.11 $&  0.60 & $-4.65 \pm 0.12$ &  \cite{simpson2011_2} \\
WASP-38	   &	51.16  &0.60  &$-5.06  \pm 0.08 $&  0.48 & $-5.01 \pm 0.10$ & \cite{barros2011_2} \\
WASP-39	   &	24.30  &0.85  &$-4.62  \pm 0.17 $&  0.82 & $-4.59 \pm 0.17$ & \cite{faedi2011}\\
WASP-40	   &	22.52  &0.96  &$-4.67  \pm 0.18 $&  1.08 & $-4.84 \pm 0.17$ & \cite{anderson2011} \\
WASP-48	   &	55.58  &0.67  &$-4.84  \pm 0.15 $&  0.66 & $-4.84 \pm 0.16$ & \cite{enoch2011} \\
WASP-52	   &	41.13  &1.01  &$-4.59  \pm 0.14 $&  0.90 & $-4.45 \pm 0.13$ & \cite{hebrard2013_2}\\
WASP-56	   &	34.76  &0.82  &$-4.77  \pm 0.13 $&  - & - &  \cite{faedi2013} \\	 
WASP-58	   &	43.68  &0.67  &$-4.70  \pm 0.19 $&  0.67 & $-4.71 \pm 0.19$ & \cite{hebrard2013_2} \\
WASP-59	   &	16.50  &1.13  &$-4.66  \pm 0.07 $&  0.92 & $-4.33 \pm 0.07$ & \cite{hebrard2013_2} \\
WASP-76	   &	8.99   &0.62  &$-4.67  \pm 0.07 $&  - & - & \cite{west2013}  \\	 
WASP-85	   &	32.08  &0.81  &$-4.66  \pm 0.07 $&  0.71 & $-4.58 \pm 0.08$ & \cite{brown2014}\\
WASP-104   &	21.58  &0.90  &$-4.73  \pm 0.20 $&  - & - &  \cite{smith2014} \\	
WASP-106   &	22.51  &0.65  &$-4.91  \pm 0.16 $&  0.83 & $-4.98 \pm 0.13$ & \cite{smith2014}\\
\hline \\
\end{tabular}}
\begin{list}{}{}
\item $^{(a)}$ Only two points gave reliable results when using the reddened $B-V$.
\end{list}
\label{tabrhk}
\end{table}

\subsection{Results}\label{resrhk}
\begin{figure}[!hbt]
\centering
\includegraphics[width = 0.6 \textwidth]{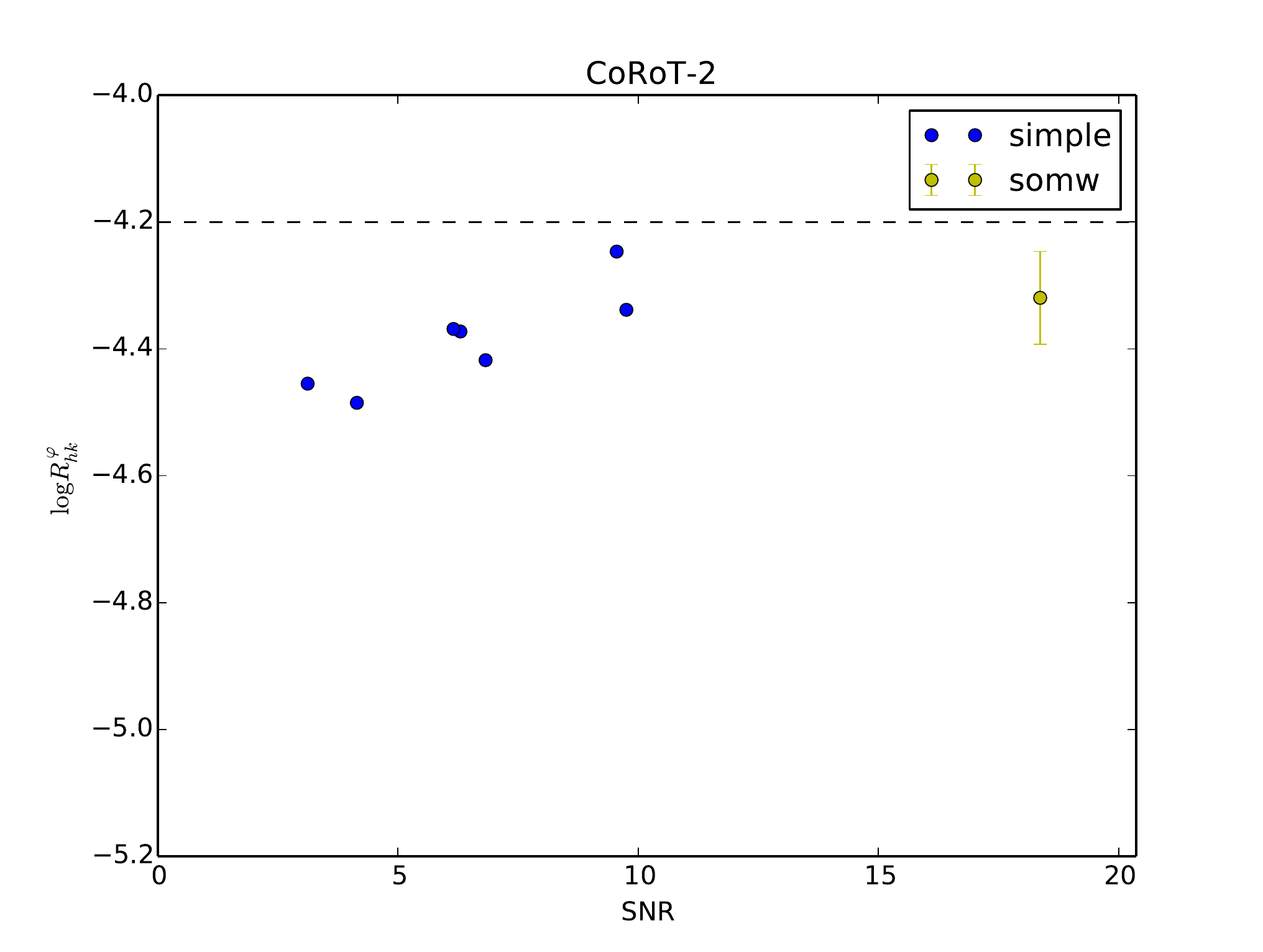}
\caption[The \logrhk\ measured on the spectra of CoRoT-2]{The \logrhk\ measured on the spectra of CoRoT-2 (blue) versus their \sn. In yellow, the value obtained on the co-added spectrum.}
\label{c2rhk}
\end{figure}

For each  star, the \logrhk\ was measured for individual exposures. Studies on solar-type stars found indexes between -5.2 (non-active star) and -4.2 (active star) \citep[e.g.][]{hall2007}. We conservatively adopted these limits, and kept only the values within this range. When several exposures of a star are available, the measured \logrhk\ increases with \sn\ until they stabilize. Figure \ref{c2rhk} shows the behaviour of the index for such a case. When possible, the averaged values were compared with those existing in literature, showing a general agreement. This behaviour suggests that a moderate degree of confidence can also be obtained for spectra with low \sn.

For each star, a co-added spectrum of the spectral window used for the calculation of the index was computed, and the \logrhk\ was derived again. This last value was adopted as the final result, and the scatter on the results from the individual spectra was adopted as the $1\sigma$ uncertainty. The results are reported in table \ref{tabrhk}.\\
The \logrhk\ values found with this study will be combined with the surface gravity of the planets, producing a plot as the one of figure \ref{hartman}.

\subsection{Discussion and perspectives}
By calculating a co-added spectrum, the evolution timescale of the activity becomes important. In some cases, spectra for the same star were recorded at some years of separation, so that there is the risk of averaging the contribution of different phases of the activity cycle. As long as the duration of the activity cycle is not known, it is difficult to quantify such a risk. \cite{knutson2010} overcame this issue by adopting the median of the \logrhk\ measured on their individual spectra. Their results were used by \cite{hartman2010}, too. However, it is likely that the variation of the index during the activity cycle is negligible with respect to the uncertainties of our results. Higher-quality spectra, such as those obtained by HARPS, are needed to compare our findings, and will be analyzed in the future.

\section{SOPHIE at low SNR}\label{lowsnr}

\begin{figure}[!hbt]
\centering
\includegraphics[width = 1.0 \textwidth]{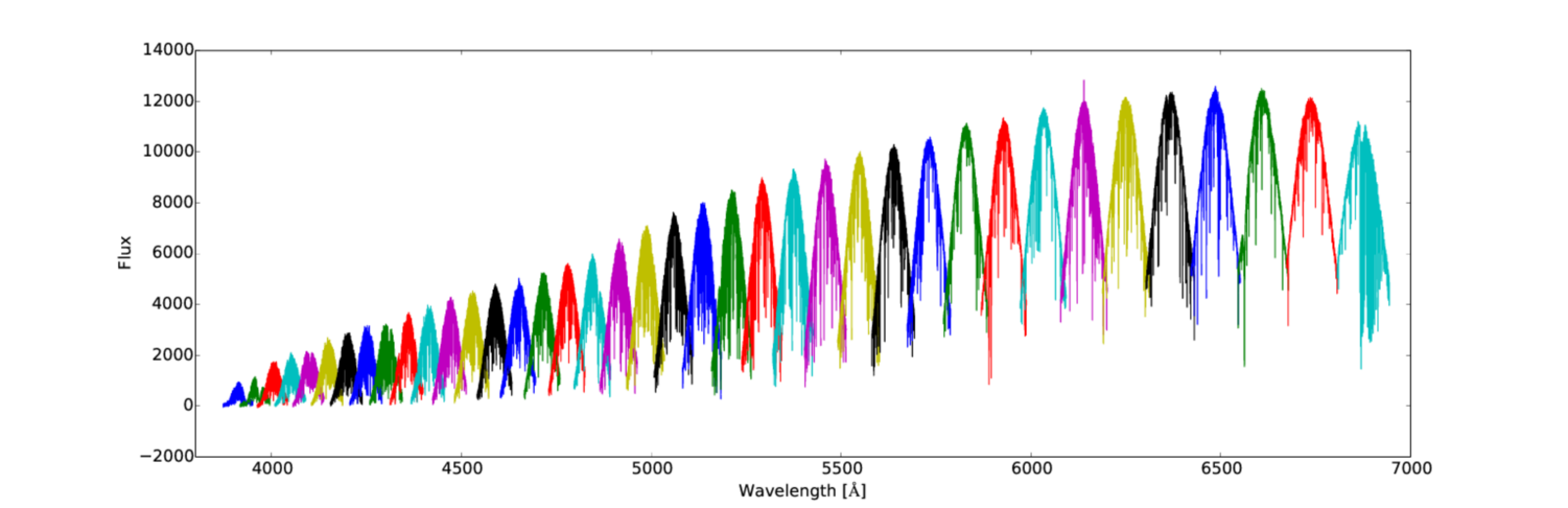}
\caption[A SOPHIE spectrum of the Sun, where the blaze function has not been corrected for yet]{A SOPHIE spectrum of the Sun, where the blaze function has not been corrected for yet.}
\label{blaze}
\end{figure}

Because of the high magnitude of some of the targets of the CoRoT and Kepler transit surveys, the spectra acquired with SOPHIE are often characterized by a very low \sn, that is, lower than 10. Experience has shown hints of a systematic underestimation of the stellar metallicity when analyzing these spectra. The cause could be an erroneous bias in the flux value in the central part of the spectral lines, introduced by a wrong correction for the blaze function on the echelle orders. Figure \ref{blaze} shows a SOPHIE spectrum of the Sun, where the blaze function has not been corrected for yet. A wrong correction would affect the EW of the spectral lines and, consequently, the derived atmospheric parameters. This risk is usually reduced by rejecting the spectra with \sn$\, <10$ and co-adding the single exposures.\\
To investigate the presence of this bias, and whether it could be due to an incomplete correction of instrumental effects, I carried out a homogeneous analysis on a high number of low-\sn\ spectra of benchmark stars. The study was meant to highlight and quantify, if present, any systematic trend in the stellar parameters as a function of the \sn, and to check whether this is maintained once several low-\sn\ spectra are co-added. Moreover, with this analysis, a quantitative information on the reliability of a spectrum, as a function of its \sn, was obtained. 

\subsection{Benchmark stars}\label{sundata}
We observed with SOPHIE a K, a G (the Sun), and an F star, whose parameters are indicated in table \ref{tabstars}. These spectral types are the main targets of RV surveys. The targets were chosen because of their strong brightness, so as to reduce the observation time to a minimum (from a few seconds to a few minutes). Every object was observed for progressively longer times, obtaining exposures with increasing \sn. The spectrum of the Sun was obtained by observing the reflection from the asteroid Hebe between April 22, 2013 and May 4, 2013, and from the Jovian satellite Europa on January 14, 2014. These targets were chosen because of their negligible contamination to the Sun's spectrum. The F and the K stars (HD016673 and HD010476, respectively) were observed on January 14, 2014. All the observations were carried in the HE mode of SOPHIE, used for faint objects, and with spectral resolution $\simeq 40000$.\\
With these observations, we secured 13 spectra for the Sun, 19 for HD16673 and 22 for HD010476, with an \sn\ going from $\simeq 4$ to $\simeq 100$.\\
The spectra of each star were processed by the SOPHIE pipeline. A standard reduction (see section \ref{spectrumtreat}) was applied to remove the cosmic rays and correct the spectra for the blaze profile. Additionally, the spectra with \sn\ $>10$ were co-added, producing a high-\sn\ spectrum for each star. No normalization was performed for any of these spectra.

\begin{table}[!ht]
\centering
\caption{Stars used as calibrators.}
\begin{tabular}{lccccc}
 \hline
Name  & Type  & $T_{eff}$ [K] & $\log g$ [cgs] & [Fe/H] [dex] & \vsini$^{(b)}$ [\kms]\\
\hline
HD010476$^{(a)}$   &  K1V &    $5229\pm33 $   &      $4.58 \pm0.06$  & $0.02 \pm0.06$ & $3.4 \pm 1.0$\\ 
Sun	   & G2V  & 5777 & 4.44 & 0.00 & 1.7 \\
HD016673$^{(a)}$   & F6V  &    $6219 \pm 46 $&     $4.29\pm0.07$   &$ -0.05\pm0.06$ & $6.8 \pm 1.0$\\ 
\hline
\end{tabular}
\begin{list}{}{}
\item $^{(a)}$: from \cite{wu2011}. $^{(b)}$: measured through the width of the SOPHIE Cross-Correlation Function \citep{boisse2010}.\
\end{list}
\label{tabstars}
\end{table}			    	

\subsection{Line selection}\label{linesel}
\begin{figure}[htb]
\centering
\includegraphics[width = 0.5 \textwidth]{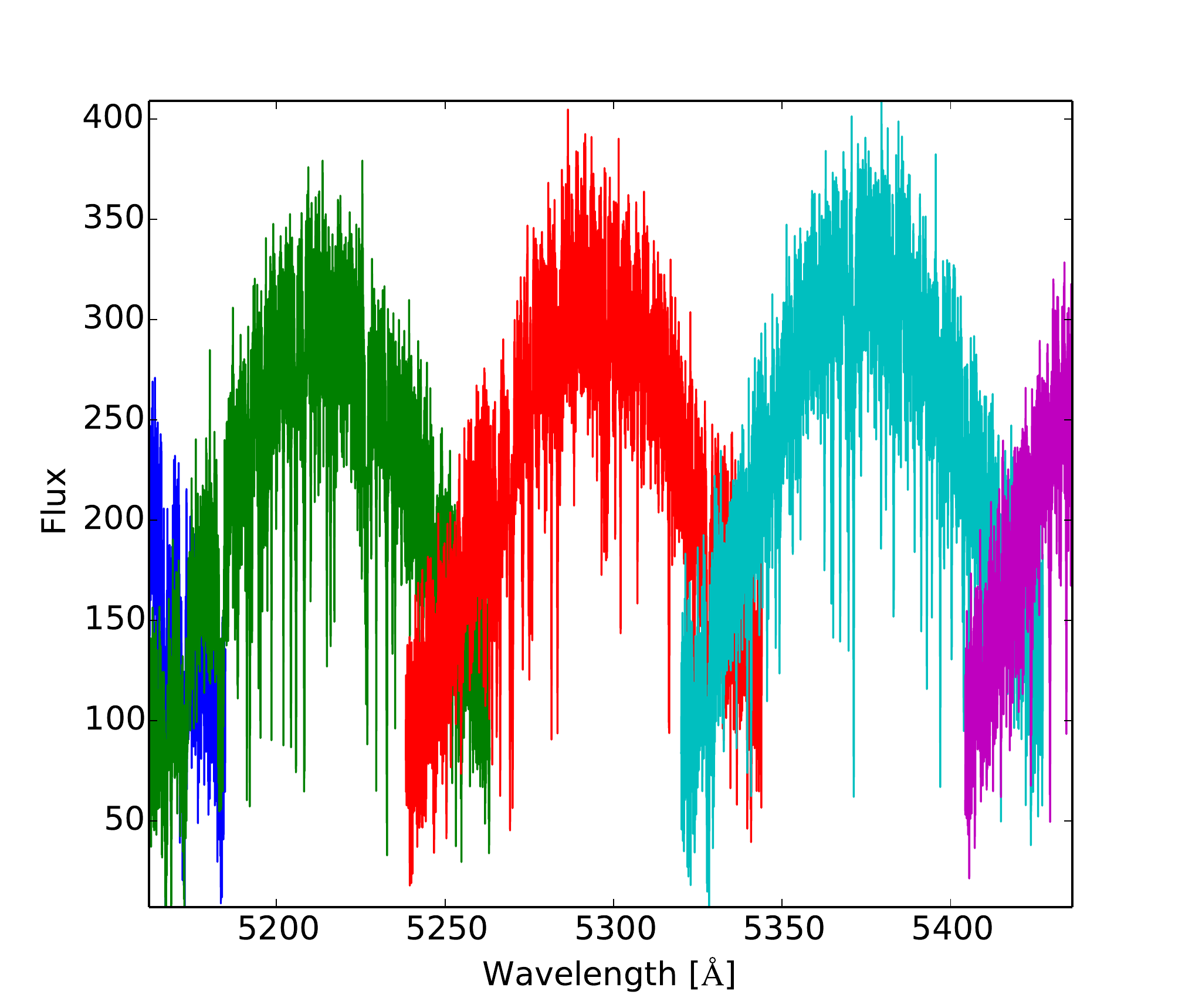}
\caption[Three adjacent orders in the a low-\sn\ spectrum of the Sun recorded with SOPHIE]{Three adjacent orders in the a low-\sn\ spectrum of the Sun recorded with SOPHIE, after the pipeline processing. Neither correction for the blaze, nor normalization have been performed yet.}
\label{blaze2}
\end{figure}

I worked on a set of 222 metallic lines of elements in the neutral or first ionization state, selected on high-\sn\ spectra. Only lines with wavelength $>5000$ \AA\ were used, to exclude the bluer and noisier part of the spectra. These lines were separated in two groups according to their position in the SOPHIE echelle orders. Using the blaze profiles  (Bouchy, private communication) they were divided between those ``at the center'' (191) and those ``at the edges'' (31) of the spectral orders. Table \ref{tabtheo}, in the appendix, reports the limits of the echelle orders and of the blaze profiles; tables \ref{tabnooverlap} and \ref{taboverlap} report the lines separated in the two groups.\\ Figure \ref{blaze2} shows three adjacent orders in a low-\sn\ spectrum of the Sun, illustrating how the spectral lines have been separated. The \sn\ at the edge of an order is approximately a factor of $\sqrt{2}$ lower than the one at the centre of that same order (Bouchy, private communication), because of the difference in flux. Moreover, the lines at the edges lie in two neighbouring orders. When co-adding a spectrum, their profile is averaged; in the present analysis, they are fitted twice, once for every spectral order they belong to, providing a double measure on a lower-\sn\ line. These two reasons make the lines at the edges less reliable than those at the center of the orders.\\

\subsection{Method}\label{rejtm}
As a large number of spectra had to be analyzed, a fast method for the determination of stellar parameters was adopted. I used the combination \ares\ - \tmcalc\  (section \ref{specmeth}), thanks to which the normalization process is skipped. As described above, the user is supposed to visually inspect the quality of the fits of the EWs and to tune the parameters of the code until a correct normalization and fit are performed. However, this approach can be automatized, at the price of introducing an uncertainty relative to the choice of the fitting function and the width of the spectral window.\\ 
In \cite{sousa2012}, the normalization parameter of \ares\ ($rejt$) was calculated as a second-degree polynomial of the \sn, calibrated on six high-\sn\ HARPS spectra. As both the instrument and the \sn\ regime were different in our analysis, a new calibration of this parameter was considered appropriate. I adopted a similar approach to \cite{sousa2012}, by examining six spectra at different \sn, two for the F star, two for the Sun, and two for the K star. As  the $rejt$ for stars of different spectral types is derived, the characteristics of different types of spectra are taken into account for the purpose of local normalization. Indeed, the different density of spectral lines, as well as their changing width and depth with \teff, affect the fitted value of the continuum for a given $rejt$ parameter.\\
For the calibration, each echelle order was analyzed separately with \ares. For each order, I adopted the \sn\ measured at 550 nm by the SOPHIE pipeline \citep{bouchy2009}. Then, the measured EWs of all the lines of a spectrum were given to \tmcalc\ to compute the \teff\ and \feh\ for that spectrum. I used spectra with \sn\ ranging from 25 to 95. Lower \sn\ were avoided, because the results become unreliable, and calibrating the method for the lowest \sn\ cases could have removed or introduced the trends I was trying to identify.

\begin{table}[!hbt]
\centering
\caption[Spectra used for the calibration of the $rejt$]{Spectra used for the calibration of the $rejt$. The \teff\ and \feh\ columns indicate the parameters given by the best choice of the $rejt$, indicated in the last column.}
\begin{tabular}{llcccc}
 \hline
 Name & Type  & \sn & \teff  [K] & \feh\ [dex] & $rejt$\\
\hline
HD010476   &  K1V &    25   & $5268\pm46$  & $0.00\pm0.18$ & 0.9710\\ 
				& & 70 & $5235 \pm 25$ & $-0.03 \pm 0.09$ & 0.9884 \\
Sun	   & G2V  & 45& $5727 \pm 42$ & $-0.06 \pm 014$ & 0.9735  \\
	&	& 68 & $5751\pm 26$ & $-0.01 \pm 0.13 $ & 0.9881 \\
HD016673   & F6V  &   51 &   $6239\pm83$   &$ -0.05\pm0.14$ & 0.9874\\
 				&   &   87 &   $6219\pm68$   &$0.00 \pm 0.12$ & 0.9890 \\
\hline
\end{tabular}
\label{tabstars2}
\end{table}

\begin{figure}[htb]
\centering
\includegraphics[width = 0.5 \textwidth]{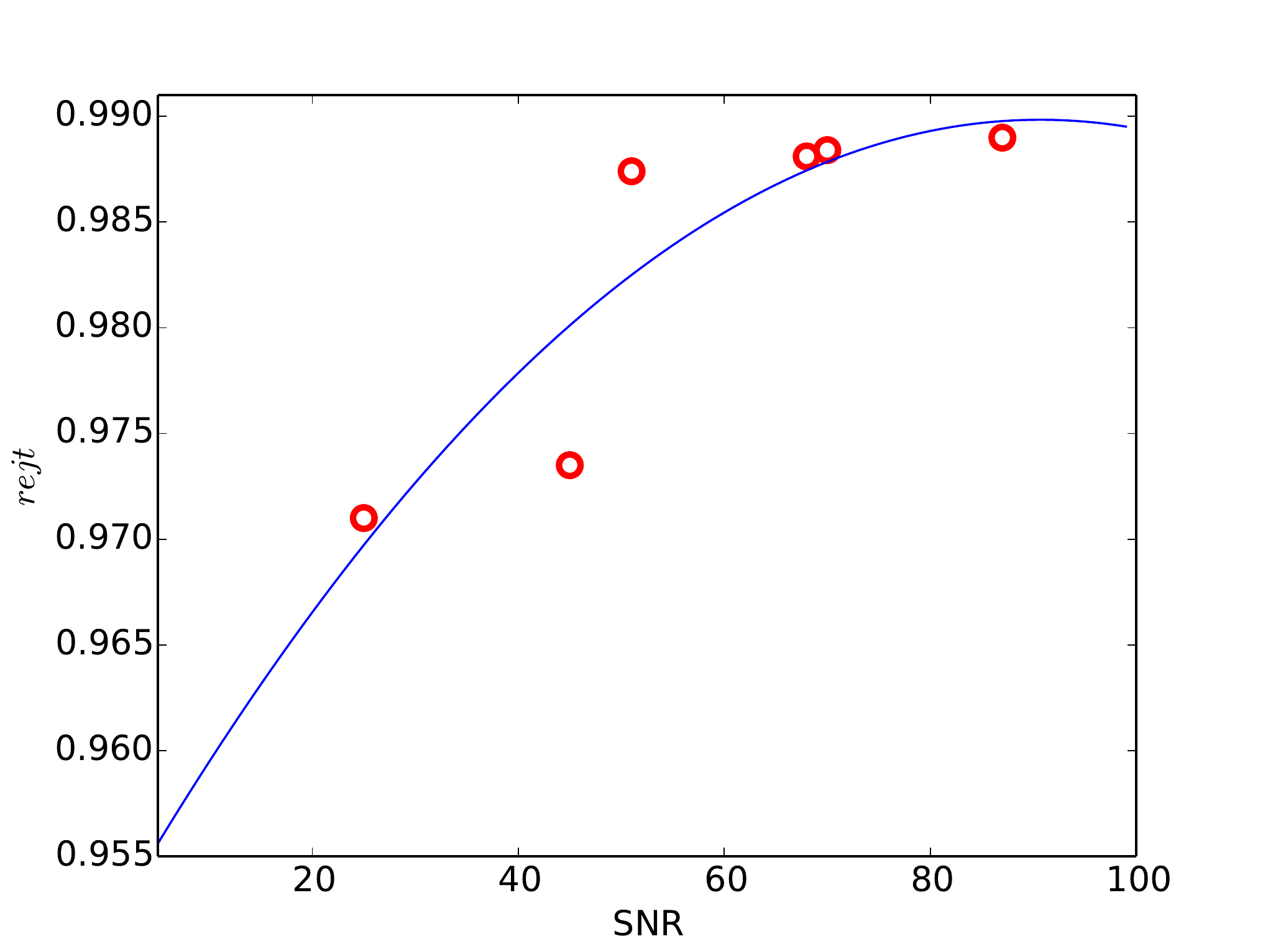}
\caption[The empirically determined $rejt$ values for the spectra used for the calibration]{The empirically determined values of the normalizing function ($rejt$) for the spectra used for the calibration (red circles), versus their \sn. In blue, the fitted second-order polynomial.}
\label{rejt_cal}
\end{figure}

For homogeneity, I fixed all the parameters of the code, excepting the normalizing function (the $rejt$). I used the conservative values of 8 pixels for the smoothing factor, 2 \AA~for the width of the spectral window, 0.1 \AA~for the minimum separation between two lines, and 5 m\AA~for the minimum reported EW. For each star, the $rejt$ was empirically adjusted until a good combination of \teff\ and \feh\ was derived. The parameters and the $rejt$ derived on the spectra are indicated in table \ref{tabstars2}.\\
Finally, the $rejt$ values were fitted with a second-degree polynomial, as a function of the \sn. This yielded
\begin{equation}
rejt =  0.9514 + 8.5450 \cdot 10^{-4}\, \rm{SNR} - 4.7663 \cdot 10^{-6} \, \rm{SNR}^2.
\end{equation}
This function, represented in figure \ref{rejt_cal}, corresponds to an increasing parabola until \sn\ $\sim 90$, that is the highest \sn\ used for the calibration. Therefore, this normalization is valid for \sn\ up to $\sim 90$: for higher ones, another calibration with SOPHIE spectra, or the one by \cite{sousa2012} (which, however, was carried out on HARPS spectra) has to be used.

\subsection{Results}

\subsubsection{Most reliable lines}
\begin{landscape}
\begin{figure}[htb]
\centering
\includegraphics[width = 0.4 \textwidth]{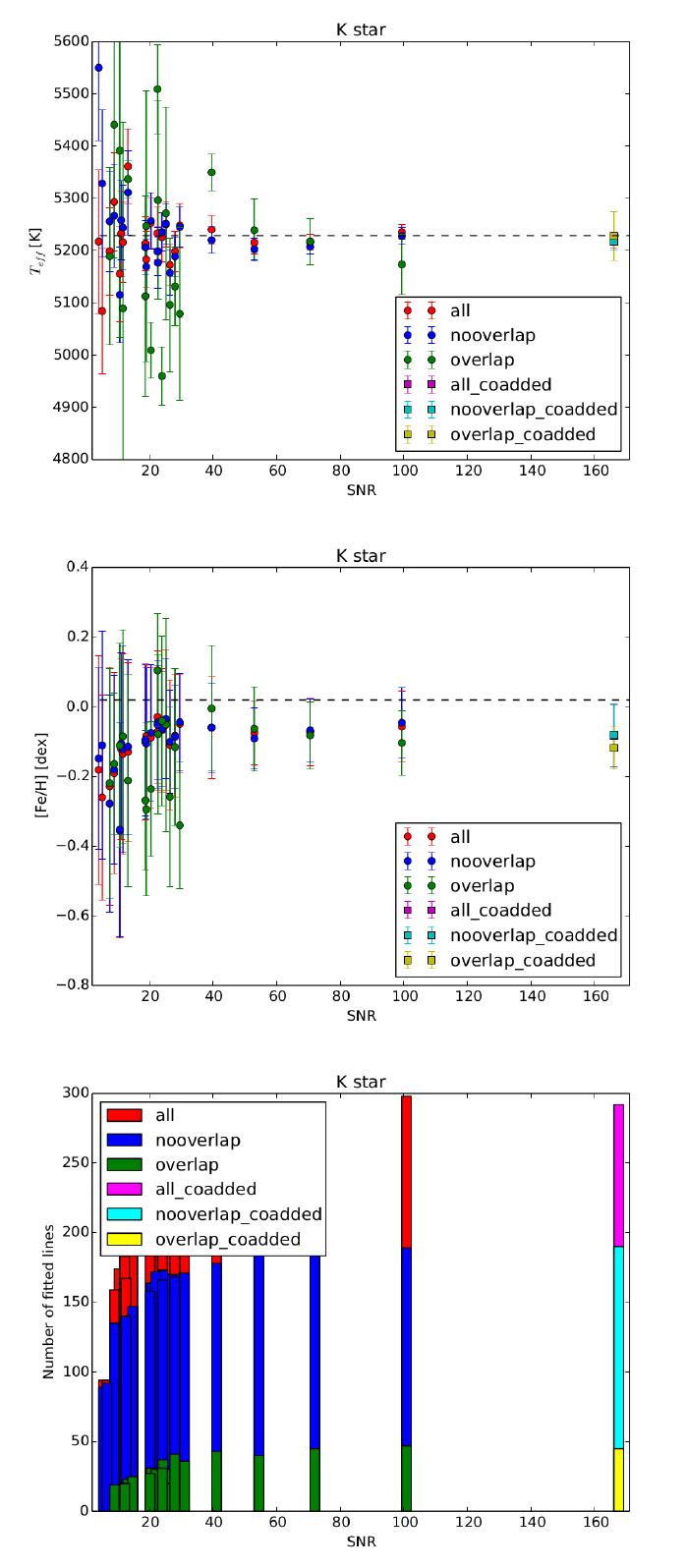}
\includegraphics[width = 0.37 \textwidth]{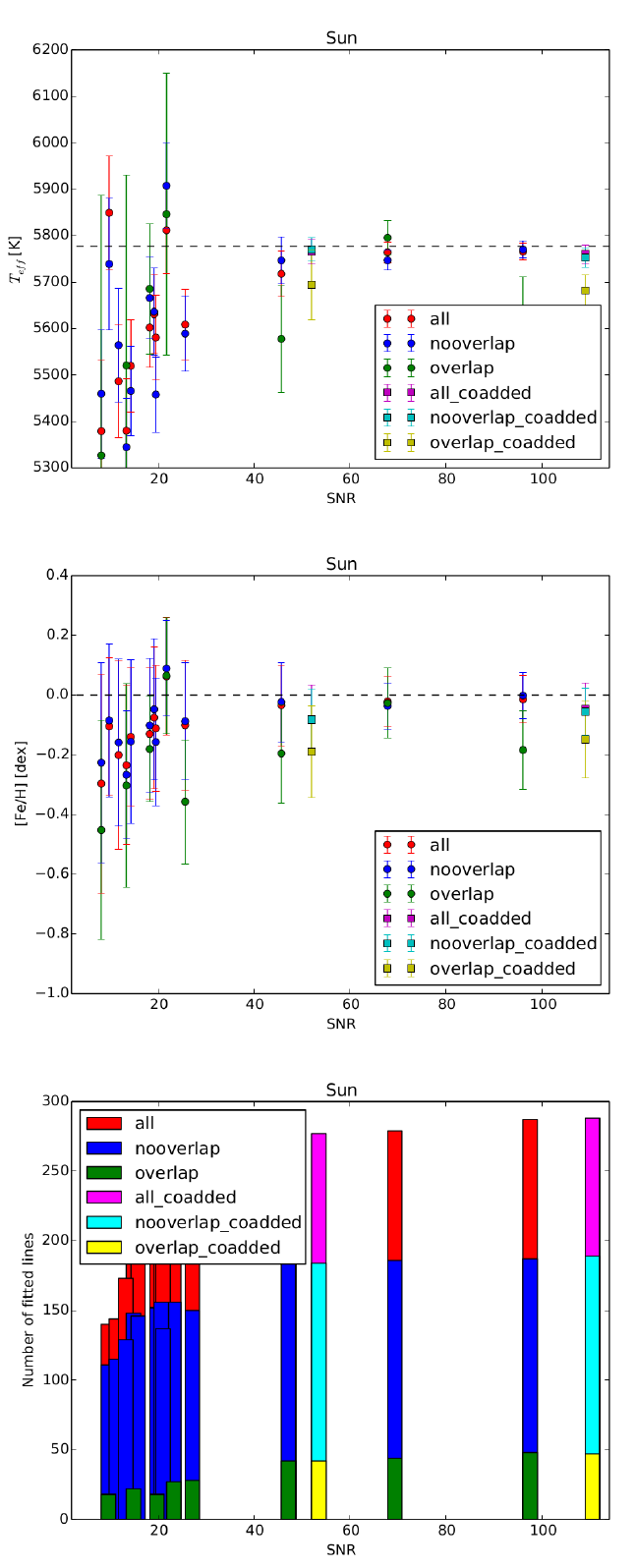}
\includegraphics[width = 0.395 \textwidth]{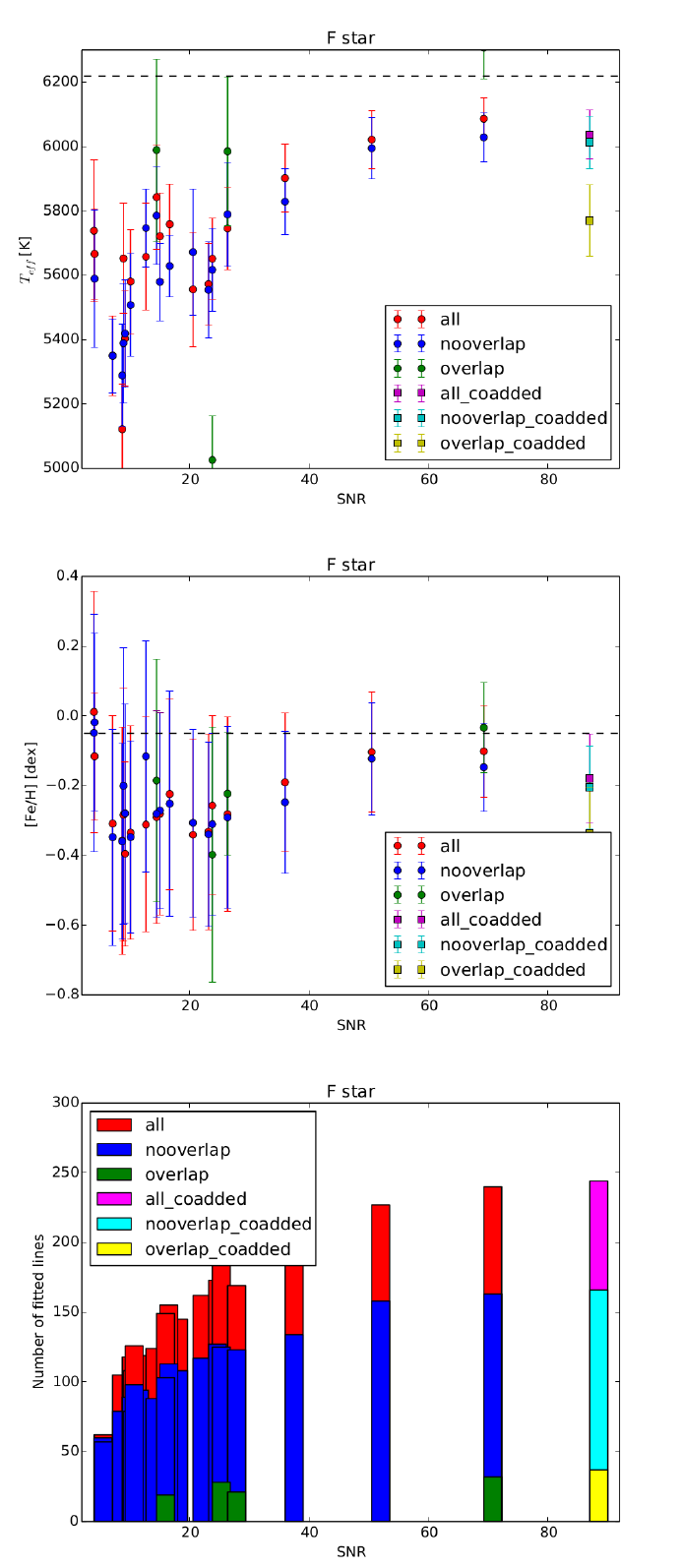}
\caption[Results from the analysis of the spectral lines at the center of the orders, at the borders, and all the lines taken together]{Results from the analysis of the lines at the center of the orders (``nooverlap'' in the legends) and at the borders (``overlap''), and all the lines taken together (``all''). The results on the co-added spectra are indicated, too. \textit{First row:} \teff\ measured on the K star, the Sun, and the F star, versus the \sn\ of the spectra. The dotted lines indicate the values from literature. \textit{Central row:} The same for \feh. \textit{Bottom row:} Number of lines successfully fitted by \ares.}
\label{resoverlap}
\end{figure}
\end{landscape}

In a first stage, the analysis was repeated for the spectral lines at the center of the orders, those at the borders, and all the lines taken together. The analysis was carried on all of the individual spectra, and on the co-added spectra. Figure \ref{resoverlap} shows that, as expected, the lines at the borders yield larger dispersion and uncertainties on the derived parameters. The \teff\ measured with the lines at the borders tends to increase in dispersion going from the K to the F star. The \feh\ measured from different sets of spectral lines, instead, agree. As discussed in section \ref{linesel}, the discrepancies were expected, mainly because of the lower \sn\ of the lines at the borders. Also, an explanation has to be looked for in the smaller number of spectral lines at the borders with respect to those at the center of the orders. Figure \ref{resoverlap} shows these results and confirms the fact that \ares\ manages to fit a larger number of lines as the \sn\ increases.\\
In a second step, \ares\ was used to create a more robust list of spectral lines, in order to obtain a smaller dispersion in the results. For this, I prepared a new list of lines whose EW does not vary importantly with changing \sn\ during the first exploratory analysis. So I selected for the K star and for the Sun only those lines whose EW does not change by more than 30\% (this resulted in 62 and 81 lines, respectively), and for the F stars by 50\% (71 lines). A compromise was found between the most reliable lines and the small number of lines satisfying the stability criterion. Being too exigent in the selection of the lines, in fact, leads to using too few of them, so that the calibration through \tmcalc\ becomes useless.\\
For all the stars $\sim90\%$ of the spectral lines selected for their smaller EW variation turned out to belong to those at the center of the orders. This new list of lines was used for the following parts of the study. 

\subsubsection{Individual spectra and noise injection}\label{indivsop}
Figure \ref{noise} shows the results for the analysis of the individual spectra with the new list of spectral lines. For \sn$\, \lesssim 30$ for the K star, and $\lesssim 50$ for the Sun and the F star, the stellar parameters derived from the individual spectra are affected by a large dispersion. 
Different trends emerge for the different spectral types. At very low \sn, \teff\ is overestimated by up to 300 K for the K star, while it is underestimated by up to several hundreds of Kelvin for the Sun and the F star. The error bars can be as large as 260 K for the K star, 170 K for the Sun, and 420 K for F star. The \teff\ measured on the F star is systematically underestimated, due to the high sensitivity of its spectrum to the normalization parameters, which was not modeled adequately by the simple $rejt$-varying approach. Moreover, this is also due to the small number of spectral lines that are used for the F star. This can be seen for the point at \sn\ $\sim 95$, as shown in the plot on the lower left. This spectrum was analyzed in more detail, by adjusting the other parameters of the code. No important improvement was obtained.\\
The \feh\ is too dispersed to identify any trend, with error bars as large as 0.3 dex for the lowest \sn\ for the three stars. The plots indicate this parameter cannot be reliably derived for \sn\ $\lesssim 50$.\\
As a sanity check, white noise was injected in the spectrum at the highest \sn\ of each star, producing spectra with \sn\ going from 5 to 50, corresponding to the most problematic \sn\ range. In figure \ref{noise}, the parameters obtained from these spectra are represented in black. It can be seen how they follow similar trends to the ``true'' spectra (red points), suggesting that the observed trends are likely related to the limitations inherent to the methods, more than to issues in the data reduction.

\subsubsection{Co-addition of low-SNR spectra}
\begin{figure}[htbp]
\centering
\includegraphics[width = 0.49 \textwidth]{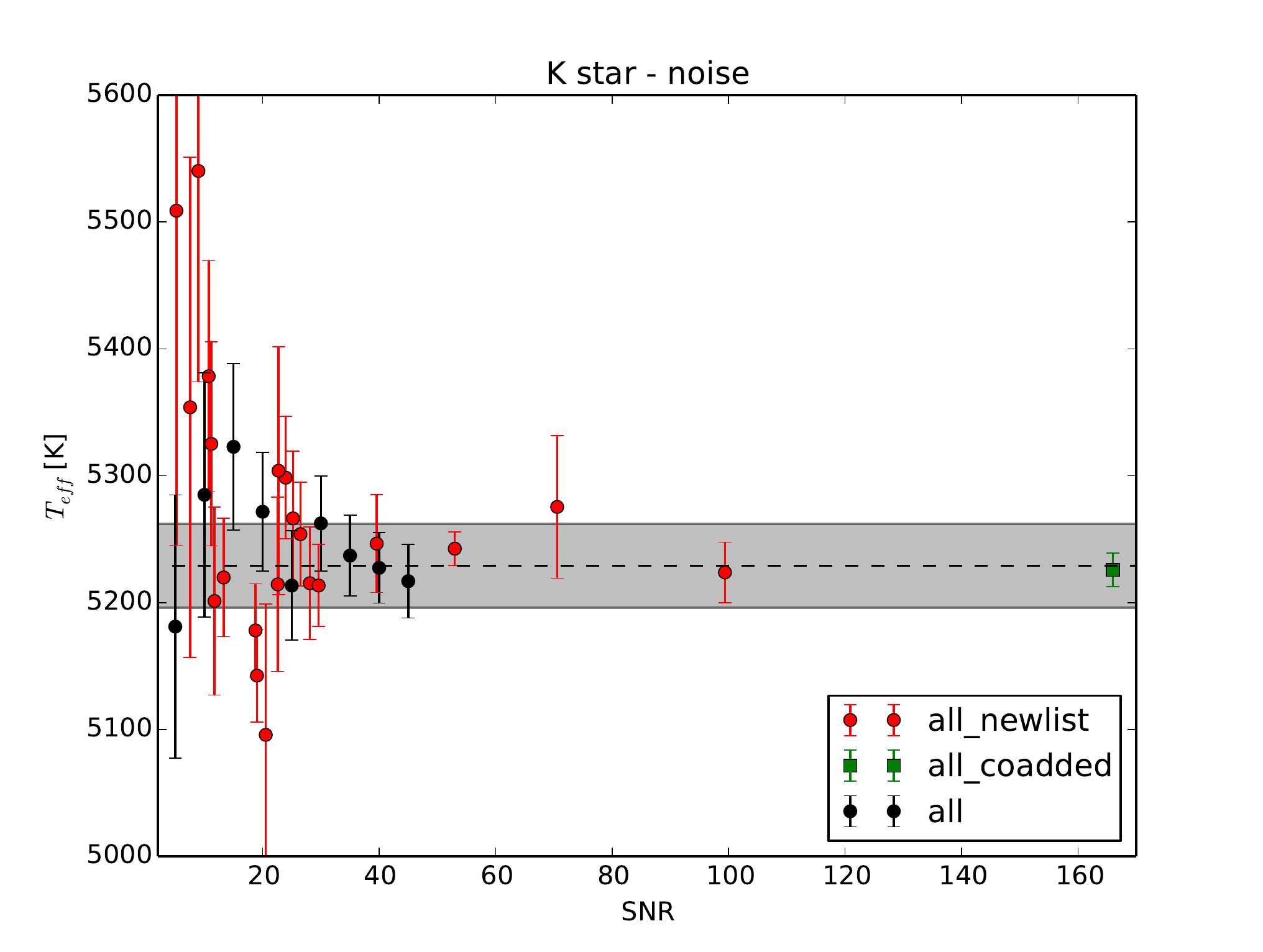}
\includegraphics[width = 0.49\textwidth]{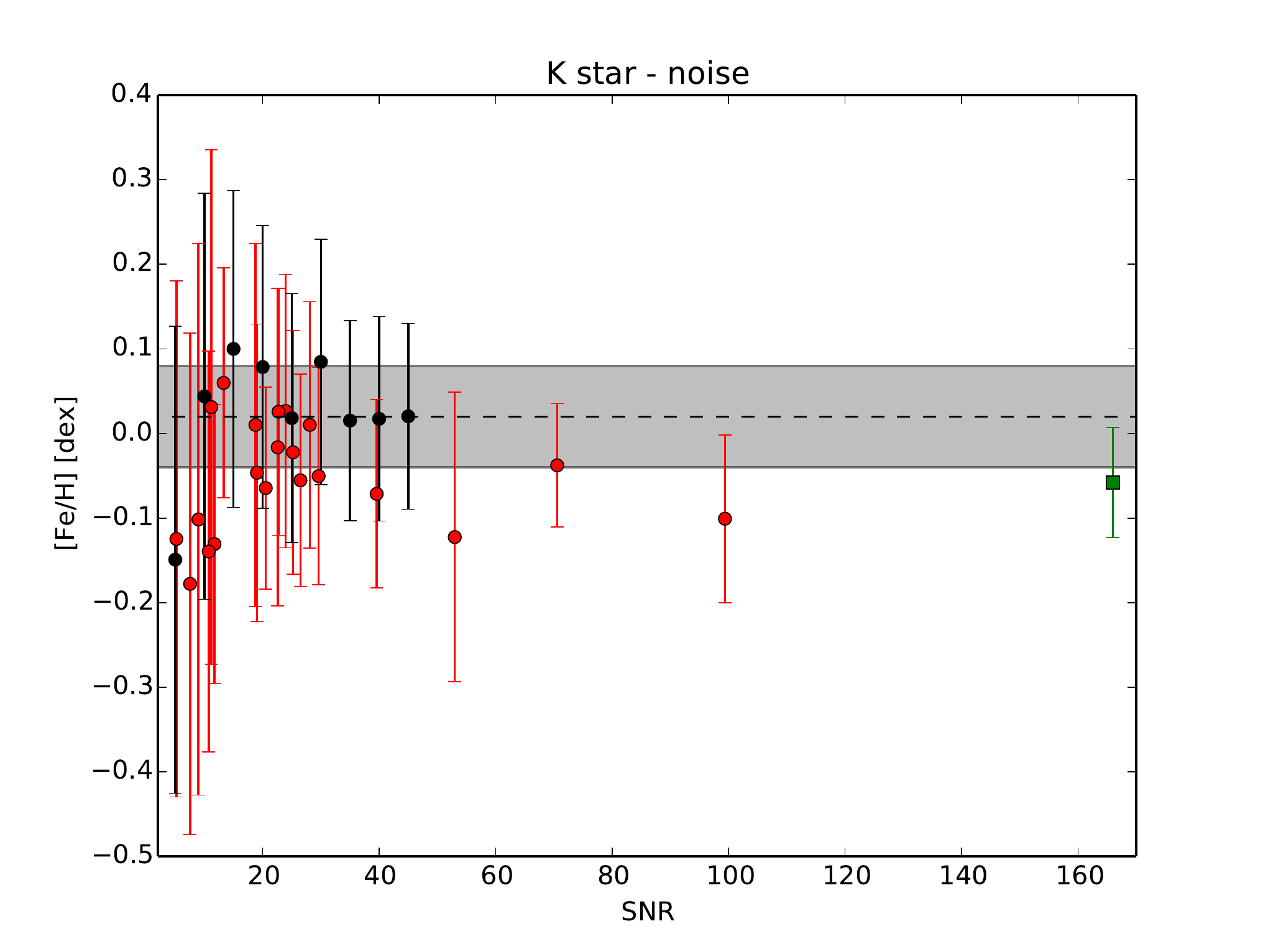}
\includegraphics[width = 0.49 \textwidth]{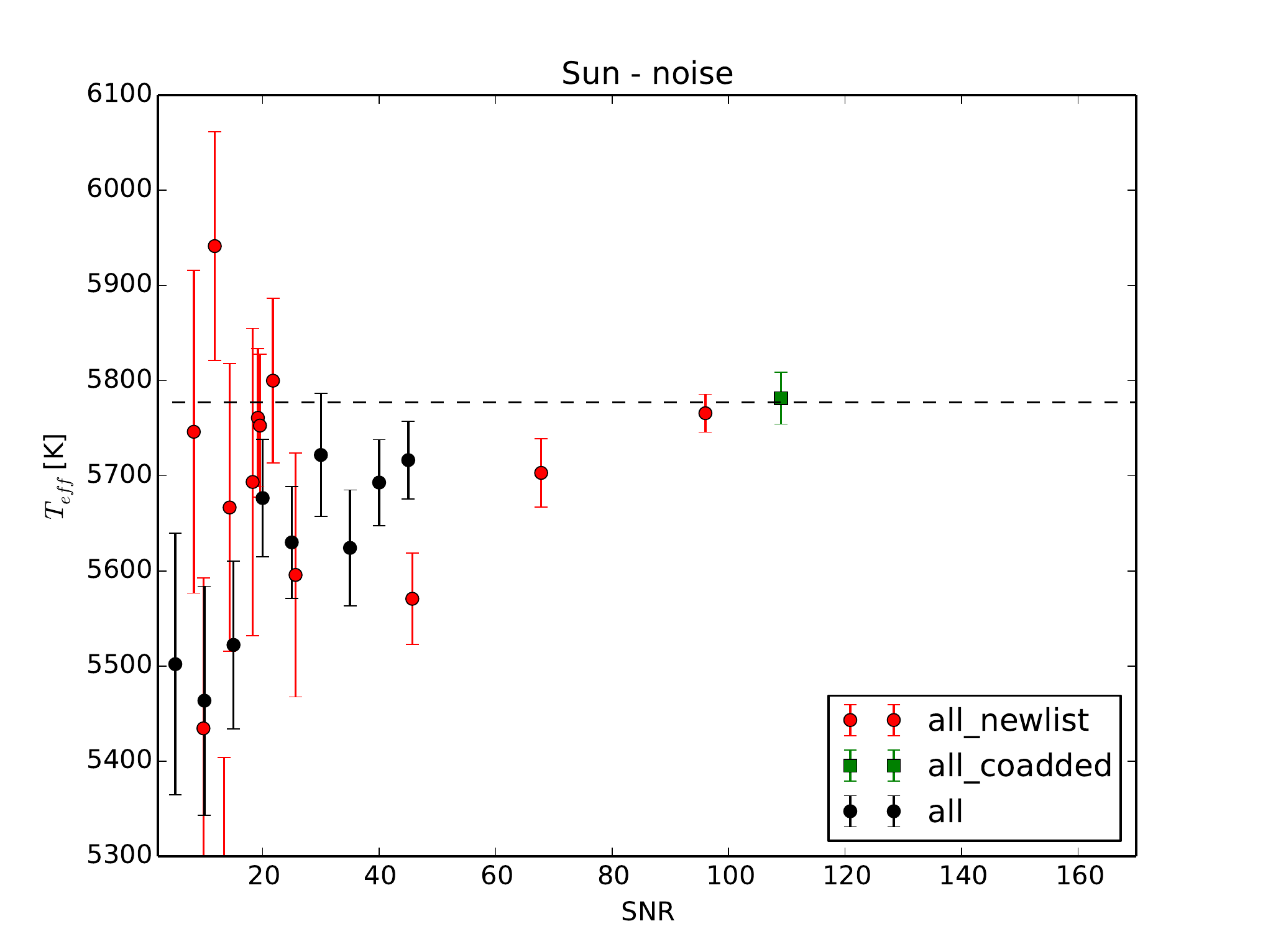}
\includegraphics[width = 0.49 \textwidth]{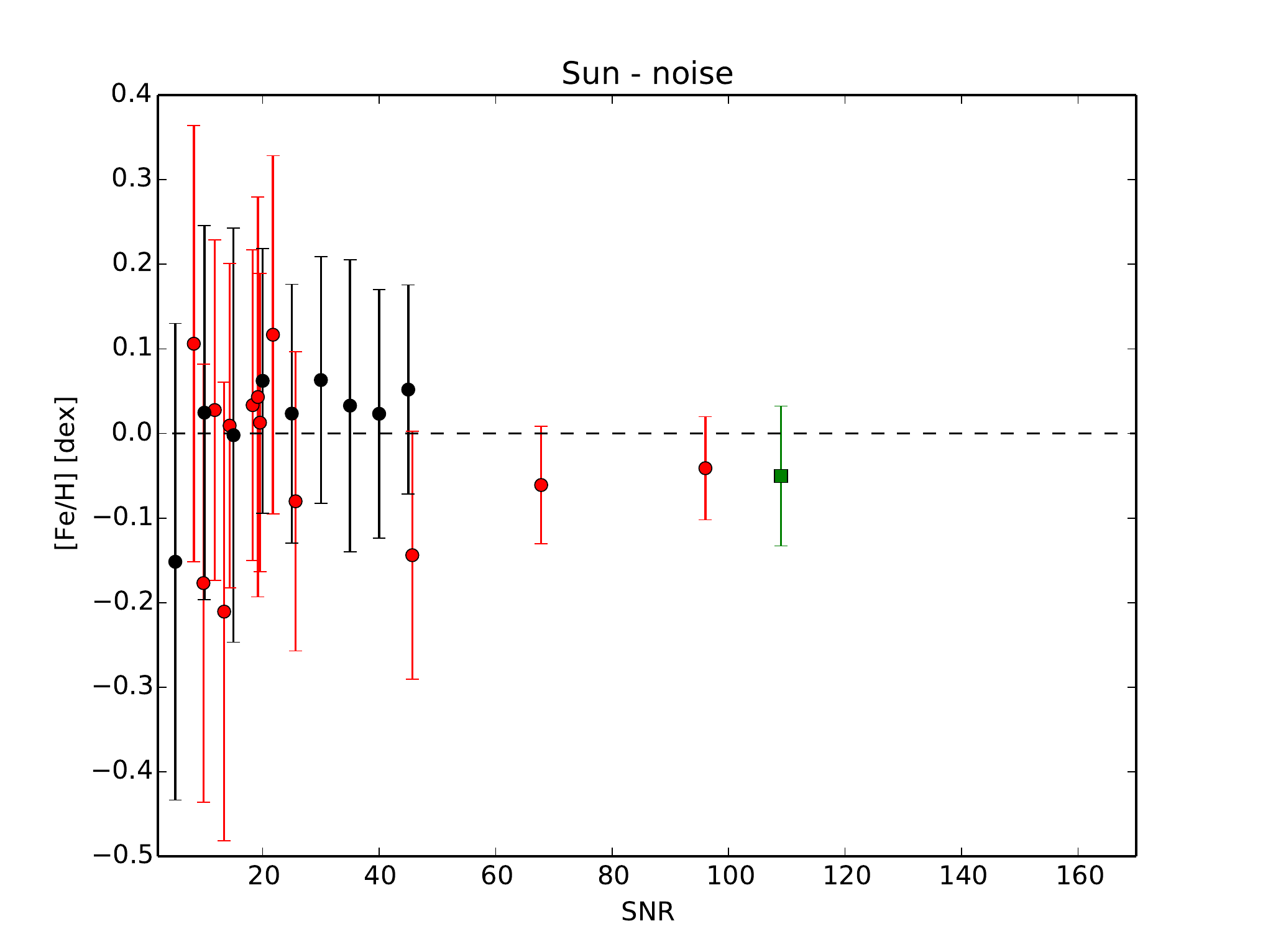}
\includegraphics[width = 0.49 \textwidth]{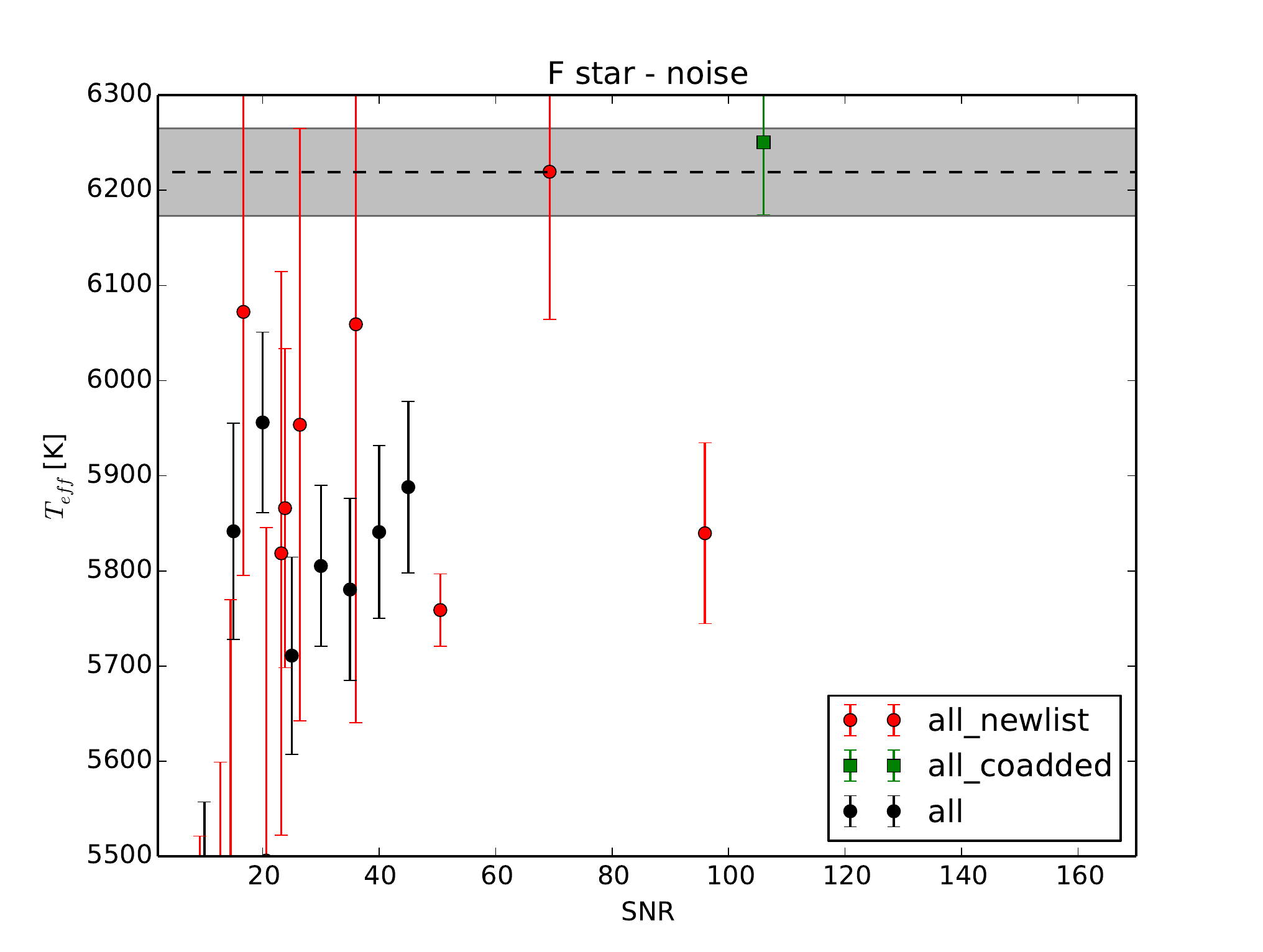}
\includegraphics[width = 0.49 \textwidth]{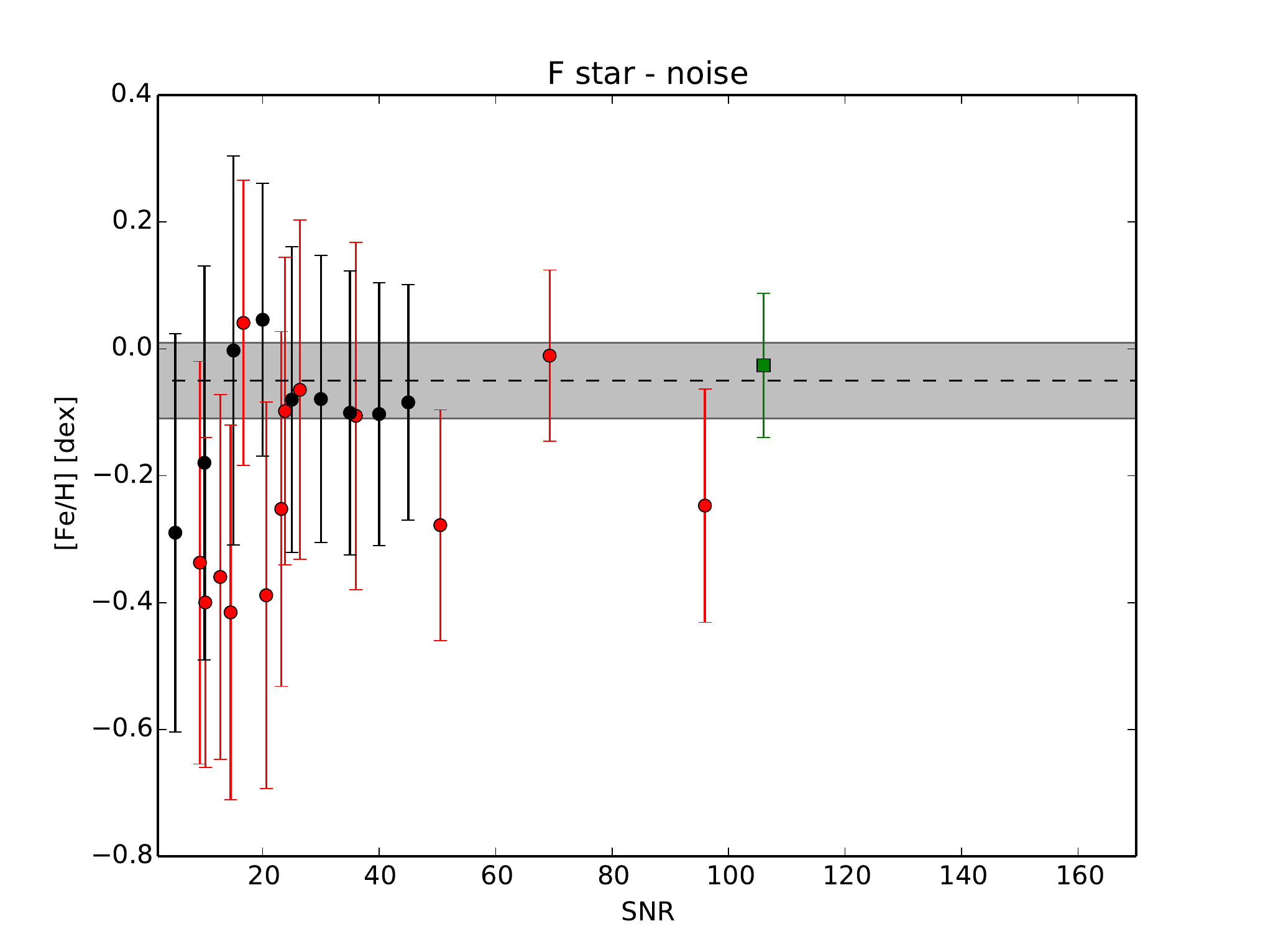}
\caption[\teff\ and \feh\ for the K star, the Sun, and the F star]{\teff\ (left) and \feh\ (right) for the K star (top), the Sun (center) and the F (bottom) star. The dashed horizontal lines indicate the mean values from the literature. For the F and K star, the literature $1\sigma$ confidence intervals of the respective parameters are colored in grey. For each plot, the results on the individual spectra are colored in red, those from the noise-added spectra are in black, and the point from the co-added spectrum is in green.}
\label{noise}
\end{figure}

In figure \ref{noise}, the results from the co-added spectra are reported. The spectra were not normalized, like the other ones. In this way, I was able to see if, disregarding the mechanism causing possible trends at low \sn, this issue could be neglected with a co-added, higher-\sn\ spectrum.\\
The co-addition produced a spectrum of \sn\ $\simeq 170$ for the K star, and $\simeq 110$ for both the Sun and the F star. As the \sn\ of the co-added spectra of the Sun and the F star are close to the limit of our $rejt$ calibration (about 90, see section \ref{rejtm}), the same $rejt$ law used for the individual spectra was applied, retrieving the correct parameters. The co-added spectrum of the K star has a larger \sn, and \ares\ was therefore tuned manually to obtain correct parameters.

\subsection{Discussion}
With this analysis, I have explored the behaviour of \teff\ and \feh\ on a set of SOPHIE spectra with different \sn, using a semi-automatic method. There were three aims to this work. The first was to recognize, if present, any problem introduced by the SOPHIE pipeline, or by the instrument, that could invalidate the parameters derived by spectral analysis. The second aim was to check the limit in \sn\ above which the atmospheric parameters derived from the spectral analysis are reliable. The third aim was to verify if the co-addition of low-\sn\ spectra gives trustworthy results.\\
The results confirm that a reliable spectral analysis requires spectra with an \sn\ of a least 30 for ``cold'' K stars, and of at least 50 for G and F stars. This is due to the varying number of spectral lines, and the difference in their profiles, which changes with \teff. The huge error bars obtained on the spectra with lowest \sn\ forbid to recognize any trend of the \feh\ with the decreasing \sn, and prevent us from using these spectra for spectral analysis.\\
The co-addition of several low-\sn\ spectra in a single high-\sn\ spectrum, though, proved to ``lose memory'' of any misleading trend, if present, so that the correct parameters could be recovered.\\
This study used the simplifying assumption that the atmospheric parameters can be correctly measured by only adjusting the normalization, while ignoring other factors, such as the blending of the spectral lines. Moreover, I tried to automatize the normalization process, in order to be able to quickly process several spectra. Finally, with \ares\ + \tmcalc\, the impact of low \sn\ on the \logg, which is probably the most severe with respect to \teff\ and \feh, could not be explored. It is likely that some improvement could be obtained by manually analyzing the spectra one by one, and by taking the \logg\ into account. However, the fundamental limitation introduced by very noisy spectra has been experienced also using other methods during this work. For example, in section \ref{lowm}, I have analyzed co-added spectra with very low \sn\, obtaining very large error bars, and sometimes not being able to derive trustworthy atmospheric parameters. Therefore, the results obtained here are likely not due to the method I adopted.

\section{Publications}
In the following, two articles of the series \textit{SOPHIE velocimetry of Kepler transit candidates} (section \ref{kepstars}) are attached. The first reports the first detection of a brown dwarf orbiting a K-type dwarf. The second paper reports the validation and characterization of three planets orbiting KOI stars: one of them is a Jupiter-like planet lying at the boundary between ``warm'' and ``hot'' Jupiters; the second one is a massive inflated hot Jupiter; the third one is one of the largest transiting planets characterized at the time of writing.\\

\cite{diaz2013}:\\
\url{http://www.aanda.org/articles/aa/pdf/2013/03/aa21124-13.pdf}

\cite{almenara2015}:\\
\url{http://www.aanda.org/articles/aa/pdf/2015/03/aa24291-14.pdf}

\clearpage
\chapter{Dynamical analysis of a multi-planetary system: Kepler-117}\label{chapttvs}
\minitoc

\bigskip
Among the Kepler systems observed with SOPHIE (section \ref{kepstars}), there is the multi-planetary system Kepler-117. This system exhibits strong transit timing variations (TTVs). TTVs give the opportunity to estimate the masses of the planets: to achieve such a goal, we needed to develop a method for modeling and fitting TTVs. In a collaborative context, I have been responsible of the analysis of this system in all its phases.\\
The results of this work have been published in an article on \aap, which is included at the end of this chapter.

\section{Context}\label{context}
\begin{figure}[!hbt]
\centering
\includegraphics[scale = 1.0]{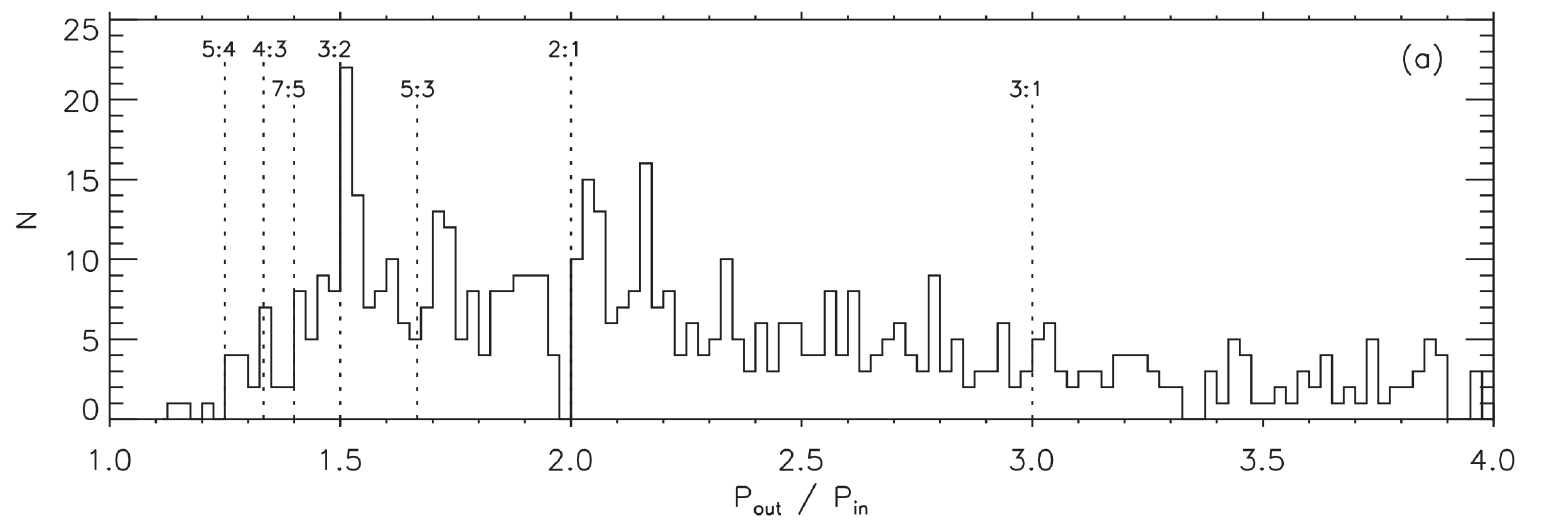}
\caption[Histogram of the period ratios of all planet pairs]{Histogram of the period ratios of all planet pairs. The main resonances are indicated with vertical dashed lines. From \cite{fabrycky2014}.}
\label{res_tr}
\end{figure}

The first detection of a multi-planetary system dates back to 1992, when \citeauthor{wolszcan1992} detected two low-mass planets around the pulsar PSR B1257+12, and found indications of the presence of at least another planet. Through the pulse timing method, and assuming a standard mass of 1.4 $M_\odot$ for the star, they found the planets to have masses of $4.3\pm0.2$ and $3.9\pm0.2 \, M_\oplus$. The first discovery of a multi-planetary system around a normal star, by radial velocities (RVs), came seven years later \citep{butler1999}. The first system known to host more than one transiting planet was Kepler-9 \citep{holman2010}.\\
The number of known multiple systems has enormously increased, mostly thanks to the detection of hundreds of them by the Kepler space telescope \citep{koch2004,borucki2008}. Systems with up to seven planets are known at the time of writing \citep{lovis2011,cabrera2014}. So far, Kepler has detected 487 multiple planet systems\footnote{http://exoplanet.eu/ \citep{wright2011}.}.\\
Multiple transiting planet systems have a low false-positive (FP) probability. Indeed, \cite{lissauer2012} established a 1.12\% probability of observing two FPs in the same system and a 2.25\% probability for a system to host a planet and show the features of a FP at the same time. They assumed that 1) FPs are randomly distributed in the Kepler population, and that 2) there is no correlation between the probability of a target to host one or more detectable planets and display false positives.\\ 
The dynamical properties of multi-planetary systems are strongly related to the period ratio between the orbits. When the ratio is equal or close to the ratio of small integers, the planets are said to be in mean motion resonance (MMR). Given two planets 1 and 2, and two integers $j$ and $k$ (with $j>k$), a MMR is found when
\begin{equation}
\frac{P_2}{P_1} \simeq \frac{j}{k}.
\end{equation}
Period commensurabilities in the Solar System, the multi-planetary system we best know about, have been shown to have a tidal origin \citep{goldreich1965}. The same can be said for exoplanet systems. RV surveys have found that planet pairs with a total mass exceeding 1 \MJ\ tend to cluster around the 2:1 MMR, as discussed by \cite{wright2011a}. According to these authors, this is unlikely a statistical fluke, and supports scenarios of planet scattering and migration.\\
The situation for transiting exoplanets is different. \cite{lissauer2011}, \cite{veras2012}, \cite{batygin2013}, and \cite{fabrycky2014} discussed Kepler multi-planetary systems, which are mainly systems with small-size planets with orbital periods shorter than one year. They found that exact first-order MMRs (as 2:1 or 3:2) are uncommon among these systems, and that there is an overabundance of systems a few percent wide of resonances, and a dearth of system just narrow of them. The distribution for higher-order resonances remains flat. This is shown in figure \ref{res_tr}, where the first MMRs are highlighted. 

This was unexpected, as the theory of planet migration predicted that gravitational interactions with the protoplanetary disk cause wide-orbit planets to move closer to the star and to be caught in resonances \citep{goldreich1980,lee2002}. A revision of the theory, including eccentricity damping \citep{goldreich2014}, or interactions with a residual protoplanetary disk \citep{chatterjee2015}, offers an explanation to this puzzle. Also, the concentration of planet pairs with period ratios slightly larger than MMR was explained as an effect of tidal dissipation that cause the orbits to diverge \citep{terquem2007,batygin2013,lithwick2012}.\\

The RV and the Kepler surveys have shown that compact systems of planets with periods shorter than a year are more often composed by Neptune-size planets or smaller, and that Jupiter-like planets are uncommon \citep{wright2009,latham2011}. For these systems, an important property in the context of formation scenarios is the mutual inclination of the orbits. Their coplanarity, as in the case of the Solar System, supports the hypothesis of the planets being formed in a flat protoplanetary disk. A discussion of individual cases can be found in the review by \cite{winn2014}. Other possible explanations are a mechanism which dampens the inclinations, and the hypothesis is that systems with large mutual inclinations could be dynamically unstable. This would lead them to destruction \citep{veras2004}.\\

\begin{figure}[htb]
\centering
\includegraphics[scale = 1.0]{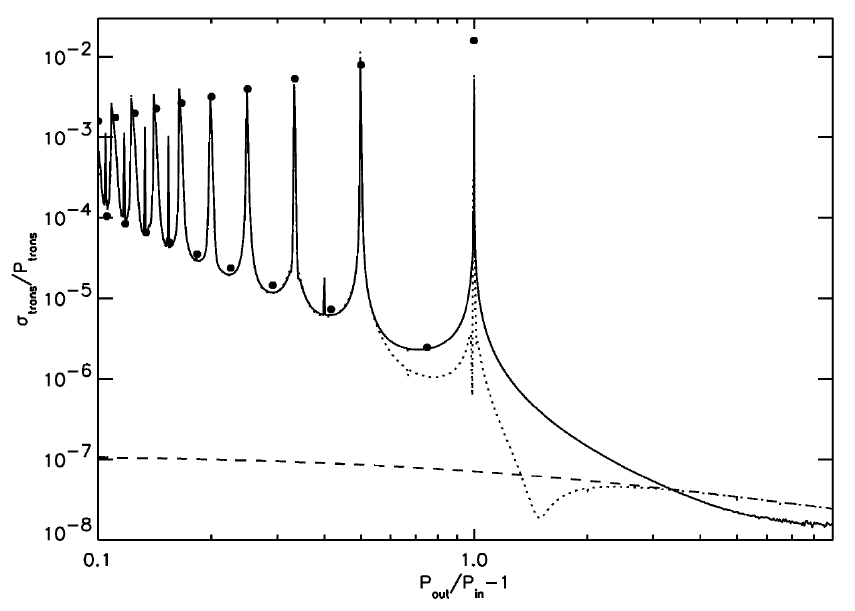}
\caption[Transit timing standard deviation for a transiting planet of mass $10^{-5} m_0$, perturbed by a planet of mass $10^{-6} m_0$]{Transit timing standard deviation for a transiting planet of mass $10^{-5} m_0$, perturbed by a planet of mass $10^{-6} m_0$. The solid (dotted) line is the numerical calculation for the inner (outer) planet averaged over 100 orbits of the outer planet. The dashed line indicated the case of planets which do not interact dynamically. The large dots represent the resonances and the halfway points between resonances. From \cite{agol2005}.}
\label{agol}
\end{figure}

Another important property of planets in compact systems is their low density ($\lesssim 1$ g cm$^{-3}$; e.g. \citealp{wu2013}). This could indicate that systems composed by small, low-density planets are dynamically stable on the long term. Indeed, \cite{ford2008} proposed that, when a closely packed system is dynamically unstable, massive planets are more likely to eject one another, while smaller planets are more likely to collide until the system stabilizes. However, the low density of planets in multiple, compact systems could also reflect a selection effect due to the main technique these systems have been characterized with, that is the transit-timing method.\\
\cite{miraldaescude2002}, \cite{holman2005}, and \cite{agol2005} showed that, in  multi-planet systems close to a MMR, gravitational interactions between the planets affect their orbital periods by a measurable amount. Transit timing variations (TTVs), that are deviations of the periods from a linear ephemeris, can be observed for these systems. Despite several ground-based efforts to detect TTVs for systems hosting Hot Jupiters \citep[e.g.][]{millerricci2008,gibson2009,barros2013}, the first observations of a system exhibiting TTVs was possible thanks to Kepler \citep{holman2010}.\\
TTVs are a powerful tool for the determination of planetary masses \citep[e.g.][]{nesvorny2013,barros2014,dawson2014}. 
The amplitude of the TTVs strongly depends on the mass ratio of the planets, their orbital period and separation, and their eccentricity. The deviation from linear ephemerides due to the dynamical interactions between planets is much stronger when planet pairs lie close to an MMR \citep{agol2005}. Figure \ref{agol} shows the transit timing deviation as a function of the period ratios of planet pairs. 

\cite{agol2005} performed the analytic derivation of the TTV strength for some extreme cases. In real cases, however, numerical integrations of Keplerian equations of motion are required. Dynamical modeling, therefore, can be used to fit the parameters determining the dynamics of a system, given its TTVs deduced from photometry.\\
The many variables involved in the dynamical fit of TTVs impose some assumptions, and often a completely consistent solution for the system parameters is not offered. Either some parameters are deduced without modeling the TTVs \citep[in particular, the transit depth: e.g.][]{sanchisojeda2012}, or fixed in the dynamical model \citep{jontofhutter2013}. Usually, a Levenberg-Marquardt algorithm is used to find a local minimum in the $\chi^2$ space and to derive the uncertainties on the system parameters \citep[e.g.][]{lissauer2013}.\\

\begin{figure}[htb]
\centering
\includegraphics[scale = 0.5]{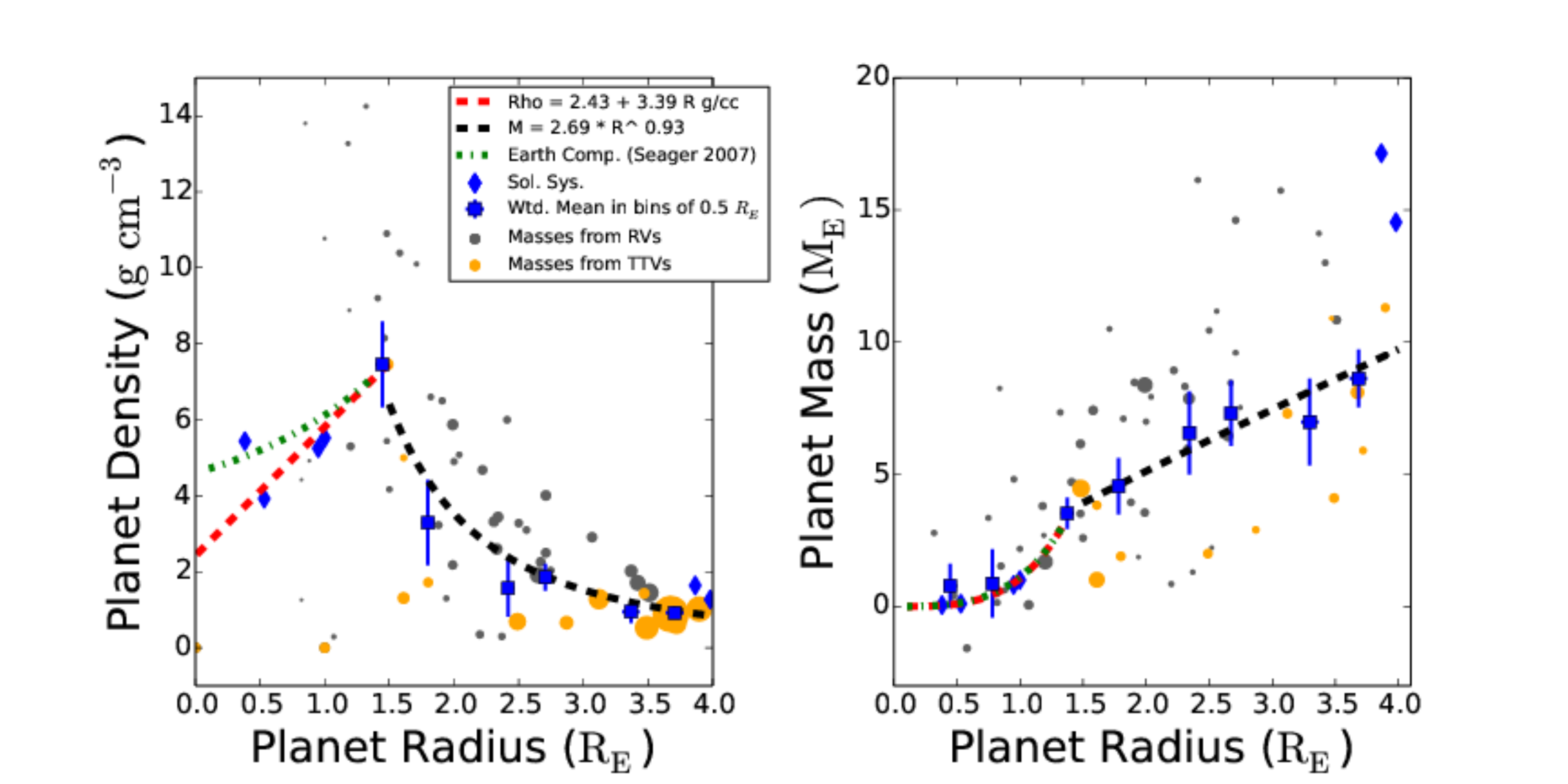}
\caption[The density-radius and mass-radius relationship for planets smaller than 4 \RE]{The density-radius (left) and mass-radius (right) relationship for planets smaller than 4 \RE. The techniques used to measure the masses, in particular RVs or TTVs, are indicated. From \cite{weiss2014}.}
\label{weiss}
\end{figure}

\cite{weiss2014} observed that planets smaller than four Earth radii with their mass measured with TTVs are systematically lower in the mass-radius diagram than those with mass derived through RVs. Figure \ref{weiss} illustrates the situation. This fact could be due to non-detected companions that might dampen the TTVs, causing a systematic underestimate of the masses, or to a lower density of the planets that show TTVs. Indeed, in the compact multiplanetary systems that are likely to produce observable TTVs, planets with lower masses for a given size are more likely to reach stable orbits \citep{jontof-hutter2014}.

Only a few of the planets known to date have their mass measured with both RVs and TTVs. Kepler 88 is the first system for which a non-transiting planet detected by TTVs \citep{nesvorny2013} was confirmed with RVs \citep{barros2014}. The RV-derived mass was found to be in agreement with the TTV-derived mass.\\

TTVs have been found to be tracers of stellar activity, as well. Stellar activity mimics the presence of non-transiting companions \citep{alonso2009,barros2013,oshagh2013}. This is one of the reasons which motivated us to develop fitting tools of stellar activity. This topic will be discussed thoroughly in chapter \ref{chapactivity}.

\section{Presentation of the Kepler-117 system}
Kepler-117 hosts the two transiting planets Kepler-117 b and Kepler-117 c. These planets were studied by \cite{steffen2010}, using only the first quarter of the Kepler photometric data. These authors predicted TTVs to be observable for this system. At that time, the photometric time coverage was not sufficient to allow this prediction to be verified.\\
Kepler-117 was presented by \cite{borucki2011} as a candidate multi-planet system. The predicted TTVs were confirmed by \cite{mazeh2013}. These authors observed that the TTVs of both the planets present a periodicity. The ratio between the periodic modulation of the TTVs of the inner planet (b) and the orbital period of the outer one (c), $P_{\rm TTVs, b}/P_c$, is $\simeq 0.997$. The similarity between the two periodicities is a strong indication that the two bodies are in the same system and thus is an argument for the validation of Kepler-117 b and c.\\
Finally, the planets were validated by \cite{rowe2014} with a confidence level of more than 99\%, while RV observations were still unavailable. After subjecting Kepler-117 b and c to various false-positive identification criteria, \citeauthor{rowe2014} used the statistical framework of \cite{lissauer2012} to promote them to \textit{bona fide} exoplanets. Kepler-117 b and c were found to have orbital periods $\simeq 18.8$ and $\simeq 50.8$ days and radii $\simeq 0.72$ and $\simeq 1.04$ \RJ.\\
The presence of TTVs for this system is in contrast with the fact that the two planets are far from both the 2:1 and the 3:1 resonance. Nonetheless, the strong harmonics of the disturbing function\footnote{This function is defined as the interaction potential for a hierarchically arranged triple system of bodies.} due to these two resonances are responsible for significant perturbations of the orbital elements, as the orbital periods \citep{almenara2015photo,mardling2013,mardling2015}.

\section{Observations and data reduction}

\subsection{Photometry}\label{photok}
\begin{figure}[!htb]
\centering
\includegraphics[scale = 0.33]{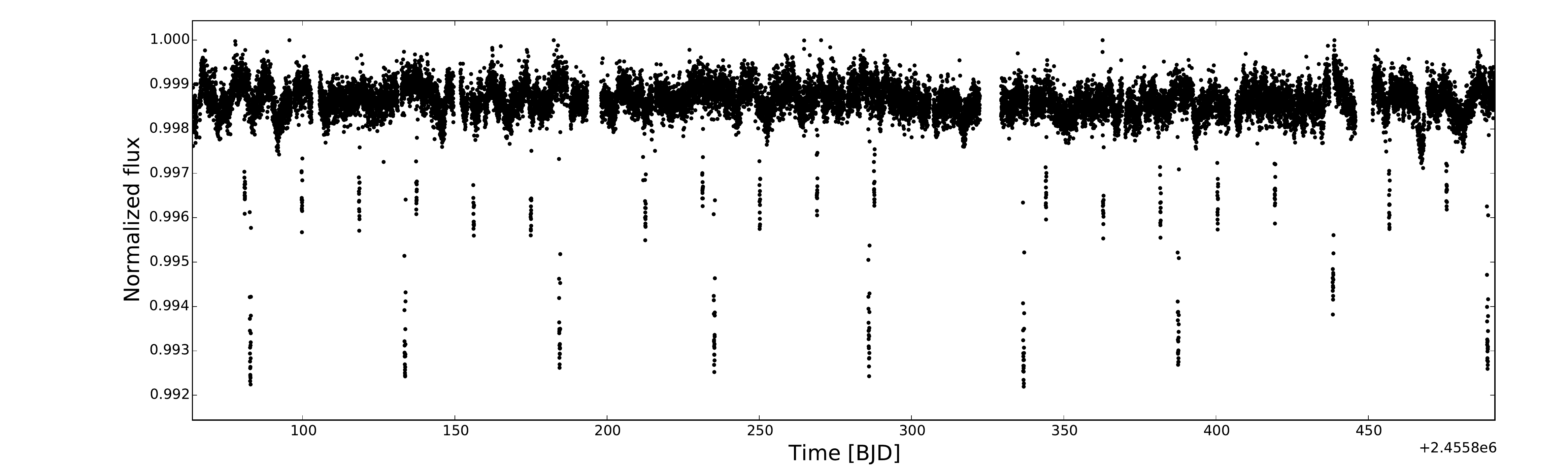}
\caption[A part of the LC light curve of Kepler-117, roughly normalized quarter by quarter]{A part of the LC light curve of Kepler-117, roughly normalized quarter by quarter.}
\label{lc_part}
\end{figure}

For this study, as for the other Kepler systems followed-up by our team, we used both photometric and spectroscopic data. Kepler observed Kepler-117 between May 2009 and May 2013, from quarter 1 to 17. We retrieved the photometric data from the Mikulski Archive for Space Telescopes (MAST)\footnote{http://archive.stsci.edu/index.html.}. There, both the long cadence (LC, sampled every 29.4 min) and short cadence (SC, sampled every 58.5 s) data is publicly available, after having been reduced with the Kepler pipeline \citep{jenkins2010}. We worked on the SC data whenever available, in reason of its higher time resolution. Figure \ref{lc_part} shows a part of the LC light curve; the lower resolution was kept to make the plot clearer. 

For all the quarters and for both the LC and the SC data, the dispersion of the contamination of the target's flux by nearby stars, corrected for by the pipeline, is lower than 1\%. The contamination value, then, was considered well-estimated, without the need to further correct it.\\
I took in charge the preparation of the photometric data. Using the preliminary estimate of the ephemeris of \cite{rowe2014}, I isolated the portion of the light curve around each transit. I discarded from the analysis the overlapping transits, i.e. the parts of the light curve where both planets happen to transit at the same time, producing a distorted transit figure. This choice was made because our transit model (described later) does not include this peculiar configuration yet. No secondary eclipse was found, as expected from the relatively long periods.\\
The light curve shows small perturbations, due to starspots, with a negligible amplitude comparing to the one of the transits ($\sim 0.1\%$). Also, no signs of occulted starspots could be observed in the transit profiles. We therefore considered that the starspots do not importantly affect the transit shape. Therefore, I conservatively normalized the transits by a second-order polynomial to the flux fitted outside of the transits. The outlier flux points were rejected through a $3\sigma$ clipping.\\

I was provided the transit times, fitted by with a procedure similar to the one presented in \cite{barros2011}, described in the following. All the transits were fitted simultaneously to constrain the transit parameters, that are the normalized separation of the planet $a/R_\star$, the ratio of planet-to-star radius $R_p /R_\star$, and the orbital inclination $i$. For each transit, the primary transit epoch $T_0$ and three normalization parameters were also fitted to account for a quadratic trend with time. The derived transit times for each planet are given in table \ref{ttvb}. The TTVs were found after removing a linear ephemeris. The TTVs exhibited by planet b are $\sim 4$ times wider than those of planet c ($\simeq 28$ min against $\simeq 7$ min). The TTV amplitude $\delta t$ is proportional to the period of the perturbed planet $P$ and to the mass $m$ of the perturbing one \citep{agol2005,holman2005}: in the limit case of two planets on circular orbits with arbitrary period ratio, but not in resonance, \cite{agol2005} showed that
\begin{equation}
\delta t_{\mathrm{in}} \sim \frac{m_{\mathrm{out}}}{M} \left( \frac{a_{\mathrm{in}}}{a_{\mathrm{in}} - a_{\mathrm{out}}} \right)^2 P_{\mathrm{in}}, 
\end{equation}
where $M$ is the total mass of the system, and the labels ``in'' and ``out'' indicate the inner and the outer planets. The labels can be inverted to obtain the $\delta t$ for the outer planet.\\
Therefore, if the two planets have a similar mass, we would expect the outer one to show stronger TTVs. The measured TTVs indicate, instead, that the outer planet is the heaviest.

\subsection{Spectroscopy}\label{rvk}
\begin{figure}[!htb]
\centering
\includegraphics[scale = 0.4]{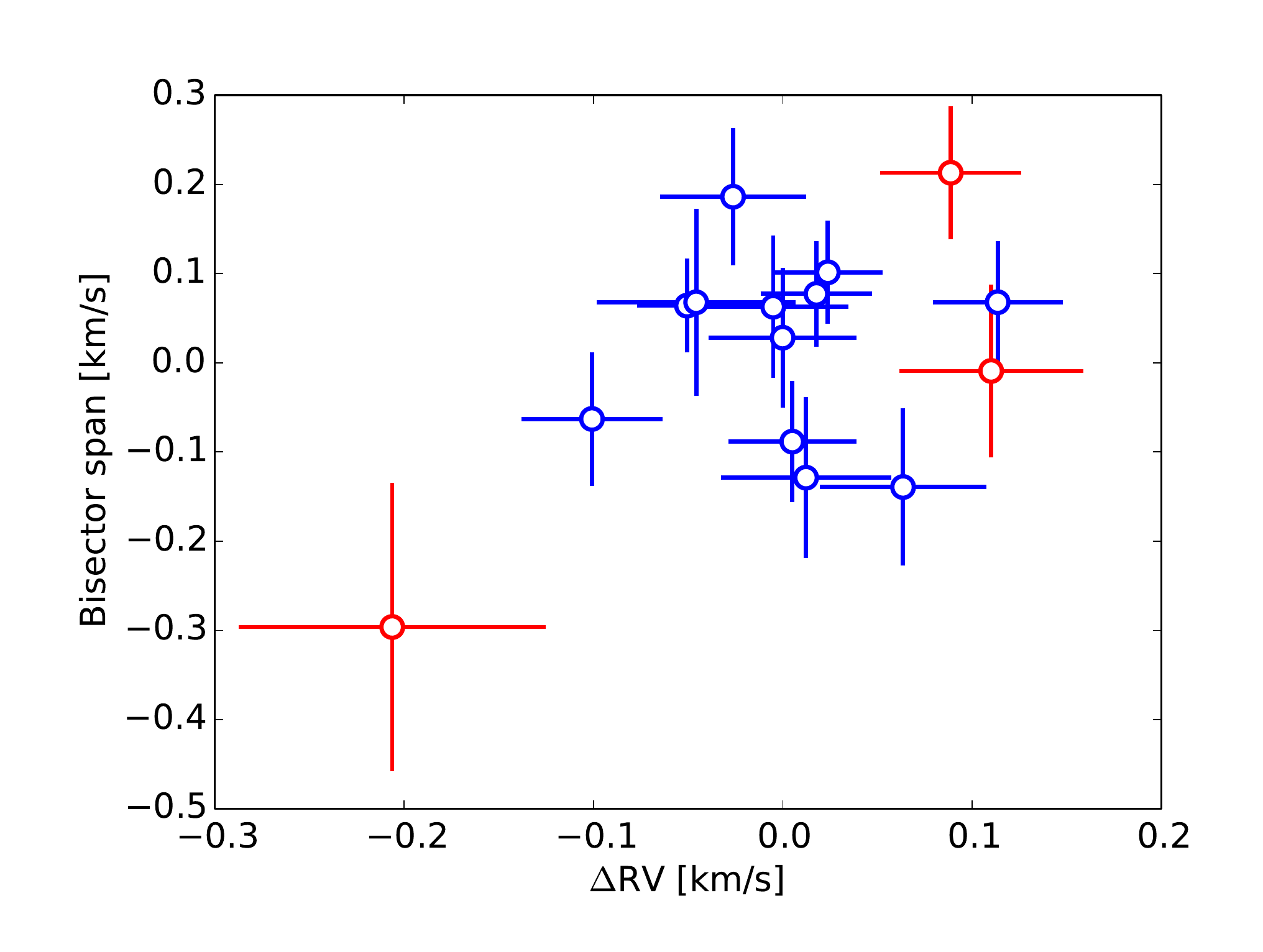}
\caption[Bisector span of the CCF plotted with respect to the RV measurements]{Bisector span of the CCF plotted with respect to the RV measurements (the mean RV has been subtracted). The red points indicate contamination by the Moon. The uncertainty on the bisector span of each point is twice the uncertainty on the RV for that point.}
\label{biss}
\end{figure}

We acquired 15 spectra of Kepler-117 during two observing seasons, between July 2012 and November 2013. In table \ref{sophiek117}, our observations are presented.\\
SOPHIE was set in HE mode (section \ref{spectrumtreat}). The exposures lasted from 1200 to 3600 s for a signal-to-noise ratio (\sn) per pixel at 550 nm between 9 and 17. The spectra were reduced using the SOPHIE pipeline \citep{bouchy2009}. In the following, I describe the reduction I carried on the RV data.\\
The RVs and their uncertainties were obtained through a Gaussian fit of the cross-correlation function (CCF) with numerical masks corresponding to the F0, G2, and K5 spectral types. The final RVs were measured with the G2 mask because of the spectral type of the star. This was proven through the analysis of the stellar spectrum, discussed later. The RV reference star HD185144 \citep{howard2010,bouchy2013,santerne2014} was used to correct the RVs for any instrumental instability of the spectrograph. This required corrections from $\sim 5$ to $\sim 30$ \ms. We corrected the three spectra affected by the moonlight for the RV of the Moon, as discussed in \cite{baranne1996}, \cite{pollacco2008}, and \cite{hebrard2008}. The charge transfer inefficiency effect was corrected for using the prescription of \cite{santerne2012}. The first three echelle orders at the blue edge of the spectrum were not used to calculate of the RVs because their low \sn\ degrades the precision of the measurements. We rejected the point at $\rm{BJD} = 2456551.49295$ because of its low \sn\ (9 at 550 nm, the lowest of all the set). Then, I proceeded with standard tests that can give hints about the planetary or false positive nature of the event causing the RV shifts. I checked for linear correlations between the bisector span of the CCF and the RVs, following \cite{queloz2001}. In figure \ref{biss}, the bisector span is plotted versus the RVs, after removal of the mean RV. If linear correlations are observed, the planetary scenario is very likely to be rejected in favor of a blend. The Spearman-rank-order correlation coefficient between the bisector span and the RVs, excluding the points contaminated by the moonlight, is $-0.08 \pm 0.32$. The p-value for this coefficient, with the null hypothesis of no correlation, is 0.98. Similarly, the Spearman correlation coefficient between the full width at half maximum of the CCFs of the spectra and their respective RV is $0.27 \pm 0.30$, with a p-value of 0.39. The two diagnostics on the CCF are clearly compatible with the planetary scenario.

\begin{table}[!htb]
\begin{center}{
\caption[Log of SOPHIE radial velocity observations]{\label {TabRV}  Log of SOPHIE radial velocity observations. The points marked with $\rightmoon$ in the date are contaminated by moonlight, while the one with $\dagger$ was discarded from the analysis.}
\begin{tabular}{|c|c|c|c|c|}
\hline
\hline
Date  & BJD    & \vrad\  & $\sigma$\vrad  & \sn\ at \\
          &      - 2450000     & [\kms]    &  [\kms]           &  550 nm$^\ast$ \\ 
\hline
2012-07-15    &     6123.54450   &  -12.842 & 0.034 &   14.12     \\
2012-07-24    &     6133.44333   &  -12.932  & 0.029  &   15.38     \\
2012-08-13    &     6153.46918   &  -13.056 & 0.037 &   13.46     \\
2012-08-22    &     6161.50353   &   -12.892 & 0.044 &   13.87     \\
2012-09-09    &     6180.46953   &  -12.943 & 0.045  &   11.83     \\
2012-09-17    &     6188.38902   &  -12.961  & 0.040 &   14.57     \\
2012-10-13    &     6214.28684   &  -12.956 & 0.039  &   12.75     \\
2013-05-08    &     6420.57184   &  -12.951 & 0.034  &   13.27     \\
2013-08-01    &     6505.52985   &  -12.982 & 0.039  &   14.76     \\
2013-08-29 $\rightmoon$    &     6533.52259   &  -12.846 &	0.049  &   11.67     \\
2013-09-16 $\rightmoon$,$\dagger$    &     6551.49295   &  -13.162 & 0.081  &   9.27      \\
2013-09-23    &     6559.30819   &  -13.006 & 0.026  &   16.44     \\
2013-10-16 $\rightmoon$ &     6582.31397   &  -12.867 &	0.037  &   16.83     \\
2013-10-27    &     6593.37762   &  -12.938 & 0.030  &   15.33     \\
2013-11-23    &     6620.24985   &  -13.001 & 0.053  &   12.47     \\
\hline
\end{tabular}
\label{sophiek117}}
\end{center}
\begin{list}{}{}
\item \small $^\ast$ Measured by the SOPHIE pipeline.
\end{list}
\end{table}

The peak-to-peak RV amplitude produced by the outer planet (planet c) is $\sim 90$ \ms, while the one produced by the inner planet (planet b) is much lower: $\lesssim 10$ \ms. This confirms the fact that planet c is the heaviest of the two. Moreover, the low RV amplitude of planet b is close to the limit of sensitivity of SOPHIE for a $\simeq 14.5$ mag$_V$ star such as Kepler-117. Indeed, it can be noticed that the RV measures indicated in table \ref{sophiek117} have $\gtrsim 30$ \ms\ uncertainties. Therefore, they will give only a poor constraint on the mass of this latter planet. This point will be discussed thoroughly in the following sections.

\section{Host star}

\subsection{Stellar parameters}\label{host}
As for the other targets that are object of a follow-up with SOPHIE, I carried out the spectral analysis. I analyzed the spectrum of Kepler-117 following the methodology described in section \ref{kepstars}. The atmospheric parameters were derived with \texttt{VWA} (section \ref{ewa}). I obtained $\rm{T}_{\rm eff} = 6260 \pm 80$ K, $\log g = 4.40 \pm 0.11$, $\rm{[Fe/H]} = 0.10 \pm 0.13$, and $v \sin i_\star = 6 \pm 2$ \kms, appropriate of an F8V-type star.\\

The addition of the TTVs to the fitting algorithm (as discussed below) allowed precisely determining the stellar \logg\ ($4.102\pm0.019$). Thanks to this, we were able to check the validity of the values derived from the spectral analysis. A first combined fit of the data sets showed a $\sim 3\sigma$ difference between the \logg\ derived from the combined fit and the spectroscopic \logg. The \logg\ is known to be a parameter whose accurate measure is problematic. We therefore compared our spectroscopic parameters with the other published ones. 
As can be seen in table \ref{tabspec}, different authors found different values for \logg, sometimes in poor agreement. 

\begin{table}[!htb]
\caption[Published spectroscopic parameters for Kepler-117 compared with those of this work]{\label {tabspec} Published spectroscopic parameters for Kepler-117 compared with those of this work. The characteristics of the corresponding spectrum are indicated. Legend for the authors: E13: \cite{everett2013}; R14: \cite{rowe2014}. $R$ indicates the spectral resolution.}
\centering
\scalebox{0.85}{\begin{tabular}{|l|c|c|c|c|c|c|}
\hline\hline
Authors & \teff [K] & \logg & \feh & \sn & $R$ & Spectral window [\AA]\\
\hline
E13 (1) & $6185\pm 75$ & $4.25 \pm 0.15$ & $-0.04 \pm 0.10$ & 72 at 4400 \AA & $\sim 3000$ & 3900-4800\\
E13 (2)$^\ast$ & $6316\pm 75$ & $4.65 \pm 0.15$ & $0.09 \pm 0.10$ & 75 at 4400 \AA  & $\sim 3000$ & 3900-4800 \\
R14    & $6169 \pm 100$ & $4.187 \pm 0.150$ & $-0.04 \pm 0.10$ & $>85^{(a)}$ at 5500 \AA & 55000 & 3600-8000\\
This work          & $6260 \pm 80$ & $4.40 \pm 0.11$ & $0.10 \pm 0.13$ & 130 at 5500 \AA & 40000 & 3972-6943 \\
\hline
\end{tabular}}
\begin{list}{}{}
\item \small $^{\ast}$ Two set of parameters are presented in E13. The fit marked with (2) is reported to have reached a parameter limit in the models. $^{(a)}$: not specified in R14.
\end{list}
\end{table}

\begin{figure}[!htb]
\centering
\includegraphics[scale = 0.6]{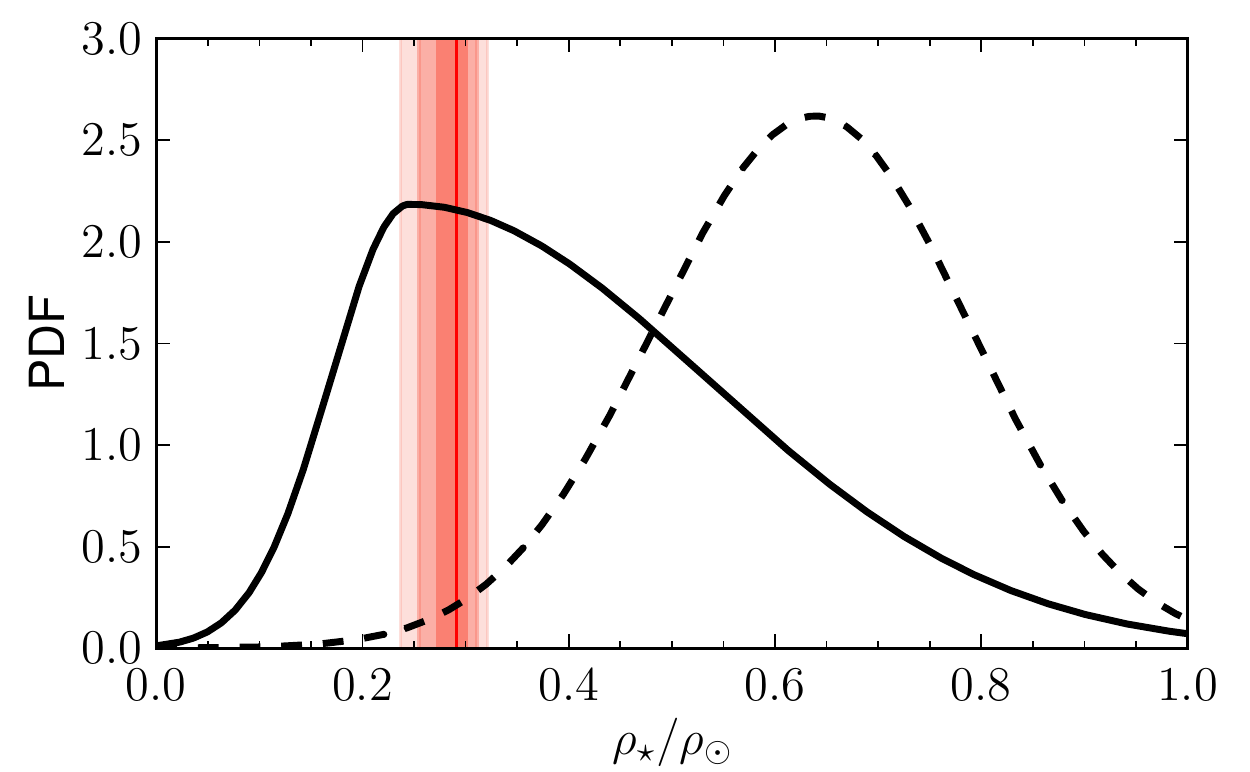}
\caption[Stellar densities derived from the spectroscopic parameters]{Stellar densities derived from the spectroscopic parameters of the SOPHIE spectrum (dashed line) and the $HIRES$ spectrum \citep[continuous line,][]{rowe2014}. In red, the posterior distribution from the \pastis\ analysis, shaded according to the 1-, 2-, and $3\sigma$ intervals.}
\label{compdens}
\end{figure}

I was unable to identify a problem in the SOPHIE spectra nor in my analysis method. However, we chose to use the published combination of \teff, \logg\, and \feh\ whose \logg\ is the closest to our posterior. We were in fact confident in the result of our MCMC, which converges to a sharp distribution even in the tail of the large spectroscopic prior of SOPHIE. This is shown in figure \ref{compdens}. The figure shows the stellar densities derived from the spectroscopic parameters obtained from the SOPHIE and the HIRES spectrum, using the Dartmouth tracks, and those from the TTVs. Hence, the final values we adopted are those of \cite{rowe2014}, who worked on a HIRES spectrum obtained at Keck. 

\subsection{Stellar activity}\label{activity}
\begin{figure}[!hbt]
\centering
\includegraphics[scale = 0.45]{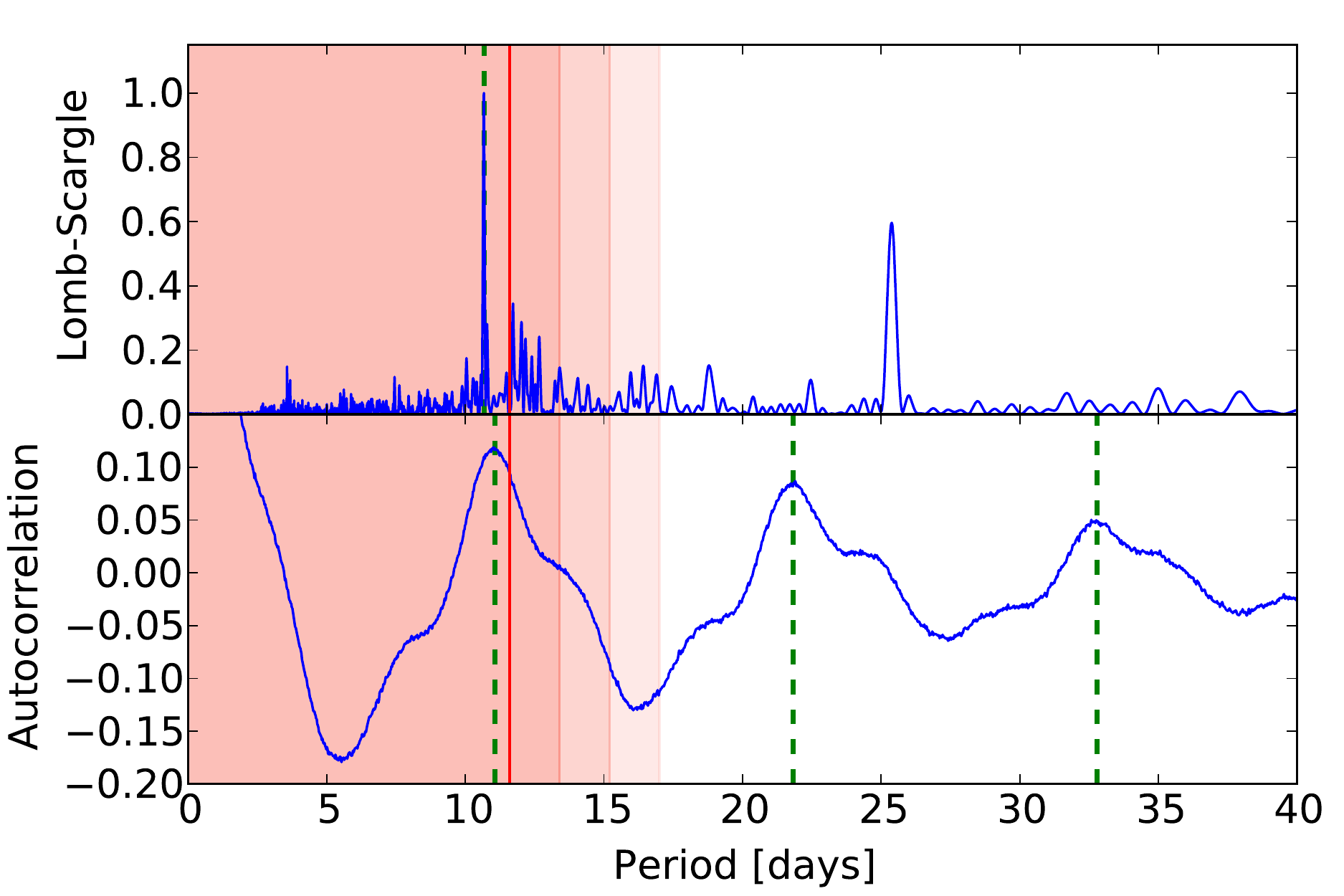}
\caption[Lomb-Scargle periodogram and autocorrelation of the light curve, after removing the transits]{\textit{Top panel}: Lomb-Scargle periodogram of the light curve after removing the transits. \textit{Bottom panel}: autocorrelation of the light curve. The green dotted lines indicate the Gaussian fit to the peaks. The red line corresponds to the maximum rotation period of the star deduced by the \vsini\ and the stellar radius. The red shadowed regions highlight the 1-, 2-, and $3\sigma$ confidence intervals for the rotation period.}
\label{perio}
\end{figure}

As discussed in section \ref{photok}, the light curve shows small periodic variations,  produced by starspots. These do not affect the transit shape importantly, but can be used to deduce some information about the stellar rotation and inclination axis. As for the stellar rotation, we removed the transits from the light curve and computed the Lomb-Scargle periodogram (LSP: \citealp{press1989}) of the light curve (figure \ref{perio}), finding a peak at $10.668 \pm 0.028$ days. The uncertainty is underestimated because it does not take into account the position of the spots on the stellar surface and the differential rotation. The $\simeq 11$-day periodicity was isolated by the autocorrelation of the light curve, too (figure \ref{perio}, bottom panel). Precisely, it was isolated by the main peak at $11.1 \pm 1.4$ days and the first two multiples at $21.8 \pm 1.7$ and $32.8 \pm 1.9$ days. The measurements are in agreement with the stellar rotation period measured from the spectroscopic \vsini\ ($6 \pm 2$ \kms) and the stellar radius $R_\star$ (section \ref{pars}), assuming that the rotation axis is perpendicular to the line of sight ($11.6\pm1.8$ days).

In conclusion, the peaks in the periodogram and the autocorrelation function can be considered as representative of the rotation of the host star, for which we conservatively adopted the photometric value with the largest uncertainty, that is, $P_\star = 11.1 \pm 1.4$ days. Comparing this with \vsini\ shows that the stellar inclination is compatible with $90^\circ$. 

\section{Combined fit of transits, RVs, and TTVs}\label{ttvsmeth}
To fit the system parameters, we developed a method for the combined fit of photometry, RVs, and TTVs. For this, we made use of the \pastis\ software \citep{diaz2014}. This code was developed at LAM to solve problems of planet validation using a Bayesian approach, by computing the Bayesian odd ratios between competing scenarios (e.g., planetary scenario or eclipsing binary). A Markov chain Monte Carlo (MCMC) algorithm is used to sample the posterior distributions of the parameters of each scenario, and can be exploited to simultaneously fit several data sets of a single scenario, as well. In particular, \pastis\ is able to simultaneously fit transits with a modified version of the \ebop\ code\footnote{Hereafter, for simplicity, \ebop.} \citep{nelson1972,etzel1981,popper1981}, and RVs with the models described in \cite{santerne2015}. The spectral energy distribution (SED) can be also fitted, solving for \teff, \logg, \feh, distance of the system and interstellar extinction.

\begin{table*}[!htbp]
\caption[Prior distributions used in the combined fit with \texttt{PASTIS}]{\label{priorsttv} Prior distributions used in the combined fit with \texttt{PASTIS}. $\mathcal{U}(a, b)$ stands for a uniform distribution between $a$ and $b$; $\mathcal{N}(\mu, \sigma)$ indicates a normal distribution with mean $\mu$ and standard deviation $\sigma$; $\mathcal{N_A}(\mu, \sigma_-, \sigma_+)$, stands for an asymmetric normal with mean $\mu$, right width $\sigma_+$ and left width $\sigma_-$; $\mathcal{S}(a, b)$ represents a sine distribution between $a$ and $b$; finally, $\mathcal{J}(a, b)$ means a Jeffreys distribution between $a$ and $b$.}
\centering
\scalebox{0.85}{\begin{tabular}{lll}
\hline\hline
\multicolumn{3}{l}{\emph{Stellar parameters}} \smallskip\\
Effective temperature \teff\ [K] & $\mathcal{N}(6169, 100)$ \\
Metallicity \feh\    & $\mathcal{N}(-0.04, 0.10)$ \\
Stellar density $\rho_{\star}$ [$\rho_\odot$]  (Dartmouth) & $\mathcal{N_A}(0.25, 0.08, 0.29)$  \\
Stellar density $\rho_{\star}$ [$\rho_\odot$]  (PARSEC) & $\mathcal{N_A}(0.25, 0.08, 0.30)$\\
Stellar density $\rho_{\star}$ [$\rho_\odot$]  (StarEvol) & $\mathcal{N_A}(0.23, 0.07, 0.25)$\\
Stellar density $\rho_{\star}$ [$\rho_\odot$]  (Geneva) & $ \mathcal{N_A}(0.24, 0.07, 0.30)$\\
Quadratic limb-darkening coefficient $u_a$     & $\mathcal{U}(-0.5, 1.2)$\\
Quadratic limb-darkening coefficient $u_b$     & $\mathcal{U}(-0.5, 1.2)$ \\
Stellar RV linear drift [\ms yr$^{-1}$]                 & $\mathcal{U}(-0.001, 0.001)$ \smallskip\\

\multicolumn{1}{l}{\emph{Planet parameters}} & \emph{Kepler-117 b} & \emph{Kepler-117 c} \smallskip\\
Orbital period $P$ [days]                   & $\mathcal{N}(18.795921, 1.3 \times 10^{-5})$ & $\mathcal{N}(50.790391, 2.3\times 10^{-5})$ \\
Primary transit epoch $T_{0}$ [BJD-2450000] & $\mathcal{N}(4978.82194, 4.7\times 10^{-4})$ & $\mathcal{N}(4968.63195, 3.1\times 10^{-4})$\\
Orbital eccentricity $e$                    & $\mathcal{U}(0, 1)$ & $\mathcal{U}(0, 1)$ \\
Argument of periastron $\omega$ [deg]       & $\mathcal{U}(0, 360)$ & $\mathcal{U}(0, 360)$\\
Orbital inclination $i$ [deg]               & $\mathcal{S}(50, 90)$ &            $\mathcal{S}(89, 91)$ \\
Longitude of the ascending node $\Omega$ [deg]       & $\mathcal{U}(135, 225)$ & - \\
Radius ratio $R_p/R_\star$   & $\mathcal{J}(0.01, 0.50)$ & $\mathcal{J}(0.01, 0.50)$ \\
Radial velocity semi-amplitude $K$ [\ms]    & $\mathcal{U}(0.0, 0.5)$ & $\mathcal{U}(0.0, 0.5)$ \smallskip\\

\multicolumn{3}{l}{\emph{System parameters}} \smallskip\\
Distance [pc]  & $\mathcal{U}(0, 1\times10^4)$  \\
Interstellar extinction $E(B - V)$ & $\mathcal{U}(0, 3)$\\
Systemic velocity (BJD 2456355), $V_{r}$ [\kms] & $\mathcal{U}(-13.1, -12.6)$ \smallskip\\

\multicolumn{3}{l}{\emph{Instrumental parameters}} \smallskip\\
Kepler jitter (LC) [ppm] & $\mathcal{U}(0, 0.0007)$\\
Kepler offset (LC) [ppm] & $\mathcal{U}(0.99, 1.01)$\\
Kepler jitter (SC) [ppm] & $\mathcal{U}(0, 0.0004)$\\
Kepler offset (SC) [ppm] & $\mathcal{U}(0.99, 1.01)$\\
TTV jitter, planet b [min]          & $\mathcal{U}(0, 9)$\\
TTV jitter, planet c [min]          & $\mathcal{U}(0, 4)$  \\
SOPHIE jitter [\ms]        & $\mathcal{U}(0, 2)$\\
SED jitter [mags]          & $\mathcal{U}(0, 1)$ \\
\\
\hline
\end{tabular}}
\end{table*}

To model the TTVs, we performed N-body dynamical simulations with the \mmercury\ code, version 6.2 \citep{chambers1999}. This code integrates the equations of motion using a variety of algorithms. We made use of a conservative Bulirsch-Stoerwe algorithm, as a compromise between accuracy of the calculations and time of execution. I contributed to the implementation of \mmercury\ in the MCMC algorithm of \pastis. At each iteration of the MCMC, a two-planet system was let evolve for the time span covered by the light curve. The transit times were calculated by interpolating the passage of the planets through the line of sight. Then, the TTVs were computed by subtracting a linear fit to the transit times.\\
As a compromise between execution time and accuracy of the TTVs with respect to the measured uncertainties, we set the simulations to cover the time span of the Kepler photometry, with a step of 0.4 days, that is, $1/47$th of the lower orbital period.\\

The combined fit was performed as follows. An exploration phase was started at random points drawn from the priors listed in table \ref{priorsttv}. From the chains computed in this phase, we used the one with the highest likelihood for the starting values of the final MCMC set.\\

In \pastis, the fits are run with four sets of stellar evolutionary tracks, to take into account the differences among models. We used four evolutionary tracks as input for the stellar parameters: Dartmouth \citep{dotter2008}, PARSEC \citep{bressan2012}, StarEvol \cite[Palacios, {priv. com.};][]{Lagarde2012}, and Geneva \citep{mowlavi2012}. For the single chains, the intrinsic uncertainties in the models were not taken into account. We ran twenty-five chains of about $10^5$ steps for each of the stellar evolutionary tracks. At each step of the MCMC, the model light curves were oversampled and then binned by a factor ten, to correct for the distortions in the signal due to the finite integration time \citep{kipping2010}. We derived the stellar density $\rho_\star$ from the spectroscopic \teff, \logg, and \feh\ and set it, together with the spectroscopic \teff\ and \feh, as a jump parameter, with normal priors for all three of them. For each planet, we used Gaussian priors for the orbital period $P$ and the primary transit epoch $T_0$ and non-informative priors for the argument of periastron $\omega$, the eccentricity $e$, and the inclination $i$. In particular, the transit modeling is degenerate with respect to the stellar hemisphere the planet covers, while the TTVs are not. In a two-planet scenario, this can lead to strong correlations between the two inclinations. We therefore constrained one of the transits in one of the hemispheres and left the other free to vary. As, from the exploration phase, the inclination of planet b was found to be lower than that of planet c, the inclination of planet b was limited to one hemisphere ($50^\circ < i < 90^\circ$) and that of planet c was left free to vary between both ($89^\circ < i < 91^\circ$).\\
We used uniform priors for the coefficients of quadratic limb darkening $u_a$ and $u_b$, for the planetary-to-stellar radius ratio $R_p/R_\star$, and for the radial velocity amplitude $K$. For Kepler-117 b, we fitted the longitude of the ascending node $\Omega$, too, for which we imposed a uniform prior. The $\Omega$ of planet c was fixed at $180^\circ$, because the symmetry of the problem allows freely choosing one of the two $\Omega$s.\\
We expressed the Kepler normalized flux offset, the systemic velocity, and the RV linear drift with uniform priors (separating LC and SC data in the photometry). As the contamination values indicated by the Kepler pipeline are self-consistent (section \ref{photok}), the contamination parameter was fixed. Finally, we modeled the instrumental and astrophysical systematic sources of error with a jitter term for Kepler, two for the TTVs (one for each planet), and one for SOPHIE. A uniform prior was assigned to all the jitter terms.\\
Every posterior distribution was thinned according to its correlation length. A combined posterior distribution was derived by taking the same number of points from each stellar evolutionary track. This combined distribution gave the most probable values and the confidence intervals for the system parameters.\\

Finally, the derived \logg\ and the posterior distributions of $R_\star$, \teff, and \feh, together with the magnitudes in table \ref{tabmag} (derived from online archives, as indicated in the table), were set as priors for another MCMC run. The magnitudes were fitted to sample the posterior distributions of the distance of the system, the interstellar extinction $E(B-V)$, and the jitter of the SED. The model SED was interpolated from the PHOENIX/BT-Settl synthetic spectral library \citep{allard2012}, scaled with the distance, the stellar radius, and the reddening $E(B-V)$, expressed through a \citet{fitzpatrick1999} extinction law. For both the distance and the reddening, non-informative priors were used.

\begin{table}[!htb]
\begin{center}{
\caption[Magnitudes of the Kepler-117 system]{\label {tabmag} Magnitudes of the Kepler-117 system.}
\begin{tabular}{|c|c|c|}
\hline\hline
Filter-Band    &    mag  &    error \\
\hline
Johnson-B $^a$ & 15.056   &  0.029	\\ 
Johnson-V $^a$ & 14.476   &  0.027 \\
SDSS-G $^a$	  & 14.688   &  0.032\\
SDSS-R $^a$	  & 14.36    &  0.026\\
SDSS-I $^a$	  & 14.227   &  0.066 \\
2MASS-J $^b$   & 13.324   &  0.026\\
2MASS-H $^b$   & 12.988  &   0.031\\
2MASS-Ks $^b$  & 13.011   &  0.031\\
WISE-W1 $^c$   & 12.946   &  0.024 \\	
WISE-W2 $^c$   & 12.992   &  0.025 \\
\hline
\end{tabular}}
\end{center}
\begin{list}{}{}
\item $^a$ APASS (http://www.aavso.org/apass); $^b$ 2MASS \citep{sktutskie2006,cutri2003}; $^c$ WISE \citep{wright2010}.
\end{list}
\end{table}

\section{Results}

\subsection{System parameters}\label{pars}
\begin{figure*}[!htbp]
\centering
\includegraphics[scale = 0.35]{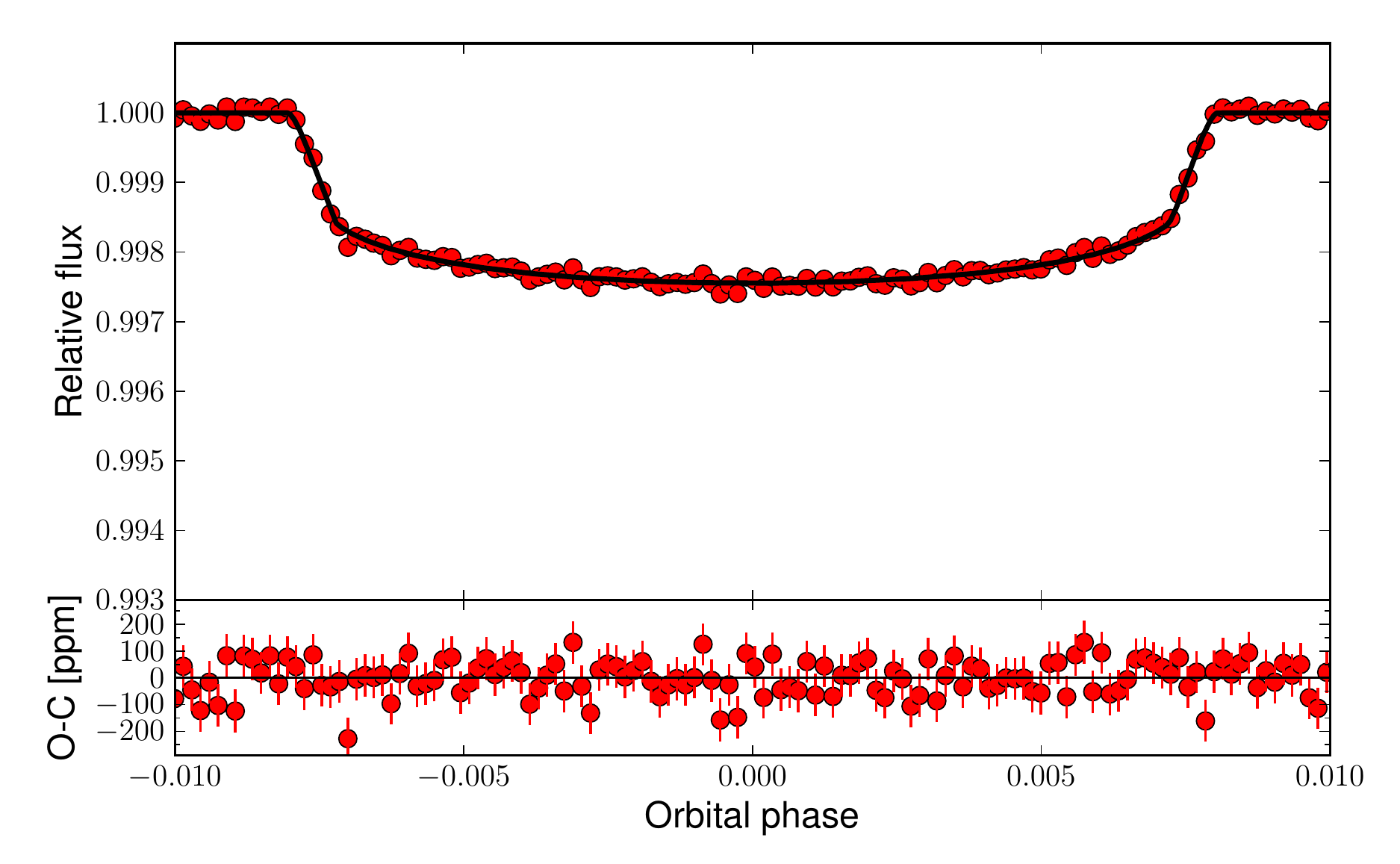}
\includegraphics[scale = 0.35]{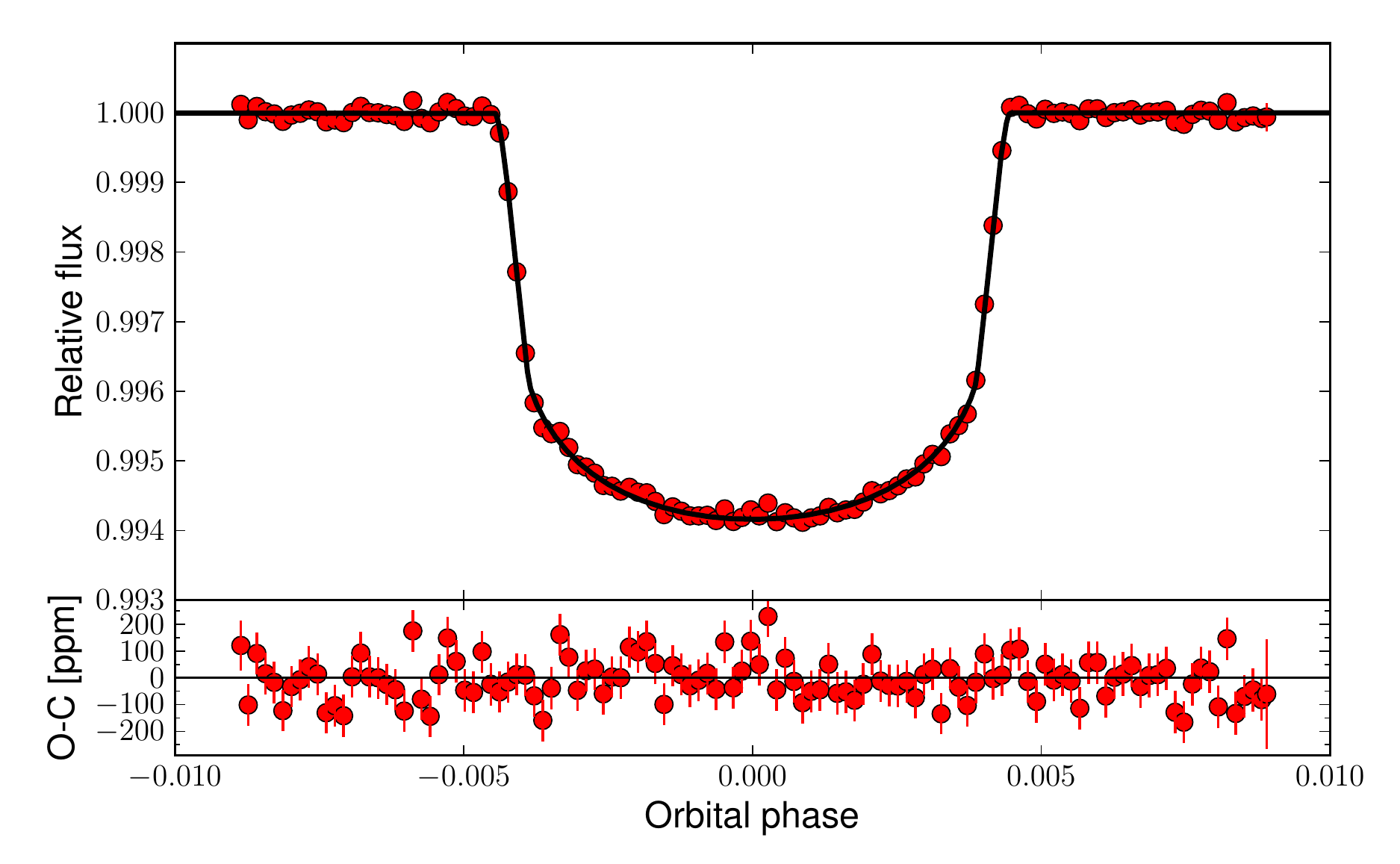}
\includegraphics[scale = 0.35]{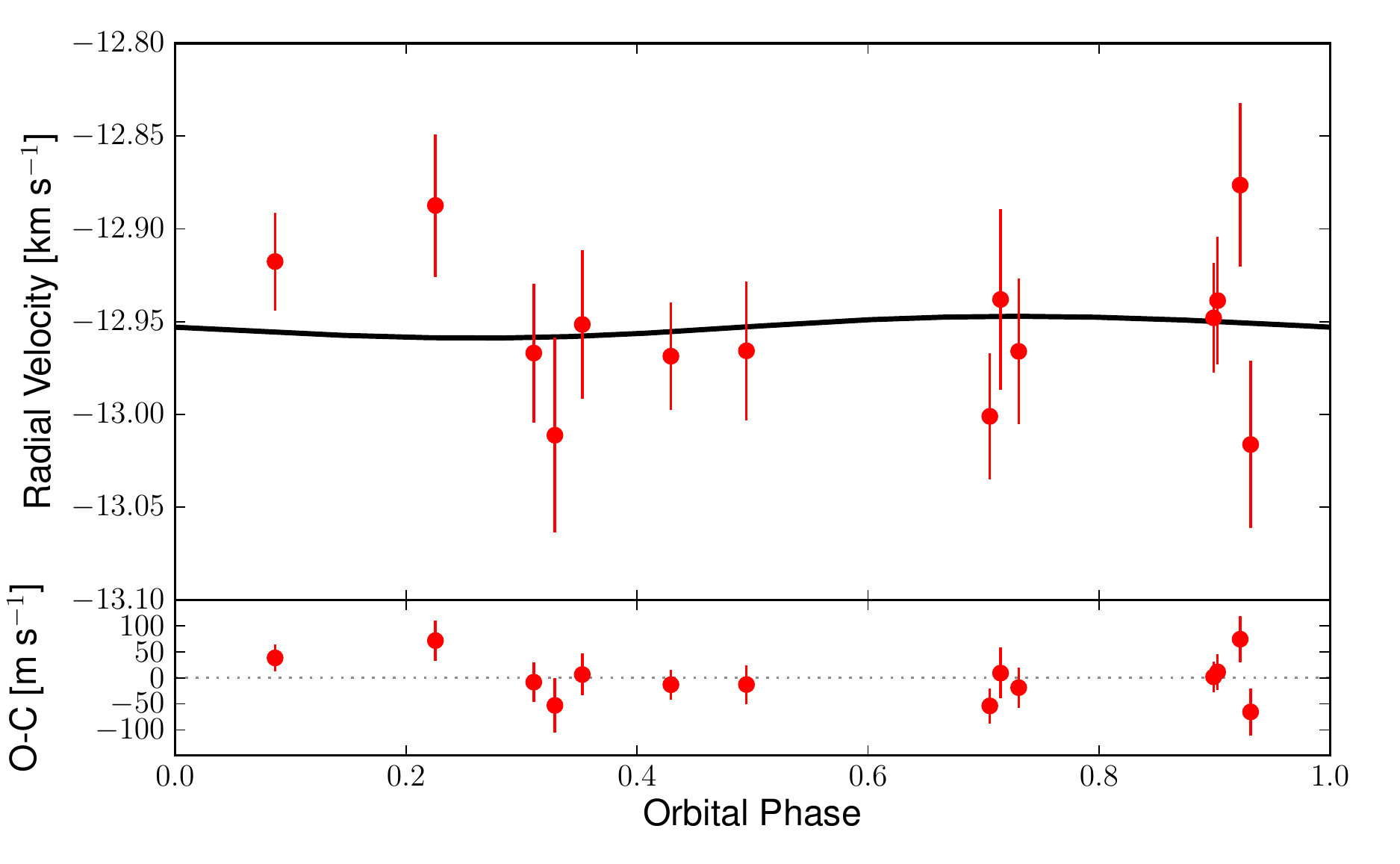}
\includegraphics[scale = 0.35]{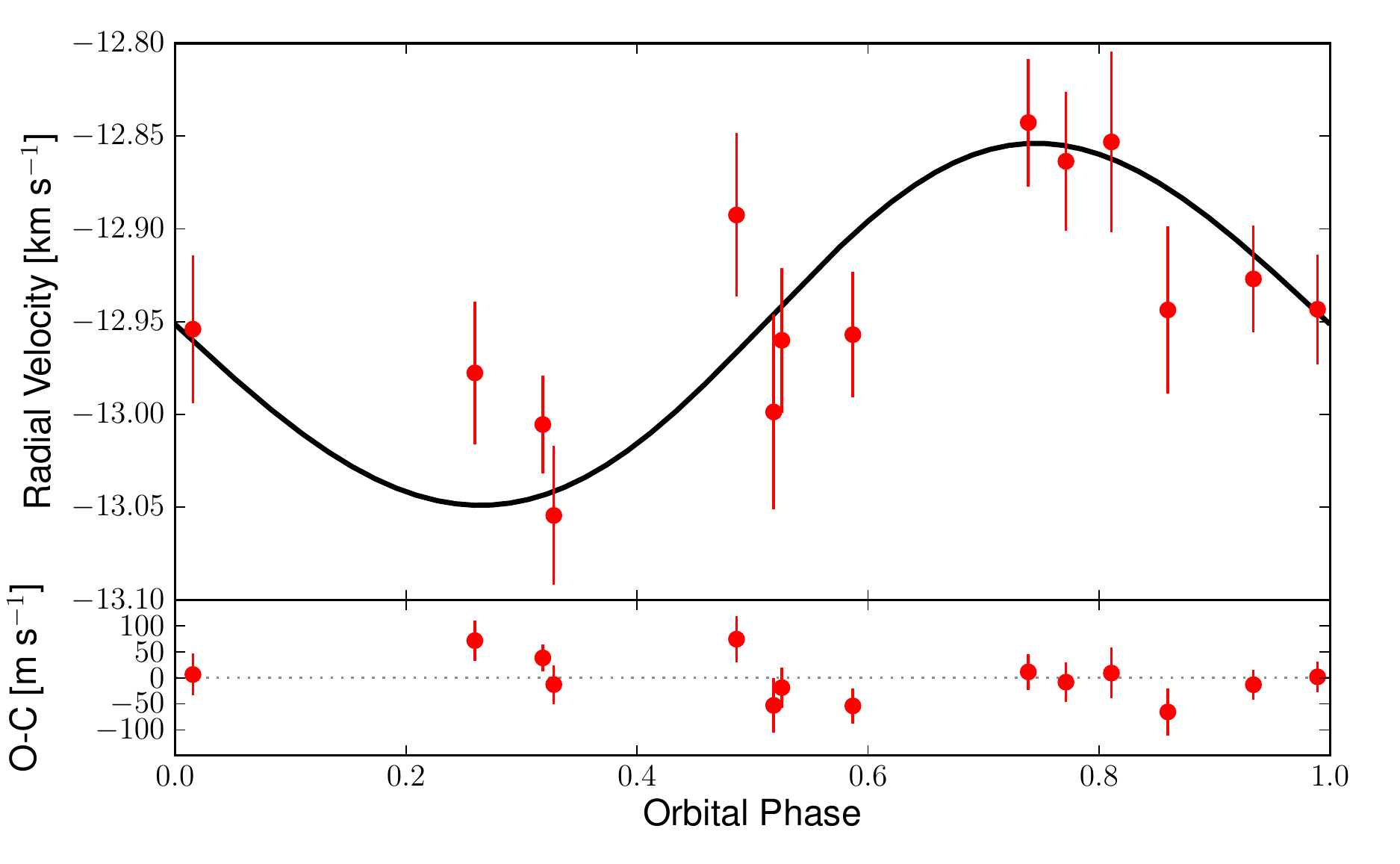}
\includegraphics[scale = 0.35]{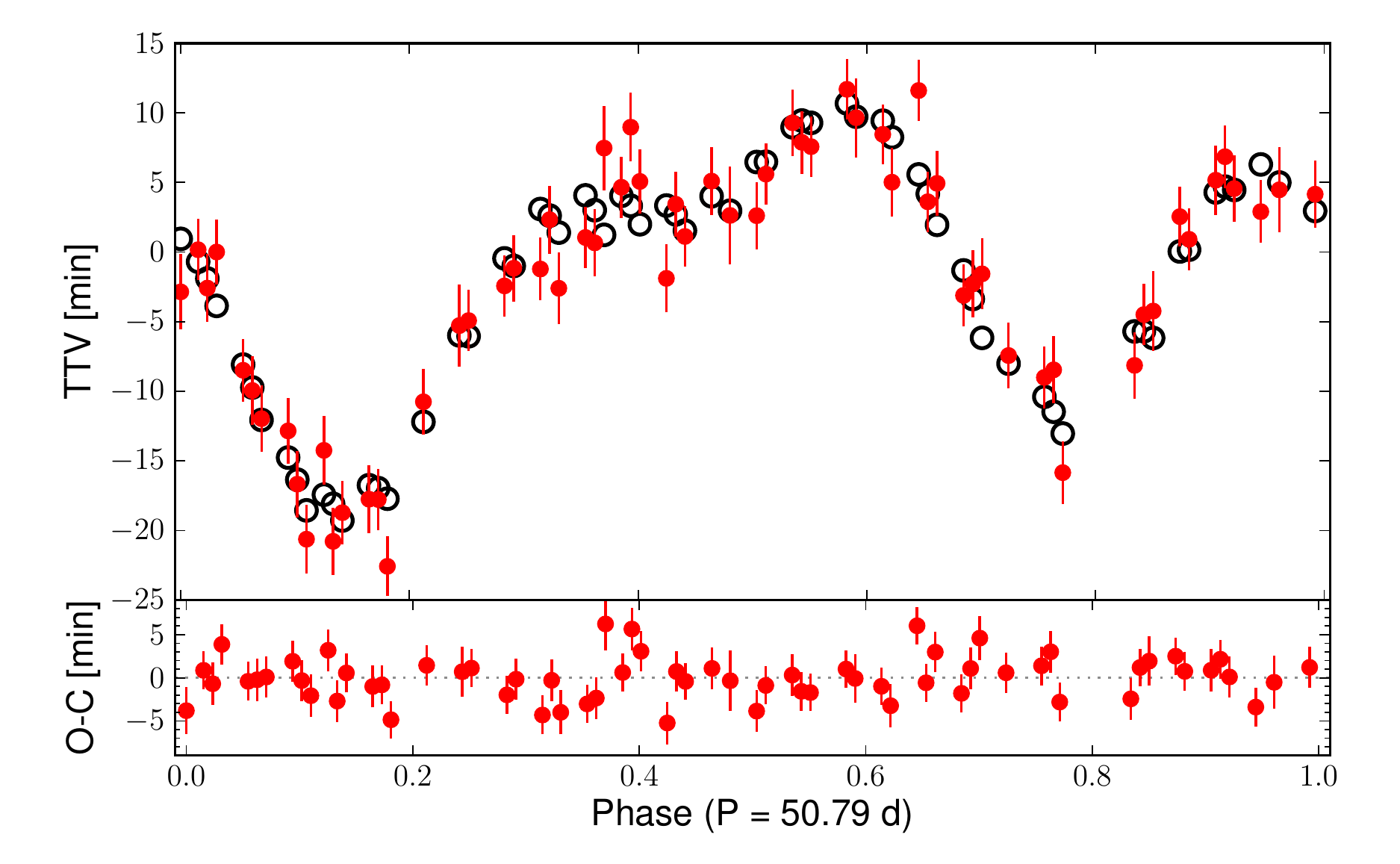}
\includegraphics[scale = 0.35]{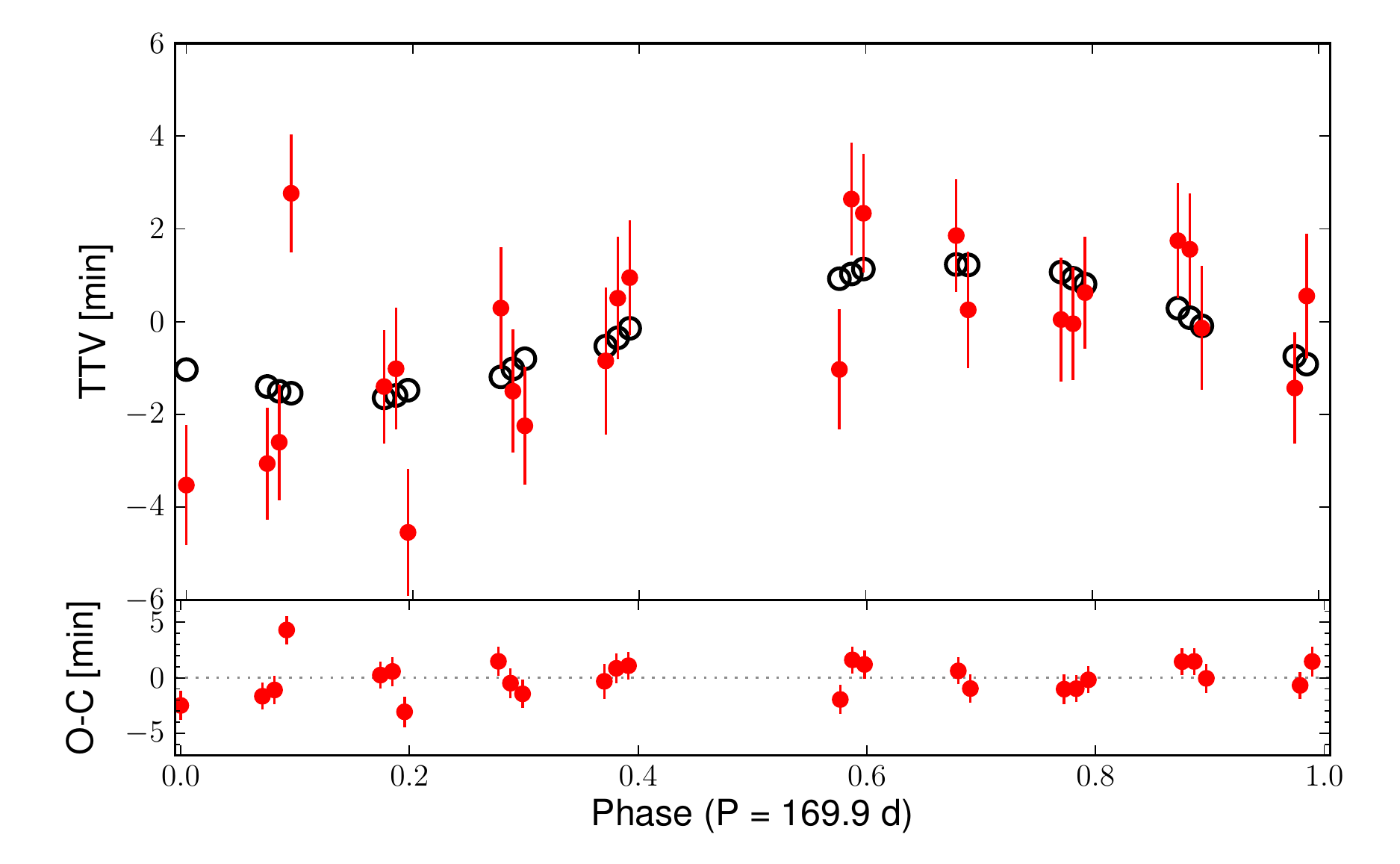}
\caption[Best-fit models of Kepler-117 plotted over the data]{\textit{Top}: phase-folded plot of the best transit model of planet b (left) and c (right), over the SC data. In black the model, in red the data binned every hundredth of orbital phase. \textit{Center}: the same for the radial velocities. \textit{Bottom}: The TTVs of planet b folded at the orbital period of planet c (left) and those of planet c folded at the first peak of its Lomb-Scargle periodogram (right, section \ref{modu}). For each plot, the lower panel shows the residuals as observed minus calculated ($O-C$) points.}
\label{modeldata}
\end{figure*}

\begin{figure}[!htbp]
\centering
\includegraphics[scale = 0.45]{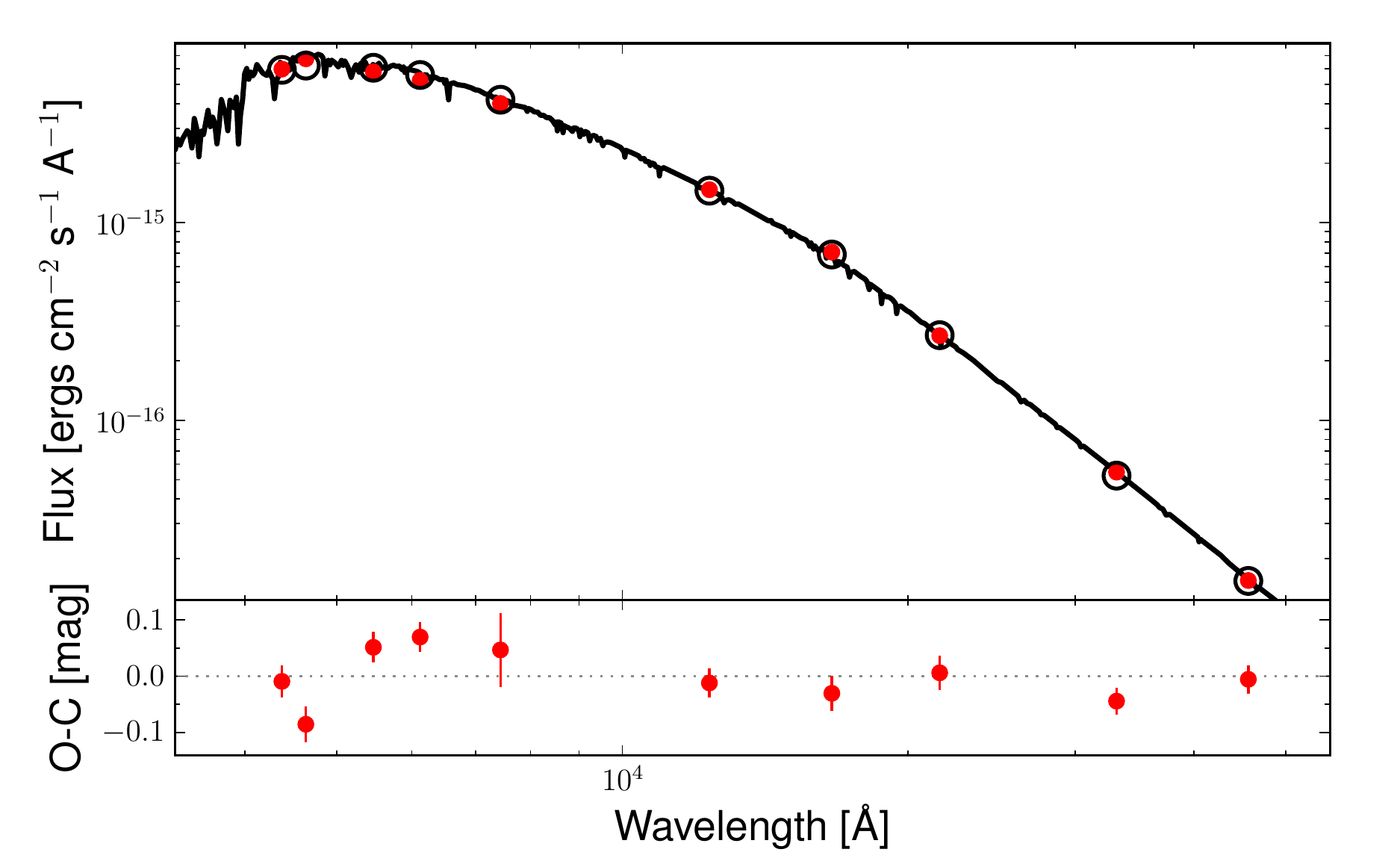}
\caption[Model SED on the photometric bands]{Model SED on the photometric bands. The residuals are shown in the lower panel.}
\label{sed}
\end{figure}

In table \ref{posteriors}, the mode and the 68.3\% equal-tailed confidence intervals of the system parameters are presented. Kepler-117 A was found to be a $\simeq 5$ Gyr old F8V star. The planets were found to be in low-eccentricity orbits ($0.0493 \pm 0.0062$ and $0.0323 \pm 0.0033$ for planet b and c), and to differ widely in their mass, but less so in their radii: $0.094 \pm 0.033$ \MJ, $0.719 \pm 0.024$ \RJ\ for planet b and $1.84 \pm 0.18$ \MJ, $1.101 \pm 0.035$ \RJ\ for planet c. The planetary radii, in particular, are in agreement with the estimate of \cite{rowe2014}: $0.72 \pm 0.14$ \RJ\ for planet b and $1.04 \pm 0.20$ \RJ\ for planet c.\\
In figure \ref{modeldata}, the solution with maximum likelihood is plotted over the various phase-folded data sets. As it can be observed from the plots of the transits of both planets, the root mean square (RMS) of the residuals of the out-of-transit flux is not significantly different from the RMS of the residuals of the in-transit flux. Therefore, as noticed in section \ref{photok}, no indication of starspots occulted by the planets is found.\\ 
The measured drift of the RVs is compatible with 0 \kms; a non-zero drift would have been an indication of a possible third companion in the system that affected the amplitude of the TTVs Figure \ref{sed} shows the maximum-likelihood SED model over the data.

\begin{table*}[!htbp]
\centering
\caption{\label{posteriors} Planetary and stellar parameters derived from the combined fit, with their 68.3\% central confidence intervals.}
\scalebox{0.85}{
\begin{tabular}{lll}
\hline\hline
\multicolumn{1}{l}{\emph{Stellar parameters from the combined analysis}} & & \smallskip\\
Stellar density $\rho_{\star}$ [$\rho_\odot$]    & 0.291$^{+0.010}_{-0.018}$ & \\
Stellar mass [\Msun]                          & 1.129$^{+0.13}_{-0.023}$  & \\
Stellar radius [\Rsun]                        & 1.606 $\pm$ 0.049      & \\
\teff [K]                                     & 6150 $\pm$ 110         & Spectroscopic value: $6160 \pm 100$ \\
Metallicity [Fe/H] [dex]                      & -0.04 $\pm$ 0.10       & Spectroscopic value: $-0.04 \pm 0.10$\\
Derived \logg [cgs]                           & 4.102 $\pm$ 0.019      & Spectroscopic value: $4.19 \pm 0.15$\\
Age $t$ [Gyr]                                 & 5.3 $\pm$ 1.4          & \\
Distance of the system [pc]                   & 1430 $\pm$ 50          & \\
Interstellar extinction $E(B - V)$            & 0.057 $\pm$ 0.029      & \\
Quadratic limb-darkening coefficient $u_a$    & 0.382 $\pm$ 0.031      & \\
Quadratic limb-darkening coefficient $u_b$    & 0.187 $\pm$ 0.056      & \\
Stellar RV linear drift [\ms yr$^{-1}$]        & 12 $\pm$ 22            & \\
Systemic velocity (BJD 2456355), $V_{r}$ [\kms]& -12.951 $\pm$ 0.013    & \smallskip\\

\multicolumn{1}{l}{} & \emph{Kepler-117 b} & \emph{Kepler-117 c} \smallskip\\
Orbital period $P$ [days]                   & 18.7959228 $\pm 7.5\e{-6}$ & 50.790391 $\pm 1.4\e{-5}$ \\
Orbital semi-major axis $a$ [AU]            & 0.1445$^{+0.0047}_{-0.0014}$   & 0.2804$^{+0.0092}_{-0.0028}$ \\
Primary transit epoch $T_{0}$ [BJD-2450000] & 4978.82204 $\pm 3.5\e{-4}$  & 4968.63205 $\pm 2.5\e{-4}$ \\
Orbital eccentricity $e$                    & 0.0493 $\pm$ 0.0062        & 0.0323 $\pm$ 0.0033 \\
Argument of periastron $\omega$ [deg]       & 254.3 $\pm$ 4.1            & 305.0 $\pm$ 7.5 \\
Orbital inclination $i$ [deg]               & 88.74 $\pm$ 0.12           & 89.64 $\pm$ 0.10$^{\dagger}$ \\
Transit impact parameter $b_{prim}$          & 0.446 $\pm$ 0.032          & 0.268$^{+0.036}_{-0.087}$$^{\ddagger}$ \\
Transit duration $T_{14}$ [h]                & 7.258 $\pm$ 0.020          & 10.883 $\pm$ 0.031 \\
Scaled semi-major axis $a/R_\star$           & 19.67 $\pm$ 0.37           & 38.18 $\pm$ 0.72 \\
Radius ratio $R_p/R_\star$                   & 0.04630 $\pm$ 0.00025      & 0.07052 $\pm$ 0.00034 \\
Radial velocity semi-amplitude $K$ [\ms]    & 6.5 $\pm$ 2.1              & 90.4 $\pm$ 7.0 \\
Longitude of the ascending node $\Omega$ [deg] & 177.9 $\pm$ 5.6         & 180 (fixed) \smallskip\\

Planet mass $M_{p}$ [M$_J$]                   & 0.094 $\pm$ 0.033         & 1.84 $\pm$ 0.18 \\
Planet radius $R_{p}$[R$_J$]                  & 0.719 $\pm$ 0.024         & 1.101 $\pm$ 0.035 \\
Planet density $\rho_p$ [$g\;cm^{-3}$]        & 0.30 $\pm$ 0.11           & 1.74 $\pm$ 0.18 \\
Planet surface gravity, $\log$\,$g_{p}$ [cgs] & 2.67$^{+0.10}_{-0.17}$       & 3.574 $\pm$ 0.041 \\
Planet equilibrium temperature, $T_{eq}$ [K]$^{\ast}$ & 984 $\pm$ 18       & 704 $\pm$ 15 \smallskip\\

\multicolumn{3}{l}{\emph{Data-related parameters}} \smallskip\\
Kepler jitter (LC)  [ppm] & 67$^{+16}_{-39}$  & \\
Kepler jitter (SC)  [ppm] & 0$^{+12}$        & \\
SOPHIE jitter$^a$[\ms]        & 0$^{+25}$         & \\ 
SED jitter [mags]          & 0.043$^{+0.026}_{-0.013}$ & \\
TTV1 jitter [min]          & 0.95 $\pm$ 0.48  & \\
TTV2 jitter [min]          & 0.90 $\pm$ 0.62  & \\
\\

\hline
\end{tabular}}
\begin{list}{}{} \footnotesize { 
\item $^{\dagger}$ From the posterior distribution, reflected with respect to $i = 90^\circ$; $^{\ddagger}$ Reflected as the inclination, with respect to $b = 0$; $^{\ast}$ $T_{eq}=T_{\mathrm{eff}}\left(1-A\right)^{1/4}\sqrt{\frac{R_\star}{2 a}}$, with $A \, \mathrm{(planet \, albedo)}\; =0$; $^a$ This value is compatible with the low level of activity observed in the photometry; \Msun $= 1.98842\e{30}$~kg, \Rsun $= 6.95508\e{8}$~m, M$_J = 1.89852\e{27}$~kg, R$_J$ = 71492000~m.
}\end{list}
\end{table*} 

\clearpage

\subsection{Modulation of the TTVs}\label{modu}

\begin{figure*}[htb]
\centering
\includegraphics[scale = 0.4]{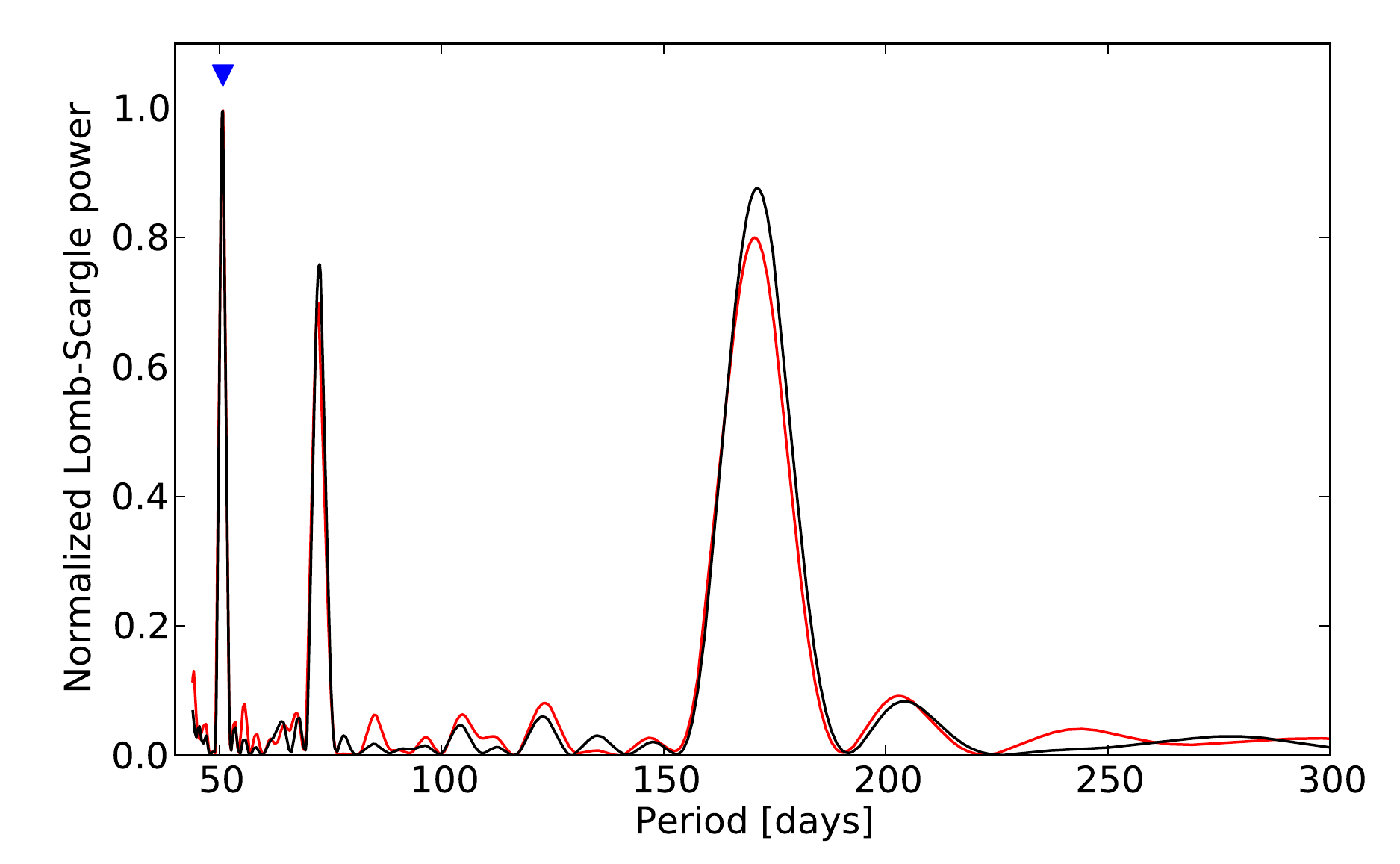}
\includegraphics[scale = 0.4]{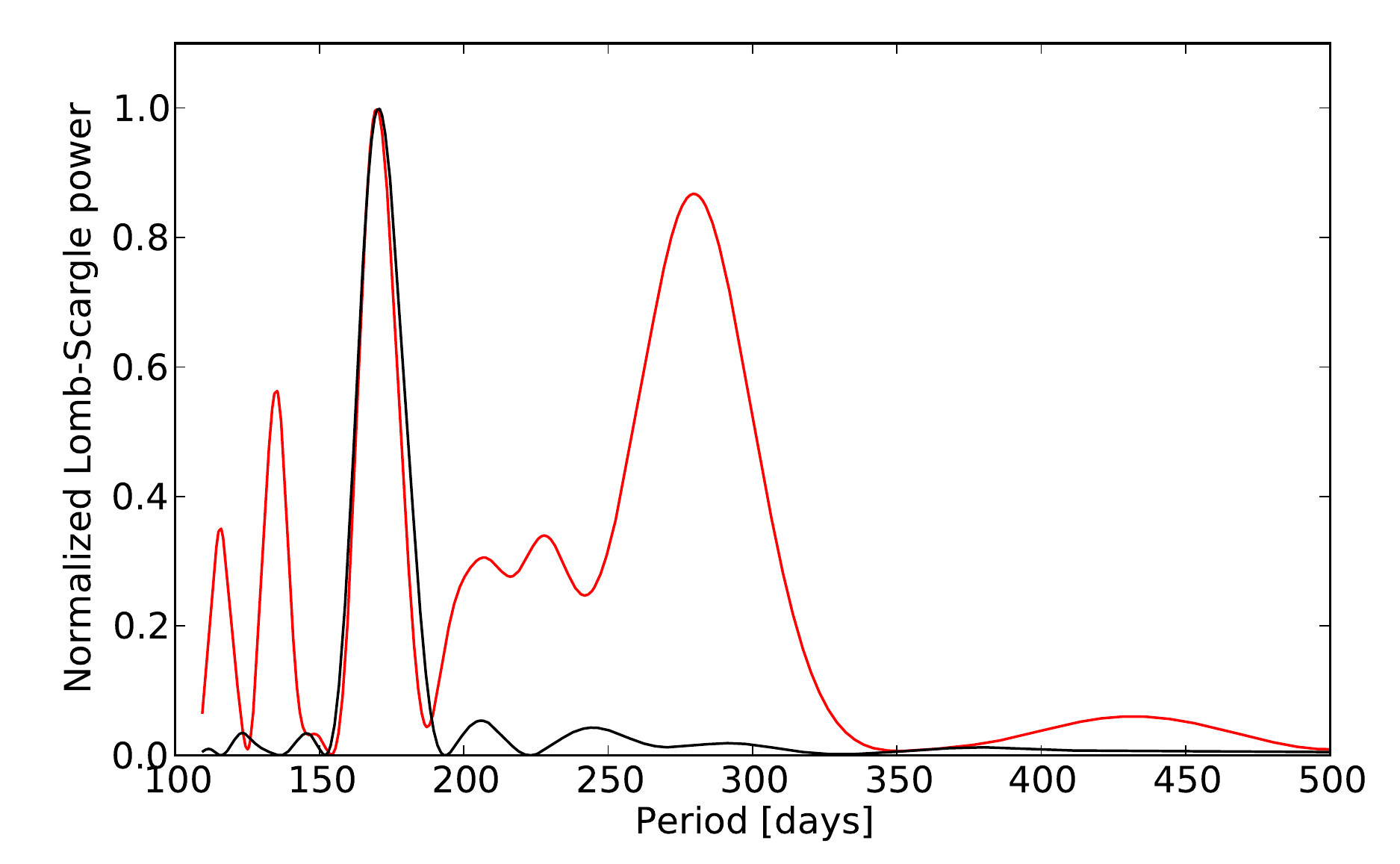}
\caption[Lomb-Scargle periodogram of the measured and modeled TTVs]{Lomb-Scargle periodogram of the measured (red) and modeled (black) TTVs for planet b (left) and c (right). The blue triangle in the plot on the left indicates the orbital period of planet c.}
\label{lsttv}
\end{figure*}

We tested the robustness of our result by inspecting the periodic modulation of the measured and the modeled TTVs. To do this, we compared their Lomb-Scargle periodograms (figure \ref{lsttv}). The main peaks coincide for both planets and also agree with the periodicities found by  \citealp{mazeh2013} \citep[see also][]{ofirtoulouse}. The periodogram of the modeled TTVs of planet b reproduces that of the measured TTVs well. Some of the peaks of planet c, on the other hand, are due to noise. This can be explained by the different amplitude of the signal in the two cases.\\
For both planets, no significant peak was recovered at the stellar rotation period of the star or at its harmonics. This suggests that the impact of possible occulted starspots on the measured TTVs is negligible.

\subsection{Discussion on the system parameters}\label{constr}

\begin{figure*}[!htb]
\centering
\includegraphics[scale = 0.5]{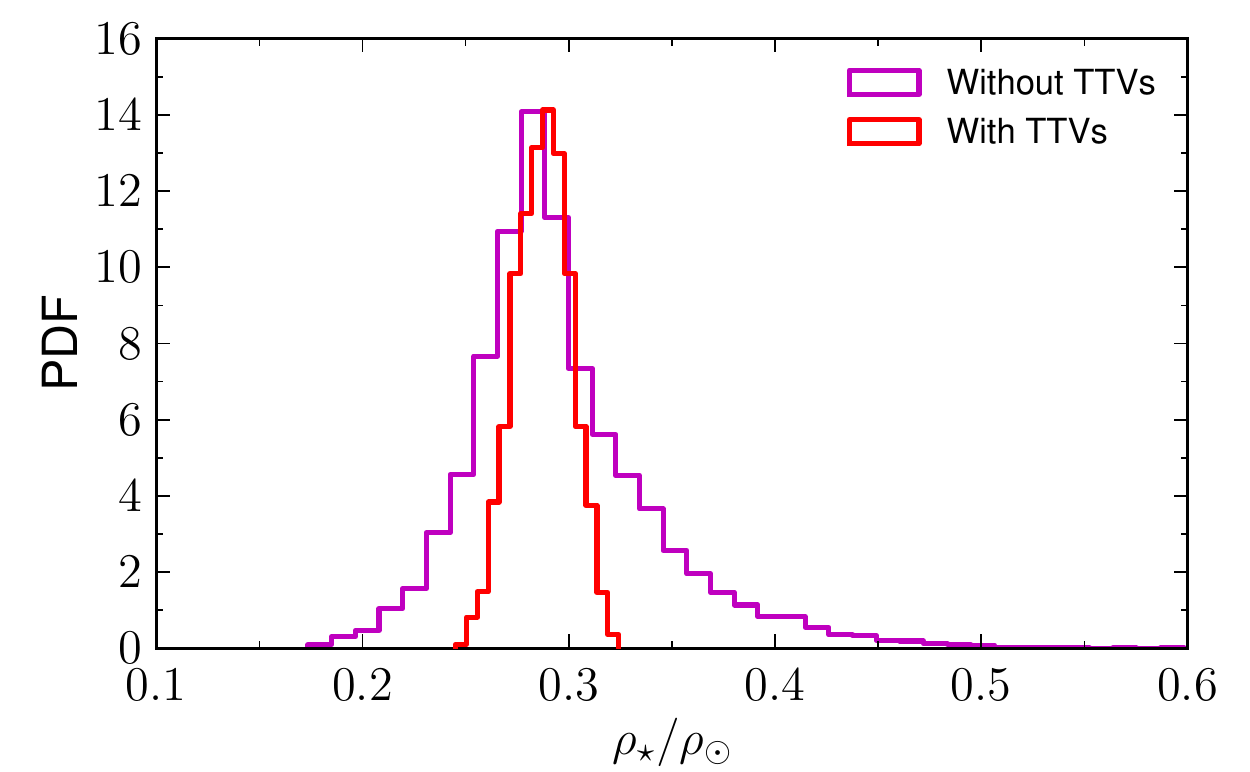}
\includegraphics[scale = 0.5]{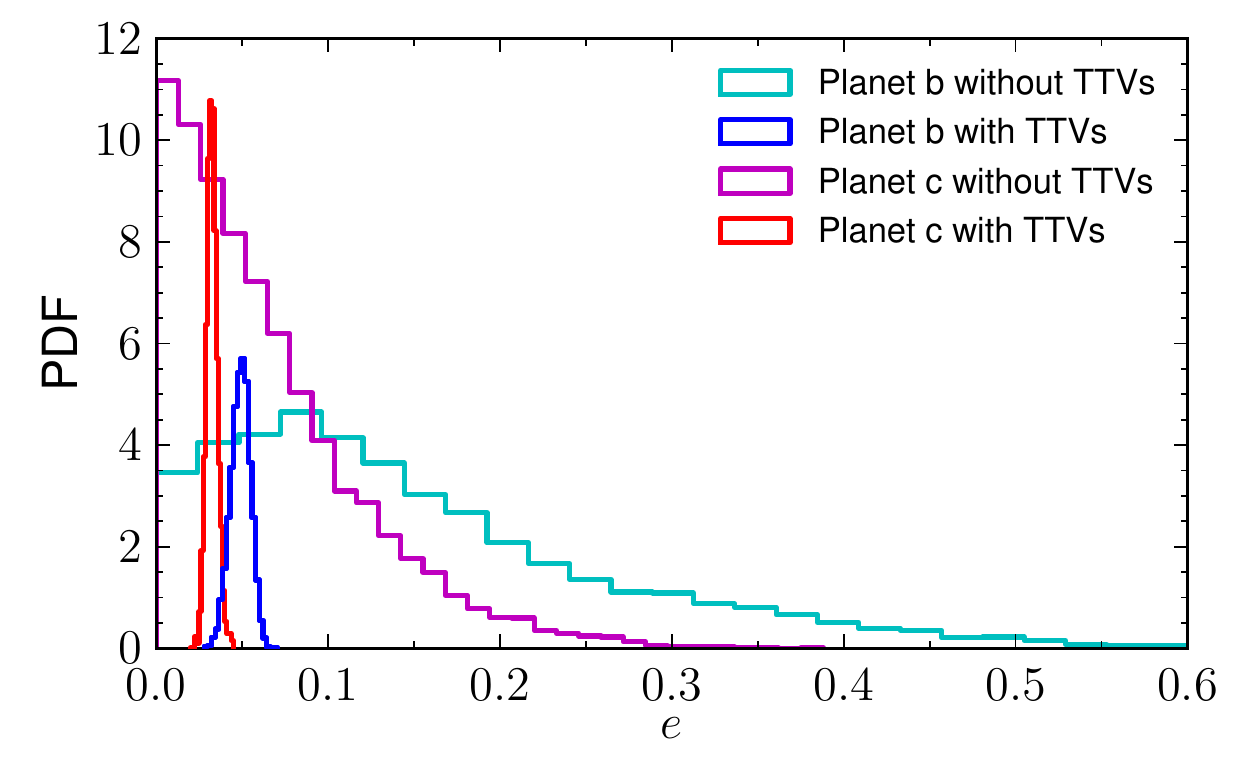}
\includegraphics[scale = 0.5]{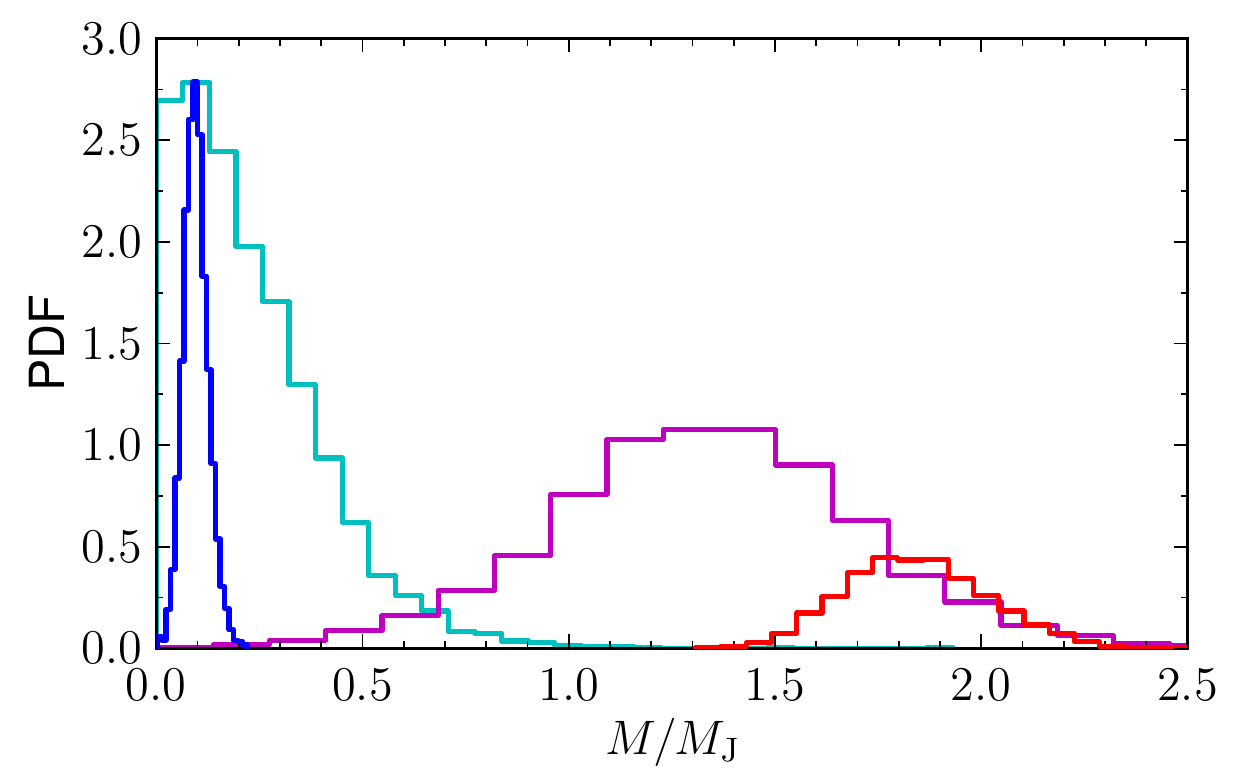}
\includegraphics[scale = 0.5]{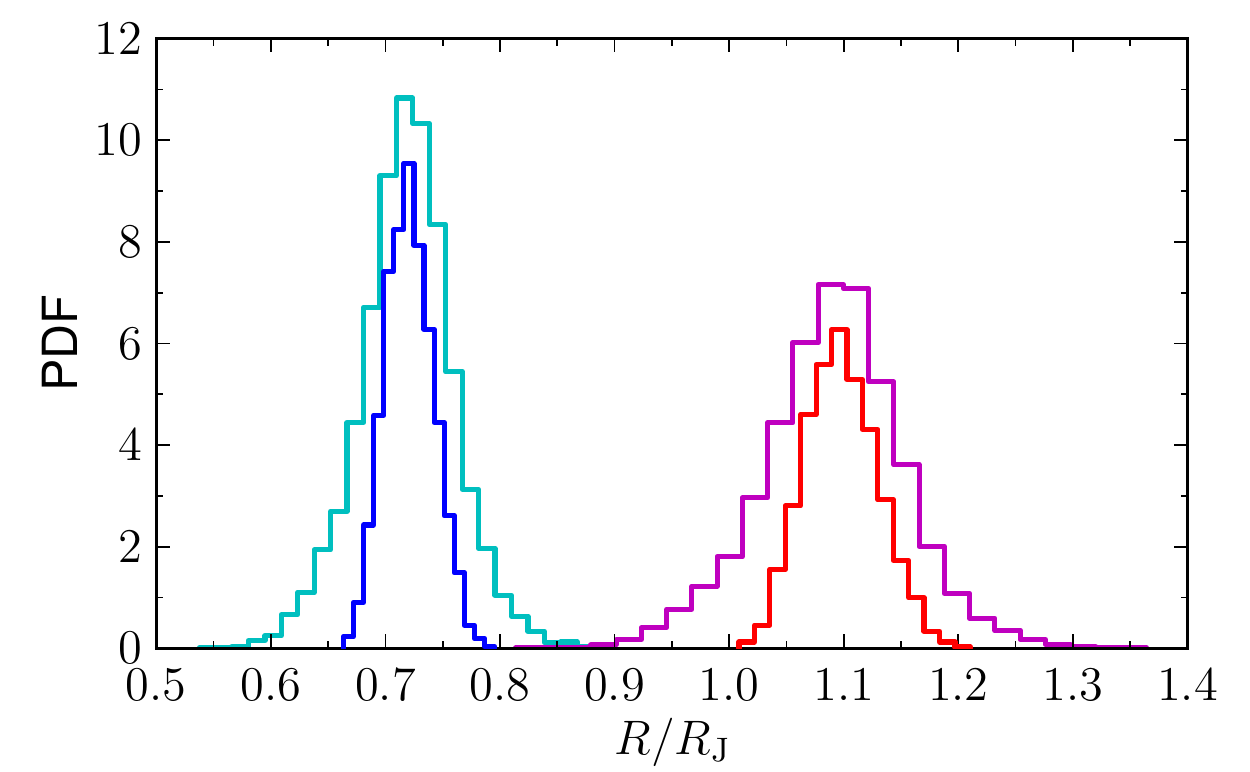}
\includegraphics[scale = 0.5]{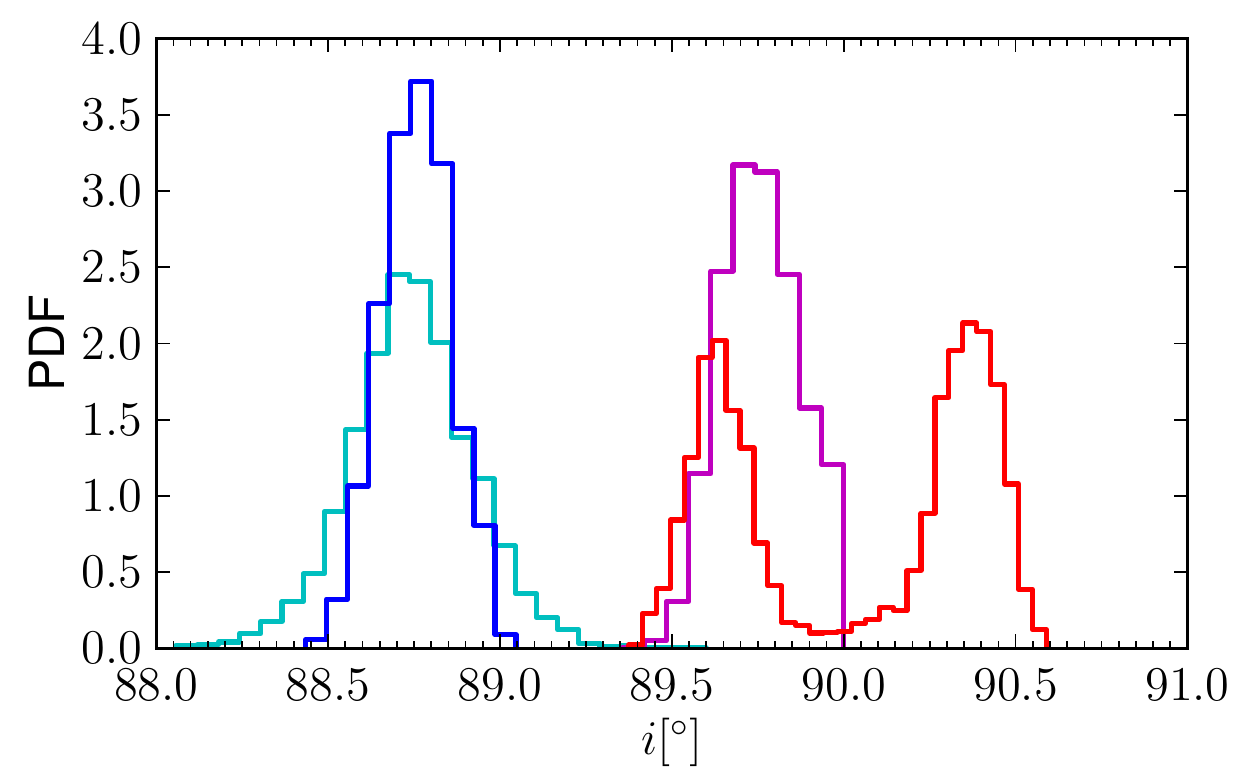}
\caption[Probability density function (PDF) of the system parameters]{\textit{Upper left:} probability density function (PDF) of the stellar density including or excluding the TTVs; to make the plot more readable, the PDF using the TTVs is divided by 2. \textit{Upper right:} the planetary eccentricities from different sets of data; the PDF using the TTVs is divided by 12. The color code is the same in the following plots. \textit{Central line:} planetary masses and radii from the fit with and without TTVs. The PDF of the masses using the TTVs is divided by 5, that of the radii by 2. \textit{Bottom:} orbital inclinations from different sets of data.}
\label{comparisons}
\end{figure*}

\begin{figure}[!hbpt]
\centering
\includegraphics[scale = 0.45]{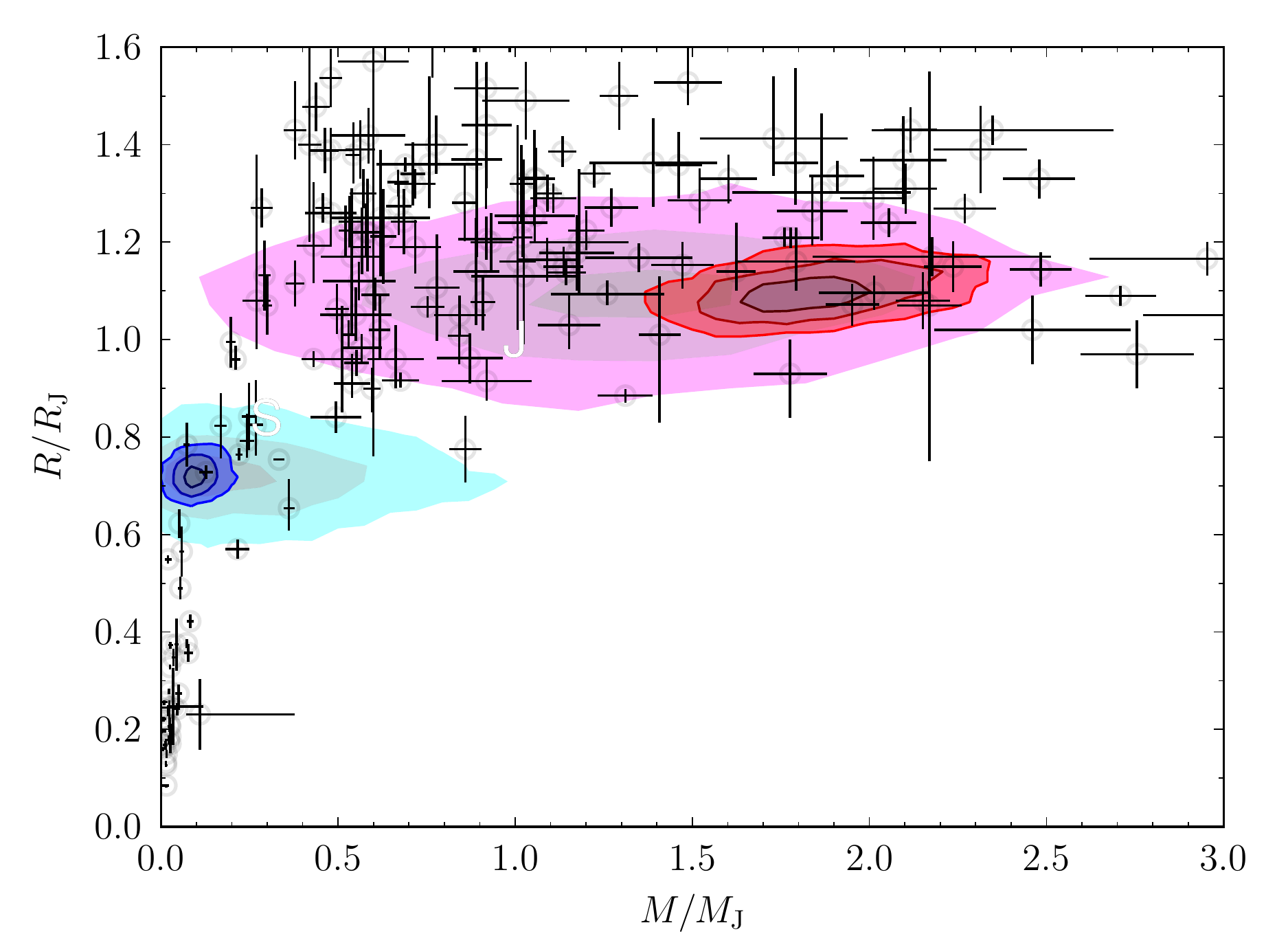}
\caption[Mass-radius diagram for the solutions without and with TTVs.]{Mass-radius diagram for the solutions without (cyan for planet b, magenta for planet c) and with (blue for planet b, red for planet c) TTVs. The blue and red solutions are those indicated in Table \ref{posteriors}. The colors, from the center to the edge of the regions, correspond to the 39.3\%, 86.5\%, and 98.9\% joint confidence intervals. Jupiter and Saturn (labeled J and S) are marked for comparison. The other planet parameters were taken from \cite{wright2011}.}
\label{mtor}
\end{figure}

The combined fit of the TTVs reduces the uncertainties on orbital separations, periods, eccentricities of the orbits and masses ratio of the planets. These parameters, in fact, determine the amplitude of the TTVs \citep{agol2005} and can be accurately measured.\\
To check the impact of the TTVs on the combined fit, we ran \pastis\ without them. The different posteriors of the most affected parameters, with or without the TTVs, are compared in figures \ref{comparisons} and \ref{mtor}. The mass of planet b presents the most important difference. The SOPHIE RVs, in fact, do not have the precision required to accurately measure this parameter, and only provide an upper limit. This is because the RV amplitude this planet produces ($6.5 \pm 2.1$ \ms) is close to the sensitivity of SOPHIE for a $\simeq 14.5$ mag$_V$ star (see section \ref{rvk}). Therefore, while without TTVs the RV fit gives to planet b an upper-limit mass of 0.28 \MJ\ at 68.3\% confidence level, with TTVs the mass is precisely measured, and the result is $0.094 \pm 0.033$ \MJ. The difference is smaller for planet c because the amplitude of the RVs is larger and better fitted. However, its uncertainty is roughly reduced by 40\%. This indicates that, when possible, including the TTVs in the combined fit is more effective than fitting them \textit{a posteriori}, using a set of orbital parameters derived without considering them.

The mass of planet c found with the RVs alone and with the TTVs are fully compatible. Therefore, the observed TTVs are completely explained by the two planets, within the data error bars. This is in agreement with a value of the RV drift compatible with 0 \kms, as found with the combined fit.\\
To determine the reliability of the fit, we ran a MCMC set without the RVs, as well. As expected, the posterior distributions are the same as with the RVs. Systems with low-mass planets presenting TTVs, which are challenging for the RV observations, would benefit from the approach used in this work.\\

Our fit allowed both hemispheres to be transited by planet c because the final inclination of its orbit is almost symmetric with respect to the stellar equator (figure \ref{comparisons}, bottom line).  The solutions with $i > 90^{\circ}$ are compatible with those without TTVs (all $< 90^\circ$) at $1\sigma$. In particular, for planet c, we found $89.64 \pm 0.10^\circ$ with TTVs and $i = 89.75 \pm 0.13^\circ$ without them.\\
Using the stellar inclination (section \ref{activity}) and the system parameters, we calculated the expected amplitude of the Rossiter-McLaughlin effects following equation 11 of \cite{gaudi2007}:
\begin{equation}
K_R \sim 0.3 K_O \left( \frac{m}{M_J} \right)^{-1/3} 
\left(\frac{P}{3~{\mathrm{days}}}\right)^{1/3} \left(\frac{v \sin i_\star}{5~\mathrm{\kms}}\right).
\end{equation}
In this equation, $K_R$ and $K_O$ are the stellar Rossiter-McLaughlin and orbital velocity amplitude, respectively, $m$ is the mass of the planet, and $P$ its period.
 For planet b, we found $10.9 \pm 3.0$ \ms\ and, for planet c, $79 \pm 13$ \ms. Measuring the spin-orbit misalignment would then be possible for planet c, but the transit duration ($\simeq 11$ hours) would require a joint effort from different locations to cover a whole transit.\\

\begin{figure}[htb]
\centering
\includegraphics[scale = 1.0]{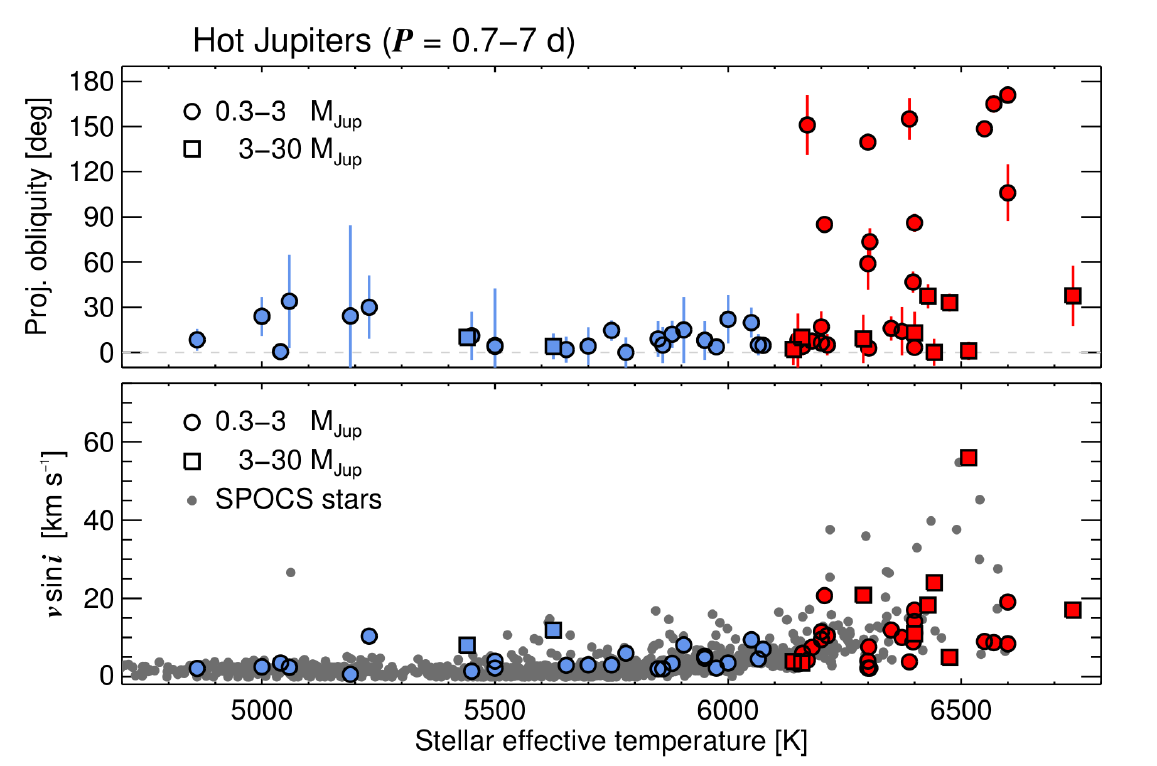}
\caption[Sky-projected stellar obliquity and \vsini\ versus \teff, for Hot Jupiters]{Sky-projected stellar obliquity and \vsini\ versus \teff, for Hot Jupiters with mass between 0.3 and 30 \MJ\ and period lower than 7 days. The colors separate the points on the left and on the right of $T_{\mathrm{eff}} =6100$ K. Grey dots indicate the \vsini\ of stars in the SPOCS catalog \citep{valenti2005spocs}. From \cite{winn2014}.}
\label{misal}
\end{figure}

In section \ref{activity}, the information coming from starspots and the stellar parameters was used to establish the alignment between the stellar equator and the planetary orbits. Verifying this property by other means would be particularly interesting for this system. In fact, its \teff\ (6150 $\pm$ 110 K) lies at the boundary between ``cool'' stars (\teff\ $ \lesssim 6100$ K), for which low obliquities are observed, and ``hot'' stars (\teff\ $\gtrsim 6100$ K)  showing a wider range of obliquities \citep{schlaufman2010,winn2010,dawson2014b}. The boundary at 6100 K coincides with the ``rotational discontinuity'', above which stars are observed to rotate significantly faster \citep{kraft1967}. A plot illustrating the behaviour of the projected obliquity of Hot Jupiters and the \vsini\ of their stars, as a function of the \teff\ of the host star, is presented in figure \ref{misal}. Cool stars have thick convective envelopes and radiative cores, causing stronger magnetic fields than in hot stars, which have a thin convective envelope and a small convective core. Observational trends suggest that magnetic fields cause magnetic brakes able to explain the difference in \vsini. The low obliquities, then, seem to be an effect of tidal forces. This picture is still not complete, and demands further modeling. \cite{valsecchi2014} and \cite{dawson2014b} tried to refine this explanation, modeling the depth of the convection zone and modeling tidal dissipation and magnetic braking at the same time.

With our model, we were able to constrain the low eccentricities of the obits ($\sim 0.05$ for planet b, and $\sim 0.03$ for planet c). An accurate measure of the eccentricities is important for testing the dynamical models of young systems with giant planets. Simulations show that the complex evolution of systems with two planets can be the result of ejection or merging in systems with three planets and can lead to stable, resonant, and low-eccentricity orbits \citep[e.g.,][]{lega2013}.\\
The difference between the resulting longitude of the ascending node $\Omega$ for planet b ($177.9 \pm 5.6^\circ$) and that of planet c (fixed to $180^\circ$) is compatible with $0^\circ$. Combined with the similar inclinations, this implies two almost coplanar orbits. As most of the Kepler planetary systems (section \ref{context}), Kepler-117 clearly has a flat configuration of the orbits.\\

\begin{figure*}[!htb]
\centering
\includegraphics[scale = 0.5]{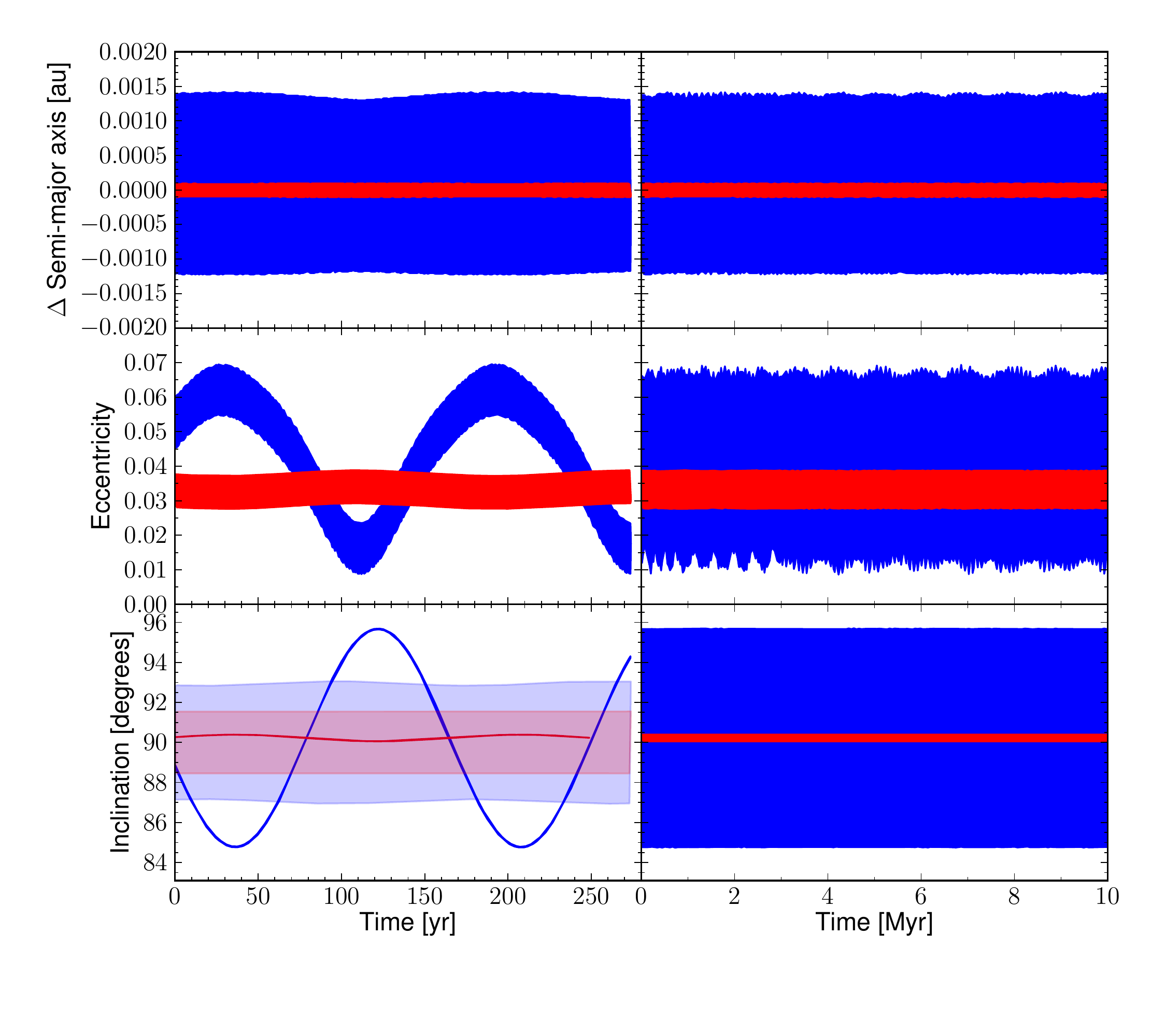}
\caption[Evolution of semi-major axes, eccentricities, and orbital inclinations over a 10 Myr simulation of the most probable solution]{Evolution of semi-major axes (top), eccentricities (center), and orbital inclinations (bottom) over a 10 Myr simulation of the most probable solution. The respective mean has been subtracted from the two semi-major axes. On the left, a zoom on the first $300$ yr; on the right, the variation intervals of the parameters. In blue planet b, in red planet c. The  shaded regions in the left panel of the inclinations correspond to the values resulting in a transit (see equation 7 of \citealp{winn2010}).}
\label{stab}
\end{figure*}

Finally, we verified that the configuration of the most probable solution is dynamically stable. We ran \mmercury\ over a time span of 10 Myr; the evolution of the semi-major axes, eccentricities, and orbital inclinations is plotted in figure \ref{stab}. These parameters oscillate over a time scale of around 200 years, but all the parameters are stable in the long term.

\subsection{A third non-transiting companion around Kepler-117?}

\begin{figure}[htb]
\centering
\includegraphics[scale = 0.4]{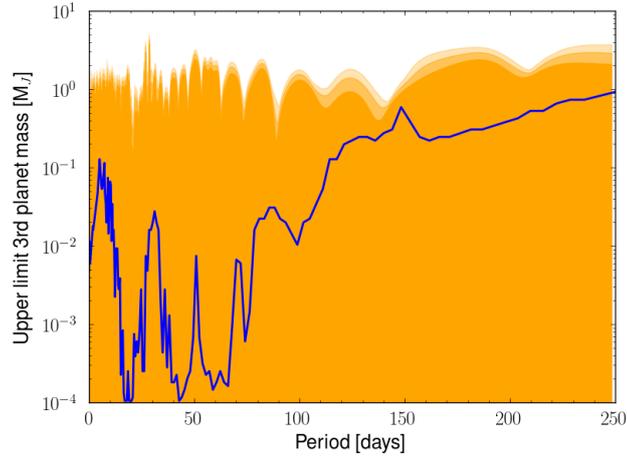}
\caption[Maximum mass of a third possible companion as a function of the orbital period]{Maximum mass of a third possible companion as a function of the orbital period, based on the RV (orange filled areas) and TTV (blue line) observations. The color, from darker to lighter, corresponds to the 68.27\%, 95.45\%, and 99.73\% confidence intervals from the RV data. The blue line represents a $3\sigma$ difference in the residuals on the TTVs from the best solution with two planets.}
\label{3rdlim}
\end{figure}

The possibility of a third non-transiting companion can be probed with the RVs and the TTVs. The absence of a systemic RV drift brings no evidence of the possible presence of a third non-transiting planet in the system. Moreover, the agreement between the mass of planet c, found with the RVs and with or without the TTVs, shows that the TTVs are not affected by a non-transiting body.\\
A more precise constraint can be obtained by subtracting the modeled RVs of the two planets from the RV measurements. We folded the residuals for several periods and fitted them with a sinusoid. The amplitude of the sinusoid and the mass of the star allow extracting the maximum mass of the possible companion. We increased the mass of the hypothetical third planet until a $3\sigma$ difference in the residuals between the fit of the TTVs with a third planet and the best solution with two was found.\\
The result is plotted in figure \ref{3rdlim} (filled regions) for the 68.27\%, 95.45\%, and 99.73\% confidence intervals. The RVs are compatible with the presence of a Jupiter-mass planet for some orbital periods. The TTVs, however, impose a stronger constraint, since including a third body with the combination of mass and period compatible with the RVs (with the simplifying assumptions of a circular orbit and $90^\circ$ inclination) would not fit the TTVs.\\
Therefore, under some simplifying assumptions, the presence of a non-detected third companion above $\sim 0.1$ \MJ\ on an orbit shorter than $\sim 100$ days, as well as that of a giant companion with an orbit shorter than $\sim 250$ days, is very unlikely.

\section{Conclusions and perspectives}\label{conclusions}

In this chapter, I presented our analysis of the Kepler-117 system and the results we achieved. Specifically, I have worked on the detailed analysis of photometric and spectroscopic data, on the implementation of the \mmercury\ code in \pastis\, and on the analysis of the results of the fit. The result of this work is the measure of the stellar, planetary, and orbital parameters of the system. In particular, the approach we developed showed that the inclusion of the TTVs in the combined fit allows tightly constraining the mass of the lighter planet, which was not precisely measured by the SOPHIE RVs. Taking into account the TTVs in the fit also allows for a better determination of the other system parameters. The planetary radii were precisely measured, as were the orbital eccentricities.\\
The measure of the planet masses with both RVs and TTVs adds important information to the controversy on the origin of the difference between the masses found with RVs and TTVs, studied by \cite{weiss2014} and discussed in section \ref{context}. Our analysis found the RV- and the TTV-measured mass of Kepler-117 c to be in agreement. The same comparison for planet b would benefit from a spectrograph with higher sensitivity than SOPHIE. In addition, under some simplifying assumptions, the TTVs almost exclude a non-detected $\sim 0.1$ \MJ\ companion with an orbit shorter than $\sim100$ days, as well as a giant companion with an orbit shorter than $\sim 250$ days. This fact, linked with the stability of the system that we proved for the next 10 Myr, is also another important information.\\

We cannot exclude the possibility that, while in this particular case the conditions are fulfilled for the combined fit to be effective, this is not the case in general. In fact, TTVs with sufficiently high amplitude are necessary. Kepler-117 exhibits significant TTVs even though the orbital period ratio of the planets is far from an exact low-order MMR, for which strong TTVs are expected \citep[e.g.,][]{lithwick2012,mardling2015}. 
Therefore, a deeper understanding of the dynamics of orbital resonances is needed to better reconstruct the history of Kepler-117, and to verify the range of applicability of our approach.\\

This system was object of a further study, carried out by \cite{almenara2015photo}. The Kepler-117 system was re-analyzed through a complete photo-dynamical model \citep{carter2011}. Such a model describes the light curve and the RV data at any time, accounting for the dynamic interactions with an N-body simulation. In principle, thanks to this technique, absolute masses and radii for the components of the system can be derived, without making use of stellar models for the determination of the stellar parameters. However, when only photometric data is available, such a derivation is not possible, because Newtonian modeling is invariant to scaling the lengths of the system by a factor and the masses by the cube of the same factor. An additional source of information is needed to break such a degeneracy. Either the time of arrival of light signals, due to the barycentric motion, or RV measurements can be used for this purpose. Alternatively, astrometric information such as the one from Gaia \citep{perryman2014} could set the scale of systems with massive planets. The light travel time is negligible for Kepler-117, because it induces differences $\lesssim 1$ s. Astrometric information is not available for this system. Therefore, the SOPHIE RVs were used.\\
The only information that is needed from RVs is the amplitude of the Doppler shift. This allowed the authors to derive the stellar radius and mass with a 20\% and 70\% precision, respectively. This precision is much lower than the one derived from the work presented here (3\% and 12\%, respectively: see sections \ref{pars} and \ref{constr}), because of the low precision of the SOPHIE RVs (with uncertainties $\gtrsim 30$ \ms). \citeauthor{almenara2015photo} showed that, with simulated RVs of 1 \ms-precision, a precision of 1\% and 2\% for stellar and planetary radii and masses, respectively, can be achieved. Therefore, they showed that photo-dynamical modeling can reduce the need of RV follow-up, of spectroscopic parameters and of stellar evolution models. This is particularly important for future missions such as PLATO 2.0 \citep{rauer2014}.

\section{Publication}
In the following, I attach the article that presents the result of the work described in this chapter. This article was published on \aap.\\

\cite{bruno2015}:\\
\url{http://www.aanda.org/articles/aa/pdf/2015/01/aa24591-14.pdf}

\clearpage
\chapter{Stellar activity}\label{chapactivity}
\minitoc
\bigskip

The increasing precision of the instruments notwithstanding, transit and radial velocity (RV) surveys are importantly limited by the signal coming from the brightness variations of the host stars. An important class of such variations is due to magnetic activity. During this PhD, I worked on the development of tools to disentangle the planetary and the activity-induced features in the data obtained through these kinds of surveys. Then, I applied the tools to three stars with different levels of activity: the Sun, CoRoT-7, and CoRoT-2.

\section{Context}\label{acticontext}
Magnetic activity is observed for a variety of stars that are cool enough to be either fully convective or to possess an outer convective envelope. Stellar magnetic fields in non-degenerate stars can have intensities from a few hundreds up to a few thousand gauss. These fields manifest via strong optical flares, coronal emissions \citep{pallavicini1981,walter1981}, chromospheric CaII and H$_\alpha$ emissions \citep[see section \ref{logrhk}]{vaughan1981,middelkoop1981,mekkaden1985}, ultraviolet line fluxes \citep{vilhu1984,simon1987}, radio emissions \citep{drake1989}, and periodic brightness variations. All these activity indicators are related to the rotation rate of the star \citep{skumanich1972}.\\
Brightness variations are due to local magnetic fields on the stellar surface, which can suppress the convective motions from the interior of the star to the surface. Regions that are cooler and darker than the average surface are produced. These regions are called starspots, and are coupled with bright structures, called faculae, that are visible in white light when they are close to the stellar limb. Hot spots can be formed, as well, that is spots brighter than the average stellar surface, but obeying the same limb darkening law as dark spots. Hot spots are visible across the whole stellar disk, oppositely to faculae\footnote{Hereafter, dark spots and faculae will be referred to as ``activity features''.}. All these active regions cross the stellar disk as the star rotates; they are distributed in groups, and vary in size, temperature, and position on the stellar disk along an activity cycle.\\
The most direct way to detect and study starspots is high-resolution spectropolarimetry, which allows one to reconstruct the topology of the magnetic field. This technique has been employed in particular for cool M and young stars \citep[e.g.][]{donati2003_2,donati2008,donati2008M} and Sun-like stars \citep{petit2008}. A variety of other techniques is used to study starspots: photometry, standard spectroscopy, interferometry, and microlensing. An exhaustive list of examples can be found in the review by \cite{berdyugina2005}.\\
More than 200 billion stars in our galaxy are estimated to be covered by starspots \citep{strassmeier2009}. These stars include: 
\begin{enumerate}[-]
\item M dwarfs, which constitute at least 80\% of the stellar population of the Milky Way \citep{kron1947,kron1952,chugainov1966,chugainov1971}; 
\item solar-type stars \citep{radick1982,radick1983b,radick1983a}; 
\item T Tauri stars, i.e.  pre-main sequence solar-mass stars, surrounded by disks of gas and dust remaining from their formation \citep{joy1945};
\item FK Com stars, i.e. G to K giants with \vsini\ between $\sim$ 50 and 150 \kms\ \citep{bopp1981,bopp1981_2,bopp1982}; 
\item close detached binaries such as RS CVn stars \citep{hall1976};
\item eclipsing binaries as W UMa stars (consisting of two solar-type stars with a common envelope and \vsini\ $\sim$ 100-200 \kms; \citealp{eggen1967,selam2004}) and Algols (consisting of a B to F main-sequence primary and a G to K evolved secondary; \citealp{hall1989}); 
\end{enumerate} 

\begin{figure}[!htb]
\centering
\includegraphics[scale = 0.83]{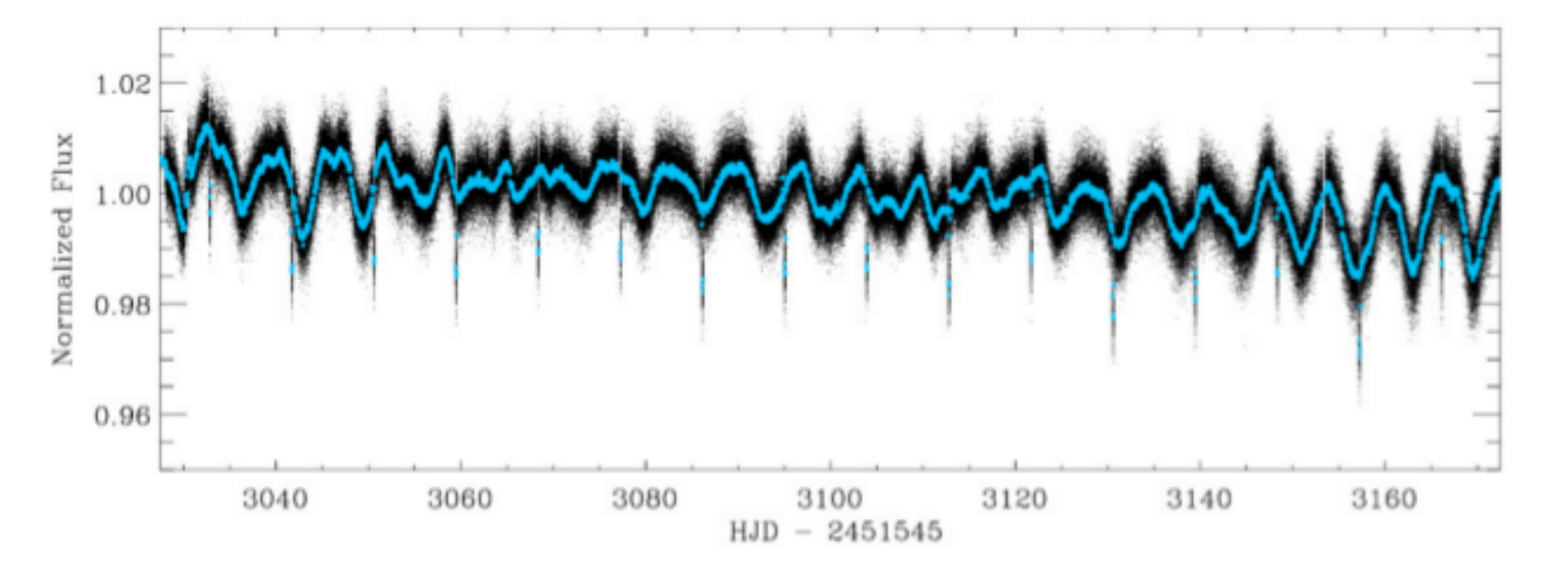}
\includegraphics[scale = 1.0]{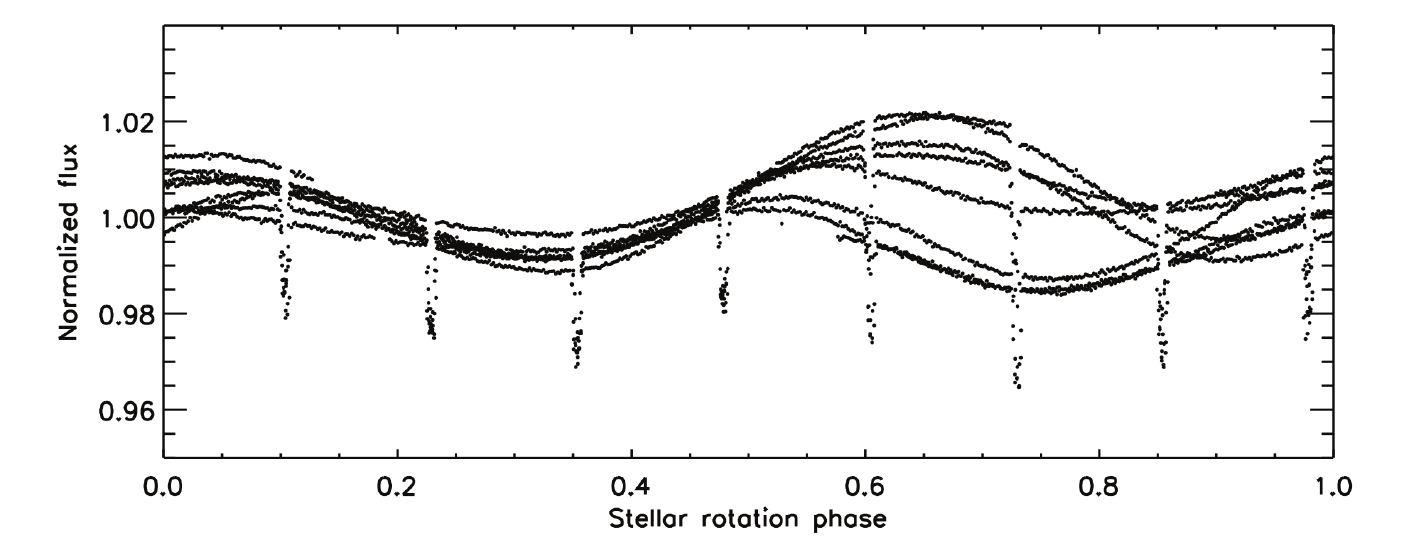}
\caption[The light curves of CoRoT-6 and Kepler-17]{$Top:$ the light curve of CoRoT-6. In black, the data binned every 32 s; in blue, the curve rebinned every 1 h. From \citealp{fridlund2010}. $Bottom:$ stellar phase folded normalized long-cadence light curves of Kepler-17 from Quarter 3. From \cite{desert2011}. The light curves show different levels of activity, rotation period, and differential rotation.}
\label{lc_comp}
\end{figure}

Despite the large number of active stars in our galaxy, only a few hundred spotted stars have been analyzed and their related observations published. Solar-type stars have local magnetic fields a few kilogauss strong; for these stars, the starspot phenomenon seems to be peaked in a \teff\ range between 4900 and 6400 K (that is, late F to early K-type). The characteristics of the activity pattern can vary widely for these stars. Figure \ref{lc_comp} shows the light curves of two active transited stars. Here, one can observe how the photometric variability can reach levels up to a few percent.\\

Magnetic activity is produced by one or several magnetic dynamo mechanisms, which were first proposed by \cite{babcock1961} to explain the solar activity cycle. A hydromagnetic dynamo is a flow that can generate and sustain the magnetic field against the effects that produce its decay \citep[for example, molecular and turbulent diffusion;][]{cattaneo1991}. It is based on a combination of rotation and convection in the stellar envelope. The modeling of a dynamo requires a numerical magneto-hydrodynamic (MHD) theory of two interface layers: the one between the convection zone of the star and its radiative interior, and the one between the stellar surface and the exterior.\\
\cite{durney1978} proposed a dimensional argument, according to which rotation can influence convection only if the convective motion is slow compared to rotation. This happens if 
\begin{equation}
v_{\mathrm{rot}} \gtrsim \frac{v_{\mathrm{conv}}}{l/R_\star},
\end{equation}
where $v_{\mathrm{rot}}$ is the rotational velocity, $v_{\mathrm{conv}}$ the characteristic convection velocity, $l$ the characteristic length scale of convection, and $R_\star$ the stellar radius.\\ 
Despite the information gathered up to now, the precise dynamo mechanism is still not precisely understood, not even for the Sun. It is not clear whether this mechanism affects stellar evolution.\\ 
Magnetic activity affects the formation of stars and of their planetary systems, as well. Indeed, it affects accretion from the protoplanetary disk to the star, as well as provokes collimated jets to be emitted from the star. This is the case for the so-called classical T Tauri stars \citep[e.g.][]{pudritz2006,bouvier2007}.\\
\cite{cuntz2000} proposed that close-in (semi-major axis $a < 0.15$ AU) massive planets interact magnetically with late type host stars. \cite{shkolnik2003} provided the first observational evidence of such an interaction, by observing the synchronous enhancement of CaII H and K emission with the short-period planetary orbit of HD 179949. In the visible band, the CoRoT space telescope provided observational evidence that close-in exoplanets trigger the emergence of magnetic fluxes from the host star, manifesting as starspots or faculae in the photosphere \citep[e.g.][]{lanza2011}. A variety of models were proposed to interpret these observations \citep[for an updated discussion, see e.g.][and references therein]{lanza2015spmi}. Further studies suggested an enhancement of coronal X-ray emission in stars hosting Hot Jupiters \citep[e.g.][]{kashyap2008,poppenhaeger2010}, although this remains a matter of debate \citep[e.g.][]{miller2015}. Possible star-planet magnetic interactions were also studied in the ultraviolet. \cite{vidal-madjar2003} and \cite{ehrenreich2008} were the first to find that stellar irradiation causes the evaporation of the planetary atmosphere, thanks to observations carried out with the Hubble Space Telescope in this band. Finally, the possibility of observing star-planet magnetic interactions in the radio domain was proposed, but confirmations are still lacking \citep[e.g.][and references therein]{see2015}.\\ 

\begin{figure}[!htb]
\centering
\includegraphics[scale = 0.38]{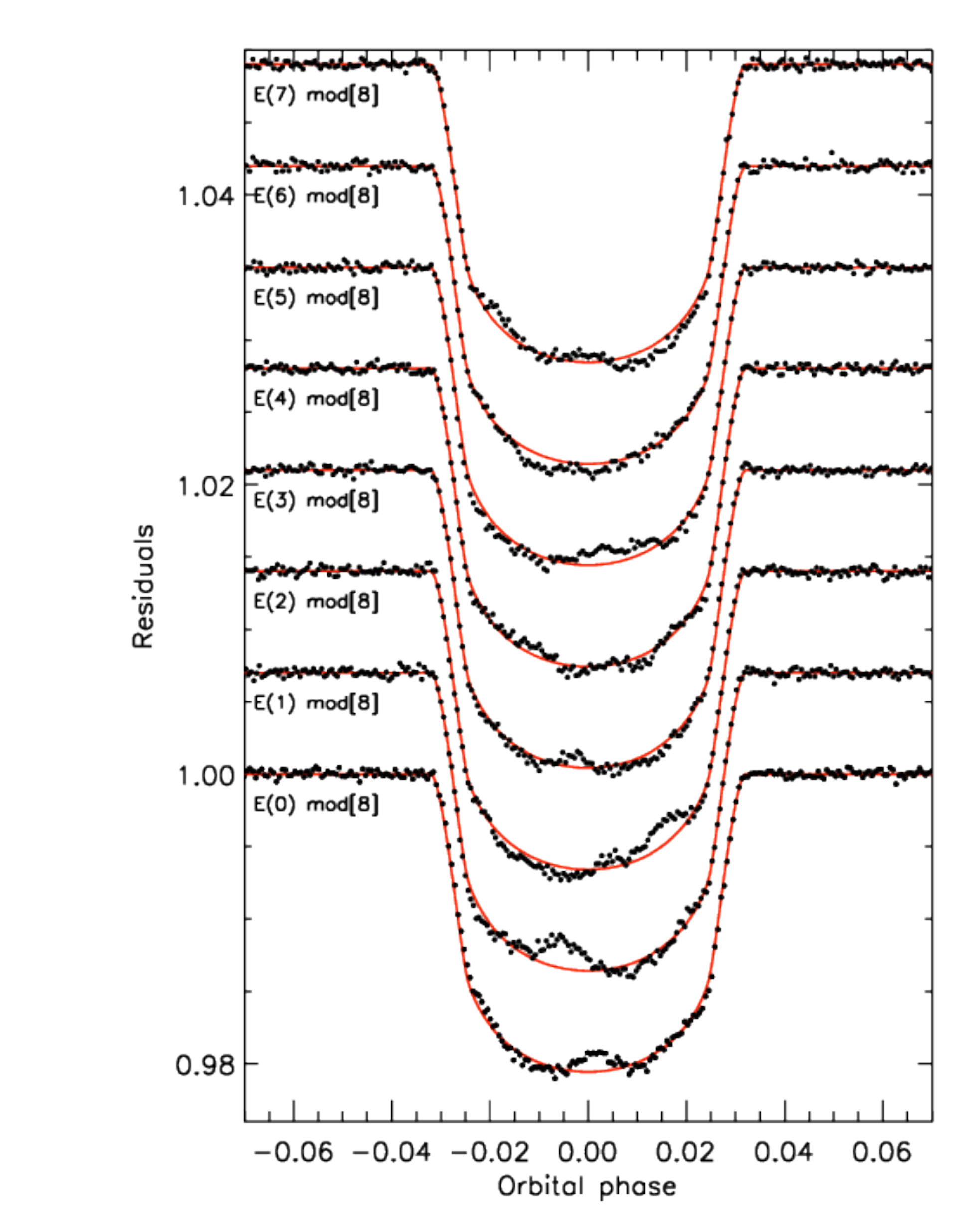}
\includegraphics[scale = 0.38]{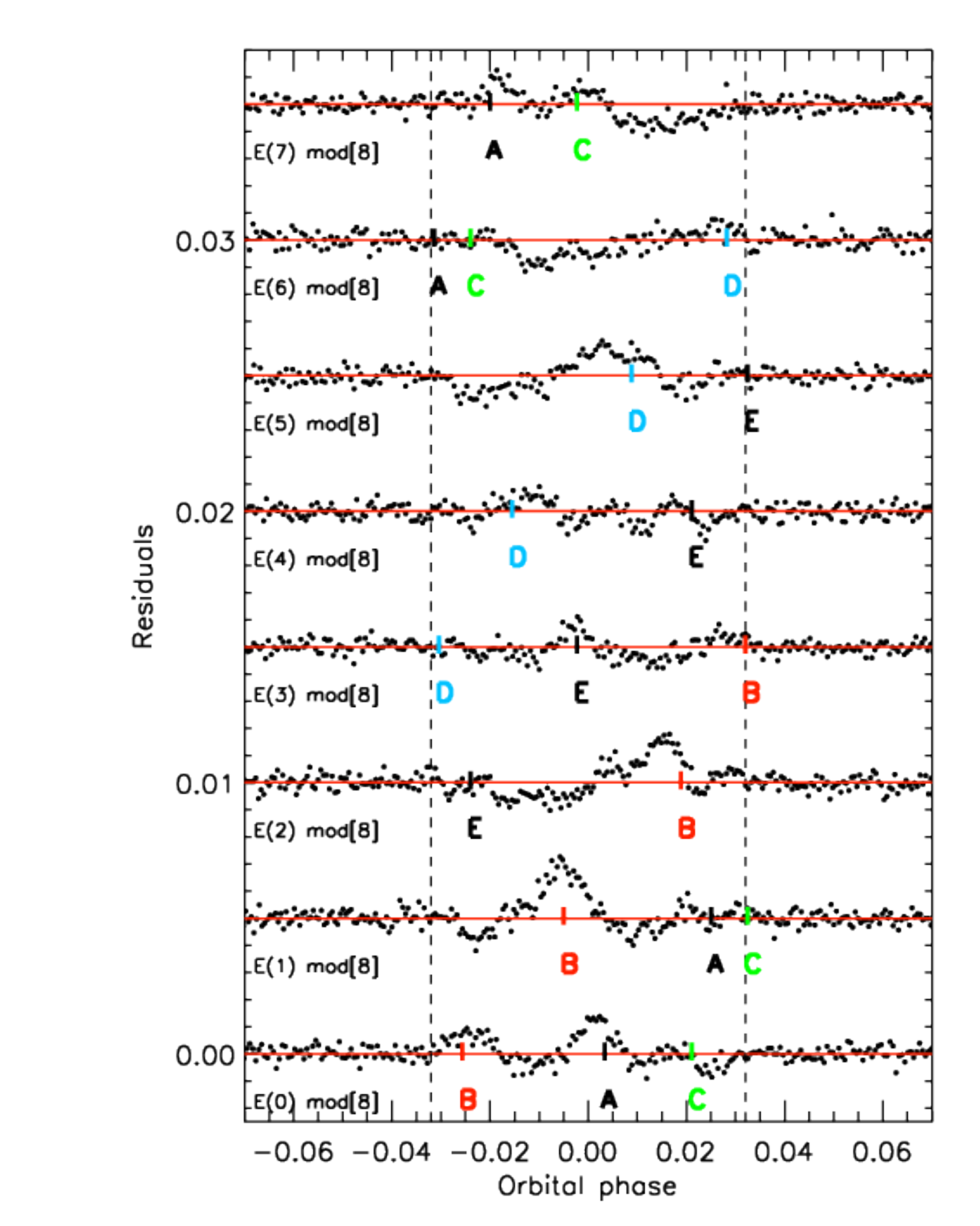}
\caption[Transits of Kepler-17 b]{$Left:$ Transits of Kepler-17 b, data (black dots) and best model (red line). Occulted spots are identifiable from the distortions of the transit profiles. $Right:$ residuals, highlighting different occulted spot, dubbed with different letters. The vertical dashed lines correspond to the beginning and to the end of the transits. From \cite{desert2011}.}
\label{desert}
\end{figure}

Starspots produce distortions in the light curve and in the RV signal. These variations affect the observables of both transit and RV methods, and therefore the derived planet parameters. In the following, I discuss the effects of spots on these two techniques.\\
Several approaches have been tested in order to disentangle the photometric signal produced by the planets and the one coming from the magnetic activity of the star. In these approaches, data is either corrected for some identified activity signature, or the activity features are masked.\\
Activity features, when they are not occulted by a planet during its transit, shape the light curve as figure \ref{lc_comp} shows.  \cite{czesla2009} showed that transit normalization is affected by non-occulted spots on the stellar disk. Indeed, transits are usually normalized according to the value of the flux immediately before and after each transit. Non-occulted spots modify the value of this flux, and therefore can introduce a time-dependent modulation of the normalized transit depth. In particular, spots darker than the average stellar surface, if not taken into account, induce an overestimate of the transit depth. \cite{czesla2009} and \cite{silva-valio2010} discussed how spots occulted during a transit act in the opposite way, producing an underestimate of the transit depth. When a planet eclipses a spot, the stellar flux increases, and the transit appears shallower. The effect is the opposite for eclipsed faculae. Figure \ref{desert} presents a case of spots occulted by a planet during its transit. The ``bumps''' inside the transits indicate the crossing of the planet over the activity features.

The smaller the planet is, the more these effects are important. To correct for them, \cite{czesla2009} proposed to adopt a different normalization than the standard one. The standard normalization consists in fitting a polynomial to the flux outside every transit. With \citeauthor{czesla2009} normalization, the flux modulations outside the transits are taken into account. After this, they propose to consider a lower envelope of the deepest transits as the closest one to the ``true'' transit.\\
The distortion in the transit profile does not only affect the planet size. Other parameters that are affected are (1) the stellar density, which starspots lead to underestimate \cite{leger2009}; (2) the limb darkening coefficients, that change from the tabulated values when the stellar surface is covered by spots \citep{csizmadia2013}; (3) the measured orbital periods, as starspots can induce apparent transit timing variations and transit duration variation \citep[e.g.][]{alonso2009,barros2013,oshagh2013}.\\

The high photometric precision of space-borne instruments has allowed for several attempts at stellar surface reconstruction using brightness variations in the light curves. Different approaches have been developed to model stellar spots. Some are based on analytical models \citep[e.g.][and references therein]{budding1977,dorren1987,kipping2012,montalto2014}, 
while others make use of numerical techniques \citep[e.g.][and references therein]{dumusque2014}. The fitting techniques also differ, going from the division of the stellar surface in segments on which a $\chi^2$ minimization is performed \citep[e.g.][]{huber2009}, to maximum entropy regularization \citep{rodono1995,colliercameron1997,lanza1998} and to modified Markov chain Monte Carlo (MCMC) algorithms to sample the spot parameter space \citep{tregloan-reed2015}. 
In particular, MCMC fitting has been proven to be an effective method to find best-fit values, uncertainties, correlations, and degeneracies for the photometric spot modeling problem \citep{croll2006}.\\
Light curve modeling is known to be an ill-posed inversion problem, because of the involved degeneracies between the many parameters. To reach convergence in the fit, strong constraints are imposed, or only part of the data is fitted \citep[e.g.][]{lanza2009,huber2009}. This is done to limit the influence of the activity evolution, which often is not modeled.\\

\begin{figure}[!bth]
\centering
\includegraphics[scale = 1.0]{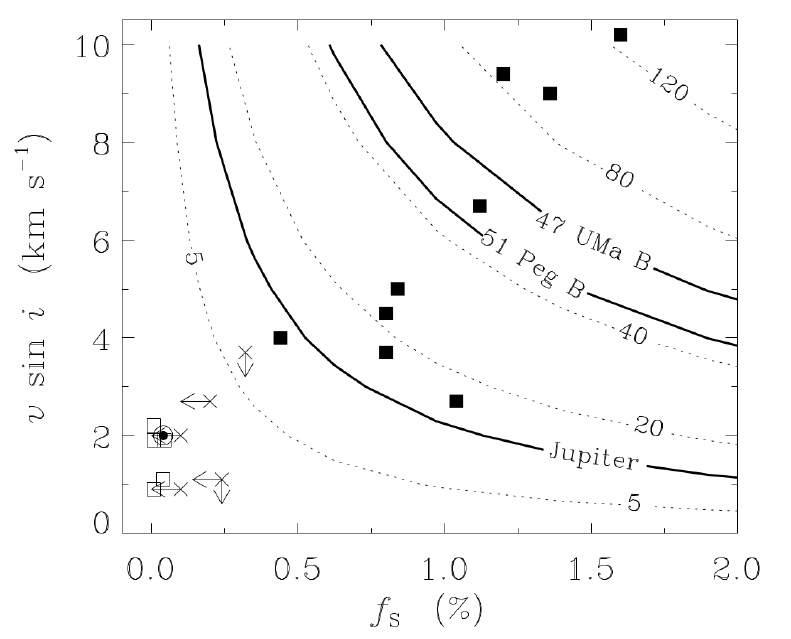}
\caption[Contours in \ms\ of expected RV shifts for several G-type stars]{Contours in \ms\ of expected RV shifts for several G-type stars (dots, arrows). The contours are function of the spot area (in percentage of the stellar surface) and the stellar \vsini. From \cite{saar1997}.}
\label{saar}
\end{figure}

In spectroscopy, starspots induce spectral line shape variations which depend on the \vsini\ of the star, the temperature of the spots, and the spectrograph resolution \citep{saar1997,hatzes1999,desort2007}. These distortions produce a time-varying jitter in the measured radial velocities (RVs) which hampers orbital parameter determination. The signal increases linearly with the spot filling factor, going form 1 \ms\ for a size comparable with large sunspots up to 20 \ms\ for filling factors of 1\% \citep{hatzes1999}. Figure \ref{saar} presents the expected RV shifts due to equatorial spots on different G stars, as a function of the spot size and \vsini\ of the star. 

\begin{figure}[!bth]

\centering
\includegraphics[scale = 0.7]{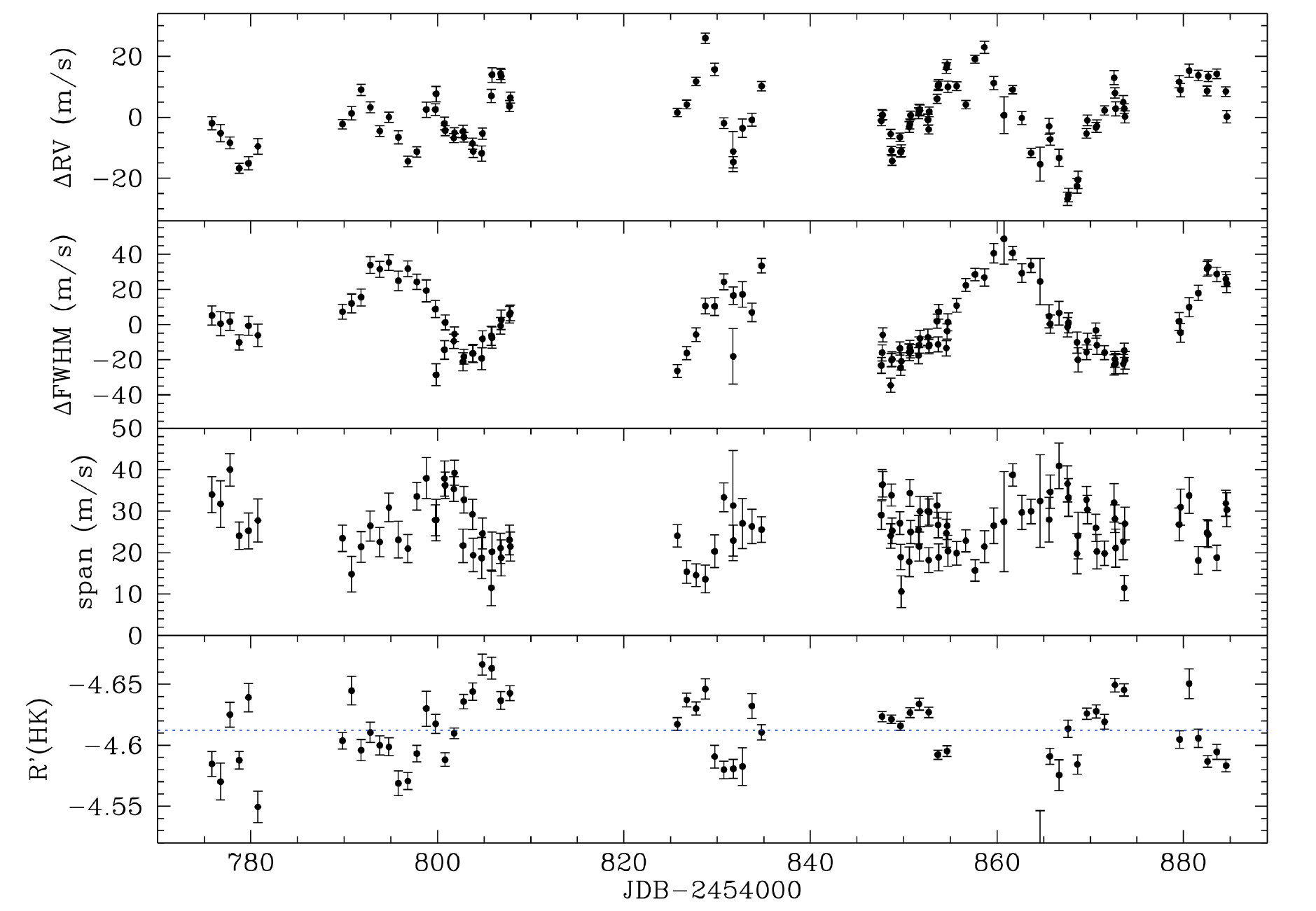}
\caption[HARPS RV observations of CoRoT-7]{HARPS RV observations of CoRoT-7 from November 2008 to February 2009, compared with other spectroscopic indicators. From \cite{queloz2009}.}
\label{queloz}
\end{figure}

\cite{desort2007} quantified the RV impact of starspots on stars of different spectral type and rotational velocity. In particular, they showed that spots covering 1\% of the stellar surface can mimic RV signals of short-period giant planets around G-K type stars with \vsini\ lower than the spectrograph resolution. Hence, starspots can even induce fake planetary signals, if their imprint on the RV signal is constant in time \citep{queloz2001,boisse2011,robertson2014}.\\
Several diagnostics are employed to detect the effect of stellar activity in RV measurements. The analysis of the bisector of the cross-correlation function (CCF) of the stellar spectrum with numerical masks and the full width at half maximum (FWHM) of the CCF is used to reveal activity-induced false positives \citep{queloz2001}. In these diagnostics, one looks for correlations between these indicators and the RV of the star: if correlations are found, the planet scenario is likely to be rejected in favour of a blend (see section \ref{rvk}). If a planet is actually present, the main derived parameter to be affected is its mass \citep[e.g.][]{queloz2009}. Indeed, the activity-induced Doppler shifts can be larger than the planetary features. Figure \ref{queloz} shows the RV pattern of an active star compared with the FWHM of the CCF, the bisector span, and the chromospheric activity index $R'_{hk}$. 

Because of these difficulties, active stars are often excluded from RV surveys. 
However, transit surveys include active stars. Therefore, it is important to develop efficient methods to disentangle the activity-induced RV shifts from the planet-induced ones, in order to measure the planet's mass. One of the proposed approaches is the removal of anti-correlations between RVs and bisector span \citep{melo2007,boisse2009}. Another approach consists in the adjustment of two Keplerian signals to the data, one of which has the stellar rotation period \citep{bonfils2007}. Similar to this, a third method proposes to adjust three sinusoids to RVs, one with the rotational period of the star $P_\star$ and two with its two first harmonics, at $P_\star/2$ and $P_\star/3$ \citep{boisse2011}. This technique can remove up to 90\% of the activity signal, but requires the knowledge of the stellar rotation period. With spectroscopy, one can only measure this parameter as a function of the stellar radius and axis inclination, i.e. with the \vsini; instead, the photometric signature of starspots allows it to be directly constrained.\\
The measure of stellar rotation is particularly important, as it is an indicator of the evolutionary state of a star. Indeed, stars are supposed to spin down with age \citep{skumanich1972}, because of angular momentum loss due to the stellar wind \citep{weber1967}. Moreover, stellar activity is a tracer of star-planet magnetic interactions (see section \ref{contrhk}). 

\section{Modeling tools}

\subsection{Photometry}\label{photmod}
Spot fitting is affected by several degeneracies and correlations \cite[e.g.][]{lanza2007,desert2011,kipping2012}. The main correlations characterizing this problem are those between the spots size and contrast (equivalent to their temperature), their longitude and evolution times, and the inclination of the stellar spin axis and the spot latitudes \cite[e.g.][]{lanza2009,desert2011,kipping2012}. If the stellar spin axis is perpendicular to both the planetary orbit and the line of sight, the only latitudinal information that can be recovered is whether a spot belongs to the belt occulted during transits or not. Without severe constraints on the spot parameters, an unique solution cannot be obtained; instead, various local minima of the $\chi^2$ space are found. However, our main objective was not surface reconstruction, but correction for the spot effects in the fit of the planet parameters. Moreover, the more realistic the model is, the more it is likely that the local minimum the fit finds is close to the global minimum. We chose not to fix any of the spot parameters, and to explore the parameter space around such a minimum, to more consistently fit the transits.\\
MCMC is an efficient method to tackle problems affected by correlated parameters. Therefore, we opted to implement photometric modeling codes in an MCMC algorithm. We adopted the one of \pastis, introduced in section \ref{ttvsmeth}. MCMC fitting for spot modeling incurs in the problems of non-uniqueness of the solution and degeneracy of the parameter space. These issues are described, together with the approaches adopted to deal with them, in sections \ref{lcsun} and \ref{evolution}. There, the discussion makes reference to the data sets analyzed for this work.\\

One of the main issues with Markov chains is the computation time. In fact, some $10^5$ or even $10^6$ iterations are often required for the chains to stabilize around a solution \citep[e.g.][]{lanza2014rot}, and to reach a stationary likelihood. 
Numerical methods define a large-resolution two-dimensional grid of the stellar surface or of its projection onto a plane, and numerically integrate over	two coordinates to obtain the light curve. This integration is computationally expensive. Numerical methods can deal with a large set of spot configurations; in principle, they allow one to compute any starspot shape, flux profile, and limb darkening law. Instead, analytic methods use analytic formulae to derive the light curve, and do not require numerical integration: therefore, they are order-of-magnitudes faster to execute. They require approximations on the spots shape. However, flux variations in the light curve represent a disk-integrated snapshot of a star, where surface features are unresolved; therefore, adding many parameters to describe complex spot shapes does not improve the quality of the fit, and a circular approximation of the spots shape works reasonably well \citep{kipping2012}. For this reason, and in order not to increase dramatically the computation time, we preferred to implement analytic over numerical methods.\\
We chose two analytic codes, which were implemented in the MCMC algorithm. Below, the codes and their main characteristics are presented.
\begin{enumerate}[-]
\item \textbf{\macula} \citep{kipping2012} is an analytically-based code, which calculates the photometric variations while activity features cross the stellar disc and evolve in size. It is based on the formalism of \cite{budding1977} and \cite{dorren1987}. While these models adopt a linear limb darkening law, \macula\ uses a four-coefficients limb darkening law \citep[as described by][]{claret2011}. A different limb darkening law can be set for the star and the features. An important reduction of the computation time is obtained through the use of a small-spots approximation ($\sin \alpha \lesssim 0.1$, where $\alpha$ is the angular size of the spot), which allows one to consider the brightness of the stellar surface as constant under the disk of a spot. This brings to a simplification of the equations.\\
Each spot is described by its coordinates (longitude and latitude), its angular radius $\alpha$ (hereafter, also referred to as its ``size''), and its contrast flux ratio with respect to the stellar surface. Differential rotation, configurations of umbra and penumbra, and the size evolution of the features are modeled, as well. Their shape is assumed to be the one of a spherical cap. The size evolution in time is modeled according to a linear growth-and-decay law, that is reported here:
\begin{equation}
\begin{split}
\frac{\alpha_k(t_i)}{\alpha_{\mathrm{max}, \, k}} & = \mathcal{I}_k^{-1}[\Delta t_1 \mathsf{H}(\Delta t_1) - \Delta t_2 \mathsf{H}(\Delta t_2)] \\
					 & - \mathcal{E}_k^{-1}[\Delta t_3 \mathsf{H}(\Delta t_3) - \Delta t_4 \mathsf{H}(\Delta t_4)], 	
\end{split}
\label{kip1}
\end{equation}
In this equation, for each spot $k$, $\alpha_k(t_i)$ represents the angular size at time $t_i$, $\alpha_{\mathrm{max}, \, k}$ is the maximum size, $\mathcal{I}_k$ and $\mathcal{E}_k$ are the ingress and egress times (i.e. time elapsing from appearance to the reaching of maximum size, and vice versa), and $\mathsf{H}$ indicate the Heaviside step function. $\Delta t_1$, $\Delta t_2$, $\Delta t_3$, and $\Delta t_4$ are defined by
\begin{equation}
\begin{aligned}
\Delta t_1 &= t_i - t_{\mathrm{max}, \, k} + \frac{L_k}{2} + \mathcal{I}_k,\\
\Delta t_2 &= t_i - t_{\mathrm{max}, \, k} + \frac{L_k}{2},\\
\Delta t_3 &= t_i - t_{\mathrm{max}, \, k} - \frac{L_k}{2},\\
\Delta t_4 &= t_i - t_{\mathrm{max}, \, k} - \frac{L_k}{2} - \mathcal{E}_k,
\end{aligned}
\label{kip2}
\end{equation}
where $t_{\mathrm{max}, \, k}$ is the time at which the spot $k$ reaches its maximum size and $L_k$ is the time during which the spot keeps its maximum size, $\alpha_{\mathrm{max}}$.\\
Therefore, four parameters are added to the description of every spot: $t_{\mathrm{max}}$, $L_k$, $\mathcal{I}_k$, and $ \mathcal{E}_k$.\\

The main limitation of \macula\ is due to the fact that the small spots approximation means that the model is limited to non-overlapping spots and faculae, as well as to features smaller than $\simeq 10^\circ$ in angular radius, above which the approximations have a non-negligible effect \citep{kipping2012}. Moreover, planetary transits are not included. The transit counterpart of \macula\ came with \texttt{spotrod} \citep{beky2014}, which operates in a semi-analytic way. However, the very short computation time required by \macula\ is its main advantage. This made it the preferred choice whenever the modeling of transits and occulted features was not needed.
\item \textbf{\ksint} \citep{montalto2014} is a code which calculates analytically a light curve presenting the signature of both planetary transits and activity features. The transits are modeled with the formalism of \cite{pal2012}. 
The star is characterized by a quadratic limb darkening law, which is the same as for the activity features. This code does not use the small-spot approximation, and therefore does not suffer of restrictions about the size and the overlap of spots and faculae. The features coordinates, size, and contrast are modeled as in \macula. The evolution of the features as a function of time, however, is not included.\\
Therefore, I introduced a simple law for spots and faculae evolution in the code. I used a linear variation of the angular size, as in \macula. I followed the prescription of \cite{kipping2012}, reported in equations \ref{kip1} and \ref{kip2}. To do this, the size parameter was translated into the maximum size reached by a spot during its evolution, $\alpha_{\mathrm{max}}$. Then, four parameters were added to the description of every spot. Using the same formalism as \macula, these are $t_{\mathrm{max}}$, $L_k$, $\mathcal{I}_k$, and $ \mathcal{E}_k$.\\
In its current version, \ksint\ needs to read files in input and write files in output. Instead, \macula\ does not need to read or write any file. This difference, together with the absence of small spot approximation in \ksint, makes the computation time required by \ksint\ larger than the one required by \macula. Therefore, \ksint\ was preferred only when the modeling of transits and occulted features was needed.
\end{enumerate}
Before the implementation in \pastis, these two codes were compared to the numerical code \soap\ \citep{boisse2012,oshagh2013,dumusque2014}. This code computes the light curve and the RV signatures of a distribution of activity features and a planet, whether transiting or not. It is based on the division of the stellar disk in a grid; every cell is assigned a flux value and a cross-correlation function (CCF). All the cells are integrated to calculate the light curve and the total CCF of the star. The contribution of each cell is weighted by the quadratic limb darkening law assigned to the star. Both spots and faculae are characterized by the same limb darkening law as the star. Because it does not use the small spot approximation, \soap\ allowed me to check the range of parameters for which I could safely use the analytic methods.

\subsection{Radial velocities}\label{rvmod}

The final purpose of this work was to develop a method for the consistent fit of activity-induced features in both photometry and RV data. For the same reasons discussed in the previous section, therefore, we selected an analytic method, with the plan of implementing it in an MCMC algorithm. This has not been possible, yet. However, a preliminary study has been carried out and a code for RV modeling has been developed and tested on observational data. Both of these stages are discussed here.\\
The method we chose was presented by \cite{aigrain2012}, and is a simple analytic model to calculate the RV-induced shift from the light curve. The method is called \ff, because it involves the use of the integrated flux ($F$) and of its time derivative ($F'$). \ff\ relies on the hypothesis of small spots and ignores limb darkening. The small spots hypothesis is similar to the one of \macula. In this way, complex shapes for the spots, their projection on the stellar disk, as well as the case of overlapping spots, do not need to be considered. The second hypothesis does not affect dark spots, but introduces uncertainties in the modeling of bright faculae, which should not be included in the modeled photometry, as they follow a different limb darkening law than dark spots\footnote{Hereafter, I will use $faculae$ and \textit{hot spots} indifferently. The codes for spot modeling described in this chapter, actually, model hot spots, and not faculae.}.\\
Dividing the activity-induced Doppler shift $\Delta RV$ in a component due to stellar rotation, $\Delta RV_{\rm{rot}}$, and one coming from convection, $\Delta RV_{\rm{c}}$, \cite{aigrain2012} proposed 
\begin{equation}
\Delta RV(t) = \Delta RV_{\rm{rot}}(t) + \Delta RV_{\rm{c}}(t).
\end{equation}
In the following, $\Psi(t)$ indicates the value of the flux at time $t$; $\Psi_0$ the flux in the absence of spots; $\dot{\Psi}$ the flux time derivative; $R_\star$ the stellar radius; $f$ the expression $2(1-c)(1- \cos \alpha)$, where $c$ is the contrast of a spot; $\delta V_{\rm{c}}$ the difference between the convective blue-shift in the unspotted photosphere and that within the spot-covered area; $\kappa$ the ratio of the unspotted-to-spotted area. \cite{aigrain2012} found
\begin{equation}
\Delta RV_{\rm{rot}}(t) = - \frac{\dot{\Psi}}{\Psi_0} \left[  1 - \frac{\Psi(t)}{\Psi_0} \right ] \frac{R_\star}{f}
\label{ff1},
\end{equation}
and
\begin{equation}
\Delta RV_{\rm{c}}(t) = \left [1 - \frac{\Psi(t)}{\Psi_0} \right ]^2 \frac{\delta V_{\rm{c}} \kappa}{f}.
\label{ff2}
\end{equation}
As it can be noticed, an important advantage of \ff\ is that it does not require the knowledge of the stellar rotation period.\\
Even if \citeauthor{aigrain2012} tested \ff\ on light curves showing the impact of many active regions, the mathematics of this method are essentially based on the case of a single spot. Using the same approach for many spots at the same time is not formally justified (Lanza, private communication), and another strategy can be suggested, once a light curve model is calculated. One can compute the RV contribution of each modeled spot considered separately, and then add all the contributions.\\
\cite{haywood2014} proposed to introduce linear coefficients for expressions \ref{ff1} and \ref{ff2}, plus an additional term, so that 
\begin{equation}
\Delta RV(t) = A \Delta RV_{\rm{rot}}(t) + B \Delta RV_{\rm{c}}(t) + \Delta RV_{\rm{additional}}.
\label{fffinal}
\end{equation}
This formulation is based on their study on the light curve and simultaneous RVs of CoRoT-7, employing a Gaussian process \citep{rasmussen2006,gibson2011}.\\
As for the photometric models, the output of the analytic method \ff\ was compared to the one of a numerical code. The numerical code \soap\ (section \ref{photmod}) was again used for this purpose. In this way, I ascertained that my implementation of the \ff\ method worked as expected on a number of simple configurations.\\
It is important to remark that \ff\ can be applied to fit only simultaneous photometry and Doppler observations. Very few cases fulfill this condition: it is the case of the data sets of the Sun and CoRoT-7 which are analyzed in the following sections.

\section{Test on the Sun}\label{maculasun}
\macula\ and \ff\ were tested on the data of the star we best know: the Sun. The Sun is a weakly active star, with $\log R'_{\mathrm{HK}}\simeq -4.81$ \citep{noyes1984}, a rotation period of $\sim 25$ days, and activity cycle $\sim 11$ years. The data set was provided by A. M. Lagrange and N. Meunier, from the Institut de Plan\'etologie et d'Astrophysique de Grenoble\footnote{UJF-Grenoble 1 / CNRS-INSU, Institut de Plan\'etologie et d'Astrophysique de Grenoble (IPAG) UMR 5274, Grenoble, F-38041, France.}. I could take advantage of the simultaneous photometry and RV of the Sun, extracted from the data sets used by \cite{meunier2010}. These data sets were generated from the magnetic features extracted from the SoHO/MDI magnetograms \citep{scherrer1995}. 
The data sets consist of two periods, corresponding to different phases in the activity cycle of the Sun: low activity (from April 24, 1996 to April 1, 1997) and high activity (from February 2, 2000 to November 1, 2000).\\

\begin{figure}[!htb]
\centering
\includegraphics[scale = 0.5]{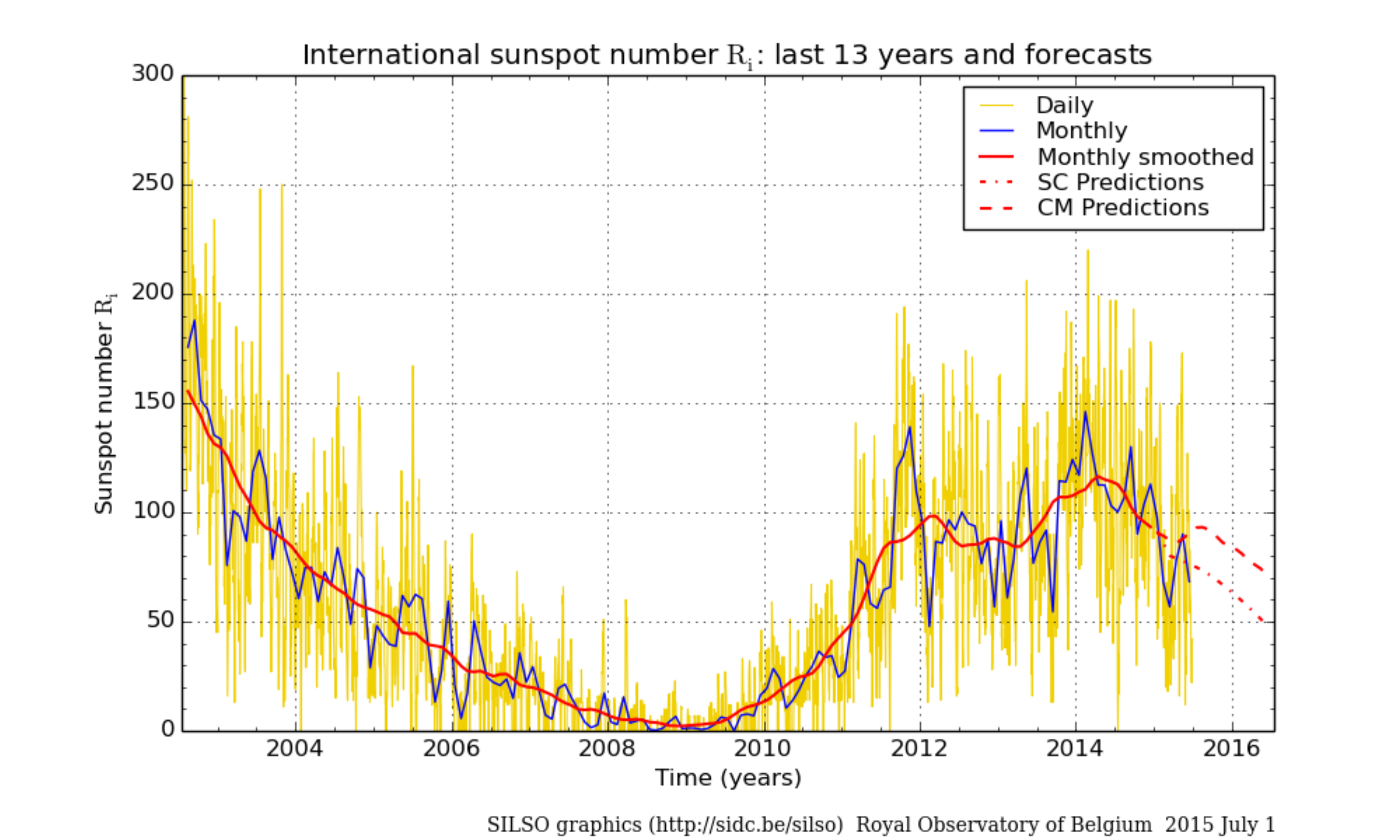}
\caption[Daily sunspot numbers]{Daily sunspot number (yellow), monthly mean sunspot number (blue), smoothed monthly sunspot number (red) for the last 13 years and 12-month ahead predictions of the monthly smoothed sunspot number. From the website of the Royal Observatory of Belgium.}
\label{spots_number}
\end{figure}

Low and high activity periods correspond to a completely different number of spots covering the solar surface. Figure \ref{spots_number} represents the number of sunspots observed during the last 13 years, in periods of different solar activity, at the Royal Observatory of Belgium\footnote{http://www.sidc.be/silso/dayssnplot.}. 

Figure \ref{spots_number} clearly indicates how the number of sunspots can go from a few units to more than one hundred. However, the number of spots in the fit, discussed in the following section, was adapted without taking into account the number of spots shown in that figure. This is the case for two reasons: 1) the number-size degeneracy, for which fewer, larger spots produce the same flux than a larger number of smaller spots with the same contrast; 2) the contrast-size degeneracy of the problem, for which decreasing (increasing) the contrast ratio of a spot allows one to reduce (increase) its size to produce the same effect. As a consequence, the spots of our model have to be considered as representative of groups of spots, and not as single spots. This allows one to use large sizes and lifetimes, without losing physical meaning. Moreover, modeling tens of spots would not be manageable with the MCMC algorithm of \pastis.\\

\begin{figure}[!htb]
\centering
\includegraphics[scale = 0.38]{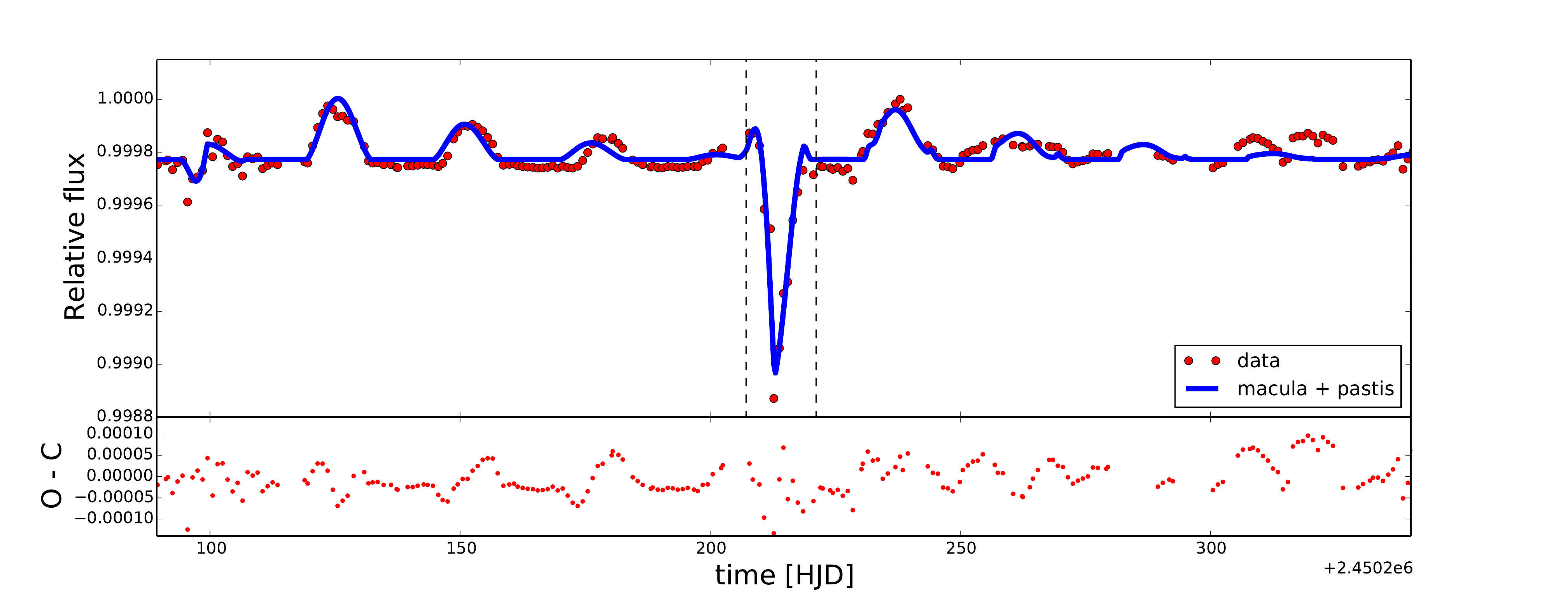}
\includegraphics[scale = 0.38]{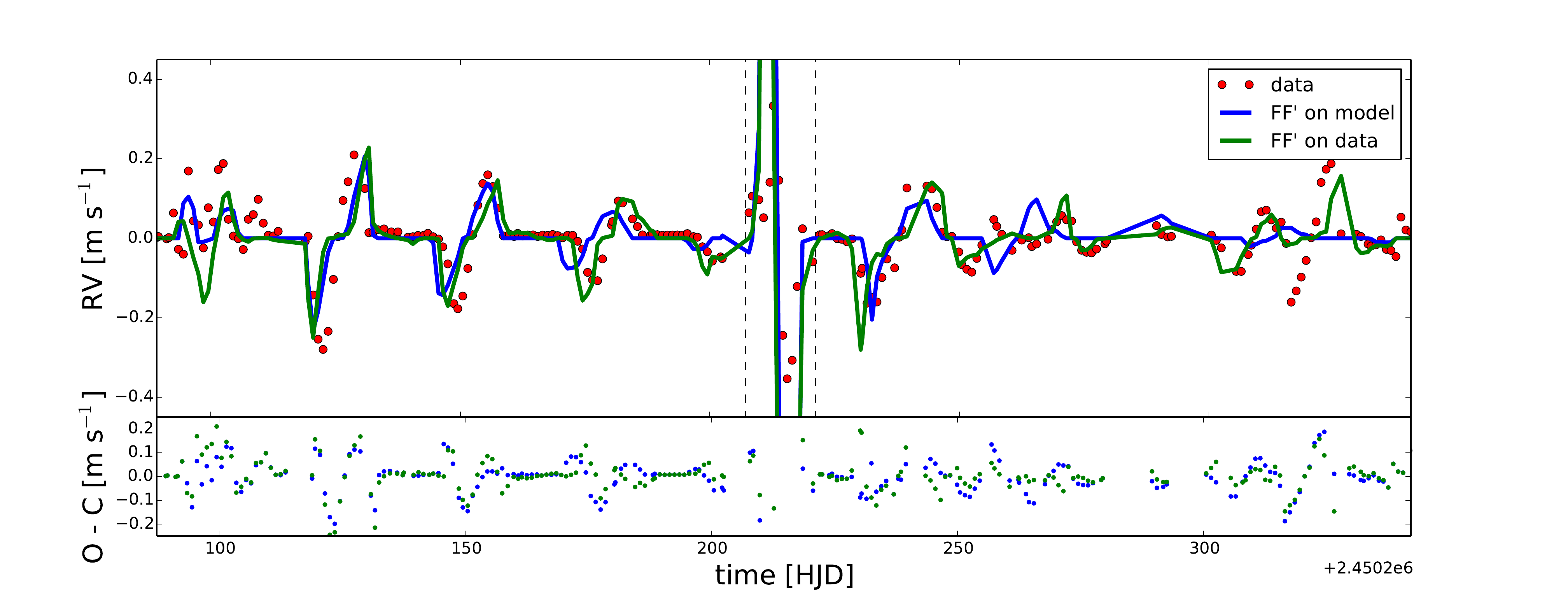}
\caption[Low-activity data set for the Sun, data and fits]{$Top:$ Synthetic light curve (red dots) and best fit with \macula\ (blue curve) for the low-activity data set. The residuals are indicated in the lower panel. The vertical dashed lines indicate the sudden decrease in flux discussed in the text. $Bottom:$ Synthetic RVs (red dots) and RVs computed on the photometric fit (blue) and on the data (green). Residuals are plotted in the lower panel, and their color correspond to the one of the photometric fit. The vertical dashed lines are the same than for the upper plot.}
\label{rvlow}
\end{figure}

Both the data sets have a sampling of a point per day. The low-activity data set is plotted in figure \ref{rvlow} (the fits are also represented: they are discussed in the following). It has a time span of about 300 days. This data set is characterized by relative flux variations of about $2 \cdot 10^{-4}$, plus a steep relative flux variation around HJD = 245041 of about $10^{-3}$, due to a large group of spots. The positive flux bumps on the sides of this sudden decrease in flux are probably the signature of faculae, which are often situated on the sides of dark spots. The corresponding RV variations have amplitude between 10 and 30 $\mathrm{cm} \, \mathrm{s}^{-1}$.

\begin{figure}[!htb]
\centering
\includegraphics[scale = 0.38]{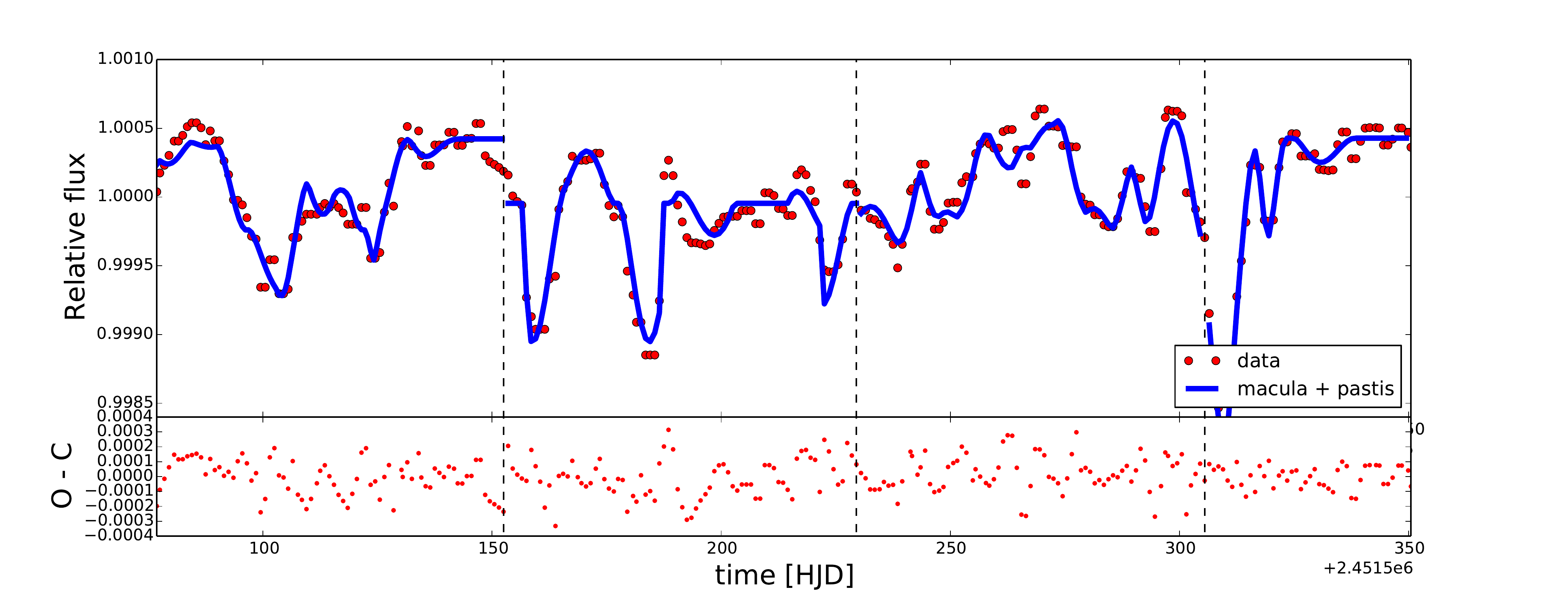}
\includegraphics[scale = 0.38]{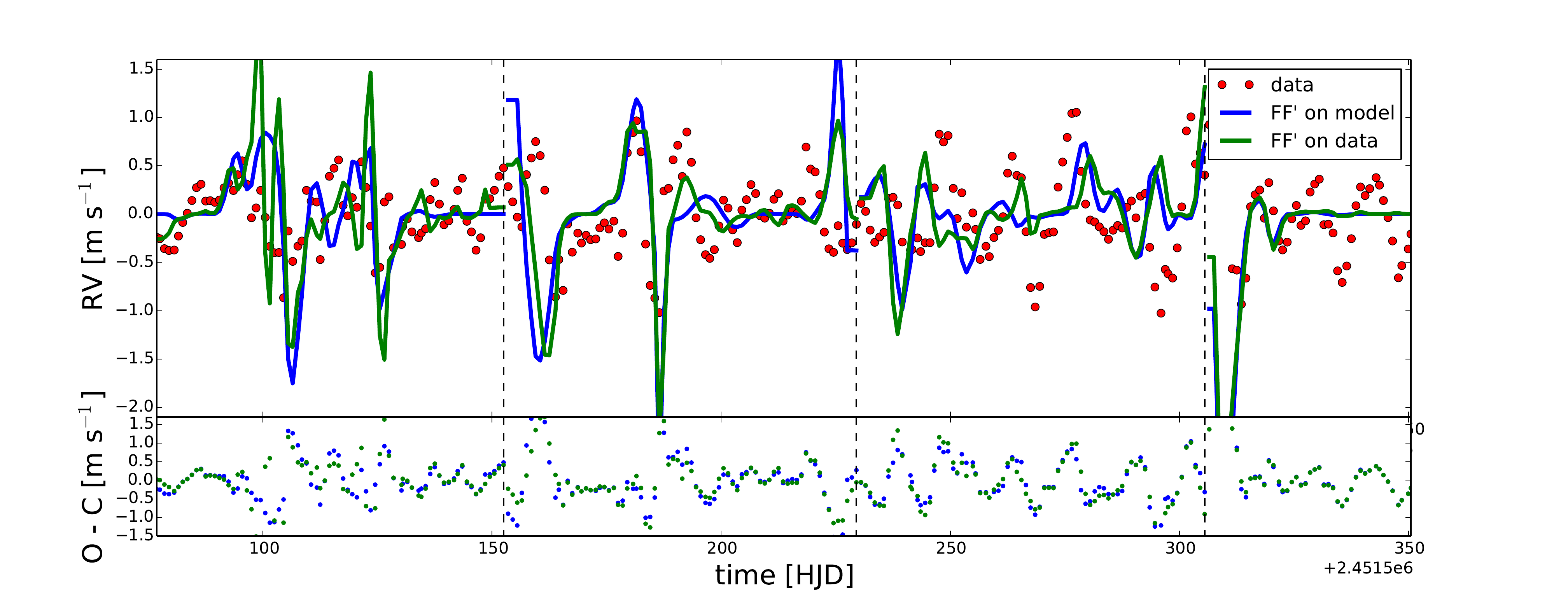}
\caption[High-activity data set for the Sun, data and fit]{The same as figure \ref{rvlow}, for the high-activity data set. The vertical lines indicate the cuts in the data set applied for the photometric fit, as discussed in the text.}
\label{rvhigh}
\end{figure}

The high-activity data set is reported in figure \ref{rvhigh} (again, the fits are plotted and discussed below). It has a duration of about 270 days. The relative flux variations measure about $1.5 \cdot 10^{-3}$, and vary on a shorter time scale than the low-activity data set. The corresponding RV peak-to-peak variation is of about 2 \ms.

\subsection{Fit of the light curve}\label{lcsun}
At first, the fit of the photometry was performed. As no transits are involved, I used \macula. To reduce the correlations and degeneracies, I fixed the stellar inclination and rotation period as well as the coefficients of differential rotation\footnote{At latitude $\phi$, the rotation period is calculated as $P_{\mathrm{eq}}/(1 - \kappa \sin^2 \phi)$, where $P_{\mathrm{eq}}$ is the rotation period at stellar equator and $\kappa$ is the quadratic differential rotation coefficient.} and limb darkening to the values reported in \cite{allen}. The values are indicated in table \ref{priorsun}.\\
Both dark spots and faculae were modeled. Uniform priors were set for every parameter of the activity features. In particular, the maximum allowed size of the features was set to $10^\circ$: as stated above, over this size the small spot approximation does not hold anymore (section \ref{photmod}). The contrast was left free in a wide range of values, between 0.3 and 1.3. Assuming spots and faculae emit as black bodies, this corresponds to a minimum \teff\ of $\sim 4300$ K for dark spots, and a maximum \teff\ of $\sim 6200$ K for faculae. To choose the coefficients, I took a wider interval than the one defined by the typical sunspot bolometric contrast of 0.67 \citep{sofia1982,lanza2004} and the bolometric facular contrast 1.115 adopted by \cite{foukal1991} and \cite{lanza2004}\footnote{\cite{lanza2004} modulated this coefficient according to the distance of the facula from the stellar limb. In this way, they accounted for the varying facular contrast with the position of the facula on the stellar disk, which is not done here.}. The observed maximum latitudes during periods of low and high solar activity were used as upper limits for the priors of the corresponding parameters: $30^\circ$ and $50^\circ$, respectively \citep[e.g.][]{allen}. The starting parameters of each Markov chain were randomly chosen within the priors.\\
\pastis\ allows one to fit the value of the flux out of transit, or photometric offset (see section \ref{ttvsmeth}). I left this as a jump parameter, to take into account the uncertainty in the  normalized ``unspotted'' flux level. To account for possible instrumental effects, I fitted an instrumental jitter as well, modeling it in the same way as the jitter of CoRoT or Kepler. Non-informative priors were set also for these two last parameters.

\begin{table}[htb]
\caption[Prior distributions for the fits on the Sun data sets]{\label{priorsun} Prior distributions used in the combined fit with \macula\ + \pastis. $\mathcal{U}(a, b)$ stands for a uniform distribution between $a$ and $b$.}
\begin{center}
\begin{tabular}{ll}
\hline\hline
\multicolumn{2}{l}{\emph{Sun parameters}} \smallskip\\
Stellar axis inclination $i_\star$ [deg] & 90.0 (fixed)\\
Stellar rotation period $P_\star$ [days] &  25.38 (fixed) \\
Quadratic differential rotation coefficient $\kappa$& 0.0145 (fixed) \\
Linear limb darkening coefficient $u_a$ & 0.44 (fixed) \\
Quadratic limb darkening coefficient $u_b$ & 0.20 (fixed) \\
\\
\multicolumn{2}{l}{\emph{Spots/faculae parameters}} \smallskip\\
Longitude [deg] & $\mathcal{U}(0, 360)$ \\
Latitude [deg] & $\mathcal{U}(-30, 30)$ (low activity) \\
			& $\mathcal{U}(-50, 50)$ (high activity)\\
Maximum size [deg] & $\mathcal{U}(0, 10.0)$ \\ 
Contrast &  $\mathcal{U}(0.3, 1.3)$ \\
Time of maximum size$^{(a)}$ [days] & $\mathcal{U}(-50, 100)$ \\
Permanence at maximum size [days] & $\mathcal{U}(0, 150)$ \\
Time of growth [days] & $ \mathcal{U}(0, 150)$ \\
Time of decay [days] &  $\mathcal{U}(0, 150) $\\
\\
\multicolumn{2}{l}{\emph{Instrumental parameters}} \smallskip\\
Offset &  $\mathcal{U}(0.9, 1.1)$ \\
Jitter & $\mathcal{U}(0.00, 0.01)$\\
\hline
\end{tabular}
\end{center}
\begin{list}{}{}
\item \small $^{(a)}$ Referred to the initial time of the data set.
\end{list}
\end{table}

For each data set, ten Markov chains were run. The posteriors of each chain were used to calculate its burn-in phase. Due to the non-uniqueness of the solution, different chains got stuck in different local minima of the $\chi^2$ space. Therefore, the chains were not merged, and I considered only the chain reaching the highest likelihood. Also, because of the large number of parameters (eight for each spot) and their mutual correlations, the number of uncorrelated points in each chain is always of the order of some units or some tens of units. This was calculated by dividing the length of the chains by their autocorrelation length. No meaningful uncertainties can be derived from such chains. Hence, no distribution was derived for the parameters of the activity features, but only the single highest-likelihood solution. To derive the uncertainties on the parameters of the features, one has to explore a tight surrounding of a local-minimum solution \citep{croll2006}. Such a study has been carried out for CoRoT-2, as described later in this chapter (section \ref{c2}).\\
The adopted methodology does not guarantee that the deepest minimum found in the $\chi^2$ space is actually the global minimum. More probably, the more Markov chains are run, the better the solution is. However, once a sufficient number of spots and faculae is included in the model, the logarithmic likelihood of the best solution of a set of chains increases by only a few units when more features are added. With this criterion, I found by trial and error the suitable number of features for the fit of each data set. A good fit for the low activity data set was found with five features. For the high activity data set, this was not possible. The larger frequency of the flux variations and the larger irregularity of their pattern, with respect to the low activity part, require a larger number of features to be modeled. Therefore, I cut the light curve in segments. This approach is often used to consider segments short enough that the size evolution of the activity features can be neglected \citep[e.g.][]{lanza2004}. In my case, I used this approach to reduce the number of spots and faculae required for the modeling. It should be reminded that this is acceptable, because the final purpose is not surface reconstruction.\\    
\cite{lanza2004} studied the same data sets, and used segments of about half the rotation period of the Sun ($\sim 14$ days). They used short parts of light curves because they did not model spot evolution; by trials, I managed to find satisfying fits on three rotation period-long windows, thanks to the modeling of the size evolution of the activity features. This resulted in four parts of the light curve, whose modeling required two or three features.\\
After this exploratory tests, I ran ten final MCMCs for each data set. I required $3 \cdot 10^5$ steps for the low activity data, and $2\cdot10^5$ steps for each part of the high activity data, as less features were needed for the modeling of the segments.\\
The top panels of figures \ref{rvlow} and \ref{rvhigh} show the best fits for the low and high activity data sets. 

\subsection{Fit of the RV signatures of the Sun}\label{rvsun}
The best solutions obtained with \macula\ + \pastis\ were used to model the RVs. As discussed in section \ref{rvmod}, \ff\ can be applied in two ways: on the whole light curve at once, or spot-by-spot. However, as the photometric solutions involved faculae, which are not properly modeled with the \ff\ method, I only applied the method on the whole light curve.\\
Figure \ref{rvlow}, bottom panel, shows the RV data and the derived RVs for the low-activity data set. The RVs were derived from both the data and the photometric model, to compare the deviation in the derived RVs. The data was first of all interpolated to increase its time resolution by a factor 3, obtaining bins of 0.33 days. A running median on a window of 7 bins (2.3 days) was then applied, to smooth the steep variations of the photometry. Seven bins were adopted as a compromise between an excessive smoothing and the quality of the fit, quantified as described below. I used a different approach from the one used by \cite{aigrain2012}. Using the notation of section \ref{rvmod}, they imposed: 
\begin{enumerate}[-]
\item $\Psi_0 = \Psi_\mathrm{max} + k \sigma$, where $\Psi_\mathrm{max}$ is the maximum value of the observed flux, $\sigma$ its dispersion, and $k$ a parameter to fit; 
\item $f = (\Psi_0 - \Psi_\mathrm{min})/\Psi_0$, where $\Psi_\mathrm{min}$ is the minimum value of the observed flux. This parameter was fixed; 
\item they fitted for $\delta V_c \kappa$; 
\item $A$ and $B$ were set equal to 1, and $\Delta RV_{\rm{additional}}$ to 0. Indeed, these parameters were added by \cite{haywood2014}.
\end{enumerate}
Instead, I fitted $\Psi_0, \, f, \, \delta V_c \kappa, \, A,$ and $B$ with a least-squares algorithm. I fixed $\Delta RV_{\rm{additional}}=0$ to reduce the number of free parameters in the fit. I used the value of the solar radius, so $R_\star = 6.9598 \cdot 10^8$ m. As remarked by \cite{aigrain2012}, the steep decrease of flux around HJD = 2450413 provokes a large computed RV shift, because of the derivative used by \ff, which has no counterpart in the RV data. To correct for this problem, \citeauthor{aigrain2012} applied a non-linear smoothing filter before the calculation of the derivative. In this case, a simpler, median filter was applied. In figure \ref{rvlow}, the time window of this steep variation is enclosed in vertical dashed lines. The variation is partly smoothed by the median filter. In this way, the RV variation calculated on the model is about a factor 2 larger than the one computed from the synthetic data, in the interval corresponding to the steep variation. This difference is out of scale in figure \ref{rvlow}, but can be noticed by the difference in the sum of the squared residuals. I computed this as $\chi^2 = \sum (y_i - d_i)^2$, where $y_i$ and $d_i$ are the modeled RV and the synthetic RV at time $i$, respectively. So, I found $\chi^2 = 50.64$ for the model-derived RVs, and 15.65 for the data-derived ones.\\
Then, I excluded from the fit the flux between HJD = 2450407 and 2450421. The fit finds the same values for the parameters, but the difference in $\chi^2$ is remarkably reduced. The fit is even better for the model-derived RVs ($\chi^2 = 0.80$) than for the data-derived RVs (0.90). The values of the $\chi^2$, whether the filter is applied or not, and the fitted parameters for \ff\ are reported in table \ref{tablow}.\\

In figure \ref{rvhigh}, bottom panel, the same kind of plot is shown for the high activity data. In this plot, the four segments in which the data set was cut are recomposed. The cuts are indicated by the vertical dashed lines. In this case, no resampling was applied to the data; a median filter on a window of three days was directly applied to smooth the signal. The parameters of \ff, as for the previous case, were fitted by least-squares minimization.\\
Quantifying the quality of the fit in this case is more difficult than in the previous one. The high-frequency brightness variations introduce noise in the calculation of the derivative. Moreover, several steep drops in the flux, as the one discussed for the low-activity data set, are present. Excluding all of them would make the fit meaningless. More adapted filters than the simple median filter would probably improve the quality of the fit. However, this issue was not addressed in this work.\\
The $\chi^2$, computed as before, and the fitted parameters on each segment are reported in table \ref{tabhigh}. As it can be noticed, for segments 1 and 3, the $\chi^2$ is lower for the fit on the modeled flux; for segments 2 and 4, it is the opposite. This can be attributed to the amplitude of the flux variations in the corresponding parts of the light curve. Larger variations imply a worse result obtained on the modeled flux. However, for each segment considered by itself, the parameters derived using the synthetic and the modeled flux are consistent.

\begin{table}[!htb]
\caption{\label{tablow} $\chi^2$ and fitted parameters for \ff\ on the low-activity data set.}
\begin{center}
\begin{tabular}{lcccccc}
\hline\hline
 & $\chi^2$ & $\Psi_0$ & $f$ &  $\delta V_c \kappa$ [\ms] & $A$ & $B$ \\
 \hline
Without filter, on model & 50.64 & 1.0002 & $ 8.7 \cdot 10^{-4}$ & 0.09 & 0.99 & 96.90 \\
Without filter, on data & 15.65 & 0.9997 & $8.1\cdot 10^{-4}$ & 0.01 & 1.00 & 1.00 \\
With filter, on model & 0.80 & 1.0002 & $8.7 \cdot 10^{-4}$ & 0.01 & 1.00 & 1.00\\
With filter, on data & 0.90 & 0.9997 &  $8.1 \cdot 10^{-4}$ & 0.01 & 1.00 & 1.00 \\
\hline
\end{tabular}
\end{center}
\end{table}

\begin{table}[!htb]
\caption{\label{tabhigh}  $\chi^2$ and fitted parameters for \ff\ on the high-activity data set.}
\begin{center}
\begin{tabular}{lcccccc}
\hline\hline
 & $\chi^2$ & $\Psi_0$ & $f$ &  $\delta V_c \kappa$ [\ms] & $A$ & $B$ \\
 \hline
Part 1, on model & 16.93 & 1.0000 & 0.00112 & 0.01 & 1.00 & 1.00 \\
Part 1, on data & 24.95 & 1.0005 & 0.00123 & 0.01 & 1.00 & 1.00 \\
Part 2, on model & 43.57 & 1.0003 & 0.00136 & 0.10 & 1.00 & 9.88 \\
Part 2, on data & 24.95 & 1.0002 & 0.00144 & 0.10 & 1.00 & 9.99 \\
Part 3, on model & 21.83 & 1.0009 &  0.00085 & 0.00 & 1.00 & 0.04 \\
Part 3, on data & 28.22 & 1.0006 & 0.00105 & 0.00 & 1.00 & 0.03 \\
Part 4, on model & 59.00 & 1.0000 & 0.00214 & 0.10 & 1.00 & 9.90 \\
Part 4, on data & 45.97 & 1.0005 & 0.00213 & 0.10 & 1.00 & 9.63 \\
\hline
\end{tabular}
\end{center}
\end{table}

These tests indicate the efficacy of \ff\ to model activity-induced RV variations from the light curve, as expected. The method also worked consistently on modeled flux variations. However, \ff\ turned out to be very sensitive to noise and high-frequency brightness variations, which worsened the quality of the fit. For highly active stars, in particular, the data needs to be smoothed with more sophisticated filters that the one employed here. In fact, \cite{aigrain2012} uses a non-linear filter for the same kind of data, together with a Gaussian process. These methods were not implemented here, as the main purpose was to test the correct behaviour of the code.

\section{Application to CoRoT-7}\label{c7}
CoRoT-7 A is a G9V-type, young (1.2-2.3 Gyr old), and active star hosting the first discovered transiting super-Earth with measured mass and radius \citep{leger2009}. It was observed by CoRoT during the LRa01 run, from October 24, 2007 to March 3, 2008. The light curve presents brightness modulations of 2\%. An analysis of the transit signal showed the presence of a 0.85 days-dimming in flux due to the transit of a $\simeq 1.7$ \RE\ planet, which was therefore called CoRoT-7 b. The strong variability of the flux affects the derived stellar density, which was found to be 14\% lower than the density derived from the spectral parameters. \cite{leger2009} found that this underestimate could be avoided by fitting subsets of 4-5 transits at a time.\\
Stellar activity severely affects the RV signal, inducing variations of about 20 \ms, larger than the variation produced by CoRoT-7 b (1.6 - 5.7 \ms). Several methods were developed to correct for this signal, yielding mass estimates going from $\simeq 2.3$ to $\simeq 8.0$ \ME, depending on the employed technique \citep{queloz2009,hatzes2010,boisse2011,pont2011,ferraz-mello2011}. Also, 3.69-days and 9.0-days periodicities in the RVs were found, leading to the proposed detection of two more planets: CoRoT-7 c, with a mass of $\simeq8.4$ \ME\ \citep{queloz2009}, and CoRoT-7 d, with a mass of $\simeq16.7$ \ME\ \citep{hatzes2010}.\\
The strive for a better understanding and characterization of this system led to a simultaneous re-observation of CoRoT-7. Simultaneous HARPS observations were carried out during the CoRoT re-observation of CoRoT-7 during the LRa06 run. The photometry observations lasted eighty days, from January 10, 2012 to March 29 \citep{barros2014c7}, while the RV observations were carried out from January 12, 2012 to February 6 \citep{haywood2014}. \cite{barros2014c7} showed that the transits in the LRa01 run are deformed, while the lower activity level during the LRa06 run allows one to disentangle the activity signal. This study required a careful selection of the transits to be fitted. \cite{haywood2014}, on the other hand, applied the \ff\ method to the RVs and needed to use Gaussian processes \citep{rasmussen2006,gibson2011} to model the activity-related noise that the \ff\ method is not able to take into account. In this way, they reduced the activity component in the RVs. They derived a mass for CoRoT-7 b ($4.73 \pm 0.95$ \ME) implying a density compatible with a rocky composition. They operated a Bayesian model selection to find the most probable number of companions in the system. In this way, they confirmed the presence of CoRoT-7 c, with a mass of $13.56 \pm 1.08$ \ME, and rejected the one of CoRoT-7 d, whose signature can be linked with the second harmonic of the stellar rotation period in a Lomb-Scargle periodogram.\\

\begin{figure}[!htb]
\includegraphics[scale = 0.4]{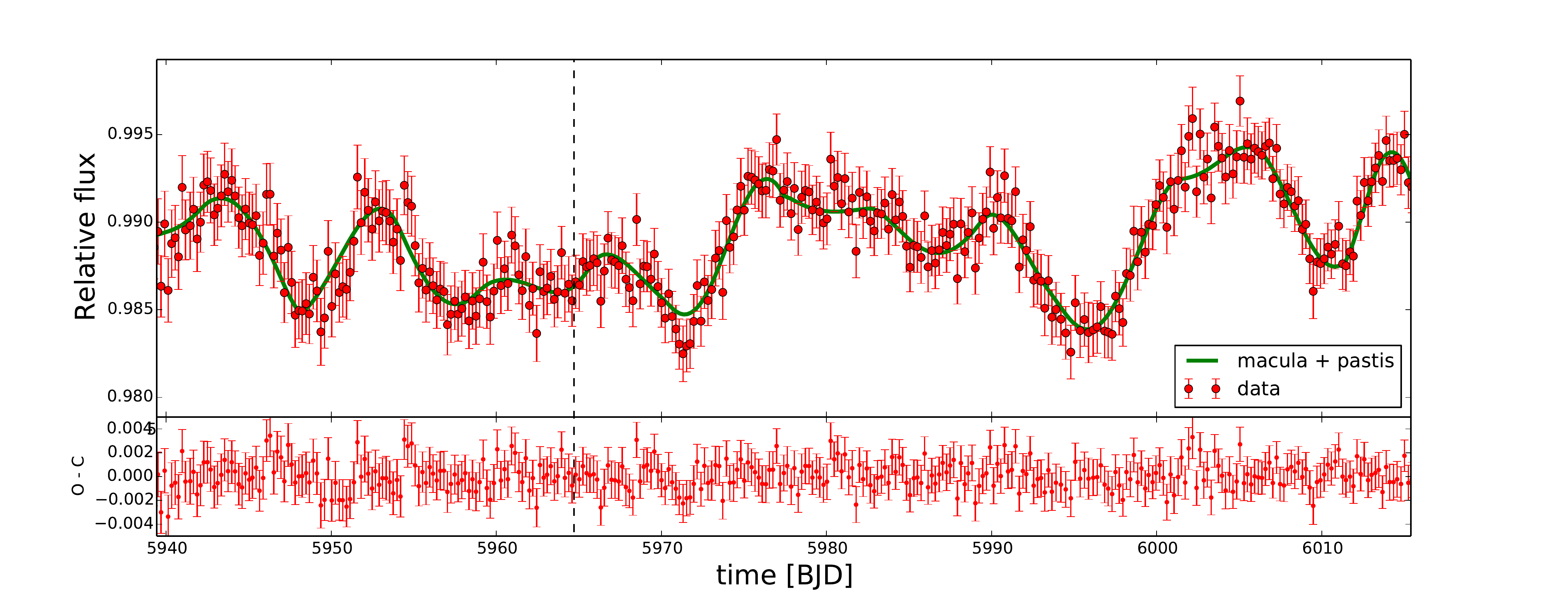}
\includegraphics[scale = 0.4]{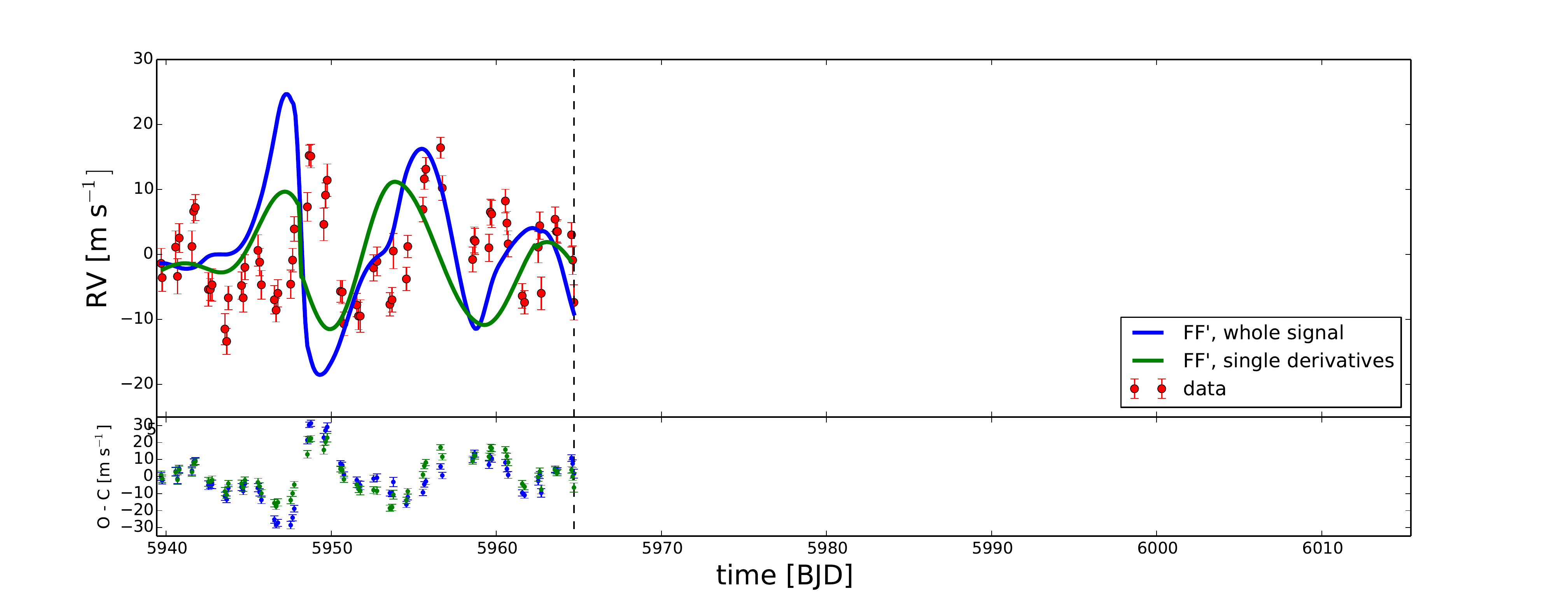}
\caption[Fits on the CoRoT-7 data]{Fits on the CoRoT-7 data. $Top:$ Light curve. $Bottom:$ RVs. The results from \ff\ applied on the modeling of the light curve as a whole (``whole signal'' in the legend) and on the spots considered singularly (``single derivatives'') are plotted. The time window of simultaneous observations on photometry and RVs is delimited by the vertical dashed lines. In both plots, residuals are indicated in the lower panel; for RV, they follow the same color code than the upper panel.}
\label{C7_ff}
\end{figure}

As discussed in section \ref{rvmod}, the \ff\ method is affected by limitations about the size of the spots, their contrast, and their simultaneous modeling. For this reason, a better insight in the RV signature produced by the spots could be gained by modeling it with a more complex tool, such as \soap\ (section \ref{photmod}). I collaborated to a study of this kind, led by M. Oshagh at CAUP\footnote{Centro de Astrofísica, Universidade do Porto (CAUP), Rua das Estrelas, 4150-762 Porto, Portugal.}. I was in charge of the fit of the starspots features in the LRa06 light curve, as simultaneous RV observations are available for this light curve. Then, the solution for the parameters of the spots would have been used to compute the activity-induced RV shift with \soap.\\
The paper relative to this study was submitted, and is currently under revision. For this reason, its results will not be discussed here. I will describe, however, the contribution I made for the fit of the photometry. For this study, I used the 32 s-cadence light curve re-binned every 500 points. This means about 260 minutes for each bin. 
The transits of CoRoT-7 b (the innermost and only confirmed planet) last about 75 min \citep{leger2009}, and their depth is about one third of the RMS of the light curve \citep{barros2014c7}. Therefore, they do not significantly affect the re-binning.\\
By avoiding the possibility of overlap of transits and spots, I could use \macula\ to model the light curve. I used the same approach as for the Sun. The inclination of the stellar axis was fixed to the value found by \cite{leger2009}, assuming it is perpendicular to the planet orbit ($i_\star = 80.1^\circ$). The limb darkening coefficients were fixed to the mean values found by \cite{barros2014c7} ($u_a = 0.515, \, u_b = 0.188$). The stellar rotation period was fixed to the value found by \cite{lanza2010}, that is 23.64 days. No differential rotation was modeled. For the spots, the same priors as for the Sun were used. However, only dark spots (i.e. spots with contrast between 0.3 and 1) were used for the fit, to be able to derive the RV signal with \ff, as described below. Their latitude was left free between -90 and $90^\circ$. Using the same method as for the Sun, seven spots were found to be necessary to model the light curve. The LRa06 light curve and its photometric best fit are plotted in figure \ref{C7_ff}, top panel.\\

I took advantage of my modeling of the CoRoT-7 light curve to test my own implementation of \ff. Thanks to the absence of faculae in the fit, I could compare the two approaches for \ff\ proposed in section \ref{rvmod}: calculation of the RVs from the whole light curve, or from the single spots.
\begin{enumerate}[-]
\item For the direct application of \ff\ on the whole light curve (solution 1 hereafter), I proceeded as for the Sun, fitting the \ff\ coefficients with a least-squares algorithm. 
\item The second approach consists in using a spot model to calculate the light curve for each spot and the corresponding RV signal with \ff; finally, all the contributions are added to derive the final RV curve. For this approach (solution 2 hereafter), I calculated the light curve corresponding to each of the spots of the photometric solution, considered separately, with \macula. I applied \ff\ to each of these light curves, and then added the resulting RV signals. In this phase, all the parameters of \ff\ were fixed, because no data set was available for the fit of a light curve with a single spot. Therefore, I set $\Psi_0 = 1, \, f = (\Psi_0 - \Psi_\mathrm{min})/\Psi_0 = 1 - \Psi_\mathrm{min}$, and $ A = 1$. Because $\delta V_c \kappa$ cannot be calculated \textit{a priori}, this parameter, as well as the $B$ coefficient, was set to 0. This approach is in contrast with the fact that for CoRoT-7, as for the Sun, the convective component is dominant over the rotational one in the RVs \citep{meunier2010plages,meunier2013conv,haywood2014}. However, fitting $\delta V_c \kappa$ and $B$ does not decrease the $\chi^2$ of the fit. I also made use of the term $\Delta RV_{\rm{additional}}$, which is useful to center the mean RV to 0 \ms. This parameter was fitted, too. I adopted a stellar radius of 0.82 \Rsun\ \citep{barros2014}.
\end{enumerate}
In figure \ref{C7_ff}, the modeled photometry and RVs are plotted over the data. The values of the fitted parameters with the two approaches are reported in table \ref{tabc7}. Both solutions retrieve the RV pattern reasonably well. The main difference is in the amplitude of the signal. However, the fits are not satisfying. Indeed, for solution 1, $\chi^2 = 46.3$, and for solution 2, $\chi^2 = 29.9$. Solution 2 is slightly shifted with respect to solution 1. This could be the effect of neglecting the convective component in the RVs. Indeed, this component causes the departure of the RV curve from an exact sinusoidal shape, given by the rotational component \citep{aigrain2012}. As discussed before, other studies have found the convective component to be dominant for slowly rotating stars. For this reason, and given the high values of the $\chi^2$ of the two solutions, it is risky to prefer solution 2 over solution 1 by relying on the value of these statistics.\\

Tests like this and the one carried out on the Sun are far from being exhaustive. They indicate that, even if the modeling of the photometric signal can help in constraining the RV signal to correct it for the activity features, it still needs to be improved. The main limitations are due to the degeneracies between the parameters, the non-uniqueness of the solution, and the limited time resolution of the RV data. \ff\ is a simple, fast, and effective method, which deserves a systematic study. For example, the simultaneous follow-up of a star observed with TESS \citep{ricker2010} (which will be launched in 2017) could be used for such a study. This would elucidate the possibilities and limitations of deducing activity-induced RV features from the analytic modeling of photometry, making it suitable for implementation in a MCMC algorithm as analytic photometric models.

\begin{table}[!bth]
\caption[$\chi^2$ and fitted parameters for \ff\ on CoRoT-7]{\label{tabc7}  $\chi^2$ and fitted parameters for \ff\ on CoRoT-7.}
\begin{center}
\scalebox{0.9}{\begin{tabular}{lccccccc}
\hline\hline
 & $\chi^2$ & $\Psi_0$ & $f$ &  $\delta V_c \kappa$ [\ms] & $A$ & $B$ &  $\Delta RV_{\rm{additional}}$ [\ms] \\
 \hline
On the whole signal &  46.3 & 0.9936 & 0.00626 & 0.1 & 0.50 & 4.97 & - \\
On single spots & 29.9 & - & - & - & 0.10 &  -  & -1.33 \\
\hline
\end{tabular}}
\end{center}
\end{table}

\section{Application to CoRoT-2}\label{c2}

\corot-2 A is a young ($< 500$ Myr old), G7V-type star observed for 152 days during the LRc01 run of the \corot\ space telescope. 78 transits were observed. The star hosts the hot Jupiter \corot-2 b \citep{alonso2008}, which has a mass of $3.31\pm 0.16 $ \MJ\ and a radius of $1.465 \pm 0.029$ \RJ. The orbit of the planet has a period of 1.74 days, and is almost aligned with the stellar equator \citep{bouchy2008}. Its radius is about 0.3 \RJ\ larger than expected for an irradiated hydrogen-helium planet of this mass. Models strive to explain a longer contraction time during the evolution of the planet \citep{guillot2011}.\\
The main characteristics of the system are listed in table \ref{presc2}, where the same notation as in chapter \ref{chapttvs} is used.

\begin{table}[!htb]
\begin{center}
\caption[Main characteristics of the CoRoT-2 system]{Main characteristics of the CoRoT-2 system. From \cite{alonso2008}.}
\begin{tabular}{lrr}
\hline\hline
& Value& Error\\
Orbital period $P$ [d]&  1.7429964&0.0000017\\
Transit duration [days] & $\sim 0.09$ \\
Transit depth $k_r$ &  0.1667&0.0006\\
Orbital inclination $i_p$ [deg]& 87.84&0.10\\
Linear limb darkening coefficient $u_a$ & 0.41&0.03\\
Quadratic limb darkening coefficient $u_b$ & 0.06&0.03\\
Semi-major axis \\
to stellar radius ratio $a/R_\star$ & 6.70&0.03 \\
Stellar density $\rho_\star$ [$\rho_\odot$] & 1.327& 0.018 \\
Eccentricity $e$ &  0 &(fixed)\\
Stellar mass $M_\star$ [$M_\odot$] & 0.97&0.06\\
Stellar radius $R_\star$ [$R_\odot$] & 0.902&0.018\\
Projected rotational velocity \vsini\ [km/s] & 11.85& 0.50\\
Effective temperature \teff\ [K] & 5625 & 120 \\
Planet mass $M_p$ [\MJ]  & 3.31&0.16\\
Planet radius $R_p$ [\RJ] & 1.465&0.029\\
\hline
\label{presc2}
\end{tabular}
\end{center}
\end{table}

\begin{figure*}[htb]
\centering
\includegraphics[scale = 0.5]{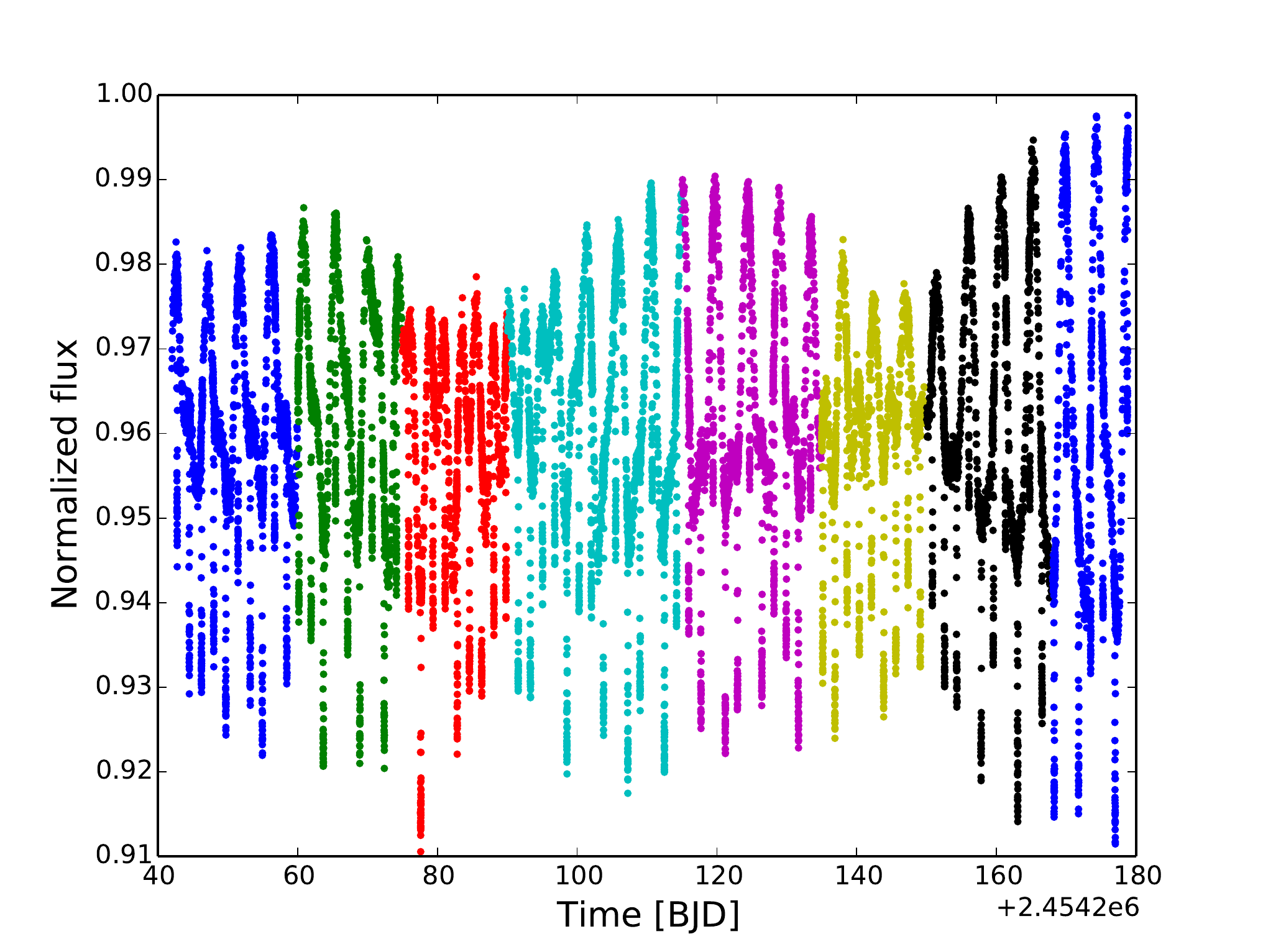}
\caption[The light curve of CoRoT-2 divided in eight segments, highlighted by color]{The light curve of CoRoT-2 divided in eight segments, highlighted by color.}
\label{slices8}
\end{figure*}

\corot-2 is an active star. Figure \ref{slices8} shows its light curve; the meaning of the colors will be explained later. The light curve exhibits peak-to-peak brightness variations between 4 and $6\%$, caused by activity features and comparable in amplitude to the transits.

The activity pattern evolves in cycles of $\sim 50-60$ days. In the light curve, about 2.5 beatings can be observed. During the activity cycle, a varying fraction of the stellar surface is covered by spots, up to tens of percent \citep{wolter2009,lanza2009,huber2009,silva-valio2010}. Moreover, the signatures of spots inside transits are clearly distinguishable in the light curve. This allows for a better quantification of the impact of starspots on the transit parameters. This is particularly important to prepare the optimal correction for active planets around active stars. For these reasons, we considered \corot-2 an ideal case for the application of \ksint\ + \pastis.\\
Different studies involving different modeling techniques and models from a few spots to several active regions have been performed. I recall hereafter the main results.
\begin{enumerate}[-]
\item \cite{wolter2009} worked on a single transit, showing an occulted spot. They constrained the size of the spot (between 4.5 and $10.5^\circ$) and its longitude with a precision of about $1^\circ$.
\item \cite{lanza2009} fitted segments of the light curve no longer than 3.2 days, because spot evolution was not allowed for in their models. They removed the transits and used both a 3-spot model and a maximum-entropy regularization method. They measured the stellar rotation period as $\sim 4.5$ days. They recovered two active longitudes on different hemispheres, and estimated the relative differential rotation of the star to be lower than $\sim 1\%$. They also measured cyclic oscillations of the total spotted area, with a period of $\simeq 29$ days. The periodicity of the oscillations was suggested to be due to star-planet magnetic interactions \citep[see also][]{lanza2011}. \cite{lanza2009} also found that the facular contribution to the optical flux variations is significantly lower than in the present Sun. 
\item \cite{huber2009} managed to fit two stellar rotations, including both the in-transit and the out-of-transit parts of the light curve. The transit parameters were fixed to the values found by previous studies, and spot evolution was not modeled. These authors managed to reproduce the photometric signal with a low-resolution surface model of 36 strips, and found that the belt occulted by the planet (close to the stellar equator) is $\sim 6\%$ darker than the average remaining surface. This study was extended to the whole light curve by \cite{huber2010}, whose results are in agreement with the previous one. Significant indications of stellar differential rotation were found. 
\item \cite{silva-valio2010} analyzed the occulted spots inside all transits, but excluded the out-of-transit part of the light curve from the fit. Up to nine spots per transit were needed for the modeling. These authors found the size of spots to be between 0.2 and 0.7 planetary radii, and a spot coverage of 10-20\% in the belt transited by the planet. For the spots, they found contrasts between 0.3 and 0.8. Finally, they found the transit depth of the deepest transit (assumed to be the less affected by spots) to be of 0.172, 3\% larger than the value found by \cite{alonso2008}. The occulted spots were modeled in a spot map for every transit by \cite{silva-valio2011}. For this map, 392 spots were used. The temporal evolution of the spot surface coverage in the transit chord was found to be between 9 and 53 days. 
\end{enumerate}
All these approaches are complementary, but none of them offers a complete modeling of the light curve. In most cases, either the out-of-transit or the in-transit part of the light curve is modeled. Even if both parts are modeled, as \cite{huber2010} did, spot evolution is not included. Moreover, the impact of the spots on the transit parameters is only partially explored.

\subsection{Data reduction}\label{phot}
I used the light curve processed by the \corot\ N2 pipeline\footnote{The technical description of the pipeline is available at \texttt{http://idoc-corotn2-public.ias.u-psud.fr/\\jsp/CorotHelp.jsp}.}. Only the part sampled every 32 s was used ($\sim 145$ over 152 days of observation), to take advantage of the full resolution inside transits. The points classified as ``bad'', including those related to the South Atlantic Anomaly, were discarded. The data were first 3$\sigma$-clipped and the transits identified with the ephemeris given by \cite{alonso2008}.\\
I prepared five data sets, to be used for an equal number of fits. The fits will be described in the next section. 
\paragraph{Data set 1.} This data set was prepared for comparison with a classical transit fit. The flux was divided by its median value; a second-order polynomial was fitted to the flux adjacent to the transits. For this, a window as long as a transit duration was considered on the sides of each transit. The flux was then divided by the fitted polynomial. 
\paragraph{Data set 2.} This data set was prepared for comparison with a standard transit fit, where occulted spots are taken into account in the data reduction phase. The normalization was performed following the prescription of \cite{czesla2009}. Indicating with $f_i$ the observed flux in the bin $i$, with $n_i$ the value of the second-order polynomial fitted to the out-of-transit local continuum in the same bin, and with $p$ the largest flux value (assumed to be the less affected by spots), the normalized flux $z_i$ in bin $i$ was calculated as
\begin{equation}
z_i = \frac{f_i - n_i}{p} + 1.
\label{czesla}
\end{equation}
\paragraph{Data set 3.} This data set was used for spot fitting. The light curve was sampled to longer time steps, to reduce the computation time of the fit. Following \cite{huber2009}, the out-of-transit flux was sampled every 2016 s (33.6 min). The out-of-transit spot pattern is not affected, as it varies on a longer time scale. The transits were sampled every 160 s. The transits being $\sim 0.09$ days long, this means about 49 points were left for each transit. This sampling keeps enough information to force some activity features to be modeled inside the transits. The uncertainty of the binned data was calculated as the standard deviation of the points in each bin, divided by the square root of the number of points in the binning window. The average uncertainty on this data set is $3.9\cdot 10^{-4}$. After resampling, the flux was divided by its maximum value and no transit normalization was performed.
\paragraph{Data set 4.} This data set was prepared for spot and transit fitting. It was prepared as data set 3, but full resolution was kept inside the transits. The average standard deviation on this data set is of $1.12 \cdot 10^{-3}$.
\paragraph{Data set 5.} The out-of-transit model fitting data set 3 was used to normalize the transits. The preparation of this data set will be described in more detail in section \ref{evolution}.

\subsection{Modeling approaches}\label{evolution}
If the transit parameters are fitted together with the spot parameters, the degeneracies and correlations between the parameters prevent the chains from converging. Almost for every chain that is run, a successful combination of the parameters is explored and produces a stable likelihood. Therefore, the fit of activity features and transits was divided in two parts. First, only the spot parameters were fitted; then, given the spot parameters, the transits were fitted. In addition to these fits, I performed three fits with a transit-only model.

\paragraph{Comparison fits (fit 1 and 2)}
Two fits were performed with a classical approach, to test the impact of transit normalization. Fit 1 was performed as a standard transit fit. \ebop, included in \pastis\ (section \ref{ttvsmeth}) was used on data set 1. The priors for this fit are listed in table \ref{tabpriors}. A Jeffreys prior was used for the transit depth $k_r$, a normal prior for the stellar density $\rho_\star$ and the orbital period $P_{\mathrm{orb}}$, a sine prior for the orbit inclination $i_p$, and uniform priors for the linear and the quadratic limb darkening coefficients $u_a$ and $u_b$. As \cite{alonso2008} found $e$ compatible with 0, $e$ and consequently also $\omega$ were fixed to 0. The flux offset and the jitter were fitted.

\begin{table*}[!tbhp]
\caption[Prior distributions used in the combined fit on CoRoT-2]{\label{tabpriors} Prior distributions used in the combined fit with \ksint\ + \pastis. $\mathcal{U}(a, b)$ stands for an uniform distribution between $a$ and $b$; $\mathcal{N}(\mu, \sigma)$ indicates a normal distribution with mean $\mu$ and standard deviation $\sigma$; $\mathcal{S}(a, b)$ represents a sine distribution between $a$ and $b$; finally, $\mathcal{J}(a, b)$ means a Jeffreys distribution between $a$ and $b$. $BV$ indicates the value of the best-likelihood solution of fit 3.}
\centering
\scalebox{0.8}{
\begin{tabular}{llll}
\hline\hline
\\
& \textit{Fit 1, 2, 5} & \textit{Fit 3} & \textit{Fit 4} \\
\multicolumn{2}{l}{\emph{Stellar parameters}} \smallskip\\
Stellar axis inclination $i_\star$ [deg]& - & 87.84 (fixed) & 87.84 (fixed) \\
Stellar rotation period $P_\star$ [days] & -&  $\mathcal{U}(4.3, 4.7)$ &  $\mathcal{U}(4.3, 4.7)^{(b)}$ \\
\\
\multicolumn{2}{l}{\emph{Spots/faculae parameters}} \smallskip\\
Longitude $\lambda$ [deg] & -& $\mathcal{U}(0, 360)$ & $\mathcal{N}(BV, 5)$\\
Latitude $\phi$ [deg]& & -$\mathcal{U}(-90, 90)$ & $BV$ (fixed)\\
Maximum size \amax\ [deg] & -& $\mathcal{U}(0, 30)$ & $\mathcal{U}(0, 30)^{(b)}$\\ 
Contrast $c$ & -&  $\mathcal{U}(0.3, 1.3)$ & $BV$ (fixed) \\
Time of maximum size \tmax$^{(a)}$ [days] & -& $\mathcal{U}(-10, 50)$ & $BV$ (fixed)\\
Permanence at maximum size \tlife\ [days]& - & $\mathcal{U}(0, 50)$ & $BV$ (fixed) \\
Time of growth \ingress\ [days] & -& $ \mathcal{U}(0, 50)$ & $BV$ (fixed)\\
Time of decay \egress\ [days] & -&  $\mathcal{U}(0, 50) $ & $BV$ (fixed)\\
\\     
\multicolumn{2}{l}{\emph{Transit parameters}} \smallskip\\
Radius ratio $k_r$   & $\mathcal{J}(0.14, 0.19)$ &  0.1667 (fixed) &$\mathcal{J}(0.14, 0.19)$\\
Stellar density $\rho_{\star}$ [g cm$^{-3}$]  - & &1.87 (fixed) &  $ \mathcal{N}(1.87, 0.5)$\\
Semi-major axis to stellar \\
radius ratio  $a/R_\star$ & $\mathcal{U}(5.0,8.0)$ &- & - \\
Orbital inclination $i_p^{(c)}$ [deg]   & $\mathcal{S}(80, 90)$ &  87.84 (fixed) &$\mathcal{S}(80, 90)$ \\
Linear limb darkening coefficient $u_a$  & $\mathcal{U}(0.0, 0.1)$ & 0.41 (fixed)  & $\mathcal{N}(0.41, 0.03)$\\
Quadratic limb darkening coefficient $u_b$  & $\mathcal{U}(0.0, 1.0)$ & 0.06 (fixed)  & $\mathcal{N}(0.06, 0.03)$\\
Orbital period $P_{\mathrm{orb}}$ [days]  & $\mathcal{N}( 1.7429964, 1.0\cdot 10^{-5})$ & 1.7429964  (fixed) & $\mathcal{N}( 1.7429964, 1.0\cdot 10^{-5})$ \\
Orbital eccentricity $e$   & 0 (fixed)  & 0 (fixed) & 0 (fixed) \\
Argument of periastron $\omega$ [deg]   & 0 (fixed)    & 0 (fixed) & 0 (fixed)  \\
\\
\multicolumn{2}{l}{\emph{Instrumental parameters}} \smallskip\\
Flux relative offset & $\mathcal{U}(0.99, 1.01)$ &  $\mathcal{U}(0.97, 1.01)$ & $\mathcal{U}(0.97, 1.01)^{(b)}$\\
Flux jitter [ppm]        & $\mathcal{U}(0, 0.01)$ & $\mathcal{U}(0, 0.01)$ & $\mathcal{U}(0, 0.01)$\\
Contamination  [\%] & $\mathcal{N}(8.81, 0.89)$ & 8.81 (fixed) &  8.81 (fixed)  \\
\\
\hline
\end{tabular}}
\begin{list}{}{}
\item \small$^{(a)}$ Referred to the initial time of a segment. $^{(b)}$ Starting from the best solution of fit 3. 
\end{list}
\end{table*}

\cite{alonso2008} used the contamination rate of CoRoT-2 based on generic point-spread functions (PSF), as calculated before CoRoT launch using a set of generic PSFs \citep[contamination level 0,][]{llebaria2006}. The contamination rate for CoRoT-2 was found to be $5.6 \pm 0.3 \%$. \cite{gardes2012} updated the values of the contamination rates for the targets observed by CoRoT, using a more realistic estimate of the CoRoT PSF (contamination level 1). They found the contamination rate of CoRoT-2 to be $8.81 \pm 0.89\%$. Therefore, I adopted this last value.\\
This fit was also used to verify the correct behaviour of our method in the case of standard transit-only fit. The fit was performed with \ksint\ + \pastis, too. The results of this fit and of the one with \ebop\ are in full agreement and are discussed in section \ref{joint}.\\
For fit 2, I used the same method as for fit 1, but on data set 2.

\paragraph{Spot fit (Fit 3).}\label{spotsonly}
This fit takes into account non-occulted activity features, and prepares the fit of the transits with the occulted ones. The light curve modeling was carried out on the data set 3. The transit parameters ($P_{\mathrm{orb}}, \, \rho_\star, \, i_p, \, k_r, \, e, \, \omega, \, u_a, \, u_b$) were fixed to the values of \cite{alonso2008}. A uniform prior was imposed for the mean anomaly $M$. The priors indicated in table \ref{tabpriors} were used. Following \cite{bouchy2008}, the stellar spin axis was assumed to be perpendicular to the planetary orbit. The planet inclination was found by \cite{alonso2008} to be of $87.84^\circ$ with respect to the plane of the sky; therefore, I fixed the stellar inclination $i_\star =  87.84^\circ$ with respect to the line of sight. A uniform prior, centered on the value found by \cite{lanza2009}, was set for the stellar period $P_\star$. Non-informative priors were used for all the spot parameters. For each spot, \amax\ was limited between 0 and $30^\circ$. The largest size was found sufficient for the modeling. Both dark spots and faculae were included in the fit, setting a prior for the contrast $c$ between 0.3 and 1.3. For each spot, a uniform prior was set for \tmax. This prior was set to be larger than the extension of the light curve, allowing \tmax\ to be both previous to the beginning of the light curve and later than its end. In this way, the fit was allowed to exclude some of the activity features if not needed. The MCMC could do this by adjusting \tmax, \ingress, and \egress, to which non-informative priors were also assigned. The contamination term was fixed to the value found by \cite{gardes2012} (8.81\%). This parameter was fixed because, from a mathematical point of view, spots and faculae act like a contamination source \citep{csizmadia2013}. A uniform prior was set for the foot and the jitter.\\

To find the optimal number of activity features needed for the fit, I manually increased it, trying to obtain a normally-distributed dispersion of the residuals, centered at zero and with a width comparable to the photometric data dispersion. By trials, it was found that several tens of features would be needed to model the entire light curve as a whole, consistent with the results of \cite{silva-valio2010} and \cite{silva-valio2011}. As our MCMC cannot handle the hundreds of resulting free parameters, the light curve was divided in shorter parts which need less features to be modeled. The initial and ending time of these segments are indicated in table \ref{tsegments}. These segments have a duration of $\sim15-25$ days (four-six stellar rotations), and can be fitted with six to nine evolving features. Longer segments tend to produce worse fits. The duration of the segments is consistent with the lifetime of individual features and active regions found by \cite{lanza2009}, that is between 20 and 50 days. In figure \ref{slices8}, introduced in the presentation of the light curve of CoRoT-2, the segments are highlighted by color. It can be noticed how the segments are related to the different phases of activity in the light curve. The brightness variations grow in amplitude, reach a maximum, and shrink again.\\
To connect the solutions of consecutive segments, the features with non-zero size at the end of a given segment were kept for the next segment. Up to eight features were kept from a segment to the following one. I used the same $\phi$, constraining $\lambda$ around the value of the solution in the previous segment with a normal prior with FWHM equal to $10^\circ$, and constraining $c$ in order to force the spot to remain darker, or brighter, than the stellar surface. The parameter \amax\ was left free in a uniform prior, starting from the best-likelihood value of the previous segment. The evolution times were left free, in order to allow for deviations from a strictly monotonic growth-and-decay law for the spot sizes. Other features were added with completely free parameters, to compensate for the fixed ones. The total number of features per segment, between the ``fixed'' and the free ones, is between six and nine.\\
The flux offset was fitted separately in every segment, starting the chains of each segment from the best value of the previous one. In this way, a possible photometric long-term trend was kept into account.\\
Chains up to $3 \cdot 10^5$ iterations-long were run for each segment. In this way, most of the chains of each segment could reach the end of their burn-in phase; a chain was considered converged after its likelihood did not increase for some thousand steps after the end of its burn-in phase. From the chain of a given segment reaching the highest likelihood, a best-likelihood solution was extracted.

\paragraph{Spot-transit fit (fit 4).}
This fit started from the solution of fit 3. With it, I took into account both non-occulted and occulted activity features. It was performed on data set 4, divided in segments as for fit 3. The transit parameters were fitted while fixing the spot parameters to their best solution of fit 3, except two. The size \amax\ and the longitude $\lambda$ of the features were fitted again. The parameter \amax\ primarily affects the transit depth; it was left free, starting from the value of the best solution of fit 3. The longitude affects the position of the spot in the transit profile; this can affect the transit duration, and therefore the limb darkening coefficients, $i_p$, and $\rho_\star$. A normal prior with FWHM $= 5^\circ$, centered on the best likelihood value, was used. Fixing the other spot parameters, the main correlations and degeneracies were removed.\\
The priors of the transit parameters are the same as for fit 1. Two parameters were set differently. Instead of $a/R_\star$, \ksint\ uses $\rho_\star$, for which I used a $0.35 \, \rho_\odot$ -FWHM Gaussian prior centered on the value of \cite{alonso2008}. Normal priors were set for the limb darkening coefficients, using the results of \cite{alonso2008}. By trials, I verified that constraining these coefficients helped the convergence of their fit.\\
In this phase, ten chains from 1.5  to $2\cdot 10^5$ steps were employed to reach convergence for each segment. For each segment, the chains were then thinned according to their correlation length and merged into a single one, giving the confidence intervals for the parameters of that segment. To be considered robust, a merged chain was required to consist of at least a thousand uncorrelated points.

\paragraph{Model-based normalization (fit 5)}
This fit consists in a standard transit fit, where transit normalization is performed via the modeling of the non-occulted spots and faculae. The best-likelihood solutions obtained from fit 3 were re-computed without the planet. They were merged in a single light curve and used to normalize the transits, obtaining data set 5 (section \ref{phot}). Equation \ref{czesla} was used, with $n_i$ indicating the model, and $p$ the flux offset value. For each segment, this parameter was fixed to the best-likelihood value obtained with fit 3. Its uncertainty, obtained with fit 4, was quadratically added to the standard deviation of the flux. In this way, the uncertainty due to the normalization technique was evaluated. This cannot be done without a spot model.\\
Once the transits were normalized, a standard transit fit was performed with \ebop. This code was used because of its higher computation speed compared to \ksint. The priors are the same as fit 1 and 2; the contamination was fitted, as well, because the activity features inside transits were not modeled any longer.

\subsection{Results}\label{res}

\begin{figure*}[htb]
\centering
\includegraphics[scale = 0.28]{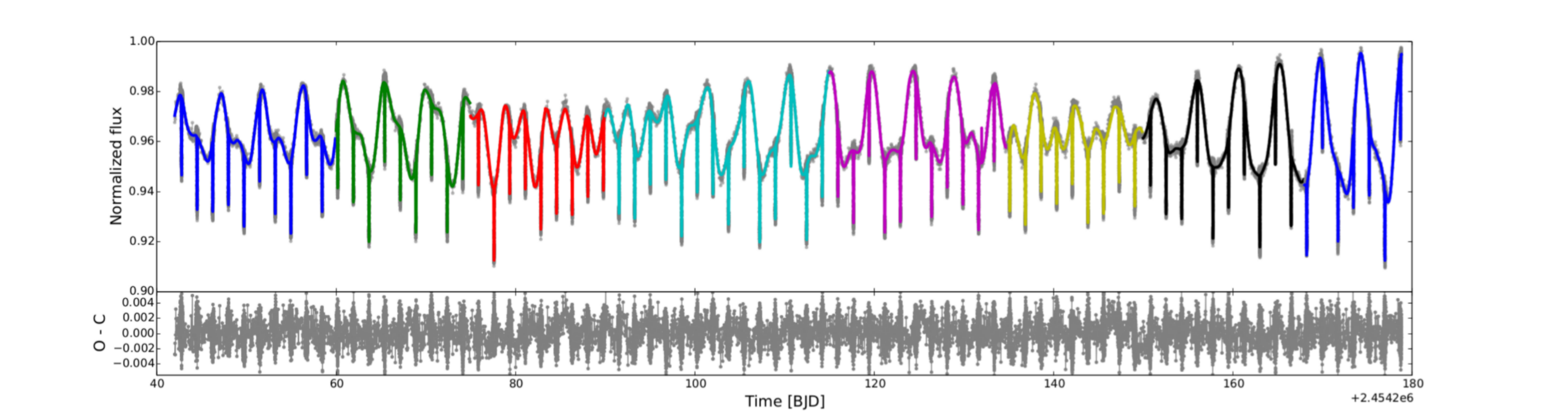}
\caption[Model light curve of CoRoT -2 from \ksint\ plotted over the data]{The model light curve from \ksint\ plotted over the data. The eight segments of the fit are divided by color. The residuals are shown in the lower panels, and error bars are not shown for clarity. The larger amplitude of the residuals in correspondence of the transits is due to the full resolution kept for the transits. The out-of-transit binning is of 2016 s, and inside the transits 32 s.}
\label{fit_all}
\end{figure*}

Figure \ref{fit_all} presents the best model of fit 4, plotted over the light curve, where the segments are highlighted by color. The spot parameters (yielded by fit 3) and the transit parameters (fit 4) are discussed separately.

\subsubsection{Spot parameters}\label{sppar}
The solutions of fit 3 correspond to the deepest local minima of the $\chi^2$ space found by the MCMC. As discussed in the previous section, six to nine activity features were needed for the fit of each segment of the light curve. This is in agreement with the results of \cite{silva-valio2010} and \cite{silva-valio2011}, who analyzed only the transits. The posterior distributions given by the best-likelihood chains give information about the spot parameters of these minima.
\begin{enumerate}[-]
\item Spot longitudes are distributed all over the stellar surface. Mean separations of $\sim 50-100^\circ$ were found between active regions. This is in contrast with the result of \cite{lanza2009}. Using a non-evolving spots-only model, they found two active longitudes separated by $\sim 180^\circ$. I compared the active longitudes of segment 1 with the result of \cite{huber2009}, who fitted a part of the light curve included in segment 1. Interestingly, \ksint\ + \pastis\ recovers an active region at $\lambda \sim 330^\circ$, as they did. No trend was observed with respect to the phase of activity (increasing brightness variations, maximum, decreasing: see section \ref{spotsonly} and figures \ref{slices8} and \ref{fit_all}). Individual features tend to move by some tens of degrees between consecutive segments, as expected from the constraint of continuity between adjacent solutions. Displacements of some tens of degrees indicate the possible need to model longitudinal migration.
\item All spot sizes are explored by the fit. Most of the features have \amax $> 5^\circ$. No net change in the mean \amax\ is observed among different activity phases (increasing, maximum, decrease).
\item Faculae (spots with $c > 1$) represent from $\sim 30\%$ to $\sim 50\%$ of the total spot coverage. 
\item In figure \ref{aeff_slice}, the effective coverage factor of the stellar surface is plotted as a function of the segment of light curve. This parameter is similar to the ratio of the facular-to-sunspot ratio, studied elsewhere. For each segment, it is defined as 
\begin{equation}
\mathcal{C} =\sum_i \alpha_{\mathrm{max},i} (1 - c_i),
\end{equation}
where the sum is run over all the features $i$ with maximum area $\alpha_{\mathrm{max}}$ and contrast $c$. For this calculation, only the most probable values of the solutions were used. A facula has $c > 1$, hence contributes negatively to the sum.

\begin{figure*}[!htb]
\centering
\includegraphics[scale = 0.5]{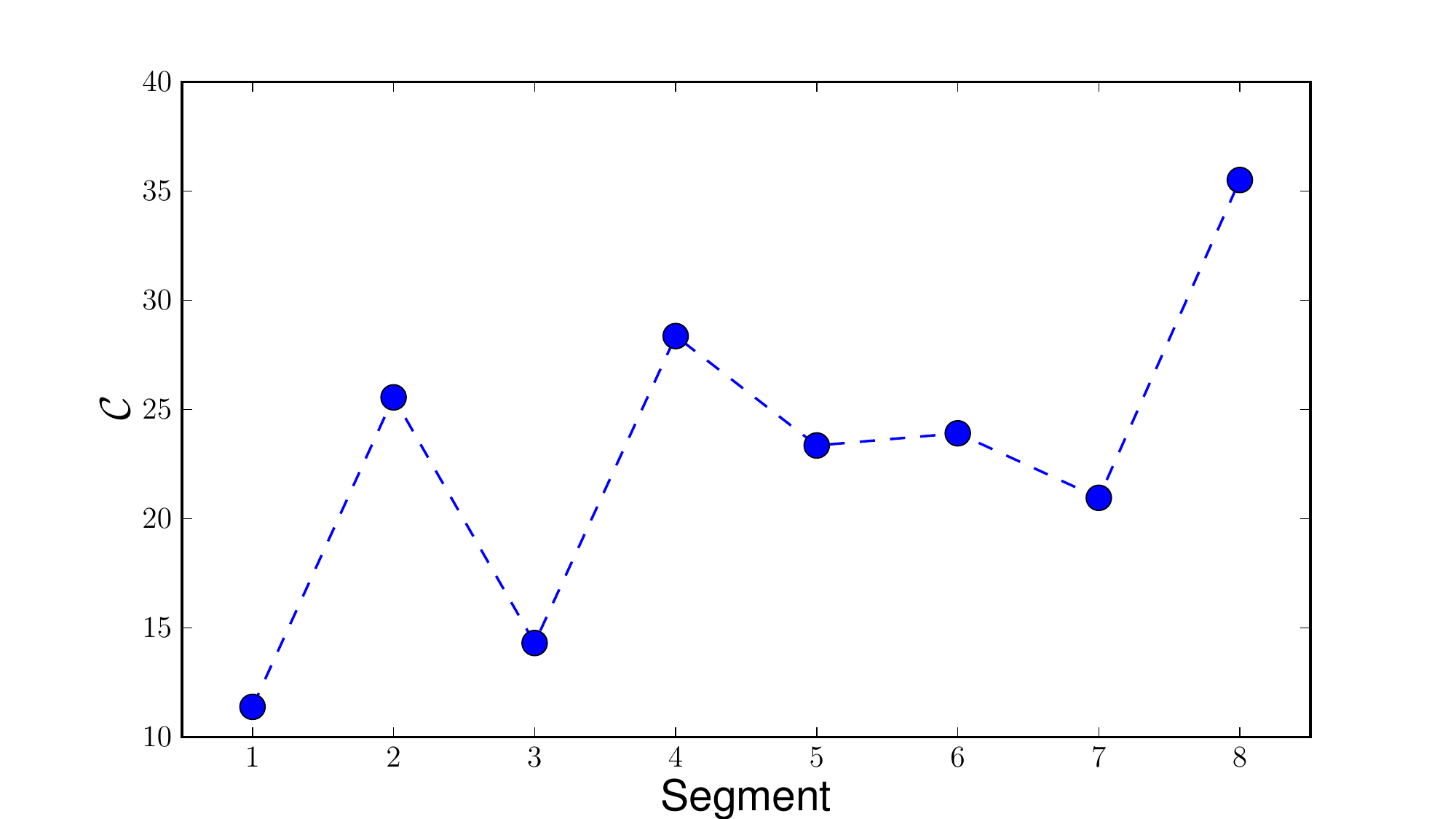}
\caption{Effective coverage factor as a function of the segment of light curve.}
\label{aeff_slice}
\end{figure*}

Figure \ref{aeff_slice} shows an increase, followed by a decrease, of $\mathcal{C}$ for the first activity cycle (segments 1 to 3, see figures \ref{slices8} and \ref{fit_all}), and indicates a similar behaviour for the third cycle (segments 7 and 8). The second cycle (segments 4 to 6) does not show this trend. This indicates a larger impact of faculae.
\item In every segment, one or two activity features (either dark spots or faculae) are found to cross the transit chord. Therefore, occulted features are a minority.
\item No trend is observed for the fitted evolution times \tmax, \tlife, \ingress, and \egress.
\end{enumerate}

\subsubsection{Transit parameters}\label{joint}
\begin{figure}[!htb]
\centering
\includegraphics[scale = 0.5]{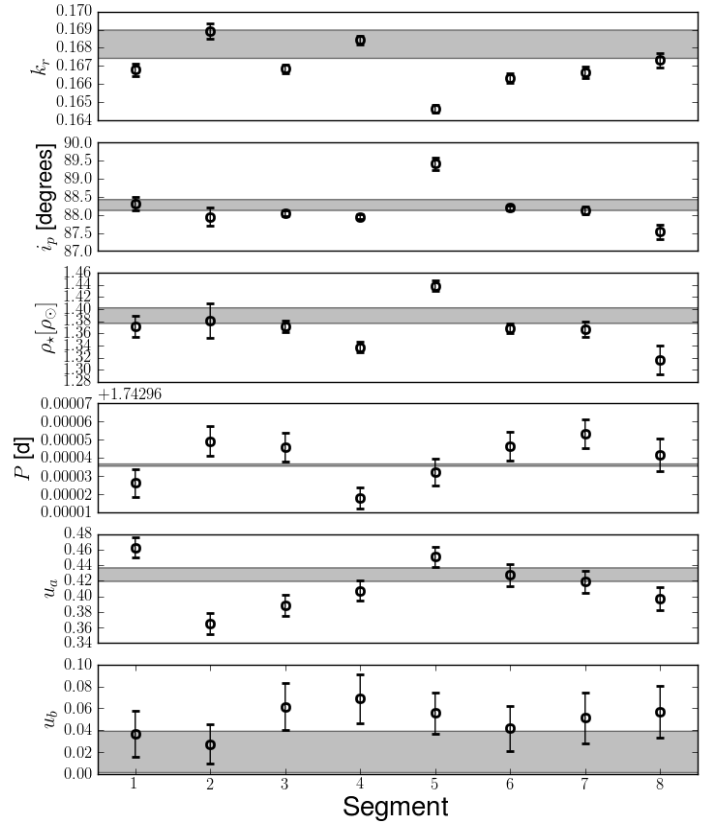}
\caption[Results of the fit on the segments of CoRoT-2 light curve]{From top to bottom: in black, values and uncertainties of $k_r$, $i_p$, $\rho_\star$, $P_{\mathrm{orb}}, u_a$, and $u_b$ for fit 4, indicated as a function of the segment of the light curve. The results from fit 5, with their 68.3\% confidence intervals, are shaded in grey.}
\label{all_post}
\end{figure}

Table \ref{tabastronorm} reports the transit parameters of all fits. The results of fit 4 are the average of all the results obtained on each segment. The values obtained for each segment are reported in table \ref{tsegments} and plotted in figure \ref{all_post}. The shaded regions correspond to the results of fit 5.

For some segments, the results of fit 4 are in poor agreement between one another. $k_r$, $i_p$, and $\rho_\star$ are determined with high precision, but low accuracy, so that they are often only at $2\sigma$ agreement, or worse. Because of this, the error bars on the average results of fit 4, in table \ref{tabastronorm}, are larger than on the other fits, whose uncertainties are similar to those found by \cite{alonso2008}. The $k_r$ on segment 5 is lower than those estimated from the other segments, and oppositely for $i_p$ and $\rho_\star$. The large contribution of faculae in this segment (see figure \ref{aeff_slice}) might indicate that faculae are more difficult to correct for than dark spots and bias the transit parameter estimate.\\
The transit depth $k_r$ was found to be in agreement between fit 4 (average $k_r = 0.1670 \pm 1.2\cdot 10^{-3}$) and fits 1, 2, and 5. Interestingly, the agreement between fits 4 and 2 ($k_r =0.16632\pm 8.0 \cdot 10^{-4}$, using \citeauthor{czesla2009} normalization) is the best. However, the low $k_r$ of segment 5 has an important role in the average $k_r$ of fit 4, bringing it close to the result of fit 2.\\
This parameter was found to depend on the normalization technique. As expected, $k_r$ is larger if a standard normalization is performed (fit 1, $k_r = 0.16917 \pm 8.5\cdot 10^{-4} $). In fact, in this case, non-occulted features are neglected and cause an overestimate of this parameter. On the other hand, normalization through equation \ref{czesla} (fit 2) produces a smaller $k_r$ ($0.16632\pm  8.0\cdot 10^{-4}$). This is because of the division by the unperturbed flux value ($p$), which is assumed to be the maximum one. For this value, without a spot model, no uncertainty is available. Therefore, an \textit{a priori} uncertainty of about half of the amplitude of the activity pattern ($\sim 3\%$) should be added to $k_r$ derived with this method. This is the main limit of this normalization. Fit 5 models non-occulted features, and therefore is to be preferred to fits 1 and 2. It yields a $k_r$ that lies between those of fits 1 and 2. The slightly larger error bars on data set 5, due to the addition of the normalization uncertainty, do not affect the error bars of this fit.\\
Besides segments 5 and 8, the results of fit 4 for $i_p$ and $\rho_\star$ are consistent between themselves. The underestimate of these parameters in segment 8 may be related to a large coverage factor, as shown in figure \ref{aeff_slice}. This aspect will be discussed in more detail in section \ref{discinc}.\\
The orbital period $P_{\mathrm{orb}}$ is found to be scattered among the segments of fit 4. This could be related to an erroneous measure of the transit duration. Indeed, an indication of apparent activity-induced TTVs in the CoRoT-2 light curve, with an amplitude of $\sim 20 s$, was presented by \cite{alonso2009}. Such TTVs would become more important once only parts of the light curve are fitted. As a consequence, the average $P_{\mathrm{orb}}$ on fit 4, in table \ref{tabastronorm}, has an uncertainty $\sim 10^2$ larger than the one measured with fit 1, 2, and 5. The measurement of TTVs needs a transit-by-transit analysis, which our code is not able to perform.\\
The limb darkening coefficients $u_a$ and $u_b$ were found to be almost the same for fits 1, 2, and 5. In fit 4, instead, $u_a$ results in disagreement among the segments, despite the imposed tight normal prior. An explanation can be the difficulty for \pastis\ + \ksint\ to recover the limb darkening coefficients if only a few transits are available. A badly fitted $P_{\mathrm{orb}}$ could contribute to this problem, as well. Another cause could be that the limb darkening coefficients actually change as a function of the varying spot coverage of the stellar surface \citep{csizmadia2013}. This possibility will be better discussed in section \ref{discinc}.\\
The following caveats need to be addressed to improve the quality of fits 3, 4 and 5.
\begin{enumerate}[-]
\item The fixed number of activity features. By manually adding features to the fit, short-lived ones inside the transits could have been missed. An automatic incremental addition of activity features, especially inside the transits, would be suitable.
\item Fitting $\lambda$ and \amax\ might not be the optimal method to propagate the uncertainties from fit 3 to fit 4. However, if more spot parameters are left free during fit 4, the chains do not converge.
\item The degeneracies between the activity features. When two or more features are in the visible stellar disk, their individual contribution to the total flux is not distinguished by the fit. This prevents the MCMC from converging. This problem has been mitigated in literature by dividing the stellar surface in fixed, non-overlapping regions \citep[e.g.][]{lanza2009,huber2009}. Another possibility is to constrain differently the parameters of the features once they come into contact, i.e. identifying them in a different way than with coordinates, or treating them as a single spot.
\item Longitudinal migration, as observed while discussing the spot parameters. 
\item Differential rotation. Once the stellar inclination is fixed, this could help the fit to be more accurate, and to constrain the spot latitudes.
\end{enumerate}

\begin{landscape}
\vspace*{.5in}~\\
\begin{table*}[!htb]
\begin{center}{
\caption[Transit parameters of CoRoT-2 with their 68.3\% confidence intervals]{\label{tabastronorm} Transit parameters with their 68.3\% confidence intervals. Results on the classic fit (fit 1), on the light curve normalized as \cite{czesla2009} (fit 2), on the segments (fits 3 and 4), and on the model-normalized light curve (fit 5). The stellar density of fit 1, 2, and 5 is derived from the respective $a/R_\star$.}
\scalebox{0.8}{\begin{tabular}{lllll}
\hline
\hline
\\
\textit{Parameter}  & \textit{Fit 1} & \textit{Fit 2} &  \textit{Fit 3-4} & \textit{Fit 5} \smallskip\\

$P_{\mathrm{orb}}$ [d]  & $1.74299628 \pm 5.8 \cdot 10^{-7}$ & $1.74299620 \pm 5.9 \cdot 10^{-7}$ & $1.742999\pm 1.2 \cdot 10^{-5}$ & $1.74299609 \pm 5.9\cdot 10^{-7}$\\
$T_0$ [BJD]   & $ 2454237.535398 \pm 2.9 \cdot 10^{-5}$ & $2454237.535399 \pm 2.8 \cdot 10^{-5}$ &  - & $2454237.535401 \pm 2.8 \cdot 10^{-5}$\\
$i_p$ [degrees]  & $88.34 \pm 0.16$ &  $88.01 \pm 0.13$  & $88.19 \pm 0.51$ & $88.27 \pm 0.15$\\
$k_r$  & $0.16917 \pm 8.5\cdot 10^{-4} $ &  $0.16632\pm  8.0\cdot 10^{-4}$ & $0.1670 \pm 1.2\cdot 10^{-3}$ & $ 0.16820\pm 7.8 \cdot 10^{-4}$\\
$\rho_\star$ [$\rho_\odot$] & $1.395 \pm 0.014$ & $1.361 \pm 0.013$ & $1.369 \pm 0.033 $ & $1.390 \pm 0.013$ \\
$u_a$  & $0.428^{+0.006}_{-0.009}$ & $0.426^{+0.007}_{-0.011}$ & $0.415 \pm 0.030$ & $ 0.428^{0.007}_{-0.011}$ \\
$u_b$ & $0.017^{+0.020}_{-0.013}$ & $0.023\pm 0.023$ &  $0.050 \pm 0.013$ & $0.020^{+0.023}_{-0.016}$ \\
Flux offset  & $1.0000473 \pm 9.0\cdot 10^{-6}$ & $1.0000459 \pm 8.6\cdot 10^{-6}$  & $0.9755\pm 4.1 \cdot 10^{-3}$  & $1.0000467 \pm  9.3\cdot 10^{-6}$\\
Flux jitter  & $0.0009171 \pm 9.1\cdot 10^{-6}$ &  $0.0009283 \pm  8.3 \cdot 10^{-6}$ & - & $0.0009387 \pm 8.9\cdot 10^{-6}$\\
Contamination [\%]  & $8.92 \pm 0.90$ & $8.81\pm 0.90 $ & - &$8.83 \pm 0.85$ \\
\\
 \hline

\end{tabular}}
}\end{center}
\end{table*}
\end{landscape}

\subsubsection{Least distorted transit}\label{individual}
\begin{figure*}[!htb]
\centering
\includegraphics[scale = 0.4]{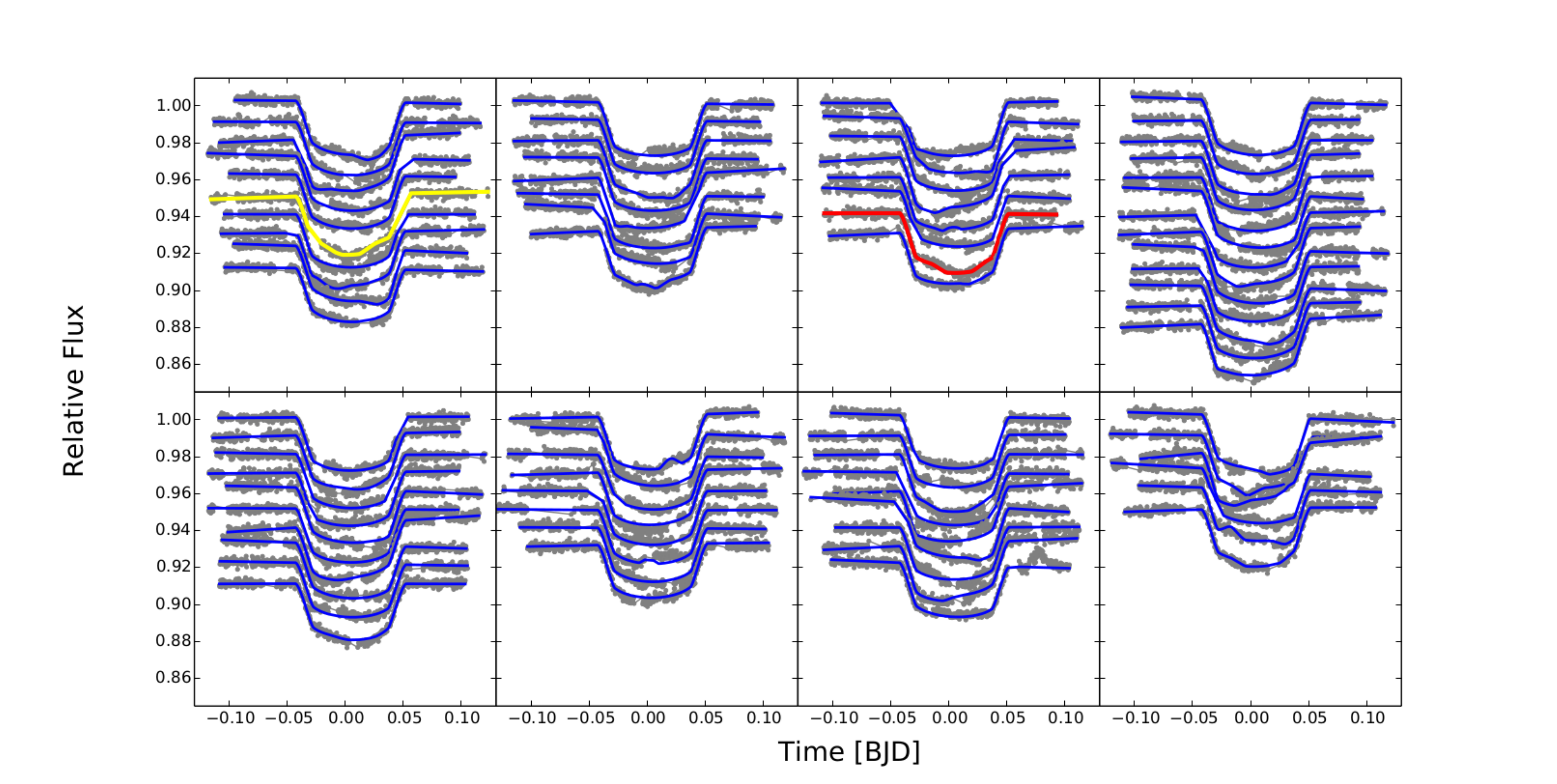}
\caption[Fit on the transits for every segment]{Fit on the transits for every segment (from left to right, and top to bottom). The deepest transit, containing an occulted facula, is shown in yellow. The transit less affected by activity features is colored in red.}
\label{transits_8segments}
\end{figure*}

According to \cite{czesla2009}, the deepest transits are those that are less affected by occulted spots; therefore, their transit depth should be closer to the true one. Working on \corot-2, they interpolated a lower envelope to an average of the deepest transits. These were found to happen at the moments where the out-of-transit flux is the largest. They fitted only $k_r$ and $i_p$, and obtained $k_r = 0.172 \pm 0.001$ and $i_p = 87.7 \pm 0.2^\circ$. Their method assumes the dominance of dark spots over faculae, as found by \cite{lanza2009} by fitting the out-of-transit light curve.\\
With our approach, the assumption of the prevalence of dark spots could be checked. I adopted a similar approach to \cite{czesla2009}, but worked on the deepest transit only, in order to avoid averaging any spot feature. In figure \ref{transits_8segments}, the best solution of fit 4 is plotted over the transits. The sixth transit (in yellow in figure \ref{transits_8segments}) was considered. 

\begin{figure*}[htb]
\centering
\includegraphics[scale = 0.25]{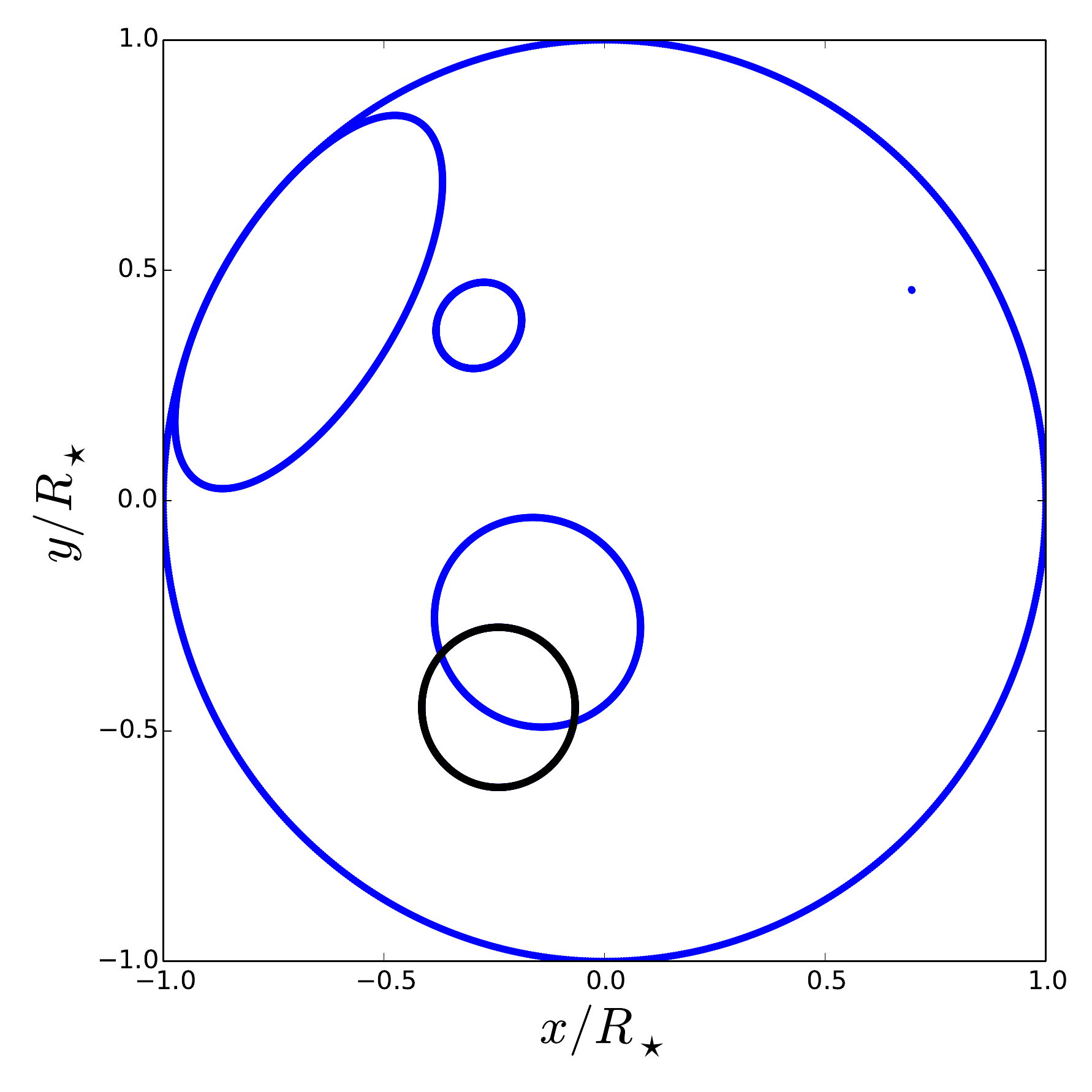}
\includegraphics[scale = 0.25]{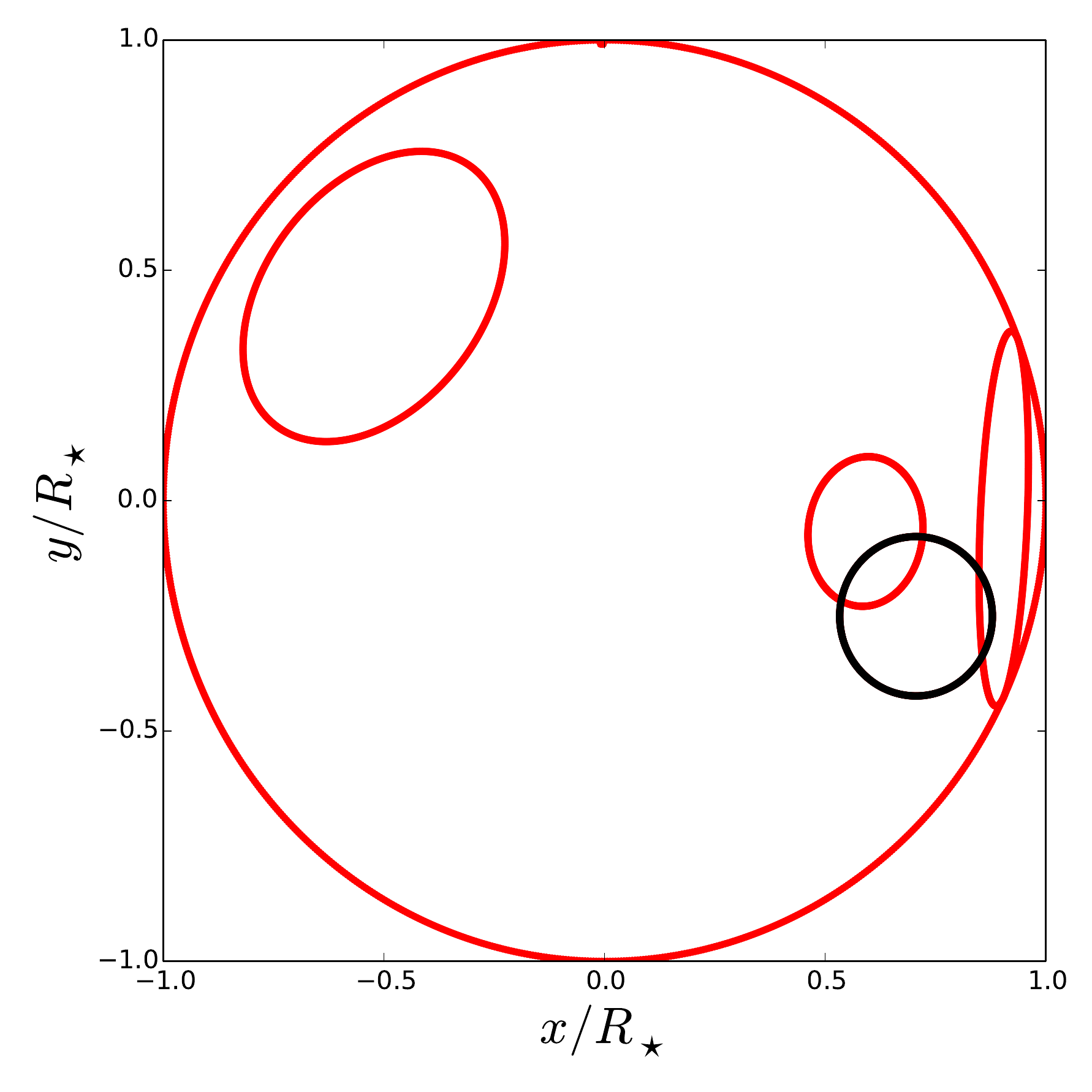}
\includegraphics[scale = 0.21]{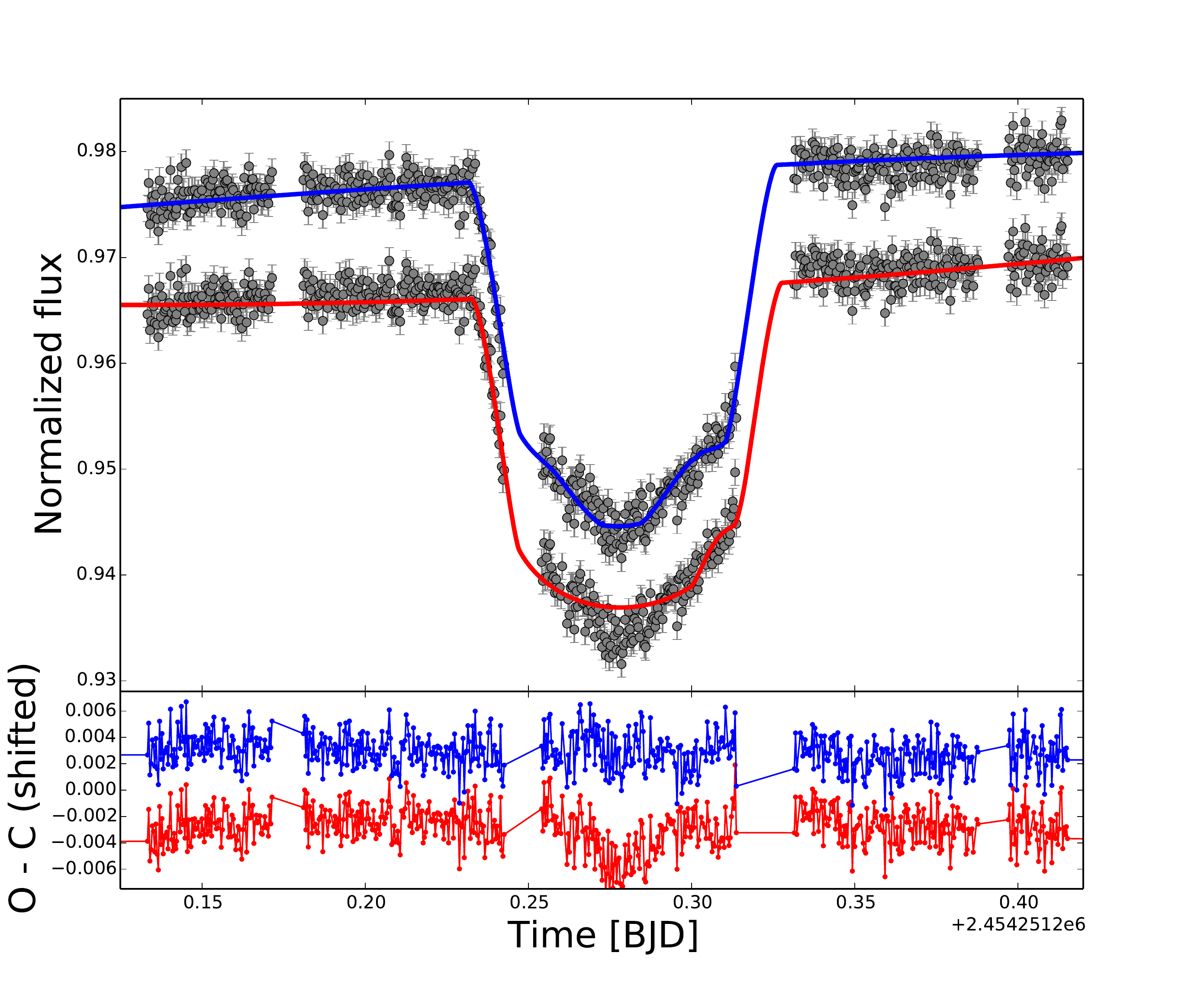}
\caption[Fits on the deepest transit of CoRoT-2 b]{\textit{Left:} Planet (black) and starspots (blue) configuration during the  deepest transit, with faculae allowed for. The faculae are crossed by the planetary disc. \textit{Center:} The solution with three dark spots, colored in red. \textit{Right:} The deepest transit fitted with the spot-facula configuration (blue) and the three-spots model (red, shifted). The residuals are shifted for clarity and use the same color code.}
\label{fitfacula6}
\end{figure*}

The transit was taken from data set 4. The spot configuration obtained with fit 4 for the corresponding segment was fixed. The transit parameters $k_r,$, $i_p$, $\rho_\star$, and the jitter were fitted. Because of the lower number of points in the transit with respect to an entire segment, the limb darkening coefficients, the foot, and the contamination value were fixed.\\
The MCMC yielded the values $k_r=0.1734^{+0.0010}_{-0.0016}$, $i_p = 86.02 \pm 0.27^\circ$, and $\rho_\star = 1.093 \pm 0.038\,  \rho_\odot$. The transit depth is in $1\sigma$ agreement with \cite{czesla2009} results. The low values of $i_p$ and $\rho_\star$, instead, have to be attributed to the distorted transit profile.\\ 
The fit finds this transit to be affected by a facula, whose position during the transit is shown on the left side of figure \ref{fitfacula6}. A facula increases the apparent $k_r$, contrarily to a dark spot. Fitting a single transit, \ksint\ + \pastis\ does not disentangle the facula from the transit profile.\\
To check whether a dark spots-only solution can be found by our model, I fixed the transit parameters to the mean values of \cite{alonso2008}, and imposed three dark spots (that is, with $c < 1$) for the fit of the deepest transit. A minimum of three spots was considered necessary. Indeed, two occulted spots at the borders of a transit can mimic a facula at the center of the transit; a third non-occulted spot is needed to generate a possible out-of-transit flux variation, if the other two are not sufficient. The contrast of the spots was fixed to the conservative solar value of 0.67 \citep{sofia1982}. The latitude of the occulted spots was fixed close to $-2.16^\circ$, to lie on the transit chord. Their longitudes were forced to lie in the visible stellar disc, to help the fit. The latitude of the non-occulted spot was set to $30^\circ$.\\
The best spots-only configuration is plotted in the center of figure \ref{fitfacula6}; this solution and the one with a facula are compared on the right side of the figure. Without a facula, the distortions of the transit profile are not recovered: the Bayesian Information Criterion (BIC, \citealp{schwarz1978}) favors the model with a facula over the one with dark spots only. Indeed, BIC$_{\mathrm{facula}}$ - BIC$_{\mathrm{dark \, spots}} \simeq -4$, corresponding to a Bayes factor $\sim e^{-4} \simeq 0.018$ between the dark spots-only and the facula model.\\
This result suggests that \cite{czesla2009} approach is flawed by the fact that faculae are not taken into account, and so lead to an overestimate of $k_r$. To constrain the presence of faculae, the modeling of activity features inside transits is necessary. The unperturbed transit profile is likely situated at a lower flux level than at the maximum peak of the light curve, as the maximum could be caused by an increased total stellar flux due to faculae.\\
Taking this into account, the twenty-sixth transit (in red in figure \ref{transits_8segments}) was considered as the closest one to the true profile. The fit was repeated in  the same way as for the deepest transit. This time, the result was $k_r = 0.1689 \pm 0.0008$, $i = 87.63 \pm0.53^\circ$, and $\rho_\star = 1.336\pm0.062 \, \rho_\odot$, in agreement with fit 5.

\subsection{Discussion}\label{disc}

\subsubsection{Correlations between spot coverage and transit parameters}\label{mostrel}
\begin{figure}[htb]
\centering
\includegraphics[scale = 0.5]{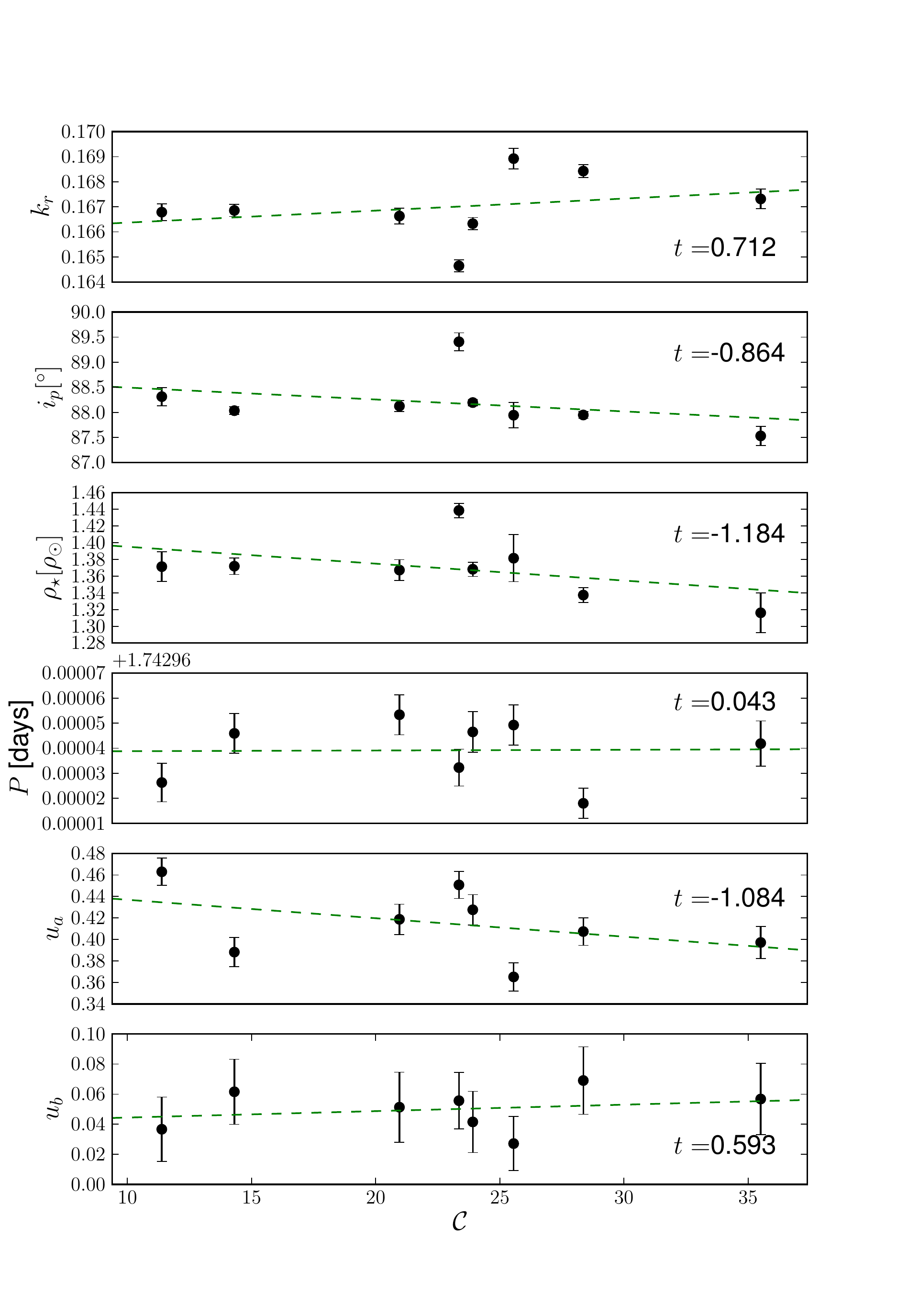}
\caption[Transit parameters as a function of the effective coverture factor]{Transit parameters as a function of the effective coverture factor, for all the segments. The values of the $t$-statistics are indicated.}
\label{correlations}
\end{figure}

The results of the modeling were used to explore the impact of the activity features on the transit parameters. In figure \ref{correlations}, all the transit parameters found with fit 4 are plotted as a function of the effective coverage factor $\mathcal{C}$ of the stellar surface, introduced in section \ref{sppar}. In each panel, the value of the Student's $t$-distribution for the plotted couple of variables is reported.

\paragraph{Transit depth and contamination.}\label{disctr}
The transit depth $k_r$ is found to increase with $\mathcal{C}$. The $t$ statistics yields 0.712. This value has small significance, suggesting that the impact of the activity features on $k_r$ is corrected for by fit 4. In that case, the scatter between segments is not due to non-disentangled spot effects.\\
The fixed value of the transit parameters imposed in fit 3 could have induced a systematic error on $k_r$. The fixed parameters may have prevented a better spot solution to be found. The scatter between the solutions of the segments of fit 4, however, suggests that the fit is not strongly dependent on the starting transit parameters. Otherwise, a smaller scatter would be expected.\\
During fits 3 and 4, the contamination term was fixed to 8.81\%, a much higher value than the level 0 value used by \cite{alonso2008} in the discovery paper (5.6\%). In addition, these authors used a ``local'' normalization, such as the one of data set 1. They found $k_r = 0.1667 \pm 0.0006$. Fit 1 yields $k_r \simeq 0.1692$. A difference of $\sim3\%$ in the contamination with the same normalization, therefore, produces a difference in $k_r$ of $\sim 1.5\%$. If the same contamination value is used, instead, choosing a classic (fit 1) or model-based (fit 5, $k_r \simeq 0.1682$) normalization yields a difference of only $\simeq 0.6\%$. This difference becomes more important if one compares the ``local'' and the \citeauthor{czesla2009}-like normalization (fit 2), yielding a $\sim 1.7 \%$ difference. This has to be taken into account for the modeling of planetary interiors and their evolution, which requires 1\% precision in planetary radii below five Earth masses \citep{wagner2011}.

\paragraph{Orbital inclination, stellar density, orbital period, and limb darkening.}\label{discinc}
A slightly larger significance is observed for the $t$-value between $\mathcal{C}$ and $\rho_\star$ (-1.184, figure \ref{correlations}), indicating a lower $\rho_\star$ measured with increasing $\mathcal{C}$. This might be linked with an underestimate of $i_p$ with increasing $\mathcal{C}$; however, the $t$-value between these parameters is not remarkably significant (-0.864). As noticed in the presentation of the results, the possible underestimate could be an effect of apparent activity-induced TTVs. The $t$-value between $P_{\mathrm{orb}}$ and $\mathcal{C}$, however, is almost null (0.043). 
An apparent lower $\rho_\star$ due to starspots was also observed for CoRoT-7 \citep{leger2009}, in comparison with the derived spectroscopic value. In the present case, it would mean that \ksint\ + \pastis\ does not effectively correct for the effect of spots and faculae on the transit edges. It has to be observed that, in these statistics, the low values of these parameters fitted on segment 8 have an important weight.\\
Another possibility to explain the slight anticorrelation of $\rho_\star$ with $\mathcal{C}$ is to associate it with the fixed value of the stellar inclination used in fit 3. This fixed parameter could affect the fit more than the fixed transit parameters during fit 3. Testing this value cannot be done with the present model. Indeed, the uncertainty on the latitude of occulted activity features, and therefore in the stellar inclination, can be estimated from the planet size \citep{desert2011}. The result is $\pm \arctan(k_r) \simeq \pm 17^\circ$, much larger than the discrepancy observed in our results.\\
The linear limb darkening coefficient $u_a$ was found to be anticorrelated with $\mathcal{C}$, oppositely to the quadratic one $u_b$ ($t= -1.084$ and 0.593, respectively). 
\cite{csizmadia2013} showed that limb darkening coefficients decrease as the fraction of the stellar surface covered by activity features increases. Our result confirms their statement, and therefore indicates that further developments in the fitting method are needed to disentangle this effect, especially when faculae are included in the fit. 

\subsubsection{Length of the segments}

The impact of the length of the segments on the fitted parameters was examined. I considered a segment half the duration of segment 1. Fits 3 and 4 were performed. This yielded $k_r = 0.1670 \pm 0.0004$, $i_p = 88.38^{+0.27}_{-0.44}$, $\rho_\star = 1.392^{+0.025}_{-0.043}$, $u_a = 0.421 \pm 0.18$, and $u_b = 0.050 \pm 0.027$, in agreement with the first segment of fit 4 (table \ref{tsegments}). Instead, longer segments than the ones used for fits 3 and 4 tend to produce worse fits for the out-of-transit flux, as indicated in the description of fit 3 in section \ref{spotsonly}. This suggests that the main improvement our method needs is related to the modeling of the time evolution of the activity features.\\

The scatter induced by fitting short segments of the light curve in a standard transit fit was quantified, too. This quantifies the errors introduced when the transit fit is performed for systems observed for a short time.

\begin{figure}[!bth]
\centering
\includegraphics[scale = 0.5]{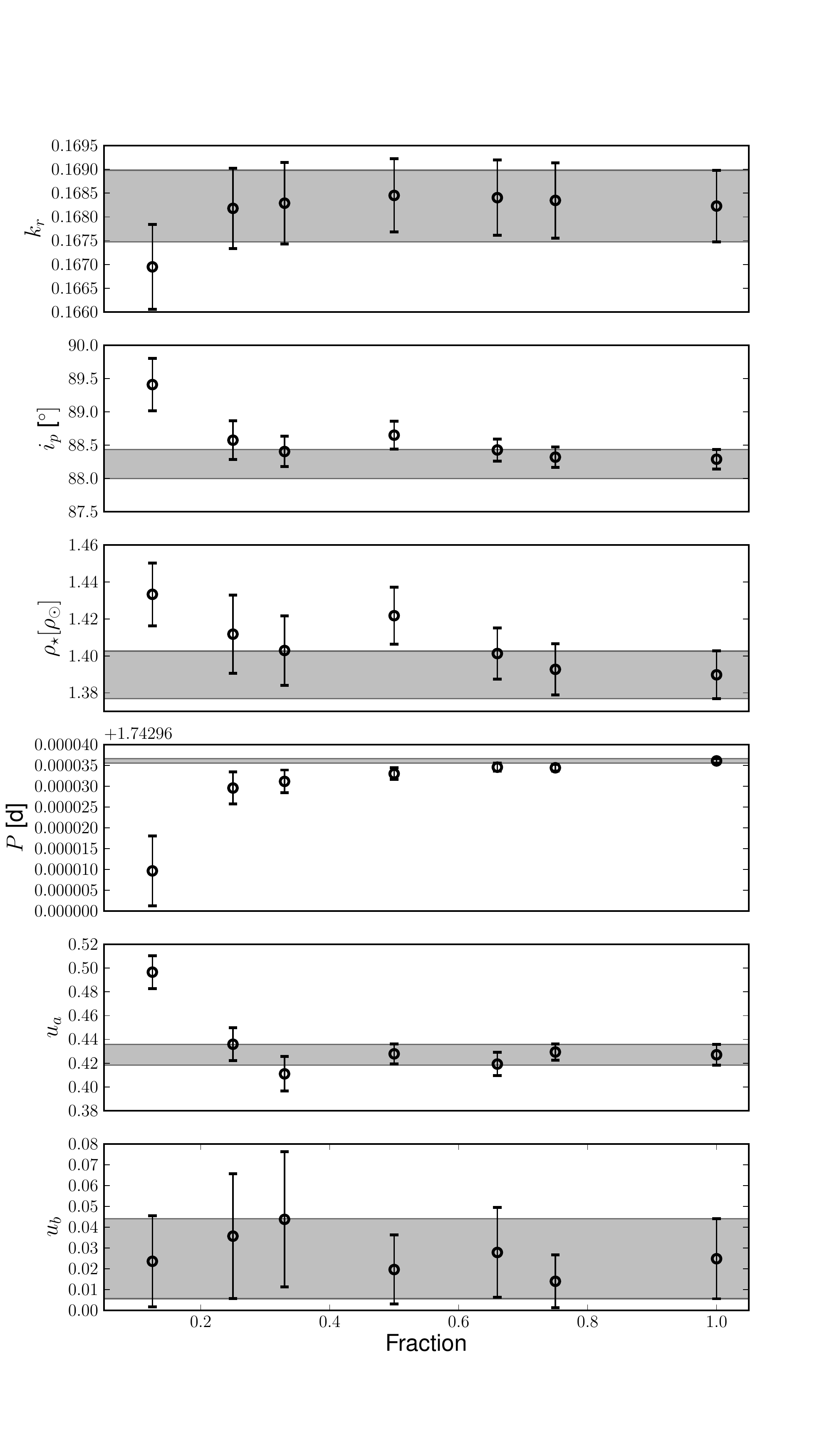}
\caption[Transit parameters derived on the model-normalized light curve with \ebop, considering increasing fractions of the data]{Transit parameters derived on the model-normalized light curve with \ebop, considering increasing fractions of the data (1 corresponds to the entire light curve). The shaded regions indicate the results and $1\sigma$ uncertainties on the entire light curve.}
\label{fractions}
\end{figure}

The model-normalized light curve (data set 5) was divided in segments of increasing duration. It was then fitted with \ebop. The derived transit parameters are listed in table \ref{tsegmentsebop}, and plotted in figure \ref{fractions}. This figure is structured as figure \ref{all_post}.\\
The plots show how the transit parameters converge after about half of the transits are used ($\sim 40$). In particular, a $1\sigma$ deviation of $i_p$ and $\rho_\star$ was found even when fitting half of the light curve. For a small number of transits, both the lower number of points to be fitted and the perturbations due to the spots and faculae make the fit less accurate. The scatter observed for fit 4, performed on segments of about one eighth of the light curve, is probably due to the same reason. Comparing the results of fit 4 on segment 1 (table \ref{tsegments}) and on one eighth of the light curve, which have a similar length, one can estimate the error induced by not modeling occulted activity features. This results in $\Delta k_r \sim -0.1\%$,  $\Delta i_p \sim 1.2\%$, and $\Delta \rho_\star \sim 4.5\%$. Because of the scatter of $k_r$ among the segments, $\Delta k_r$ can be considered a lower limit.

\subsubsection{Planet radius}
Our results indicate that a fit like the combination of fits 3 and 4 has the potential to measure the transit parameters in the most unbiased way. Indeed, it is not affected by the transit normalization, and takes into account the occulted activity features. However, it is limited by the need to cut the light curve in parts, which introduces a scatter in the results because of the lower number of points to be fitted. Until a model such as \ksint\ + \pastis\ manages to fit an entire light curve, fit 5 can be adopted as a good compromise. Despite neglecting occulted spots and faculae, this fit is neither affected by the normalization, nor by the chopping of the light curve into segments.\\ 
The $k_r$ and $\rho_\star$ derived from fit 5 were used to calculate the radius of \corot-2 b. I used the Geneva stellar evolutionary tracks \citep{mowlavi2012} and the updated stellar atmospheric parameters of \cite{torres2012} (\teff$=5575\pm 70$ K, \feh$=-0.04 \pm 0.08$). The fit yielded $R_P = 1.424 \pm 0.021$ \RJ. This value is lower by $\sim3\%$ than the one found by \cite{alonso2008} ($1.465\pm0.029$ \RJ), but the two results are compatible. The inflated radius of the planet is confirmed.

\subsection{Conclusions}\label{conclusions}
I presented a method for the fit of transit photometry which takes into account the impact of starspots and faculae in the transit parameters. This allowed me to explore the impact of the activity features on the calculation of the transit (and therefore planetary) parameters. This approach is based on an analytic modeling of activity features and transits, which includes both non-occulted and occulted features. The method was applied to the light curve of \corot-2. Two main types of fit were performed: one considering both non-occulted and occulted features, and one considering only the non-occulted ones.\\
I compared the results of \pastis\ + \ksint\ with other modeling approaches presented in the literature, as well as with other ways of correcting for starspots and faculae in the data reduction. The uncertainties introduced by the type of transit normalization were quantified. It was confirmed that a standard ``local'' transit normalization causes an overestimate of the transit depth. I quantified the underestimate induced by \cite{czesla2009} normalization, due to neglecting the role of faculae. With the method presented here, non-occulted features are taken into account, therefore avoiding these problems. Despite these improvements, the transit parameters are not importantly modified. The inflated radius of \corot-2 is confirmed.\\
The fit of the transits required faculae. This result could not have been achieved with the fit of the non-occulted features only. Our result, therefore, warns against the use of a lower envelope to fit the transits, as faculae are likely to be neglected.\\
A slight anticorrelation was found between the spot filling factor and the stellar density. An apparent lower stellar density due to the impact of starspots was also observed for \corot-7 \citep{leger2009}. The discrepancies could be related to the prior given to the stellar inclination, or to apparent activity-induced TTVs that our fit is not able to detect. The presence of apparent activity-induced TTVs was suggested by \cite{alonso2009}; a detailed analysis of such TTVs would benefit from a dedicated study.\\
Our modeling of the evolution of the activity features needs to be improved to fully correct for the impact of spots and faculae on limb darkening coefficients.\\
By including the temporal evolution of the features in the code, it was possible to model longer parts of the light curve with respect to other attempts presented in literature. Despite this, it was necessary to cut the light curve into segments and to fit them separately to achieve a correct fit. This suggests that the most important developments needed by our method are related to the modeling of spot and faculae evolution and migration. Indeed, the fit of a large number of transits (at least 40) is needed to be confident about the derived transit parameters. Because of the low number of transits in each segment, instead, the transit parameters fitted on different segments are in poor agreement between each other.\\
This study is a first step towards a consistent modeling of activity and transit features in transit photometry, and requires further investigation on cases of active stars. To begin with, a study on synthetic data could quantify the efficacy of \pastis\ + \ksint\ in recovering the correct parameters. Secondly, \pastis\ + \ksint\ should be applied to other real cases. An interesting candidate is CoRoT-7. In section \ref{c7}, it was shown how its light curve can be modeled without the need of dividing it into segments. \cite{barros2014c7} showed that different transit parameters are found in the fit of the LRa01 and the LRa06 data sets. The two light curves show different activity levels; therefore, the application of \ksint\ + \pastis\ on the LRa01 and LRa06 light curves would allow for the quantification of how much the activity features can be disentangled from the transit profiles with respect to a standard transit fit, on an entire light curve.\\ 

Important improvements are needed to model data sets several tens or hundreds of days long. This is the usual time span of observations of present-day transit surveys. The consistent modeling of long-duration light curves affected by stellar activity would be an advantage for future missions such as PLATO 2.0 \citep{rauer2014}. On the same line, one of the aspects that cannot be neglected is the one of computation time. Even if an analytic model was employed, some days of calculations were needed for the fits. This is due to the calculations needed to model the overlap of activity features and transits, and to technical details of the code, which need to be optimized. At the present time, for example, \pastis\ + \ksint\ needs to read and write files in input and output.

\section{Publication}
In the following pages, I attach the submitted paper about the modeling of the light curve of \corot-2.

\clearpage
\includepdf[pages={-}]{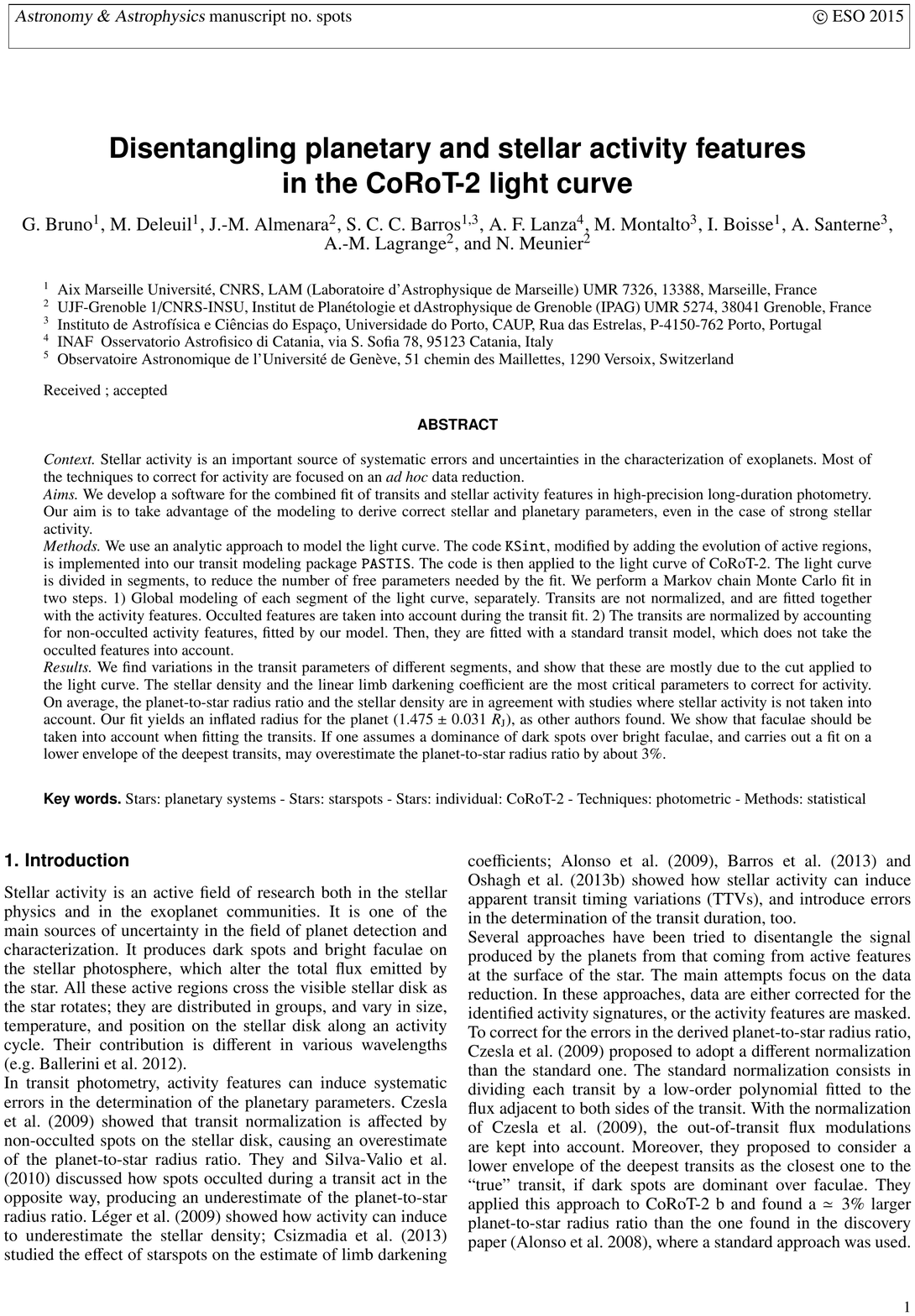}
\clearpage
\chapter{Conclusions and perspectives}

\section{Summary}
During this PhD, I worked on various problematics related to the characterization of exoplanet host stars and their planets.\\
I participated in a campaign of follow-up and characterization of Kepler giant planets. I performed the spectral analysis of nine stars of various spectral types, observed with SOPHIE and other spectrographs. This resulted in the determination of the mass and the radius of the stars and of their planets. Despite belonging to the broad family of giant planets, these bodies have a variety of properties, and foster the development of the theories of planet and brown dwarf formation and of planet interiors.\\
I have taken part in an on-going study aiming at improving the statistics of low-mass stars.  I derived the atmospheric parameters of twenty-one CoRoT and Kepler stars, which are the primary members of long-period binary systems whose secondary is a low-mass star. With these parameters, the mass and radius of the secondary object will be derived. This will allow for a better constraint on the mass-radius relationship of low-mass stars in long-period binaries.\\
I have participated also in an on-going study related to magnetic star-planet interactions. I derived the magnetic activity index \logrhk\ for a sample of thirty-one CoRoT, HAT, and WASP stars observed with SOPHIE, extending the sample of \cite{hartman2010} and \cite{figueira2014}. These measures will be used to further constrain the correlation between the magnetic activity of host stars and the surface gravity of Hot Jupiters. These two latter studies are in phase of development, and their results are planned to be submitted in the short term.\\
I have also explored problematics related to the behavior of the SOPHIE pipeline in low-\sn\ regime. I examined low-\sn\ spectra of benchmark stars, looking for trends in the stellar parameters introduced by a possible wrong correction of the blaze function. With such a study, I determined the \sn\ above which the stellar parameters derived from SOPHIE spectra can be considered reliable, as a function of the stellar spectral type. I verified that, if several low-\sn\ exposures are available, they are not reliable individually, but can be safely co-added in order to avoid the problems due to the high level of noise in the individual exposures.\\
An important part of this project was dedicated to the study of a two-planet Kepler system which is part of the SOPHIE follow-up program, Kepler-117. This system required the development of a method for the fit of transit timing variations (TTV). To that purpose, a code for planet dynamic simulations was implemented in a Bayesian-oriented software for planet validation and characterization developed by members of the exoplanet team of the Astrophysics Laboratory of Marseille (PASI). This allowed for an increase in the precision on the system parameters \citep{bruno2015}. In particular, it allowed us to constrain the mass of the inner and lighter planet of the system. Indeed, the small RV amplitude of this latter only allows for putting an upper limit to its mass. This is due to both the low mass of the planet and to the faint brightness of the host star. Our result was achieved by consistently taking advantage of the high photometric precision of Kepler. The technique we developed is applicable to other similar cases.\\
Among the many problems affecting the precise characterization of exoplanets and of their host stars, I focused on stellar magnetic activity. I developed a method for the fit of starspots and faculae in transit photometry, that exploits two analytic models. The first method does not include transit modeling, and was applied to the Sun and \corot-7. This first approach allowed us to benchmark the method presented by \cite{aigrain2012} to derive activity-induced RVs from the corresponding simultaneous photometric features. The second method involves the modeling of transits and both non-occulted and occulted activity features. This latter was applied to the light curve of \corot-2, which was studied in detail. With such a study, I explored an approach aimed at correcting for the signal of starspots and faculae in the light curve, by taking into account its full extent. I was able to put constraints on the planet and activity features parameters that a fit without spot and faculae modeling would not have allowed for. Strong points and drawbacks of the method were examined and discussed, and key aspects to develop the technique were identified. In particular, the method presented here would take advantage of refinements of the analytic models it is based on, and of optimizations in the way the model is implemented in the fitting algorithm. 
   
\section{Future developments}
In the last twenty years, our knowledge and understanding of exoplanet systems have enormously improved. However, many questions remain unanswered. What is the distribution of the bulk properties of exoplanets? How are these correlated with their atmospheric properties? How do planetary systems evolve with age? How do planet properties and their frequency correlate with stellar metallicity and type, orbital distance, and the properties of the protoplanetary disk? What are the statistical properties of terrestrial planets, and how are they related to their formation environment? How unique is our Solar System, and in particular our Earth?\\
These are just a few of the questions that present and future research needs to address. These questions are the motivation for future missions and surveys. The problematics addressed in this work will remain important also in the perspective of such future programs. Spectral analysis will remain a necessity for host star characterization. Short and middle-term future transit missions such as TESS \citep{ricker2010}, CHEOPS \citep{broeg2013}, and PLATO 2.0 \citep{rauer2014}, as well as future RV instruments such as ESPRESSO \citep{pepe2014}, will be based on the same methodology as previous surveys for the determination of planet parameters. Ongoing and future projects also rely on the precise spectral characterization of the host stars in other bands than the visible. Future RV instruments focused on planets orbiting M dwarfs as the Habitable Zone Planet Finder \citep{mahadevan2012}, CARMENES \citep{quirrenbach2012}, and SPIROU \citep{carmona2013} will require to move spectral characterization to the near-infrared.\\
The large number of stars monitored in the search for exoplanets would benefit from the development of automatic spectral analysis techniques. After the need of high-\sn\ spectra and the sensitivity to data reduction, probably the main limitation of spectral analysis is the correlation between \teff, \logg\, and \feh\ \citep{torres2012}. This and slight differences in the methodologies are at the origin of discrepancies between the atmospheric parameters derived with different automatic methods \citep[][and references therein]{blancocuaresma2014}. Therefore, a list of previously selected spectral lines is needed for these methods to be effective \citep[e.g.][]{mucciarelli2013}, or the methods need to be trained for the instrument \citep[e.g.][]{gazzano2010}. Constraints from asteroseismology can be used to remove the degeneracies between the atmospheric parameters \citep[e.g.][]{bruntt2012,morel2012,thygesen2012,huber2013a}. 
To derive the stellar density, surface gravity, mass, radius, and age of a star, though, asteroseismology requires an estimation of \teff\ and \feh, and therefore has to be coupled with spectroscopy.\\
Spectroscopically-derived parameters, as well as stellar mass and radius, depend on the accuracy of model atmospheres, synthetic spectra, and stellar evolutionary tracks. As of model atmospheres and the related synthetic spectra, 3D model atmospheres \citep[e.g.][]{ludwig2009} need to be developed for a wide range of atmospheric parameters. Such models are expected to model non-thermal velocity fields more accurately than present-day 1D model atmospheres. Indeed, the introduction of the microturbulence parameter, correlated with \teff, is due to the fact that convective motions in the photosphere are generally not well described by one-dimensional, static model atmospheres. As of stellar evolutionary tracks, the effects of magnetic activity and stellar rotation are still unclear. Moreover, a clear picture of star-planet interactions for close-in giant planets is still lacking. Such interactions may modify the evolution of host stars with respect to the one of stars without planets. The advancement of planet characterization will depend on developments in this field, as well.\\
In the lack of RVs, the fit of TTV can constrain the planetary mass and size down to a few percent \citep{carter2012}. Photo-dynamical modeling is a very efficient (yet highly time-consuming) method to exploit TTV and to characterize multiple transiting systems. This technique takes full advantage of the information contained in the varying transit shape along the light curve. If the light curve is affected by occulted spots and faculae, however, the precise measurement of the transit times is seriously affected. This, together with the need to improve ``standard'' transit and RV fit, is another reason which motivates in fostering activity modeling.\\
One of the main limitations in activity modeling, as presented in this work, is the \textit{ad hoc} adaptation of the fitting method to the data set of a single target. Conversely, it is desirable to obtain a method that can be routinely employed to the fit of any light curve. The importance of developing standard approaches for stellar activity will become even more important for the analysis of the light curves that PLATO 2.0 will provide. These light curves will have a time span of two-three years \citep{rauer2014}, much longer that the one studied here for CoRoT-2 (150 days). This highlights the importance of improving the modeling of the activity features evolution, which will severely affect the light curves of future surveys. Activity features modeling is already able to constrain important information about the transit parameters; unluckily, however, we are still far from its application in routine. Its development would benefit from a study on synthetic light curves and on other systems with an active star.\\ 
Also, and no less important, the time span of the light curves of future surveys will need an optimization in the computation times of the models, if activity modeling is to be of any practical use. Indeed, the fit of the light curve of CoRoT-2 required some days of computation.\\
As explored during this work, the modeling of the photometric signal can help in correcting the RV signal from activity-induced features. However, this method needs to be developed further. It would benefit from studies on other cases than the ones explored here. As it needs simultaneous photometry and RV data, dedicated observations are required for this.\\
Finally, an important help for transit fit can come from simultaneous and near-infrared observations, which have proven to be useful to discriminate between planetary signals and activity-induced flux variations \citep{jha2000,ballerini2012}. It would be interesting to extend the fitting method developed here to different passbands, and to perform simultaneous fits. This may also help to constrain the transit depth variations determining the shape of transmission spectra. Indeed, \cite{oshagh2014} showed that occulted stellar plages can mimic the signature of atmospheric scattering.\\

The techniques exploited and developed throughout this work are driven by the need for precise stellar and planetary parameters in the understanding of the exoplanet and host star populations. Without these, current and new technologies will not be able to express their full potential. To conclude, one can then go back to figure \ref{yearR}, which shows the improvement of the measurement accuracy of planet masses and radii throughout the years. Thanks to this improvement, it has been possible to detect planets of ever smaller masses and radii. Keeping the same trend will require an effort in the direction this work has tried to follow.

\clearpage
\begin{appendix}
\chapter{Additional tables}\label{appA}

\begin{table}[!ht]
\centering
\caption[SOPHIE echelle orders]{SOPHIE echelle orders: from left to right-column, echelle order, shortest CCD and blaze wavelengths
($\lambda^{\rm{CCD}}_{min}, \, \lambda^{\rm{blaze}}_{\rm{min}}$), central wavelength
($\lambda_{\rm{c}}$), and largest blaze and CCD wavelengths ($\lambda^{\rm{blaze}}_{\rm{max}}, \,\lambda^{\rm{CCD}}_{max}$).}
\scalebox{0.91}{
\begin{tabular}{|rccccc|}
 \hline
Order & $\lambda^{\rm{CCD}}_{\rm{min}}$ & $\lambda^{\rm{blaze}}_{\rm{min}}$ & $\lambda_{\rm{c}}$ &
$\lambda^{\rm{blaze}}_{\rm{max}}$ & $\lambda^{\rm{CCD}}_{\rm{max}}$\\  
\hline			      				     
  0 & 3872.41 & 3887.91 & 3910.00 & 3932.34 & 3950.24 \\
  1 & 3916.75 & 3932.34 & 3954.94 & 3977.80 & 3995.48 \\
  2 & 3962.12 & 3977.80 & 4000.93 & 4024.33 & 4041.78 \\
  3 & 4008.56 & 4024.33 & 4048.00 & 4071.95 & 4089.17 \\
  4 & 4056.10 & 4071.95 & 4096.19 & 4120.72 & 4137.69 \\
  5 & 4104.80 & 4120.72 & 4145.54 & 4170.67 & 4187.38 \\
  6 & 4154.68 & 4170.67 & 4196.10 & 4221.84 & 4238.28 \\
  7 & 4205.80 & 4221.84 & 4247.90 & 4274.29 & 4290.44 \\
  8 & 4258.19 & 4274.29 & 4301.00 & 4328.05 & 4343.91 \\
  9 & 4311.92 & 4328.05 & 4355.44 & 4383.19 & 4398.73 \\
 10 & 4367.02 & 4383.19 & 4411.28 & 4439.74 & 4454.96 \\
 11 & 4423.56 & 4439.74 & 4468.57 & 4497.78 & 4512.65 \\
 12 & 4481.59 & 4497.78 & 4527.37 & 4557.35 & 4571.86 \\
 13 & 4541.16 & 4557.35 & 4587.73 & 4618.52 & 4632.65 \\
 14 & 4602.35 & 4618.52 & 4649.73 & 4681.36 & 4695.08 \\
 15 & 4665.22 & 4681.36 & 4713.43 & 4745.93 & 4759.23 \\
 16 & 4729.84 & 4745.93 & 4778.89 & 4812.31 & 4825.16 \\
 17 & 4796.28 & 4812.31 & 4846.20 & 4880.57 & 4892.95 \\
 18 & 4864.62 & 4880.57 & 4915.43 & 4950.79 & 4962.68 \\
 19 & 4934.94 & 4950.79 & 4986.67 & 5023.07 & 5034.43 \\
 20 & 5007.34 & 5023.07 & 5060.00 & 5097.48 & 5108.30 \\
 21 & 5081.89 & 5097.48 & 5135.52 & 5174.14 & 5184.36 \\
 22 & 5158.71 & 5174.14 & 5213.33 & 5253.13 & 5262.74 \\
 23 & 5237.89 & 5253.13 & 5293.54 & 5334.57 & 5343.53 \\
 24 & 5319.56 & 5334.57 & 5376.25 & 5418.58 & 5426.85 \\
 25 & 5403.81 & 5418.58 & 5461.59 & 5505.28 & 5512.82 \\
 26 & 5490.79 & 5505.28 & 5549.68 & 5594.80 & 5601.56 \\
 27 & 5580.62 & 5594.80 & 5640.66 & 5687.27 & 5693.21 \\
 28 & 5673.45 & 5687.27 & 5734.67 & 5782.86 & 5787.92 \\
 29 & 5769.42 & 5782.86 & 5831.87 & 5881.71 & 5885.84 \\
 30 & 5868.71 & 5881.71 & 5932.42 & 5984.00 & 5987.14 \\
 31 & 5971.49 & 5984.00 & 6036.49 & 6089.91 & 6092.00 \\
 32 & 6077.94 & 6089.91 & 6144.29 & 6199.64 & 6200.60 \\
 33 & 6188.26 & 6199.64 & 6256.00 & 6313.40 & 6313.16 \\
 34 & 6302.67 & 6313.40 & 6371.85 & 6431.40 & 6429.89 \\
 35 & 6421.40 & 6431.40 & 6492.08 & 6553.91 & 6551.02 \\
 36 & 6544.70 & 6553.91 & 6616.92 & 6681.17 & 6676.82 \\
 37 & 6672.84 & 6681.17 & 6746.67 & 6813.47 & 6807.55 \\
 38 & 6806.11 & 6813.47 & 6881.60 & 6951.11 & 6943.51 \\
\hline				     
\end{tabular}}
\label{tabtheo}
\end{table}

\begin{table}[htb]
\caption[Lines at the center and at the edges of the SOPHIE echelle orders]{Lines at the center (left) and at the edges (right) of the SOPHIE echelle orders.}
\begin{minipage}[b]{0.8\linewidth}
\centering
\scalebox{0.82}{
\begin{tabular}{|cc|cc|cc|cc|}
\hline
$\lambda$ [\AA] & Element & $\lambda$ [\AA] & Element & $\lambda$ [\AA] & Element & $\lambda$ [\AA] & Element \\
\hline			   
5044.22	 &  FeI   &   5649.99  &  FeI  &    6111.08  &  NiI    &    6322.17  &  NiI   \\
5054.65	 &  FeI   &   5651.47  &  FeI  &    6111.65  &  VI     &    6322.69  &  FeI   \\
5109.65	 &  FeI   &   5652.32  &  FeI  &    6119.76  &  NiI    &    6330.13  &  CrI   \\
5187.91	 &  FeI   &   5653.87  &  FeI  &    6120.25  &  FeI    &    6358.68  &  FeI   \\
5197.57	 &  FeI   &   5661.35  &  FeI  &    6125.02  &  SiI    &    6378.26  &  NiI   \\
5219.70	 &  TiI   &   5667.52  &  FeI  &    6126.22  &  TiI    &    6392.54  &  FeI   \\
5223.19	 &  FeI   &   5670.85  &  VI   &    6127.91  &  FeI    &    6416.93  &  FeI   \\
5225.53	 &  FeI   &   5671.82  &  ScI  &    6128.98  &  NiI    &    6481.88  &  FeI   \\
5228.38	 &  FeI   &   5694.99  &  NiI  &    6130.14  &  NiI    &    6498.94  &  FeI   \\
5234.63	 &  FeI   &   5701.11  &  SiI  &    6135.36  &  VI     &    6598.60  &  NiI   \\
5264.81	 &  FeI   &   5703.59  &  VI   &    6142.49  &  SiI    &    6599.12  &  TiI   \\
5288.53	 &  FeI   &   5715.09  &  FeI  &    6145.02  &  SiI    &    6608.03  &  FeI   \\
5295.32	 &  FeI   &   5727.05  &  VI   &    6149.25  &  FeI    &    6609.12  &  FeI   \\
5300.75	 &  CrI   &   5731.77  &  FeI  &    6151.62  &  FeI    &    6625.02  &  FeI   \\
5376.83	 &  FeI   &   5737.07  &  VI   &    6156.02  &  CaI    &    6627.55  &  FeI   \\
5379.58	 &  FeI   &   5738.24  &  FeI  &    6157.73  &  FeI    &    6635.13  &  NiI   \\
5386.34	 &  FeI   &   5739.48  &  TiI  &    6159.38  &  FeI    &    6703.57  &  FeI   \\
5395.22	 &  FeI   &   5741.85  &  FeI  &    6160.75  &  NaI    &    6705.11  &  FeI   \\
5398.28	 &  FeI   &   5748.36  &  NiI  &    6165.36  &  FeI    &    6710.32  &  FeI   \\
5432.95	 &  FeI   &   5752.04  &  FeI  &    6173.34  &  FeI    &    6713.05  &  FeI   \\
5436.30	 &  FeI   &   5753.64  &  SiI  &    6175.37  &  NiI    &    6713.74  &  FeI   \\
5436.59	 &  FeI   &   5793.92  &  FeI  &    6176.82  &  NiI    &    6721.85  &  SiI   \\
5441.34	 &  FeI   &   5797.87  &  SiI  &    6177.25  &  NiI    &    6725.36  &  FeI   \\
5461.55	 &  FeI   &   5805.22  &  NiI  &    6186.72  &  NiI    &    6726.67  &  FeI   \\
5464.28	 &  FeI   &   5806.73  &  FeI  &    6187.99  &  FeI    &    6732.07  &  FeI   \\
5466.99	 &  FeI   &   5809.22  &  FeI  &    6204.61  &  NiI    &    6733.15  &  FeI   \\
5473.17	 &  FeI   &   5814.81  &  FeI  &    6216.35  &  VI     &    6739.52  &  FeI   \\
5481.25	 &  FeI   &   5827.88  &  FeI  &    6220.79  &  FeI    &    6741.63  &  SiI   \\
5490.16	 &  TiI   &   5852.22  &  FeI  &    6223.99  &  NiI    &    6750.16  &  FeI   \\
5517.54	 &  SiI   &   5853.15  &  FeI  &    6224.51  &  VI     &    6786.86  &  FeI   \\
5520.50	 &  ScI   &   5855.08  &  FeI  &    6226.74  &  FeI    &    6793.26  &  FeI   \\
5522.45	 &  FeI   &   5856.09  &  FeI  &    6229.24  &  FeI    &    6820.37  &  FeI   \\
5538.52	 &  FeI   &   5902.48  &  FeI  &    6230.10  &  NiI    &    6828.60  &  FeI   \\
5543.94	 &  FeI   &   5916.26  &  FeI  &    6237.33  &  SiI    &    6837.01  &  FeI   \\
5546.51	 &  FeI   &   5927.79  &  FeI  &    6238.39  &  FeI    &    6839.84  &  FeI   \\
5547.00	 &  FeI   &   5929.68  &  FeI  &    6240.65  &  FeI    &    6842.69  &  FeI   \\
5553.58	 &  FeI   &   5934.66  &  FeI  &    6243.82  &  SiI    &    6843.66  &  FeI   \\
5560.22	 &  FeI   &   5956.70  &  FeI  &    6244.48  &  SiI    &    6855.72  &  FeI   \\
5618.64	 &  FeI   &   5996.73  &  NiI  &    6247.56  &  FeI    &    6857.25  &  FeI   \\
5619.60	 &  FeI   &   6005.55  &  FeI  &    6251.83  &  VI     &    6858.15  &  FeI   \\
5621.60	 &  SiI   &   6027.06  &  FeI  &    6258.11  &  TiI    &    6861.94  &  FeI   \\ \cline{7-8}
5625.32	 &  NiI   &   6034.04  &  FeI  &    6261.10  &  TiI    &    \\ 
5628.35	 &  NiI   &   6035.34  &  FeI  &    6270.23  &  FeI    &    \\
5635.83	 &  FeI   &   6039.73  &  VI   &    6274.66  &  VI     &    \\
5636.70	 &  FeI   &   6054.08  &  FeI  &    6285.17  &  VI     &    \\
5638.27	 &  FeI   &   6056.01  &  FeI  &    6290.98  &  FeI    &    \\
5641.44	 &  FeI   &   6064.63  &  TiI  &    6297.80  &  FeI    &    \\
5641.88	 &  NiI   &   6094.38  &  FeI  &    6315.81  &  FeI    &    \\
5643.08	 &  NiI   &   6096.67  &  FeI  &    6318.72  &  MgI    &    \\
5645.61	 &  SiI   &   6098.25  &  FeI  &    6319.24  &  MgI    &    \\
\cline{1-6}
\end{tabular}}
\label{tabnooverlap}
\end{minipage}
\hspace{0.3cm}
\begin{minipage}[b]{0.15\linewidth}
\centering
\scalebox{0.82}{
\begin{tabular}{|cc|}
\hline
$\lambda$ [\AA] & Element \\
\hline
5010.94	 &  NiI   \\
5180.06	 &  FeI   \\
5243.78	 &  FeI   \\
5247.06	 &  FeI   \\
5250.21	 &  FeI   \\
5253.47	 &  FeI   \\
5406.78	 &  FeI   \\
5417.04	 &  FeI   \\
5491.83	 &  FeI   \\
5587.58	 &  FeI   \\
5679.03	 &  FeI   \\
5680.24	 &  FeI   \\
5682.20	 &  NiI   \\
5684.49	 &  SiI   \\
5690.43	 &  SiI   \\
5772.15	 &  SiI   \\
5775.08	 &  FeI   \\
5778.46	 &  FeI   \\
6078.49	 &  FeI   \\
6079.01	 &  FeI   \\
6081.45	 &  VI	  \\
6082.72	 &  FeI   \\
6086.29	 &  NiI   \\
6089.57	 &  FeI   \\
6090.21	 &  VI	  \\
6091.18	 &  TiI   \\
6195.46	 &  SiI   \\
6200.32	 &  FeI   \\
6303.46	 &  FeI   \\
6806.85	 &  FeI   \\
6810.27	 &  FeI   \\
\hline     
\end{tabular}}
\label{taboverlap}
\end{minipage}
\end{table}

\begin{table}[!htb]
\caption[Transit timing variations for Kepler-117 b and c]{\label{ttvb} Transit timing variations for Kepler-117 b (left) and Kepler-117 c (right).}
\begin{minipage}[b]{0.45\linewidth}
\centering
\scalebox{0.7}{
\begin{tabular}{|c|c|c|c|}
\hline\hline
Epoch & Mid-transit time  & TTV & Uncertainty \\
    & [BJD - 2450000] & [min] & [min] \\
\hline
0   & 4978.82211 &  -2.9 & 2.7  \\
1   & 4997.62519 &   7.5 & 3.0  \\
3   & 5035.19749 & -20.6 & 2.5  \\
4   & 5054.00955 &   2.6 & 3.5  \\
5   & 5072.80069 &  -4.2 & 2.9  \\
7   & 5110.40214 &   9.6 & 2.8  \\
8   & 5129.19445 &   4.5 & 3.0  \\
9   & 5147.98545 &  -2.6 & 2.6  \\
10  & 5166.78207 &  -1.6 & 2.5  \\
11  & 5185.57076 & -12.0 & 2.4  \\
12  & 5204.37575 &   1.1 & 2.2  \\
14  & 5241.95112 & -22.6 & 2.2  \\
15  & 5260.76796 &   7.6 & 2.2  \\
16  & 5279.56177 &   4.6 & 2.4  \\
17  & 5298.35371 &  -1.2 & 2.4  \\
18  & 5317.15385 &   4.9 & 2.4  \\
19  & 5335.94633 &   0.0 & 2.3  \\
20  & 5354.74575 &   5.1 & 2.3  \\
21  & 5373.52714 & -15.8 & 2.3  \\
22  & 5392.32105 & -18.7 & 2.3  \\
23  & 5411.13384 &   5.6 & 2.2  \\
24  & 5429.92650 &   0.9 & 2.2  \\
25  & 5448.71835 &  -4.9 & 2.2  \\
26  & 5467.52116 &   5.0 & 2.5  \\
27  & 5486.31647 &   4.2 & 2.4  \\
28  & 5505.10995 &   0.7 & 2.4  \\
30  & 5542.68973 & -16.7 & 2.4  \\
32  & 5580.29000 &  -4.5 & 2.2  \\
33  & 5599.08156 & -10.8 & 2.3  \\
34  & 5617.89305 &  11.7 & 2.2  \\
36  & 5655.47836 &   2.3 & 2.4  \\
37  & 5674.27106 &  -2.3 & 2.4  \\
38  & 5693.06165 &  -9.9 & 2.5  \\
39  & 5711.86685 &   3.4 & 2.3  \\
41  & 5749.44394 & -17.8 & 2.2  \\
42  & 5768.25766 &   7.9 & 2.2  \\
43  & 5787.05285 &   6.8 & 2.2  \\
44  & 5805.84231 &  -2.4 & 2.2  \\
45  & 5824.64242 &   3.6 & 2.2  \\
46  & 5843.43402 &  -2.6 & 2.4  \\
47  & 5862.23795 &   9.0 & 2.5  \\
48  & 5881.02176 &  -8.5 & 2.4  \\
49  & 5899.80911 & -20.8 & 2.4  \\
50  & 5918.62126 &   2.6 & 2.4  \\
51  & 5937.41712 &   2.5 & 2.2  \\
52  & 5956.20759 &  -5.3 & 2.9  \\
53  & 5975.01304 &   8.4 & 2.1  \\
55  & 6012.59970 &   1.0 & 2.2  \\
56  & 6031.38973 &  -7.4 & 2.4  \\
57  & 6050.18187 & -12.8 & 2.3  \\
58  & 6068.99024 &   5.1 & 2.4  \\
59  & 6087.77696 &  -8.1 & 2.4  \\
62  & 6144.17234 &   2.9 & 2.3  \\
63  & 6162.96539 &  -1.2 & 2.3  \\
64  & 6181.75998 &  -3.1 & 2.2  \\
65  & 6200.55215 &  -8.5 & 2.3  \\
66  & 6219.35264 &  -1.9 & 2.5  \\
68  & 6256.93344 & -17.8 & 2.4  \\
69  & 6275.74812 &   9.3 & 2.4  \\
70  & 6294.54117 &   5.2 & 2.5  \\
72  & 6332.13746 &  11.6 & 2.2  \\
73  & 6350.92542 &   0.2 & 2.2  \\
74  & 6369.72444 &   4.6 & 2.2  \\
75  & 6388.51087 &  -9.0 & 2.2  \\
76  & 6407.30314 & -14.2 & 2.4  \\
\hline
\end{tabular}}
\label{ttvb}
\end{minipage}
\hspace{0.5cm}
\begin{minipage}[b]{0.45\linewidth}
\centering
\scalebox{0.7}{
\begin{tabular}{|c|c|c|c|}
\hline\hline
Epoch & Mid-transit time  & TTV & Uncertainty \\
    & [BJD - 2450000] & [min] & [min] \\
\hline
0   &  4968.63008   &    -3.5 &  1.3 \\
1   &  5019.42134   &    -2.2 &  1.3 \\
2   &  5070.21489   &     2.3 &  1.3 \\
3   &  5121.00355   &    -0.1 &  1.3 \\
4   &  5171.79086   &    -4.5 &  1.4 \\
6   &  5273.37520   &     0.6 &  1.2 \\
7   &  5324.16706   &     2.8 &  1.3 \\
8   &  5374.95617   &     0.9 &  1.2 \\
9   &  5425.74606   &     0.3 &  1.2 \\
10  &  5476.53664   &     0.6 &  1.3 \\
11  &  5527.32559   &   -1.5 &  1.3 \\
12  &  5578.11883   &     2.6 &  1.2 \\
13  &  5628.90845   &     1.6 &  1.2 \\
14  &  5679.69704   &   -1.0 &  1.3 \\
16  &  5781.27846   &   0.0 &  1.2 \\
17  &  5832.06706   &    -2.6 &  1.2 \\
18  &  5882.85959   &     0.5 &  1.3 \\
19  &  5933.65090   &     1.9 &  1.2 \\
20  &  5984.43899   &    -1.4 &  1.2 \\
21  &  6035.23056   &     0.3 &  1.3 \\
22  &  6086.02001   &    -1.0 &  1.3 \\
23  &  6136.81231   &     1.7 &  1.2 \\
24  &  6187.60050   &    -1.4 &  1.2 \\
26  &  6289.18225   &     0.0 &  1.3 \\
27  &  6339.97047   &    -3.1 &  1.2 \\
28  &  6390.76238   &    -0.8 &  1.6 \\
\hline
\end{tabular}}
\end{minipage}
\end{table}

\begin{landscape}
\begin{table}[thb]
\begin{center}{
\caption[Transit parameters for CoRoT-2 on segments of light curve]{\label{tsegments} Transit parameters of CoRoT-2 b with their 68.3\% confidence intervals for each segment of fit 4. $t_i$ and $t_f$ stand for the initial and final time of each segment.}
\scalebox{0.8}{\begin{tabular}{ccccccccc}
\hline
\hline
\\
Segment & $t_i - 2454000$ [BJD]& $t_f - 2454000$ [BJD]& $k_r$ & $i_p [^\circ]$ &  $\rho_\star [\rho_\odot]$ & $P_{\mathrm{orb}}$ [days] & $u_a$ & $u_b$  \smallskip\\
\hline\\
1	&	242.009666	&	259.936338&	$0.16679\pm0.00033$	&	$88.31\pm0.18$	&	$1.371\pm0.018$	&	$1.742986\pm8\cdot 10^{-6}$	&	$0.463\pm0.013$	&	$0.037\pm0.021$ \\
2	&	260.019699	&	274.929539&	$0.16892\pm0.00042$	&	$87.94\pm0.25$	&	$1.381\pm0.028$	&	$1.743009\pm8\cdot 10^{-6}$	&	$0.365\pm0.013$	&	$0.027\pm0.018$ \\
3	&	275.011778	&	289.962123&	$0.16685\pm0.00024$	&	$88.03\pm0.08$	&	$1.372\pm0.01$	&	$1.743006\pm8\cdot 10^{-6}$	&	$0.388\pm0.014$	&	$0.062\pm0.022$ \\
4	&	290.025447	&	314.952613&	$0.16842\pm0.00025$	&	$87.95\pm0.07$	&	$1.337\pm0.009$	&	$1.742978\pm6\cdot 10^{-6}$	&	$0.407\pm0.013$	&	$0.069\pm0.022$ \\
5	&	315.032994	&	334.999139&	$0.16465\pm0.00024$	&	$89.41\pm0.18$	&	$1.439\pm0.009$	&	$1.742992\pm7\cdot 10^{-6}$	&	$0.451\pm0.013$	&	$0.056\pm0.019$ \\
6	&	335.00025	&	349.938381&	$0.16633\pm0.00025$	&	$88.2\pm0.06$	&	$1.368\pm0.009$	&	$1.743007\pm8\cdot 10^{-6}$	&	$0.428\pm0.014$	&	$0.042\pm0.02$ \\
7	&	350.013573	&	367.946176&	$0.16663\pm0.00032$	&	$88.12\pm0.11$	&	$1.367\pm0.013$	&	$1.743013\pm8\cdot 10^{-6}$	&	$0.419\pm0.014$	&	$0.051\pm0.023$ \\
8	&	368.022108	&	378.821162&	$0.16731\pm0.00039$	&	$87.53\pm0.19$	&	$1.316\pm0.024$	&	$1.743002\pm9\cdot 10^{-6}$	&	$0.397\pm0.015$	&	$0.057\pm0.024$ \\

\hline
\\
\end{tabular}}}
\end{center}
\end{table}

\begin{table}[!bhtp]
\begin{center}{
\caption[Transit parameter for CoRoT-2 on segments of light curve of increasing length]{\label{tsegmentsebop} Mean and standard deviations of the transit parameters of CoRoT-2 b on segments of the light curve with increasing length. The first column indicates the fitted fraction of the light curve; $t_i$ and $t_f$ stand for the initial and final time of each segment.}
\scalebox{0.8}{\begin{tabular}{ccccccccc}
\hline
\hline
\\
Fraction & $t_i - 2454000$ [BJD]& $t_f - 2454000$ [BJD]&  $k_r$ & $i_p [^\circ]$ &  $\rho_\star [\rho_\odot]$ & $P_{\mathrm{orb}}$ [days] & $u_a$ & $u_b$  \smallskip\\
\hline\\
0.125	&	242.618596	&	255.102084&	$0.16695\pm0.00089$	&	$89.41\pm0.39$	&	$1.433\pm0.017$	&	$1.74297\pm8\cdot 10^{-6}$	&	$0.497\pm0.014$	&	$0.024\pm0.022$ \\
0.25	&	242.618596	&	276.007638&	$0.16818\pm0.00085$	&	$88.57\pm0.29$	&	$1.412\pm0.021$	&	$1.74299\pm4\cdot 10^{-6}$	&	$0.436\pm0.014$	&	$0.036\pm0.03$ \\
0.33	&	242.618596	&	287.999862&	$0.16829\pm0.00086$	&	$88.4\pm0.23$	&	$1.403\pm0.019$	&	$1.742991\pm3\cdot 10^{-6}$	&	$0.411\pm0.015$	&	$0.044\pm0.033$ \\
0.5	&	242.618596	&	310.877715&	$0.16845\pm0.00077$	&	$88.65\pm0.21$	&	$1.422\pm0.015$	&	$1.742993\pm1\cdot 10^{-6}$	&	$0.428\pm0.008$	&	$0.02\pm0.017$ \\
0.66	&	242.618596	&	331.793366&	$0.1684\pm0.00079$	&	$88.43\pm0.17$	&	$1.401\pm0.014$	&	$1.742995\pm1\cdot 10^{-6}$	&	$0.419\pm0.01$	&	$0.028\pm0.022$ \\
0.75	&	242.618596	&	343.994937&	$0.16835\pm0.00079$	&	$88.32\pm0.15$	&	$1.393\pm0.014$	&	$1.742994\pm1\cdot 10^{-6}$	&	$0.429\pm0.007$	&	$0.014\pm0.013$ \\
1.0	&	242.618596	&	378.814125&	$0.16823\pm0.00075$	&	$88.29\pm0.15$	&	$1.39\pm0.013$	&	$1.742996\pm1\cdot 10^{-6}$	&	$0.427\pm0.009$	&	$0.025\pm0.019$ \\

\hline
\\
\end{tabular}}}
\end{center}
\end{table}
\end{landscape}

\end{appendix}

\clearpage

\bibliographystyle{aa}
\setlength{\bibsep}{0.0pt}
\fancyhead[LO]{\itshape Bibliography}
\fancyhead[RE]{\itshape Bibliography}
\bibliography{biblio}
\pagestyle{empty}
\clearpage


\section*{Supplementary material}

# Acknowledgments

This should be the page of a dissertation written with the biggest pleasure and fun, and actually it's well the case. While writing these concluding lines, I can finally look with relief at all the pages I managed to fill (it didn't sound obvious at all, some time ago!), and be happy and surprised about the fact that these pages are there. They exist, and they are almost already part of the past.

And while looking at them, I think they represent a very important period in my life. In first place, for having followed me during these years of education to research, with patience when I was têtu, support in moments of stress, and openness to discussion when I didn't know where I was going, I warmly thank my supervisor Magali Deleuil. I thank her, also, for the time she spent in reading these pages and suggesting me on how to improve them.

I wish to thank all the members of the jury, who will travel until Marseille to participate to my discussion. And I am grateful to my two referees, who have read this work and provided me with helpful comments. Taking a leap to the future, I hope they will be kind during the defense...

There are many people at LAM I would like to thank. Some of them are not at LAM anymore, but somewhere else in the world. Some others are "new entries". I will remember all the members of the PASI team for having been a great group to work and to grow up with. Cilia, Susana, and Benjamin, thank you for having helped me in clearing my thoughts when they were fuzzy - especially those about life - and for having taken care of me when I needed. By the way, Ben, thank you for your corrections on my French, you have been my first French French teacher (with no repetition). José, thank you for your patience when we worked together, thank you for helping during the trips to IKEA (!), and thank you for the difficult times. They helped. And thank you to all those I have not space to mention here. I wish to all of you guys the best of luck and I really hope to meet you again.

There are many other people at LAM that I won't forget. There I met friends, people who love discussing about the past, the present, and the future, and people who had an important role in making my days warm and human. Thibaud, it's been a long time since that ride to Château Gombert, hein? Brian, I hope you keep... being Brian, "dude" (that one is for Matvey). And Pol, keep playing music. Just to name a few.

I wish to thank the people who helped me understanding the French paper system - a very interesting universe, I confess -, especially in the moments where my ability in doing things last-minute was at its best. So, thank you Frédérique, Agnès, and Françoise. Thank you Gaby and Christine for being so welcoming every day I meet you. And oh, thank you Vincent for not throwing me and my computer out of the window!

I won't forget the telescope operators at the OHP, who were friendly and made my observing runs more enjoyable. Merci Jean Paul for having driven me from Manosque to the OHP and back many times, for having come stop a false fire alarm, and for a drive to the hospital in the middle of the night, always with a smile.

It's difficult not to think that I'm not in my country of origin. So, a thought of gratitude is also for the country which is hosting me, and for the people and institutions who - despite all what doesn't work in these difficult days - make this and similar experiences possible. Walking back along the karmic chain just for a little bit, my memory goes to my professor of astronomy lab at university, Giuseppe Gavazzi, who put me in touch with the exoplanet community of Marseille. And to my professor of stellar astrophysics Monica Colpi, who supported me during my first Erasmus experience at LAM.

And then there's people outside LAM, who were so important for these years I spent in Marseille. I am mostly grateful to my flatmates Francesca and Luca, who have been Friends, with a capital F, during this journey. Thank you Alice, it was great fun being both house and office mates. I thank all the people who spent some time living at the "17, rue Caisserie" and contributed in making of it one of the best places in Marseille, even though not one of the most known.

I thank the people I met in this city (except those who stole my bike), and the places I had the luck to discover. I want to thank my friends Adam and Ceci, who, among many other things, offered me a repair for my first homeless weeks at the beginning of this adventure. Special thanks to Will, who was so kind to read (him, too) a large part of these pages and correct all the mistakes of English. He hasn't read these last two pages, so, if there are mistakes, tant pis !
I thank the people I met going from, and coming back to, Marseille. And those who are in the foggy countryside where I come from. They made my life fresher.

(By the way, I like fog. And also countryside.)

Last, but of course not least, I am enormously grateful to my parents and family. Without them, I wouldn't be writing these happy, concluding words.



\end{document}